%% file: tesis.tex
\documentclass[a4paper,11pt,twoside]{report}

    \makeatletter 
    \def\cleardoublepage{\clearpage\if@twoside \ifodd\c@page\else%
    \hbox{}%
    \thispagestyle{empty}
    \newpage%
    \if@twocolumn\hbox{}\newpage\fi\fi\fi} 
    \makeatother

\usepackage{amsmath,amsfonts,amssymb}
\usepackage{amsthm}
\usepackage{graphicx}
\usepackage{a4wide}
\usepackage[bf,small,center]{titlesec}
\usepackage{epsfig}

\usepackage[spanish]{babel} 
\usepackage[latin1]{inputenc}
\usepackage{mathrsfs}
\usepackage{latexsym}
\usepackage{psfrag}

\usepackage{Insert} 

\DeclareMathOperator{\tr}{Tr} 
\newcommand{\ls}{\ensuremath{{l_s}}} 
\newcommand{\gs}{\ensuremath{{g_s}}} 
\newcommand{\ket}[1]{\lvert #1 \rangle} 
\newcommand{\gym}{{\ensuremath{g_{\text{YM}}^2}}} 
\newcommand{\hd}[1]{\ensuremath{\phantom{ }^{\star_{#1}}}} 

\newcommand{\talpha}{{\tilde{\alpha}}}
\newcommand{\tpsi}{{\tilde{\psi}}}

\newcommand{\bea}{\begin{eqnarray}}
\newcommand{\eea}{\end{eqnarray}}
\newcommand{\be}{\begin{equation}}
\newcommand{\ee}{\end{equation}}
\newcommand{\cleqn}{\setcounter{equation}{0}}
\newcommand{\ii}{{{\'\i}}}

\def\tr{{\rm Tr}}
\def\det{{\rm det}}

\def\cn{{\cal N}}
\def\lbldef#1#2{\expandafter\gdef\csname #1\endcsname {#2}}
\def\IZ{{\mathbb Z}}

\begin{document}

 \renewcommand{\listtablename}{{\'I}ndice de tablas}
 \renewcommand{\tablename}{Tabla}

\include{intro_logo}

\pagestyle{plain}
\pagenumbering{roman}

\include{resumen}

\newpage

\include{agradecimientos}

\newpage
\include{dedicatoria}


\tableofcontents
\listoffigures  


\newpage

\pagenumbering{arabic}

\include{chamuyo}

\include{dbranes_mau}

\include{general_ads_cft}


\include{sol_nocritica}
\include{sym}

\include{conclusiones}

\appendix  


\include{appendI}

\include{appendII}

\include{appendIII}

\include{4mesc}

\cleardoublepage
\bibliography{tesis}

\end{document}

%% file: intro_logo.tex

\thispagestyle{empty}


\begin{center}

\vspace{4cm}

\begin{figure}[!ht]
\centering
\includegraphics[scale=0.2,angle=0]{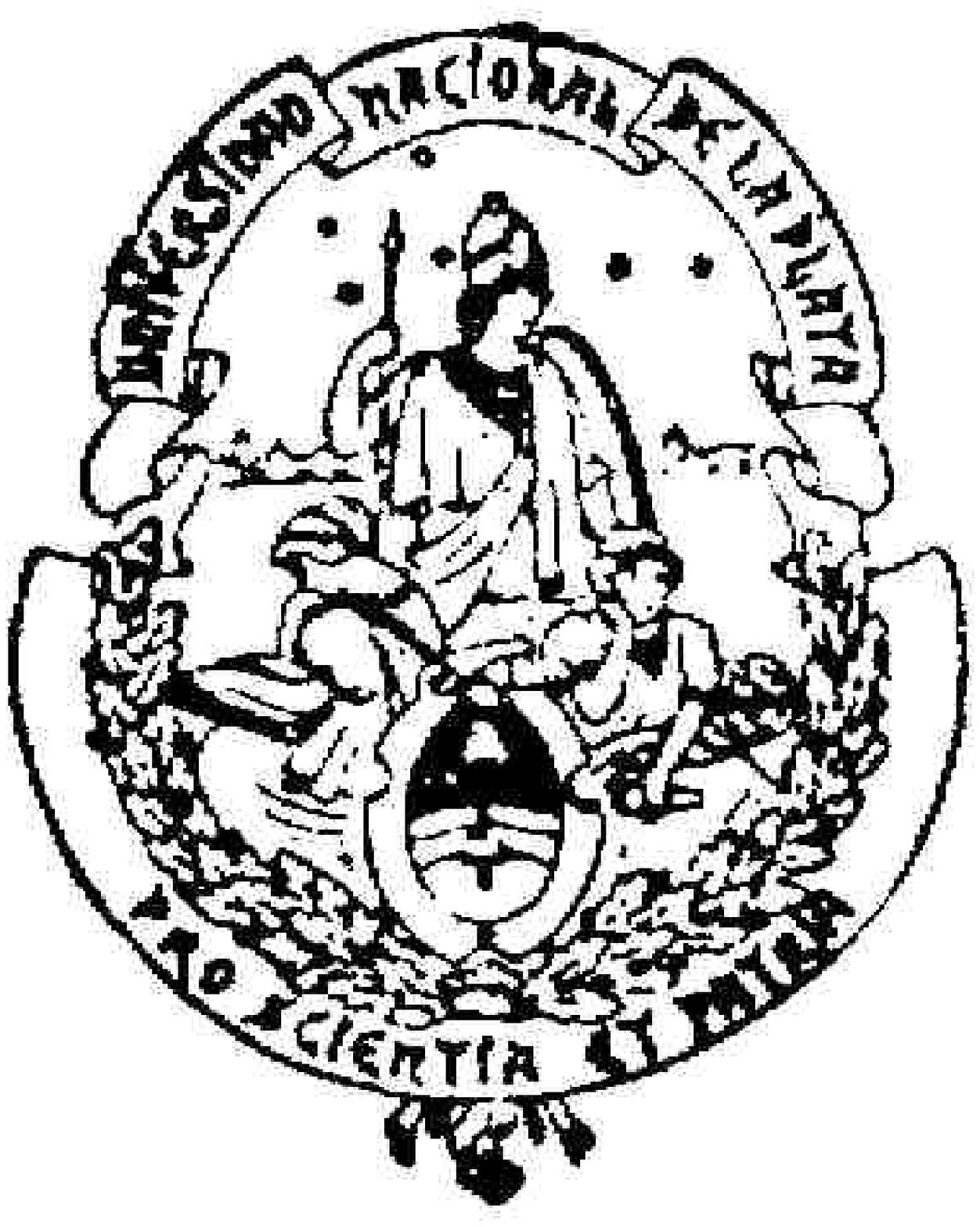}
\end{figure}

{\bf \Large UNIVERSIDAD NACIONAL DE LA PLATA}

{\bf \large FACULTAD DE CIENCIAS EXACTAS}

{\bf   \large DEPARTAMENTO DE FISICA}


\vspace{2cm}


---------------------------------------------------------------------------------------------------------------\par

\vspace{0.5cm}

{\large Trabajo de Tesis Doctoral}

\vspace{0.5cm}

{\Large \bf F\ii sica de $D$-branas en Teor\ii as de Cuerdas No Cr\ii ticas}\\

\vspace{0.5cm}

---------------------------------------------------------------------------------------------------------------\par


\vspace{2cm}

{\Large Mauricio Bernardo Sturla}

\vspace{1cm}

{\Large Director: Dr. Adri\'an R. Lugo}

\vspace{2cm} 

{\Large 2009}






\end{center}


\newpage

%% file: resumen.tex

\begin{center}
{\bf Resumen}
\end{center}
\vspace{1cm}

En esta tesis hemos abordado el estudio de la f\ii sica de las teor\ii
as de cuerdas no cr\ii ticas, en la aproximaci\'on de baja energ\ii a. 

Luego de introducir los conceptos fundamentales de cuerdas
supersim\'etricas y D$_p$-branas, presentamos brevemente la conjetura
de Maldacena \cite{Maldacena:1997re} \cite{Witten:1998qj} y su
generalizaci\'on a teor\ii as no supersim\'etricas
\cite{Witten:1998zw}, el estudio de lazos de Wilson y el c\'alculo del
espectro de glueballs. Seguidamente, y tras una breve introducci\'on a
teor\ii as de cuerdas no cr\ii ticas, presentamos las soluciones a su
acci\'on de baja energ\ii a. El estudio de estas soluciones est\'a
dividido en tres grupos; soluciones de vac\ii o, soluciones cargadas
NSNS y cargadas frente R-R. En el primer caso, hemos sido capaces de
encontrar todas las posibles soluciones de vac\ii o recuperando, como
casos particulares, las previamente conocidas. En el caso de
soluciones de fondo cargadas NSNS, hemos encontrado la soluci\'on que
representa la cuerda fundamental doblemente localizada en el vac\ii o
Minkowski $\times$ el cigarro. Tambi\'en hemos podido solucionar en
forma completa, el problema de encontrar soluciones que llenen todo
Minkowski. En particular, hemos encontrado las que pueden ser
identificadas con las soluciones de cuerda fundamental, tanto en el
vac\ii o del dilat\'on lineal como en el vac\ii o del cigarro. En el
caso de R-R, hemos encontrado algunas familias de soluciones, en
particular las que podr\ii an corresponder al l\ii mite de horizonte
cercano de soluciones de D$_p$-branas embebidas en el vac\ii o del
dilat\'on lineal. En este caso, tambi\'en hemos sido capaces de
recuperar algunas soluciones previamente conocidas, presentadas en
\cite{Kuperstein:2004yk}.

Con las soluciones de fondos cargados R-R, en particular, con una
familia de soluciones a dilat\'on constante, hemos estudiado las
teor\ii as de gauge duales, obteniendo condiciones de fase confinante
a trav\'es del c\'alculo de lazos de Wilson, que se traducen en este
caso en condiciones sobre el espacio de par\'ametros que caracteriza
nuestras soluciones. De esta manera, es posible obtener ``familias'' de
teor\ii as de gauge duales en su fase confinante. Hemos estudiado las
teor\ii as duales YM en $D=3$ y YM en $D=4$, obteniendo los espectros de
las masas de los glueballs correspondientes, para distintos puntos del
espacio de par\'ametros. Finalmente, estudiamos la dependencia de los
espectros con estos par\'ametros, y encontramos valores de los mismos
que llevan a un muy buen acuerdo con los resultados predichos por {\it
  Lattice QCD}.

%% file: agradecimientos.tex
{\bf Agradecimientos:}

\vspace{1cm}

Quisiera agradecer, en primer lugar, a mi director Adri\'an R. Lugo
por la oportunidad brindada de realizar este trabajo de tesis en un
marco cordial y ameno, por haberme ense\~nado casi todo lo que s\'e de
F\ii sica, pero fundamentalmente por el apoyo constante en cada paso
de este trabajo. A mis amigos de la Facultad, Carlos, Daniel, Diego
M., Diego R., Guille, Juan, Marcelo, Nico, Pablo, Raulo,... por lo
gratos momentos compartidos. En particular a Diego M., Gaston G.,
Guille, Nico, Marcelo, Pablo y Ra\'ul, por las innumerables
discusiones de F\ii sica de las que aprend{\ii} tanto. A mis
companeros de oficina, Silvana, Marcos, Mariano y Mario, con quienes
compart{\ii} tantos momentos gratos. Tambi\'en quiero agradecer a mi
familia por el apoyo constante, y muy especialmente a mi esposa
Claudia, a quien est\'a dedicada esta tesis.

%% file: dedicatoria.tex
\hspace{10cm}$\mathfrak{Dedicada\;\;a\;\; Claudia...}$

%% file: chamuyo.tex

\chapter{Introducci\'on}\label{c:introduction}

La descripci\'on de la Naturaleza a nivel microsc\'opico, en la
actualidad, es entendida y verificada experimentalmente en t\'erminos
de teor\ii as de campos cu\'anticos. Esta descripci\'on supone que
todas las part\ii culas son excitaciones de alg\'un campo, que son
puntuales e interact\'uan entre s{\'\i} localmente. En particular, es
bien conocido el \'exito del Modelo Est\'andar. Este modelo es una
teor\ii a de gauge cu\'antica; es decir, se describe como una teor\ii
a de campos cu\'anticos con simetr\ii a de gauge local. Las teor\ii as
de campos son capaces de describir la Naturaleza a las escalas que
podemos medir hoy en d\ii a (1 TeV). Sin embargo, existen fuertes
indicaciones de nueva f\ii sica subyacente a distancias extremadamente
peque\~nas, del orden de la longitud de Planck ($\sim 10^{-35}m$), que
no puede ser descripta en t\'erminos de estas teor\ii as. La raz\'on
fundamental es que, a distancias tan peque\~nas, los efectos
cu\'anticos de la gravedad no pueden despreciarse.  Hasta el momento
no se ha podido construir una teor\ii a de gravedad cu\'antica
satisfactoria siguiendo los procedimientos usuales. Sin embargo,
abandonando la idea de part\ii culas puntuales y suponiendo en su
lugar que las piezas fundamentales son objetos extendidos
unidimensionales, {\it cuerdas} \cite{Polchinski:1998}, parece ser
posible obtener una descripci\'on consistente de gravedad a peque\~nas
distancias.

Estas teor\ii as de cuerdas, que originalmente surgieron para explicar
las teor\ii as de interacciones fuertes, incluyen de manera natural en
su espectro una part\ii cula sin masa de esp\ii n 2 que puede ser
identificada con el gravit\'on. Por este motivo, se considera a las
teor\ii as de cuerdas como fuertes candidatos para una teor\ii a
cu\'antica de la gravedad. Por otra parte, existe un consenso
generalizado de que todas las interacciones conocidas en la Naturaleza
podr\ii an ser unificadas utilizando modelos de cuerdas
supersim\'etricas ({\it supercuerdas}) en interacci\'on
\cite{Polchinski:1998}. Por consistencia, estas teor\ii as requieren
ser formuladas sobre espacio-tiempos de dimensiones cr\ii ticas;
$D=26$, en el caso de teor\ii as de cuerdas bos\'onicas, o $D=10$, en
el caso de supercuerdas sobre espacio-tiempos planos. Naturalmente,
este alto n\'umero de dimensiones aparece como un obst\'aculo si se
pretende que estas teor{\'\i}as describan las interacciones de un
mundo que, al menos para los experimentos que se han llevado a cabo
hasta el momento, parece lucir cuadridimensional.

Sin embargo, en la misma formulaci\'on del problema parece estar su
soluci\'on. Est\'a claro que los experimentos realizados pueden
considerarse todos por debajo de alguna escala de energ\ii a dada. En
consecuencia, nada impide que las dimensiones extras aparezcan a
escalas de energ\ii as superiores a las puestas en juego hasta el
momento. Es decir, podemos pensar que las teor\ii as de cuerdas
realmente describen nuestro Universo y que a escalas de energ\ii as
suficientemente bajas se comportan de manera efectiva como teor\ii as
cuadrimensionales.  Un mecanismo que permite remover a bajas energ\ii
as las dimensiones extras es el de compactificaci\'on.  Esto significa
que las dimensiones extras pueden ser consistentemente ``enrolladas''
sobre variedades compactas (toros generalizados) llamadas variedades
de Calabi-Yau. De esta manera, las dimensiones compactas ya no resultan
observables a escalas de energ\ii as menores que la inversa del
radio de compactificaci\'on\footnote{Los modos de Kaluza-Klein que se
  propagan sobre estas direcciones compactas tienen masas del orden de
  $1/r_{\rm comp}$, por lo que para poder observar estas dimensiones
  extras es necesario que las energ\ii as puestas en juego en los
  experimentos sean de este orden.}.

Una alternativa a la compactificaci\'on surge si tenemos en cuenta que
la necesidad de formular las teor\ii as de cuerdas en las llamadas
dimensiones cr\ii ticas proviene del requerimiento de que la teor\ii a
en la hoja de mundo sea invariante de escala, y que, por lo tanto, su
funci\'on $\beta$ deba anularse. En el caso de cuerdas bos\'onicas,
esto sucede cuando la dimensi\'on del espacio-tiempo es $D=26$, y en
el caso de cuerdas supersim\'etricas, cuando $D=10$. Sin embargo, esta
situaci\'on puede modificarse si se permite considerar campos de
fondo, como campos de Kalb-Ramond, campos dilat\'onicos, o si se
consideran fondos curvos. En este caso, las funciones $\beta$
(calculadas orden por orden en $\alpha'$) naturalmente cambian,
permitiendo ahora ser anuladas por configuraciones de campo no
triviales en dimensiones diferentes de las cr\ii ticas. Estas teor\ii
as de cuerdas se conocen como teor\ii as de cuerdas no cr\ii ticas.
Se conocen varios ejemplos de estas teor\ii as; los m\'as renombrados
son las soluciones de dilat\'on lineal y la soluci\'on del
cigarro. Ambas soluciones corresponden a teor\ii as de campos
conformes exactas. Es decir, a\'un cuando estas soluciones pueden
obtenerse a partir de las ecuaciones de movimiento para los campos que
surgen de anular las funciones $\beta$ al orden $\alpha'$, puede
demostrarse que \'estas constituyen por s{\'\i} mismas teor\ii as
conformes, por lo que la invarianza de escala est\'a satisfecha en
forma exacta, no s\'olo a orden $\alpha'$.  \\ \\ Resulta esencial,
para 
interacciones fundamentales, entender el comportamiento de las teor\ii
as con grupo de gauge $SU(N)$ para grandes valores del par\'ametro
$N$. La comprensi\'on de estas teor\ii as ofrece tal vez la
posibilidad m\'as clara de abordar QCD (Quamtun Chromo Dynamics) en el
r\'egimen de acoplamiento fuerte \cite{'tHooft:1973jz}. Se sospecha,
desde hace d\'ecadas, que el comportamiento a $N$ grande de estas
teor\ii as podr\ii a ser descripto en t\'erminos de una teor\ii a de
cuerdas. Hace ya diez a\~nos que Maldacena \cite{Maldacena:1997re}
parece haber dado la clave para relacionar estas teor\ii as a trav\'es
de la dualidad AdS/CFT\footnote{AdS:Espacio Anti de Sitter\\ CFT:
  Teor\ii a de Campos Conforme}, o {\it Conjetura de Maldacena}. En
esta conjetura, Maldacena propone que el l\ii mite $N$ grande de una
teor\ii a de gauge conforme $SU(N)$ en $D$ dimensiones puede ser
descripto en t\'erminos de excitaciones de una teor\ii a de cuerdas
sobre un fondo de supergravedad producto de un espacio AdS $(D +
1)$-dimensional, por una variedad compacta, que en el caso
maximalmente supersim\'etrico corresponde a una esfera. El ejemplo
m\'as claro que sustenta esta conjetura es el de la dualidad entre
${\mathcal N} = 4$ Super Yang Mills $SU(N)$ en 4 dimensiones y
supercuerdas tipo IIB en $AdS_5 \times S^5$.  Sin embargo, a la hora
de intentar describir teor\ii as de gauge m\'as realistas, con la
esperanza de llegar a describir QCD, nos encontramos con serios
inconvenientes dado que la conjetura, en su nivel m\'as b\'asico,
relaciona teor\ii as de gauge que son supersim\'etricas y conformes,
con cuerdas sobre un espacio $AdS$. Mucho trabajo se ha hecho en los
ultimos a\~nos y muchos son los resultados para tratar de extender
esta dualidad en el contexto de teor\ii as de cuerdas cr\ii ticas. Sin
embargo, no se ha avanzado mucho en el contexto de cuerdas no cr\ii
ticas. En particular, el estudio de la dualidad en el contexto no
cr\ii tico podr\ii a proveer algunas soluciones a problemas t\ii picos
de las teor\ii as cr\ii ticas. Por ejemplo, las teor\ii as de gauge
confinantes, duales a teor\ii as cr\ii ticas ($D=10$), sufren de
varias limitaciones; entre otras, los espectros hadr\'onicos que
\'estas predicen est\'an seriamente contaminados por modos de
Kaluza-Klein provenientes de la necesidad de reducir dimensiones
compactificando las direcciones extras, y adem\'as, los duales
obtenidos suelen resultar conformes.  En el caso de cuerdas no cr\ii
ticas, estas dimensiones extras pueden ser evitadas de manera natural
y, al mismo tiempo, las soluciones admiten en general dilatones no
triviales que llevan a teor\ii as duales no conformes. Sin embargo, el
hecho de que las teor\ii as no cr\ii ticas deben ser formuladas sobre
fondos curvos, hace que debamos tratar las soluciones obtenidas a
partir de las ecuaciones provenientes de las funciones $\beta$ a orden
$\alpha'$ con cautela.  \\~\\ En esta tesis estudiaremos en
profundidad las soluciones de baja energ\ii a de las teor\ii as no
cr\ii ticas y analizaremos las teor\ii as de gauge duales que se
desprenden de algunas de sus soluciones.  \\~\\ En el cap\ii tulo 2,
introduciremos los conocimientos b\'asicos de teor\ii as de cuerdas
que nos permitir\'an dar la descripci\'on de los objetos extendidos en
varias dimensiones, conocidos como branas. En el cap\ii tulo 3,
daremos una breve descripci\'on de la dualidad $AdS/CFT$, su
generalizaci\'on a teor\ii as no supersim\'etricas y los fundamentos
del c\'alculo de lazos de Wilson y espectro de glueballs en la teor\ii
a dual. En el cap\ii tulo 4, presentaremos un estudio exhaustivo de la
acci\'on de baja energ\ii a de teor\ii as de cuerdas no cr\ii ticas y
de algunas de sus soluciones. En el cap\ii tulo 5, construiremos las
teor\ii as de gauge duales a una de nuestras familias de soluciones no
cr\ii ticas, estudiaremos sus reg\ii menes de confinamiento y
calcularemos los espectros de glueballs asociados. Finalmente, en el
cap\ii tulo 6 presentaremos las conclusiones.

\vspace{0.5cm}

Esta tesis est\'a basada en los trabajos
\begin{itemize}
\item ``The fundamental non critical string,'' A. R. Lugo and M. B. Sturla.\\
  Phys.\ Lett.\ B {\bf 637}, 338 (2006)
  [arXiv:hep-th/0604202] \cite{Lugo:2005yf}.

\item ``Space-time filling branes in non critical (super) string
  theories'', A.~R.~Lugo and M.~B.~Sturla.\\
 Phys.\ Lett.\ B {\bf 637}, 338 (2006)
  [arXiv:hep-th/0604202] \cite{Lugo:2006vz}.

\item ``Gauge invariant
  perturbation theory and non-critical string models of Yang-Mills
  theories'', A.~R.~Lugo and M.~B.~Sturla.\\
  A ser enviado \cite{adrian}
\end{itemize}

%% file: dbranes_mau.tex

\chapter{D-branas en Teor\ii a de Cuerdas}\label{c:dbranes}

En este cap\ii tulo, introducimos los rudimentos de la teor\ii a de
cuerdas y supergravedad, que ser\'an necesarios para comprender las
ideas de branas asociadas con las condiciones de contorno de cuerdas
abiertas, y de su descripci\'on ``dual'' como soluciones de
teor\'{\i}as de supergravedad.  El tratamiento aqu{\'\i} presentado se
basa en el review de Emiliano Imeroni~\cite{Imeroni:2003jk}.

\section{Aspectos Perturbativos de Teor\ii a de Cuerdas}\label{s:strings}
\subsection{Cuerda Cl\'asica}\label{clasica_cuerda}

Primero, estudiaremos algunos resultados de las teor\'{\i}as de
supercuerdas a nivel perturbativo. Las referencias usuales son
~\cite{Polchinski:1998,Green:1987} y 
~\cite{DiVecchia:1999rh,DiVecchia:2003ne}.

La acci\'on de hoja de mundo que describe la propagaci\'on de una
supercuerda en el espacio-tiempo plano en el gauge superconforme est\'a dada
por:

\begin{equation}\label{polyact}
	S = -\frac{T}{2}\int_{\mathcal{M}}  d\tau d\sigma \left( \eta^{\alpha\beta}
		\partial_\alpha X^\mu \partial_\beta X_\mu
		- i \bar{\psi}^\mu \rho^\alpha \partial_\alpha \psi_\mu \right)\,.
\end{equation}
donde
\begin{itemize}
\item $T$ es la tensi\'on de la cuerda, que est\'a relacionada con su
  longitud caracter\ii stica via $T = \tfrac{1}{2\pi\ls^2}$;
\item $\mathcal{M}$ es el volumen de mundo de la cuerda, parametrizada
  por las coordenadas $\xi^\alpha=(\tau,\sigma)$, donde $\sigma$ toma
  valores en $\sigma\in[0,s]$, y la m\'etrica plana
  $\eta^{\alpha\beta}$ tiene signatura $(-,+)$;
\item $X^\mu$, $\mu=0,\ldots,9$ desde el punto de vista de la hoja de
  mundo, son diez campos escalares, que corresponden a las coordenadas
  cartesianas del espacio blanco;
\item $\psi^\mu$, $\mu=0,\ldots,9$ son espinores de Majorana en la
  hoja de mundo, y las matrices $\rho^\alpha$ son una representaci\'on
  del \'algebra Clifford
  $\{\rho^\alpha,\rho^\beta\}=-2\eta^{\alpha\beta}$.
\item La acci\'on es invariante ante transformaciones de
  supersimetr\ii as en la hoja de mundo
\begin{equation}\label{wssusy}
	\delta X^\mu = \bar{\epsilon}\psi^\mu\,,\qquad
	\delta \psi^\mu = -i \rho^\alpha \partial_\alpha X^\mu \epsilon\,,\qquad
	\delta \bar{\psi}^\mu = i \bar{\epsilon} \rho^\alpha \partial_\alpha X^\mu,
\end{equation}
donde el par\'ametro de la transformaci\'on $\epsilon$ es un espinor
de Majorana que satisface $\rho^\beta\rho^\alpha \partial_\beta
\epsilon = 0$.
\end{itemize}

Comencemos con las ecuaciones de
movimiento y las condiciones de borde para los campos escalares que se derivan
de ~\eqref{polyact}:
\begin{align}
	\partial^\alpha \partial_\alpha X^\mu &= 0\,,\label{steom}\\
	\partial_\sigma X \cdot \delta X \big\rvert_{\sigma=s}
		- \partial_\sigma X \cdot \delta X \big\rvert_{\sigma=0} &= 0\,.\label{stbc}
\end{align}
Hay varias posibilidades para satisfacer las condiciones de contorno~\eqref{stbc}:
\begin{itemize}
\item Condici\'on de \emph{cuerda abierta de Neumann} (N). En este caso, elegimos $s=\pi$ e imponemos:

\begin{equation}
	\partial_\sigma X^\mu \big\rvert_{\sigma=0,\pi} = 0\,.
\end{equation}
La soluci\'on m\'as general de las ecuaciones de movimiento~\eqref{steom} con condiciones de contorno de Neumann en ambos extremos es:
\begin{equation}
	X^\mu_{\text{NN}} (\tau,\sigma) = x^\mu + 2\ \ls^2 p^\mu \tau
		+ i \sqrt{2}\ \ls \sum_{n\neq 0} \left( \frac{\alpha_n^\mu}{n}
		e^{-in\tau} \cos n\sigma \right)\,.
\end{equation}

\item Condici\'on de \emph{cuerda  abierta de Dirichlet} (D). Elegimos nuevamente  $s=\pi$ e imponemos:
\begin{equation}
	\delta X^\mu \big\rvert_{\sigma=0,\pi} = 0\,.
\end{equation}
La soluci\'on m\'as general de la ecuaci\'on con condiciones de contorno de Dirichlet resulta:
\begin{equation}
	X^\mu_{\text{DD}} (\tau,\sigma) = \frac{c^\mu (\pi-\sigma) + d^\mu \sigma}{\pi}
		- \sqrt{2}\ \ls \sum_{n\neq 0} \left( \frac{\alpha_n^\mu}{n}
		e^{-in\tau} \sin n\sigma \right)\,.
\end{equation}
\item Condici\'on de contorno de \emph{Cuerda cerrada}. Elegimos ahora
  $s=2\pi$ e imponemos:
\begin{equation}
	X^\mu (\tau,0) = X^\mu(\tau,2\pi)\,.
\end{equation}
La soluci\'on general de la ecuaci\'on~\eqref{steom} es:
\begin{equation}\label{Xcl}
	X^\mu_{\text{closed}} (\tau,\sigma) = x^\mu + \ls^2 p^\mu \tau
		+ i \frac{\ls}{\sqrt{2}} \sum_{n\neq 0} \frac{1}{n}
		\left( \alpha_n^\mu e^{-in(\tau-\sigma)} 
		+ \talpha_n^\mu e^{-in(\tau+\sigma)}\right)\,.
\end{equation}
\end{itemize}

En el caso de cuerdas abiertas, uno puede satisfacer las condiciones de
contorno ~\eqref{stbc} de varias maneras. Es decir, es posible elegir
condiciones de contorno mixtas (en un extremo una condici\'on, en el otro
extremo la otra), pudiendo elegir diferentes combinaciones para los
diferentes $\mu$.

Es interesante notar que las condiciones de contorno de Dirichlet no
preservan la simetr\ii a de Poincar\`e. Veremos m\'as adelante que esta
condici\'on puede ser impuesta asumiendo que describe objetos extendidos
en el espacio-tiempo.

Para tratar los grados de libertad fermi\'onicos, resulta \'util
considerar las llamadas coordenadas del cono de luz $\xi_{\pm} = \tau
\pm \sigma$. En estas coordenadas, las ecuaciones resultan:
\begin{align}
	\partial_+ \psi_-^\mu = \partial_- \psi_+^\mu &= 0\,,\label{psieom}\\
	\left( \psi_+ \delta \psi_+ - \psi_- \delta \psi_-\right)\big\rvert_{\sigma=0,s} &= 0\,,
\end{align}
donde $\psi^\mu_\pm = \tfrac{1 \mp \rho^0 \rho^1}{2} \psi^\mu$. Para el
caso de la cuerda abierta, las condiciones de contorno se satisfacen
imponiendo:
\begin{equation}
	\psi^\mu_- (\tau,0) = \eta_1 \psi^\mu_+ (\tau,0)\,,\qquad
	\psi^\mu_- (\tau,\pi) = \eta_2 \psi^\mu_+ (\tau,\pi)\,,
\end{equation}
donde $\eta_{1,2}=\pm 1$. En consecuencia, obtenemos dos sectores diferentes para el espectro de cuerda abierta:

\begin{itemize}
\item $\eta_1 = \eta_2$: sector \emph{Ramond} (R);
\item $\eta_1 = -\eta_2$: sector \emph{Neveu--Schwarz} (NS).
\end{itemize}

Las soluciones generales que satisfacen~\eqref{psieom} con estas
condiciones de contorno son:
\begin{equation}
	\psi_{\pm}^\mu \sim \sum_r \psi_r^\mu e^{-ir(\tau\pm\sigma)}\,,\qquad
	\text{donde}\qquad \begin{cases}
		r\in\mathbb{Z}\quad \text{sector R}\\
		r\in\mathbb{Z}+\frac{1}{2}\quad \text{sector NS}
	\end{cases}
\end{equation}

En el caso de cuerda cerrada, los campos $\psi_\pm$ son independientes
y pueden ser tanto peri\'odicos como antiperi\'odicos:

\begin{equation}
	\psi^\mu_- (\tau,0) = \eta_3 \psi^\mu_- (\tau,2\pi)\,,\qquad
	\psi^\mu_+ (\tau,0) = \eta_4 \psi^\mu_+ (\tau,2\pi)\,,
\end{equation}

Tenemos, por lo tanto, cuatro sectores diferentes:
\begin{itemize}
\item $\eta_3 = \eta_4 = 1$: sector \emph{Ramond-Ramond} (R-R);
\item $\eta_3 = \eta_4 = -1$: sector \emph{Neveu--Schwarz-Neveu--Schwarz}
  (NS-NS);
\item $\eta_3 = -\eta_4 = 1$: sector \emph{Ramond-Neveu--Schwarz} (R-NS);
\item $\eta_3 = -\eta_4 = -1$: sector \emph{Neveu--Schwarz-Ramond} (NS-R).
\end{itemize}

La soluci\'on general resulta:
\begin{subequations}
\begin{align}
	\psi_{-}^\mu &\sim \sum_r \psi_r^\mu e^{-ir(\tau-\sigma)}\,,\quad
	\text{donde}\quad \begin{cases}
		r\in\mathbb{Z}\quad \text{sector R (left-moving)}\\
		r\in\mathbb{Z}+\frac{1}{2}\quad \text{sector NS (left-moving)}
	\end{cases}\\
	\psi_{+}^\mu &\sim \sum_r \tpsi_r^\mu e^{-ir(\tau+\sigma)}\,,\quad
	\text{donde}\quad \begin{cases}
		r\in\mathbb{Z}\quad \text{sector R  (right-moving)}\\
		r\in\mathbb{Z}+\frac{1}{2}\quad \text{sector NS (right-moving)}
	\end{cases}
\end{align}
\end{subequations}

\subsection{Cuantizaci\'on Can\'onica de la Cuerda}\label{cuarda_cauntica}

Como siempre, la idea es promover a operadores de creaci\'on y
destrucci\'on los coeficientes del desarrollo de Fourier de las
soluciones antes estudiadas: $\alpha_n^\mu$, $\talpha_n^\mu$,
$\psi_r^\mu$ y $\tpsi_r^\mu$. \'Estos act\'uan como operadores de
aniquilaci\'on y destrucci\'on sobre el espacio de Fock.  Las
relaciones de conmutaci\'on can\'onicas entre los campos y sus momentos
conjugados se traducen, para los osciladores, en las
siguientes relaciones de conmutaci\'on:
\begin{equation}
	[x^\mu, p^\nu] = i \eta^{\mu\nu}\,,\quad
	[\alpha_m^\mu,\alpha_n^\nu]
		= m \delta_{m,-n} \eta^{\mu\nu}\,,\quad
	\{\psi_r^\mu,\psi_s^\nu\}
		= \delta_{r,-s} \eta^{\mu\nu}\,,
\end{equation}

Lo mismo ocurre  para los osciladores tilde. Las dem\'as relaciones de
conmutaci\'on se anulan.

Comencemos por la cuerda abierta: $\ket{0,k}$ representa al vac\ii o de
momento $k$ y esta determinado por las siguientes ecuaciones:
\begin{equation}
	p^\mu \ket{0,k} = k^\mu \ket{0,k}\,,\qquad\qquad \alpha_n^\mu
        \ket{0,k} = \psi_r^\mu \ket{0,k} = 0\,,\qquad \forall n,r >
        0\,.
\end{equation}

Mientras que el estado fundamental del sector NS es
\'unico, en el sector de R hay varios modos cero que satisfacen el
\'algebra $\{ \psi_0^\mu,\psi_0^\nu\}=\eta^{\mu\nu}$, que pueden
ser representados por  matrices $\Gamma$ (Gamma) 10-dimesionales  $32\times32$.

Esto implica que el estado fundamental del sector de R transforma como
un spinor de Dirac de 32 dimensiones. Para describirlo, es conveniente
tomar una nueva base de osciladores:
\begin{equation}
	d_i^\pm = \begin{cases}
		\frac{1}{\sqrt{2}} (\psi_0^1 \mp \psi_0^0)\qquad &i=0\\
		\frac{1}{\sqrt{2}} (\psi_0^{2i} \pm \psi_0^{2i+1})\qquad &i=1,2,3,4
		\end{cases}
\end{equation}
En t\'erminos de esta base, el \'algebra deviene $\{ d_i^+ , d_j^- \}=
\delta_{ij}$, donde $d_i^\pm$ act\'uan como operadores escalera en el
vac\ii o de Ramond de 32 dimensiones, que podemos denotar como:
\begin{equation}\label{Rgs}
	\ket{\mathbf{s}} = \ket{s_0,s_1,s_2,s_3,s_4}\,,
\end{equation}
donde  $s_i=\pm\tfrac{1}{2}$. La acci\'on de un  $d_i^+$ sube $s_i$ de $-\tfrac{1}{2}$ a  $\tfrac{1}{2}$ y:
\begin{equation}
	d_i^- \ket{-\tfrac{1}{2},-\tfrac{1}{2},-\tfrac{1}{2},-\tfrac{1}{2},-\tfrac{1}{2}} = 0\,.
\end{equation}

Debido a la signatura Minkowski de la m\'etrica, el espacio de Fock
que hemos definido contiene estados de norma negativa, que resultan
f\ii sicamente inaceptables. Para seleccionar los estados f\'{\i}sicos,
analizamos el tensor de energ\'{\i}a-momento y las supercorrientes y
observamos su acci\'on sobre los estados. En las coordenadas del
cono de luz, el tensor de energ\'{\i}a-momento tiene las
siguientes componentes no nulas:

\begin{equation}
	T_{++} = \partial_+ X \cdot \partial_+ X + \frac{i}{2} \psi_+
        \cdot \partial_+ \psi_+\,,\qquad T_{--} = \partial_- X \cdot
        \partial_- X + \frac{i}{2} \psi_- \cdot \partial_- \psi_-\,,
\end{equation}
mientras que las supercorrientes de  N\"oether asociadas a la supersimetr\ii a
 de hoja de mundo (ver ecuaci\'on .~\eqref{wssusy})  son:
\begin{equation}
	J_+ = \psi_+ \cdot \partial_+ X\,,\qquad
	J_- = \psi_- \cdot \partial_- X\,.
\end{equation}
A partir de la expansi\'on de $X^\mu$ y $\psi^\mu$ en modos de Fourier
que escribimos arriba, se deriva la expansi\'on del tensor $T$ y las
corrientes $J$. Las componentes del desarrollo de Fourier del tensor
de energ\'{\i}a-momento y los generadores del \'algebra de Virasoro en
orden normal resultan:

\begin{equation}
	L_n = \frac{1}{2} \sum_m \alpha_{-m} \cdot \alpha_{n+m} +
        \frac{1}{2} \sum_r \left(\frac{n}{2}+r\right) \psi_{-r} \cdot
        \psi_{r+n}\,,
\end{equation}

donde $m,n\in\mathbb{Z}$,  $r\in\mathbb{Z}$ en el sector de R, y
$r\in\mathbb{Z}+\tfrac{1}{2}$ en el sector de NS. En el caso de $L_0$,
una vez resueltos los inconvenientes provenientes del orden normal, se obtiene:
\begin{equation}
	L_0 = \ls^2 p^2 + \sum_{n=1}^\infty \alpha_{-n} \cdot \alpha_n
		+ \sum_{r>0} r\ \psi_{-r} \cdot \psi_r\,.
\end{equation}
Las componentes de Fourier de las supercorrientes vienen dadas por:
\begin{equation}
	G_r = \sum_{n\in\mathbb{Z}} \alpha_{-n} \cdot \psi_{r+n}.
\end{equation}
Estos modos satisfacen el siguiente super\'algebra de Virasoro :
\footnote{El \'algebra ~\eqref{virasoro} se satisface en el sector NS,
  y en el de R con la siguiente redefinici\'on $L_0\to L_0 +
  \tfrac{5}{8}$.}
\begin{equation}\label{virasoro}
\begin{split}
	[ L_m, L_n ] &= (m-n) L_{m+n} + \tfrac{5}{4}(m^3-m)\delta_{m,-n}\,,\\
	\{ G_r, G_s \} &= 2L_{r+s} + \tfrac{5}{4} (4r^2 - 1)\delta_{r,-s}\,,\\
	[ L_m, G_r ] &= \tfrac{1}{2} (m-2r) G_{m+r}\,.
\end{split}
\end{equation}
Los estados f\'{\i}sicos de la teor\ii a son aquellos que satisfacen las
siguientes condiciones:
\begin{equation}
	(L_0 - a) \ket{\psi} = 0\,,\qquad\qquad L_n \ket{\psi} = G_r
  \ket{\psi} = 0\,,\qquad \forall n>0, r\ge 0\,,
\end{equation}
donde en el sector de NS $a=\tfrac{1}{2}$ y en el de R  $a=0$.

El espectro de masas de la teor\ii a es obtenido desarrollando el hamiltoniano en t\'erminos de los osciladores:
\begin{equation}
	M^2 = \frac{1}{\ls^2} \Big[ \sum_{n,r>0} \left( \alpha_{-n}
          \cdot \alpha_n + r\ \psi_{-r} \cdot \psi_r \right) - a
          \Big]\,.
\end{equation}
Notar que el primer estado del espectro de NS es taqui\'onico, $M^2 =
- \tfrac{1}{2\ls^2}$.

Consideremos ahora cuerdas cerradas. El an\'alisis en este caso sigue
en forma directa el de cuerdas abiertas, con la precauci\'on de
considerar dos conjuntos de osciladores, dos copias del \'algebra de
Virasoro, dos conjuntos de condiciones f\ii sicas, etc.  En
particular, $L_0$ resulta diferente que en el caso anteriormente
considerado:
\begin{equation}
	L_0 = \frac{\ls^2}{4} p^2 + \sum_{n=1}^\infty \alpha_{-n}
        \cdot \alpha_n + \sum_{r>0} r\ \psi_{-r} \cdot \psi_r\,,\quad
        \tilde{L}_0 = \frac{\ls^2}{4} p^2 + \sum_{n=1}^\infty
        \talpha_{-n} \cdot \talpha_n + \sum_{r>0} r\ \tpsi_{-r} \cdot
        \tpsi_r\,.
\end{equation}
 El espectro de masa resulta:
\begin{equation}
	M^2 = \frac{2}{\ls^2} \Big[ \sum_{n,r>0} \left( \alpha_{-n} \cdot \alpha_n + \talpha_{-n} \cdot \talpha_n
		+ r\ \psi_{-r} \cdot \psi_r + r\ \tpsi_{-r} \cdot \tpsi_r \right) - a - \tilde{a} \Big]\,,
\end{equation}
Vemos nuevamente que el estado de energ\'{\i}a m\'as bajo corresponde a  un
  taqui\'on con $M^2=-\tfrac{2}{\ls^2}$. Adem\'as, se debe imponer una {\it level-matching condition} sobre los estados f\'{\i}sicos:
\begin{equation}
	\left( L_0 - \tilde{L}_0 - a + \tilde{a} \right) \ket{\psi} = 0\,.
\end{equation}

El espectro obtenido arriba, tanto para cuerdas abiertas como para
cuerdas cerradas, contiene taquiones y no es supersim\'etrico en
espacio-tiempo. Para obtener un espectro supersim\'etrico y libre de
taquiones, es necesario truncar el espectro de forma consistente
aplicando la llamada {\it proyecci\'on GSO} (Gliozzi, Scherk, Olive).
Esta proyecci\'on consiste en retener en la teor\ii a s\'olo los
estados con un n\'umero par $F$ de fermiones en la hoja de mundo. Por
ejemplo, en el sector de NS para la cuerda abierta podemos definir el
siguiente proyector:
\begin{equation}\label{GSONS}
	P_{\text{NS}} = \frac{1+ (-1)^{F_{\text{NS}}}}{2}\,,\qquad
        F_{\text{NS}} = \sum_{r=1/2}^{\infty} \psi_{-r} \cdot \psi_r -
        1\,.
\end{equation}
Usando este proyector ~\eqref{GSONS}, vemos que el estado fundamental
taqui\'onico es removido del espectro. En el sector de R, sobre los
estados diferentes de modos cero, la proyecci\'on act\'ua de manera
totalmente an\'aloga, mientras que para los modos cero, act\'ua como
proyecci\'on de quiralidad, conservando los estados~\eqref{Rgs} con:
\begin{equation}
	\sum_{i=0}^4 s_i = \text{par (impar).}
\end{equation}

Examinemos el primer nivel de cuerda abierta que sobrevive a la
proyecci\'on.  Del sector de NS, tenemos los siguientes estados no
masivos:
\begin{equation}
	\epsilon_\mu(k) \psi^\mu_{-1/2}\ket{0,k}\,,\qquad k^2 =
        0\,,\qquad k \cdot \epsilon = 0\,.
\end{equation}
Se trata de un campo vectorial de gauge con polarizaci\'on transversa,
con 8 grados de libertad on-shell. En el caso del sector de R, el estado  es
\begin{equation}
	u_{\mathbf{s}} (k) \ket{\mathbf{s}}\,,\qquad
	k^2 = 0\,,\qquad k\cdot \Gamma_{\mathbf{s'}\mathbf{s}} u_{\mathbf{s}} = 0\,.
\end{equation}

Por acci\'on de la proyecci\'on GSO, este estado resulta en un espinor
de Mayorana-Weyl, con 8 grados de libertad on-shell en 10
dimensiones. Vemos entonces que la condici\'on necesaria para tener
supersimetr\'ia (es decir, que el n\'umero de grados de libertad
bos\'onicos y fermi\'onicos coincida a cada nivel de masas), se cumple
a nivel massless. Se puede mostrar que el espectro completo de la
cuerda luego de la proyecci\'on GSO es supersim\'etrico.

\subsection*{Teor\ii as Tipo II}

En el caso de cuerdas cerradas, tenemos cuatro sectores diferentes y
es necesario realizar la proyecci\'on GSO en cada uno de ellos. En
particular, surge la posibilidad de realizar la misma proyecci\'on o la
opuesta en las partes left-moving y right-moving del sector
R. Dependiendo de la elecci\'on, tendremos dos teor\ii as no equivalentes
con los siguientes sectores, donde el signo denota la paridad frente
$(-)^F$:
\begin{center}
\begin{tabular}{lcccc}
Type IIB (quiral): & (NS+,NS+) & (R+, NS+) & (NS+,R+) & (R+,R+) \\
Type IIA (no quiral): & (NS+,NS+) & (R+, NS+) & (NS+,R$-$) & (R+,R$-$) \\
\end{tabular}
\end{center}

Analizaremos entonces las partes no masivas del espectro de las
teor\ii as de cuerdas tipo II. El sector NS-NS posee el mismo contenido
en ambas teor\ii as, y el estado massless correspondiente es:
\begin{equation}\label{NSmassless}
	\psi_{-1/2}^\mu \tpsi_{-1/2}^\nu \ket{0,k}\,,\qquad k^2 = 0\,.
\end{equation}
Saturando este estado con un tensor sim\'etrico sin traza, obtenemos:
\begin{equation}
	\epsilon_{\mu\nu}^{(h)} = \epsilon_{\nu\mu}^{(h)}\,,\qquad
	\epsilon_{\mu\nu}^{(h)}\eta^{\mu\nu}=0\,,\qquad
	k^\mu\epsilon_{\mu\nu}^{(h)} = 0\,,
\end{equation}
esto da un estado f\'{\i}sico que puede ser identificado con el
gravit\'on $h_{\mu\nu}$ en el espacio-tiempo. Alternativamente, podemos
saturar~\eqref{NSmassless} con un tensor antisim\'etrico:
\begin{equation}
	\epsilon_{\mu\nu}^{(B)} = -\epsilon_{\nu\mu}^{(B)}\,,\qquad
	k^\mu\epsilon_{\mu\nu}^{(B)} = 0\,.
\end{equation}
Esto da lugar a una 2-forma que corresponde a un campo de gauge
$B_2$. Finalmente, podemos obtener una campo escalar, el dilat\'on
$\Phi$, saturando~\eqref{NSmassless} con:
\begin{equation}
	\epsilon_{\mu\nu}^{(\Phi)} = \frac{1}{\sqrt{8}} (\eta_{\mu\nu}
        - k_\mu \ell_\nu - k_\nu\ell_\mu)\,,\qquad \ell^2 = k^2 =
        0\,,\quad \ell\cdot k = 1\,.
\end{equation}
En el sector NS-R , tenemos espinor-vector no masivo:
\begin{equation}
	u_{\mu \mathbf{s}} \psi_{-1/2}^\mu \ket{0,\mathbf{\tilde{s}},k}\,,\qquad
	k^2 = k^\mu \tilde{u}_{\mu\mathbf{\tilde{s}}} 
		= k \cdot \Gamma_{\mathbf{\tilde{s}'}\mathbf{\tilde{s}}}
		\tilde{u}_{\mu \mathbf{\tilde{s}}}=0\,,
\end{equation}
\'este es reducible Lorentz. Su descomposici\'on da lugar a un
gravitino de espin $3/2$ y un dilatino de espin $1/2$, con quiralidades
opuestas. El mismo an\'alisis puede ser repetido para el sector R-NS,
y la teor\'{\i}a tiene entonces dos gravitinos. Estas teor{\'\i}as se
llaman de Tipo II. Por \'ultimo, analizamos el sector R-R. El estado
massless gen\'erico tiene la forma:
\begin{equation}\label{Rmassless}
	u_{\mathbf{s}}\tilde{u}_{\mathbf{\tilde{s}}}
        \ket{\mathbf{s},\mathbf{\tilde{s}}}\,,\qquad k \cdot
        \Gamma_{\mathbf{s}'\mathbf{s}} u_{\mathbf{s}} = k \cdot
        \Gamma_{\mathbf{\tilde{s}'}\mathbf{\tilde{s}}}
        \tilde{u}_{\mathbf{\tilde{s}}}=0\,.
\end{equation}

Para poder analizar este caso, conviene pasar al gauge conforme. La
teor\ii a de cuerdas en el gauge superconforme resulta ser una teor\ii a de
campos bidimensional superconforme. Recordemos que en teor\ii as de
campos conformes~\cite{Polchinski:1998}, un estado $\ket{\psi}$ de
la teor\ii a es dual a un operador de v\'ertice, tal que $\ket{\psi} =
\lim_{z,\bar{z}\to 0} \mathcal{V}^{(\psi)} (z,\bar{z})\ket{0}$, donde
$z = e^{i(\tau-\sigma)}$ y $\bar{z}=e^{i(\tau+\sigma)}$ son las
coordenadas complejas de la hoja de mundo. En nuestro caso, el vac{\'\i}o
de R-R es obtenido a partir del de NS-NS via la acci\'on de \emph{spin fields}:

\begin{equation}
  \ket{\mathbf{s},\mathbf{\tilde{s}}} = \lim_{z,\bar{z}\to 0}
  S^{\mathbf{s}} (z) \tilde{S}^{\mathbf{\tilde{s}}} (\bar{z})
  \ket{0}\,.
\end{equation}
usando esto junto con ~\eqref{Rmassless}, podemos escribir la siguiente descomposici\'on:
\begin{multline}\label{Rdec}
	\lim_{z,\bar{z}\to 0}
        u_{\mathbf{s}}\tilde{u}_{\mathbf{\tilde{s}}} S^{\mathbf{s}}
        (z) \tilde{S}^{\mathbf{\tilde{s}}} (\bar{z}) \ket{0}\\ =
        \frac{1}{32} \sum_{n=0}^{10} \frac{(-1)^{n+1}}{n!}
        u_{\mathbf{s}} (\Gamma_{\mu_1\dotsm \mu_n} C^{-1})
        \tilde{u}_{\mathbf{\tilde{s}}} \lim_{z,\bar{z}\to 0}
        S^{\mathbf{s}} (z) (C \Gamma^{\mu_1\dotsm \mu_n})
        \tilde{S}^{\mathbf{\tilde{s}}} (\bar{z}) \ket{0}\,,
\end{multline}
donde $C$ denota la matriz de conjugaci\'on de carga. Se puede mostrar
que, en el caso de la teor\ii a tipo IIB, s\'olo t\'erminos con $n$
impar contribuyen a la suma~\eqref{Rdec}, mientras que para la teor\ii
a IIA se da la situaci\'on inversa, es decir, s\'olo los t\'erminos con $n$
par sobreviven. Es decir que los campos
\begin{equation}
	F_{\mu_1\dotsm\mu_n} = u \Gamma_{\mu_1\dotsm\mu_n} C^{-1} \tilde{u}\,,
\end{equation}
existen s\'olo si $n$ es impar en la teor{\'\i}a tipo IIB, y s\'olo si
$n$ es par en la teor{\'\i}a tipo IIA.

 De la expresi\'on anterior y de ~\eqref{Rmassless}, se tiene que $F$
 satisface las ecuaciones de movimiento y las identidades de Bianchi
 apropiadas para la $n$-forma intensidad de campo (en lugar de la del
 potencial de gauge):
\begin{equation}
	dF_n = d (\hd{} F)_{10-n} = 0.
\end{equation}

\begin{table}
\begin{center}
\begin{tabular}{|c|c|c|}
\hline
& sector NS-NS  & sector R-R \\
\hline
Tipo IIB & $G_{\mu\nu}$, $B_2$, $\Phi$ & $C_0$, $C_2$, $C_4$ \\
Tipo IIA & $G_{\mu\nu}$, $B_2$, $\Phi$ & $C_1$, $C_3$ \\
\hline
\end{tabular}
\end{center}
\caption{{\small Espectro bos\'onico no masivo de Teor{\'\i}as de Supercuerdas Tipo II.}}
\label{t:spectrum}
\end{table}

\subsection*{Acci\'on de baja energ\ii a}

Otro ingrediente importante en esta descripci\'on es  
la acci\'on de baja energ\ii a de las supercuerdas tipo II. Cuando
hablamos de acci\'on de baja energ\ii a nos referimos a una acci\'on
cl\'asica de campos que describe la din\'amica de los estados no
masivos en el l{\'\i}mite $\ls\to 0$, donde la cuerda se reduce a un
punto. Los campos que aparecen en la acci\'on efectiva son los que se
listan en la tabla~\ref{t:spectrum}, y esta acci\'on efectiva se
construye de forma tal que sea capaz de reproducir al m\'as bajo orden
las amplitudes de dispersi\'on o las funciones beta de un modelo sigma
para la cuerda.

En cualquier caso, uno obtiene la acci\'on de \emph{type II
  supergravity} (SUGRA).  La parte bos\'onica de la acci\'on de
supergravedad tipo IIB en el string frame (donde la m\'etrica aparece
tal como en la acci\'on de hoja de mundo) resulta:
\begin{multline}\label{IIBst}
	S_{\text{IIB}}^{\text{(st)}} = \frac{1}{2\kappa^2} \Bigg\{
        \int d^{10}x \sqrt{-\det G}\ e^{-2\Phi}R - \frac{1}{2} \int
        \Big[ - 8 e^{-2\Phi} d\Phi \wedge \hd{} d\Phi + e^{-2\Phi} H_3
          \wedge \hd{}H_3\\ + F_1 \wedge \hd{}F_1 + \tilde{F}_3 \wedge
          \hd{}\tilde{F}_3 + \frac{1}{2} \tilde{F}_5 \wedge
          \hd{}\tilde{F}_5 - C_4\wedge H_3 \wedge F_3 \Big] \Bigg\}\,.
\end{multline}
donde $\kappa$ est\'a relacionada con las cantidades de la cuerda como:
$\kappa = 8\pi^{7/2}\gs\ls^4$ y:
\begin{equation}
	H_3 = dB_2\,,\qquad F_n = dC_{n-1}\,,\qquad
	\tilde{F}_3 = F_3 + C_0 \wedge H_3\,,\qquad
	\tilde{F}_5 = F_5 + C_2 \wedge H_3\,.
\end{equation}
La acci\'on~\eqref{IIBst} no toma en cuenta la autodualidad del $\tilde{F}_5$, que deber\'a ser impuesta on-shell.

La parte bos\'onica de la acci\' on de SUGRA tipo IIA en el string frame es:
\begin{multline}\label{IIAst}
	S_{\text{IIA}}^{\text{(st)}} = \frac{1}{2\kappa^2} \Bigg\{
        \int d^{10}x \sqrt{-\det G}\ e^{-2\Phi}R - \frac{1}{2} \int
        \Big[ - 8 e^{-2\Phi} d\Phi \wedge \hd{} d\Phi + e^{-2\Phi} H_3
          \wedge \hd{}H_3\\ - F_2 \wedge \hd{}F_2 - \tilde{F}_4 \wedge
          \hd{}\tilde{F}_4 + B_2\wedge F_4 \wedge F_4 \Big] \Bigg\}\,,
\end{multline}
donde
\begin{equation}
	H_3 = dB_2\,,\qquad F_n = dC_{n-1}\,,\qquad
	\tilde{F}_4 = F_4 - C_1 \wedge H_3\,.
\end{equation}

La dependencia no est\'andar con el dilat\'on $\phi$ en la
acci\'on~\eqref{IIBst} puede ser expresada en la forma usual al pasar
al Einstein frame, usando la definici\'on
\begin{equation}
	ds^2_{\text{(E)}} = e^{-{\frac{4}{D-2}}\phi} ds^2_{\text{(st)}}\,.
\end{equation}
donde $D$ es la dimensi\'on del espacio-tiempo ($D=10$ en el caso usual).
La acci\'on de SUGRA tipo II en el Einstein frame, est\'a dada por:

\begin{multline}\label{IIBE}
	S_{\text{IIB}}^{\text{(E)}} = \frac{1}{2\kappa^2} \Bigg\{ \int
        d^{10}x \sqrt{-\det G}\ R - \frac{1}{2} \int \Big[ d\Phi
          \wedge \hd{} d\Phi + e^{-\Phi} H_3 \wedge \hd{}H_3\\ +
          e^{2\Phi} F_1 \wedge \hd{}F_1 + e^{\Phi} \tilde{F}_3 \wedge
          \hd{}\tilde{F}_3 + \frac{1}{2} \tilde{F}_5 \wedge
          \hd{}\tilde{F}_5 - C_4\wedge H_3 \wedge F_3 \Big] \Bigg\}\,,
\end{multline}
y
\begin{multline}\label{IIAE}
	S_{\text{IIA}}^{\text{(E)}} = \frac{1}{2\kappa^2} \Bigg\{ \int
        d^{10}x \sqrt{-\det G}\ R - \frac{1}{2} \int \Big[ d\Phi
          \wedge \hd{} d\Phi + e^{-\Phi} H_3 \wedge \hd{}H_3\\ -
          e^{3\Phi/2} F_2 \wedge \hd{}F_2 - e^{\Phi/2} \tilde{F}_4
          \wedge \hd{}\tilde{F}_4 + B_2\wedge F_4 \wedge F_4 \Big]
        \Bigg\}\,.
\end{multline}

\section{D-branas y cuerdas abiertas}\label{s:dbranes}

En esta secci\'on, presentaremos brevemente las caracter\ii sticas
esenciales de unos objetos de la teor\ii a de cuerdas llamados
\emph{branas de Dirichlet} o D-branas. En la teor\ii a de cuerdas
abiertas, estos objetos pueden ser definidos como hipersuperficies
donde los extremos de las cuerdas pueden moverse libremente. Es decir,
hiperplanos determinados por las condiciones de contorno de Dirichlet
~\cite{Johnson:2003gi}.

\subsection*{Cuerdas en un C\ii rculo}

Comencemos considerando la compactificaci\'on de una cuerda cerrada en
un c\ii rculo. La soluci\'on m\'as general de las ecuaciones de
movimiento~\eqref{steom} para una cuerda cerrada pueden escribirse
como siguen:
\begin{equation}
	X^\mu (\tau,\sigma) = x^\mu +
        \frac{\ls}{\sqrt{2}}(\alpha_0^\mu+\talpha_0^\mu) \tau -
        \frac{\ls}{\sqrt{2}} (\alpha_0^\mu-\talpha_0^\mu) \sigma + i
        \frac{\ls}{\sqrt{2}} \sum_{n\neq 0} \frac{1}{n} \left(
        \alpha_n^\mu e^{-in(\tau-\sigma)} + \talpha_n^\mu
        e^{-in(\tau+\sigma)}\right)\,.
\end{equation}
y el momento est\'a dado por:
\begin{equation}
	p^\mu = \frac{1}{\sqrt{2}\ \ls} (\alpha_0^\mu + \talpha_0^\mu)\,.
\end{equation}
Como vimos en el caso no compactificado, los dos modos cero deben
estar identificados $\alpha_0^\mu = \talpha_0^\mu =
\tfrac{\ls}{\sqrt{2}} p^\mu$, debido a la condici\'on de periodicidad
ante $\sigma\to\sigma+2\pi$, y de aqu\'i uno obtiene el desarrollo en
modos~\eqref{Xcl}, en t\'erminos del momento $p^\mu$.

Supongamos que compactificamos una sola direcci\'on del espacio-tiempo
de fondo en un c\ii rulo de radio $R$. Llamemos a esta direcci\'on
$X$ sin {\'\i}ndices:
\begin{equation}
	X \simeq X + 2\pi R\,.
\end{equation}
En este caso, el momento a lo largo de esta direcci\'on debe estar cuantizado como:
\begin{equation}\label{quantmom}
	p = \frac{n}{R}\,,\qquad n\in\mathbb{Z}\,.
\end{equation}
Si adem\'as tenemos en cuenta que una cuerda cerrada puede enroscarse
alrededor de la direcci\'on compactificada $X$, ante
$\sigma\to\sigma+2\pi$, $X$ no necesita ser monovaluada, y como
consecuencia tenemos:
\begin{equation}\label{wind}
	X(\tau,\sigma+2\pi) \simeq X(\tau,\sigma)+ 2\pi R w\,,
\end{equation}
donde $w$ es el n\'umero de winding (enroscamiento) alrededor de
$X$. Tomando las condiciones~\eqref{quantmom} y~\eqref{wind},
obtenemos la siguiente ecuaci\'on:
\begin{equation}
	\alpha_0 + \talpha_0 = \sqrt{2}\ \ls \frac{n}{R}\,,\qquad
	\alpha_0 - \talpha_0 = \frac{\sqrt{2}}{\ls} wR\,,
\end{equation}
que implica:
\begin{equation}
	\alpha_0 = \frac{\ls}{\sqrt{2}} \left(\frac{n}{R}+\frac{wR}{\ls^2}\right)\,,\qquad
	\talpha_0 = \frac{\ls}{\sqrt{2}} \left(\frac{n}{R}-\frac{wR}{\ls^2}\right)\,.
\end{equation}
La parte bos\'onica de los modos cero de los generadores de Virasoro se modifican como sigue:
\begin{equation}
\begin{aligned}
	L_0 &= \frac{\ls^2}{4}p^2 + \frac{\ls^2}{4}\left( \frac{n}{R}
        + \frac{wR}{\ls^2} \right)^2 + \sum_{n=1}^{\infty}
        \alpha_{-n}\cdot \alpha_n\,,\\ \tilde{L}_0 &=
        \frac{\ls^2}{4}p^2 + \frac{\ls^2}{4}\left( \frac{n}{R} -
        \frac{wR}{\ls^2} \right)^2 + \sum_{n=1}^{\infty}
        \talpha_{-n}\cdot \talpha_n\,,
\end{aligned}
\end{equation}
donde $p$ ahora denota el momento de las direcciones no
compactificadas. Usando estas expresiones, se ve que la parte bos\'onica
del operador de masa tiene la forma siguiente:
\begin{equation}\label{masscomp}
	M^2 = \left(\frac{n}{R}\right)^2 + \left(\frac{wR}{\ls^2}\right)^2
		+ \text{parte oscilatoria}\,.
\end{equation}
donde vemos que aparecen dos nuevos tipos de estados  en el espectro de
la teor{\'\i}a compactificada. \'Estos son, por un lado, modos de Kaluza-Klein
(KK) que contribuyen a la energ\ii a como $\tfrac{n}{R}$, y por otro
lado aparecen las excitaciones debidas a los modos de winding en torno
a la direcci\'on compactificada.

\subsection*{T-dualidad}

La f\'ormula~\eqref{masscomp} para el espectro de masa de la cuerda
cerrada compactificada en un c\ii rculo nos permite hacer varias
observaciones interesantes. Por ejemplo, en el l\ii mite $R \to \infty$,
vemos que los modos $K K$ se vuelven muy livianos, mientras que los
modos de winding se vuelven infinitamente masivos, desacopl\'andose de la
teor\ii a, como ser{\'\i}a de esperar, 
ya que \'este corresponde al l\ii mite de descompactificaci\'on de la 
direcci\'on $X$. Sin embargo, cuando miramos el l\ii mite $R \to 0$, 
donde los modos $KK$ son los que se
desacoplan, los modos de winding tienden al continuo y observamos que
el espectro resultante es exactamente el obtenido en el l{\'\i}mite
anterior. En efecto, ~\eqref{masscomp} es invariante ante la
transformaci\'on: 
\begin{equation}\label{RtoTR}
	n \leftrightarrow w\,,\qquad
	R \to \hat{R} = \frac{\ls^2}{R}\,,
\end{equation}
y entonces resulta claro que, al menos en lo que se refiere a la parte
bos\'onica del espectro, los l\ii mites $R\to0$ y $R\to\infty$ son
f\ii sicamente id\'enticos: El espectro de cuerdas luce exactamente igual
en el caso compactificado que en el caso descompactificado. Esta
simetr\ii a del espectro bos\'onico es conocida como \emph{T-dualidad}. A\'un
m\'as, intercambiar el n\'umero de winding con el momento es equivalente
a la siguiente acci\'on sobre los modos cero bos\'onicos:
\begin{equation}
	\alpha_0 \to \alpha_0\,,\qquad
	\talpha_0 \to -\talpha_0\,,
\end{equation}
Esta operaci\'on se extiende a los modos no cero, de forma tal que
podemos resumir la acci\' on de esta simetr\ii a en la coordenada $X$
como sigue. Escribamos el desarrollo de la coordenada original $X$
correspondiente a la direcci\'on compacta as\ii:

\begin{equation}
	X (\tau,\sigma) = X_L (\tau-\sigma) + X_R (\tau+\sigma) \,,
\end{equation}
donde
\begin{equation}\label{XLR}
\begin{aligned}
	X_L (\tau-\sigma) &= \frac{x^\mu}{2} + \frac{\ls}{\sqrt{2}}
        \alpha_0^\mu (\tau-\sigma) + i \frac{\ls}{\sqrt{2}}
        \sum_{n\neq0} \frac{\alpha_n^\mu}{n}
        e^{-i(\tau-\sigma)}\,,\\ X_R (\tau+\sigma) &= \frac{x^\mu}{2}
        + \frac{\ls}{\sqrt{2}} \alpha_0^\mu (\tau+\sigma) + i
        \frac{\ls}{\sqrt{2}} \sum_{n\neq0} \frac{\alpha_n^\mu}{n}
        e^{-i(\tau+\sigma)}\,.
\end{aligned}
\end{equation}
De la condici\'on anteriormente establecida, se ve que la coordenada
$\hat{X}$ que ser\'a usada en la descripci\'on T-dual debe satisfacer:
\begin{equation}\label{NtoD}
	\partial_\tau \hat{X} = - \partial_\sigma X\,,\qquad
	\partial_\sigma \hat{X} = - \partial_\tau X.
\end{equation}
Esto nos permite identificar la coordenada T-dual como:
\begin{equation}\label{TdX}
	\hat{X} (\tau,\sigma) =  X_L (\tau-\sigma) - X_R (\tau+\sigma)\,,
\end{equation}
en t\'erminos de la expresi\'on~\eqref{XLR}.

Hasta aqu{\'\i}, hemos sido capaces de identificar la acci\'on de la
simetr\ii a de T-dualidad sobre la parte bos\'onica del espectro, pero
debemos ver que es lo que ocurre cuando tratamos de establecer la
acci\'on de esta transformaci\'on sobre la parte fermi\'onica.  Esta
acci\'on puede ser deducida requeriendo invarianza
superconforme sobre la hoja de mundo, que implica: 

\begin{equation}
	\psi_+ \to \psi_+\,,\qquad
	\psi_- \to - \psi_-\,,
\end{equation}
o, en t\'erminos de los osciladores:
\begin{equation}
	\psi_r \to \psi_r\,,\qquad
	\tpsi_r \to -\tpsi_r\,.
\end{equation}

Es decir, esta transformaci\'on invierte la quiralidad del estado
fundamental de los modos right-moving del sector R-R. Esto implica
directamente que dado que el espectro bos\'onico es invariante ante
T-dualidad, y el sector right-moving del R-R cambia de quiralidad, si
comenzamos con una teor\ii a IIA en un c\ii rculo de radio $R$ y
hacemos una T-dualidad, obtendremos una teor\ii a IIB en un c\ii rculo
de radio $\tfrac{\ls^2}{R}$, y viceversa. Este resultado se extiende
naturalmente a T-dualidades que envuelven m\'as de una direcci\'on. Si
comenzamos con un tipo de teor\ii a tipo II, un n\'umero impar de
T-dualidades nos llevar{\'\i}a a otro tipo de teor\ii a, mientras que
un n\'umero par nos dejar\ii a de nuevo dentro de la misma teor\ii a tipo
II. En ese sentido, se dice que las teor\ii as tipo II son el l\ii
mite de decompactificaci\'on de un \'unico espacio de teor\ii as
compactas.  En efecto, la relaci\'on~\eqref{RtoTR} implica que
en una teor\ii a compactificada podemos limitarnos a la regi\'on en la
que $R\ge \ls$, y es por esta raz\'on que podemos decir que $\ls$ es la
longitud m\ii nima de la teor\ii a.

Dado que, por T-dualidad, pasamos de teor\ii as tipo IIA a teor\ii as
tipo IIB y viceversa, los espectros completos de las teor\ii as IIA y
IIB deben mapearse uno en el otro. En particular, veamos como los
campos no masivos se transforman entre s{\'\i} para las acciones de
supergravedad~\eqref{IIBst}
y~\eqref{IIAst}~\cite{Bergshoeff:1995as}. Se observa que los
potenciales de rango impar en el sector de R-R son mapeados en los de
rango par y viceversa. El siguiente ejemplo corresponde a T-dualizar
s\'olo en la direcci\'on $X^9$, mientras que los {\'\i}ndices $\mu$ y
$\nu$ denotan las direcciones que no fueron afectadas por la
tranformaci\'on. Los campos que tiene sombrero ($\,\hat{}\,$) corresponden
a los transformados:
\begin{equation}\label{Tduality}
\begin{gathered}
	\hat{G}_{99} = \frac{1}{G_{99}}\,,\qquad e^{2\hat{\Phi}} =
        \frac{e^{2\Phi}}{G_{99}}\,,\qquad \hat{G}_{\mu 9} =
        \frac{B_{\mu 9}}{G_{99}}\,,\qquad \hat{B}_{\mu 9} =
        \frac{G_{\mu 9}}{G_{99}}\,,\\ \hat{G}_{\mu\nu} = G_{\mu\nu} -
        \frac{G_{\mu 9}G_{\nu 9}-B_{\mu 9}B_{\nu 9}}{G_{99}}\,,\qquad
        \hat{B}_{\mu\nu} = B_{\mu\nu} - \frac{B_{\mu 9}G_{\nu
            9}-G_{\mu 9}B_{\nu 9}}{G_{99}}\,,\\
\begin{aligned}
	(\hat{C}_{p})_{\mu\cdots\nu\alpha9} &=
  (C_{p-1})_{\mu\cdots\nu\alpha} - (p-1)
  \frac{(C_{p-1})_{[\mu\cdots\nu|9}G_{|\alpha|9}}{G_{99}}\,,\\ (\hat{C}_{p})_{\mu\cdots\nu\alpha\beta9}
    &= (C_{p+1})_{\mu\cdots\nu\alpha\beta9} + p
    (C_{p-1})_{[\mu\cdots\nu\alpha}B_{\beta]9} + p(p-1)
    \frac{(C_{p-1})_{[\mu\cdots\nu|9}B_{|\alpha|9}G_{|\beta]9}}{G_{99}}\,.
\end{aligned}	
\end{gathered}
\end{equation}

\subsection*{T-dualidad Para Cuerdas Abiertas: D-branas}

Resulta natural preguntarse si es posible extender la T-dualidad a
teor\ii as de cuerdas abiertas.  Como vimos, las cuerdas abiertas no
satisfacen ninguna condici\'on de periodicidad, y claramente no pueden
ser enrolladas alrededor de una direcci\'on peri\'odica. Por lo tanto, si
bien encontramos modos $KK$, no encontraremos modos de winding. Esto
significa que el l\ii mite $R \to 0$ ser\'a realmente diferente del
l\ii mite de decompactificaci\'on $R\to\infty$.

Ahora bien, si tenemos en cuenta que las teor\ii as con cuerdas
abiertas contienen cuerdas cerradas, parece que algo no trivial est\'a
sucediendo. La pregunta es: > C\'omo es posible que en el l\ii mite de
compactificaci\'on $R\to 0$ las cuerdas cerradas perciban todas las
direcciones como no compactas, mientras que las cuerdas abiertas ven
que una direcci\'on efectivamente desaparece al desacoplarse los modos
de $KK$?. Dado que las cuerdas abiertas y cerradas son localmente
indistinguibles unas de las otras, s\'olo puede haber diferencias en sus
extremos, entonces, es posible salir de esta aparente paradoja
pensando que los extremos de las cuerdas deben estar pegados a
 hipersuperficies del espacio-tiempo.

Para clarificar, consideremos el siguiente ejemplo. Supongamos que
tenemos cuerdas abiertas y cerradas viviendo en un espacio
$D$-dimensional, en el que nos disponemos a compactificar una
direcci\'on en un c\ii rculo de radio $R$. Ahora queremos hacer el
l\ii mite $R \to 0$, y para eso hacemos una T-dualidad en la
direcci\'on considerada. La nueva teor\ii a es \'esta compactificada
en un c\ii rculo de radio $\hat{R}=\tfrac{\ls^2}{R}$. En esta teor\ii a,
el l\ii mite que nos interesaba corresponde a
$\hat{R}\to\infty$. Resulta f\'acil ver que ahora la condici\'on de
contorno de Neumann para $X$ se ha transformado en una condici\'on de
contorno de Dirichlet para $\hat{X}$.
  
Esto significa que, en la teor\ii a T-dual, los extremos de la cuerda
abierta fueron obligados a vivir sobre una hipersuperficie
$p$-dimensional, donde $p=D-1$ es la dimensi\'on de la
hipersuperficie.  Queda claro entonces por qu\'e, cuando miramos el
l\ii mite en cuesti\'on, el espectro de la cuerda abierta ha perdido
un grado de libertad. Estos hiperplanos toman el nombre de
\emph{D$p$-branas}~\cite{Polchinski:1995mt}. Resulta claro por otro
lado que, dado que una transformaci\'on de T-dualidad intercambia
condiciones de contorno \emph{N} por condiciones de contorno \emph{D},
una nueva T-dualidad realizada en una direcci\'on longitudinal a la
D$p$-brana dar\'a como resultado una D$(p-1)$-brana, mientras que una
T-dualidad en una direcci\'on transversa lleva de una D$p$-brana a una
D$(p+1)$-brana.

Hasta aqu{\'\i}, hemos tratado a las D$p$-branes como hiperplanos r\ii
gidos donde terminan las cuerdas, pero las excitaciones de las cuerdas
hacen que las D$p$-branas sean objetos din\'amicos en la teor\ii a de
cuerdas. Entre estas excitaciones, las no masivas tiene la importante
propiedad de no cambiar la energ\ii a de la D-brana y pueden ser
entendidas, por lo tanto, como las coordenadas colectivas de la
misma. Consideremos, entonces, el espectro no masivo de las cuerdas
abiertas pegadas a una D$p$-brana. En lo que sigue, representaremos
las configuraciones de las D-branas a trav\'es de las tablas como la
que siguen, donde los s\ii mbolos $-$ y $\cdot$ representan,
respectivamente, direcciones longitudinales y transversas al volumen
de mundo de la D$p$-brana.

\begin{center}
\begin{tabular}{|c|ccc|ccc|}
\hline
&0&$\cdots$&$p$&$p$+1&$\cdots$&9\\
\hline
D$p$ &$-$&$-$&$-$&$\cdot$&$\cdot$&$\cdot$\\
\hline
\end{tabular}
\end{center}
Separemos  las 10 coordenadas del espacio-tiempo como sigue:
\begin{itemize}
\item $x^\alpha\,,\quad \alpha=0,\ldots,p\,:$ direcciones
  pertenecientes al volumen de mundo de la D$p$-brana;
\item $x^i\,,\quad i=p+1,\ldots,9\,:$ direcciones transversas al
  volumen de mundo de la D$p$-brana. La brana puede encontrarse, por
  ejemplo, en $x^i=0$.
\end{itemize}
Los estados no masivos de cuerda abierta est\'an dados por:
\begin{center}
\begin{tabular}{|c|l|}
\hline
\multicolumn{2}{|c|}{estados NS} \\
\hline
$\psi_{-1/2}^\alpha \ket{0,k}$ &
$\to$ $1$ vector $A^\alpha$\\
$\psi_{-1/2}^i \ket{0,k}$ &
$\to$ $9-p$ escalares reales $\Phi^i$\\
\hline
\end{tabular}
\qquad
\begin{tabular}{|c|l|}
\hline
\multicolumn{2}{|c|}{estados R} \\
\hline
$\ket{s_0,s_1,s_2,s_3,s_4}$ &
$\to$ $16$ fermiones\\
\hline
\end{tabular}
\end{center}

Estos estados conforman precisamente el multiplete vectorial de una
teor\ii a de gauge $(p+1)$-dimensional, con grupo de gauge $U(1)$ y 16
supercargas. \'Este es un hecho clave de las D$p$-branas: La din\'amica
de baja energ\ii a de los estados de cuerdas abiertas m\'as bajos
pegados a una D$p$-brana, se describen a trav\'es de una teor\ii a de
gauge $(p+1)$-dimensional. Los escalares $\Phi^i$ describen la forma
de la hipersuperficie y, en efecto, est\'an relacionados con las
coordenadas del embedding como:
\begin{equation}
	X^i = 2\pi\ls^2 \Phi^i\,.
\end{equation}
El an\'alisis anterior es f\'acilmente generalizado al caso de varias
D$p$-branas paralelas en el mismo punto. Consideremos, por ejemplo, el
caso de dos branas en el mismo punto $x^i=y^i$ del espacio transverso
a ambas. Adem\'as de las cuerdas que empiezan y terminan en la misma
brana, podemos tener cuerdas que empiecen en una y terminen en la otra,
con las dos posibles orientaciones. Estas configuraciones pueden ser
representadas introduciendo matrices de $\lambda$ Chan-Paton de $2
\times 2$, que etiquetan los estados de una cuerda abierta.

\begin{equation}
	\lambda \otimes \text{``osciladores''}\ \ket{0,k}\,,\qquad
	\lambda = \begin{pmatrix}
		\text{D}p-\text{D}p & \text{D}p-\text{D}p' \\
		\text{D}p'-\text{D}p & \text{D}p'-\text{D}p' \\
		\end{pmatrix} \,,
\end{equation}

Repitiendo el an\'alisis anterior, vemos que ahora los estados no
masivos de cuerdas abiertas forman un multiplete vectorial
correspondiente a una teor\ii a de gauge en $p+1$ dimensiones con
grupo de gauge $U(2)$. Si las branas fueran separadas en alguna
direcci\'on del espacio transverso, los estados de cuerda abierta con
elementos no diagonales en $\lambda$ se volver\ii an masivos, es
decir, representar\ii an bosones $W$ masivos y la simetr\ii a del
grupo se romper\ii a en $U(1)\times U(1)$. Cuando las dos branas
coinciden, los bosones $W$ se vuelven no masivos y la simetr\ii a
$U(2)$ es recuperada.

Siguiendo con este razonamiento, podemos considerar $N$ D-branas. La
teor\ii a de baja energ\ii a correspondiente a un conjunto de $N$
D$p$-branas coincidentes, es una teor{\'\i}a de Super Yang-Mills
($p+1$)-dimensional con grupo de gauge $U(N)$ y 16 supercargas.

Cuando una de estas branas es separada del resto, digamos del punto
$x^i=0$ a cualquier otro punto del espacio transverso, el valor de
expectaci\'on del escalar $\Phi^i$ en el multiplete vectorial de la
teor\ii a de gauge toma un valor no nulo, $x^i = 2\pi\ls^2 \Phi^i$, y
la simetr\ii a de gauge se rompe como $U(N-1)\times U(1)$ via
mecanismo de Higgs.

\begin{figure}
\begin{center}
\includegraphics[scale=.7]{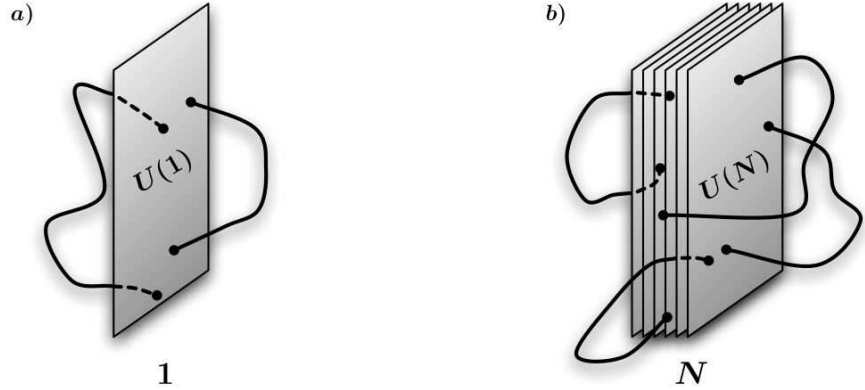}
\caption[D-branas como hiperplanos donde terminan las
  cuerdas abiertas.] {Las D-branas son hiperplanos din\'amicos donde
  terminan las cuerdas abiertas. a) La din\'amica de baja energ{\'\i}a
  de los estados no masivos de la cuerda abierta sobre una \'unica
  D-brana da una teor{\'\i}a de gauge $U(1)$; b) La teor{\'\i}a que
  vive en un stack de $N$ D-branas coincidentes tiene grupo de gauge
  $U(N)$.}
%
\label{f:stack}
\end{center}
\end{figure}

\subsection*{D-branas y Cargas de Ramond-Ramond}

Hemos visto hasta aqu{\'\i} que las D-branas pueden considerarse como
objetos f\ii sicos de la teor\ii a de cuerdas, donde las cuerdas
abiertas pueden tener sus extremos. As\ii mismo, hemos visto que las
fluctuaciones de las cuerdas abiertas sobre la brana tienen una
din\'amica de baja energ\ii a descripta por una teor\ii a de gauge
supersim\'etrica. Desde el punto de vista de las teor\ii as de gauge,
resulta posible mostrar que estos objetos, las D$p$-branas, son
objetos BPS y que por lo tanto preservan la mitad de las supercargas,
que en este caso (D=10) corresponden a 16, la mitad de las 32
posibles. Una propiedad importante de los estados BPS de teor\ii as
supersim\'etricas es que ellos deben estar cargados. En efecto, existe
s\'olo un tipo de carga capaz de acoplarse a una D$p$-brana y resulta
ser una carga antisim\'etrica de
Ramond-Ramond~\cite{Polchinski:1995mt}. Un volumen de mundo
$(p+1)$-dimensional $\mathcal{M}_{p+1}$ de una D$p$-brana se acopla
naturalmente a $(p+1)$-forma potencial $C_{p+1}$ en el sector de R-R
via el siguiente acoplamiento m\ii nimo:
\begin{equation}\label{mincoup}
	 \mu_p \int_{\mathcal{M}_{p+1}} C_{p+1}\,,
\end{equation}
donde $\mu_p$ es el acoplamiento que debe ser determinado en t\'erminos
de cantidades de la teor\ii a de cuerdas. Como vemos, no son las cuerdas
las que act\'uan como fuentes de los campos de $R-R$, sino las D$p$-branas, que tambi\'en son objetos contenidos en la teor{\'\i}a. 
 En este punto es interesante notar que el hecho de que los
 potenciales $C_p$ de $RR$ no existan para cualquier $p$ en las
 teor\ii as tipo II, restringe el espectro de las D$p$-branas en las
 teor\ii as de supercuerdas tipo IIA o tipo IIB (ver
 tabla~\ref{t:branes}).

En la secci\'on que sigue, estudiaremos con m\'as detalle la relaci\'on
existente entre campos viviendo  sobre una D$p$-brana y los campos
de cuerdas cerradas, al derivar la acci\'on de volumen de mundo de las
D-branas.

\begin{table}
\begin{center}
\begin{tabular}{|l|ccccc|ccccc|}
\hline
&\multicolumn{5}{|c|}{Tipo IIA}&\multicolumn{5}{|c|}{Tipo IIB}\\
\hline
D-brana & D0 & D2 & D4 & D6 & D8 & D$(-1)$ & D1 & D3 & D5 & D7 \\
potencial R-R  & $C_1$ & $C_3$ & $C_5$ & $C_7$ & $C_9$ 
& $C_0$ & $C_2$ & $C_4$ & $C_6$ & $C_8$ \\
\hline
\end{tabular}
\end{center}
\caption{El espectro de D-branas en teor{\'\i}as tipo II. Notar que la
  D9-brana no est\'a presente en la teor{\'\i}a est\'andar de tipo
  IIB, dado que implicar{\'\i}a anomal{\'\i}as de carga, pero puede
  existir en sobre orientifolds. La D$(-1)$-brana IIB, acoplada
  con el escalar R-R $C_0$, es un ``D-instanton'', ya que es una
  configuraci\'on localizada en el tiempo (euclideano).}
\label{t:branes}
\end{table}

\section{La acci\'on del volumen de mundo de la D-brana}\label{s:wv}

Hemos visto que los estados no masivos de las cuerdas sobre una
D$p$-brana forman un multiplete vectorial de una teor\ii a de gauge en
$(p+1)$ dimensiones, con 16 supercargas. Resulta natural entonces
preguntarnos como es la din\'amica del volumen de mundo de la
D$p$-brana y como \'esta experimenta  la influencia de
los campos de fondo del espacio-tiempo.  Hay varias maneras de derivar
esta acci\'on. Por razones de simplicidad, aqu{\'\i} derivaremos la
acci\'on en forma heur\ii stica, usando T-dualidad como
en~\cite{Polchinski:1998}.

Comencemos con el acoplamiento de la m\'etrica con el dilat\'on. Una
D$p$-brana tiene un volumen de mundo $(p+1)$-dimensional, y la
acci\'on de baja energ\ii a debe ser la que reproduzca la amplitud de
dispersi\'on de disco que va como $g_s^{-1}$, por lo que esperamos entonces
\begin{equation}\label{wv1}
	S_{\text{D}p} = - \tau_p \int_{\mathcal{M}}
        d^{p+1}\xi\ e^{-\Phi} \sqrt{-\det \hat{G}_{ab}} + \ldots\,,
\end{equation}
donde $\xi^a$, $a=0,\ldots,p$ parametrizan las direcciones del volumen
de mundo, $\tau_p$ es la tensi\'on de la brana que incluye el factor
$g_s^{-1}$, y $\hat{G}_{ab}$ es el pullback de la m\'etrica del
espacio-tiempo en el volumen de mundo de la brana:
\begin{equation}\label{pullback}
	\hat{G}_{ab} = \frac{\partial x^\mu}{\partial \xi^a}
        \frac{\partial x^\nu}{\partial \xi^b} G_{\mu\nu}\,.
\end{equation}

El siguiente paso es considerar el acoplamiento al campo de $B$ con el
campo de gauge del volumen de mundo $A_a$. Para determinarlo, consideremos el ejemplo
  de una D2-brana extendida a lo largo de $x^1$ y $x^2$, donde
  encendemos el tensor intensidad de campo $F_{12}$ en el volumen de
  mundo y elegimos el gauge particular en el que $A_2 = x^1 F_{12}$.
Si realizamos una T-dualidad a lo largo de la direcci\'on $x^2$,
obtendremos una D1-brana con:
\begin{equation}
	x^2 = 2\pi\ls^2 A_2 = 2\pi\ls^2 x^1 F_{12}\,.
\end{equation}
Podemos interpretar este resultado diciendo que la D1-brana est\'a inclinada un \'angulo:
\begin{equation}
	\theta = \tan^{-1} \left(2\pi\ls^2 F_{12}\right)\,,
\end{equation}
con respecto a la direcci\'on  $x^2$. 
De~\eqref{pullback} tenemos que:
\begin{equation}
	\int dx^1 \sqrt{1-(\partial_1 x^2)^2} = \int dx^1 \sqrt{1 +
          (2\pi\ls^2 F_{12})^2}\,.
\end{equation}

En el caso general, $F_{ab}$ se puede llevar a la forma diagonal en
bloques, y el t\'ermino tiene la forma:
\begin{equation}\label{wv2}
	\int d^{p+1}x \sqrt{\det (\eta_{ab} + 2\pi\ls^2 F_{ab})}\,.
\end{equation}

Bien, ahora por un lado debemos covariantizar el t\'ermino~\eqref{wv2}
para poder incorporar el acoplamiento~\eqref{wv1}. Por el otro lado,
debemos ser cuidadosos con la invarianza de gauge espacio-temporal. Es
decir, no cualquier combinaci\'on de $F$ y $B_2$ preserva esta
invarianza. En efecto, una transformaci\'on de la forma $\delta B_2 =
d\zeta$, donde $\zeta$ es una 1-forma arbitraria, da lugar a un
t\'ermino de borde en el modelo sigma que describe el background. Esta
variaci\'on debe ser cancelada por una transformaci\'on de gauge sobre
el campo $A$, que puede ser $\delta A = - \zeta / 2\pi\ls^2$. De esta
manera, la combinaci\'on invariante de gauge considerada ser\'a $B +
2\pi\ls^2 F$. Todo esto sugiere que la forma de la acci\'on debe ser
la siguiente:
\begin{equation}\label{DBI}
	S_{\text{DBI}} = -\tau_p \int_{\mathcal{M}_{p+1}}
        d^{p+1}\xi\ e^{-\Phi} \sqrt{\det (\hat{G}_{ab} + \hat{B}_{ab}
          + 2\pi\ls^2 F_{ab})}\,,
\end{equation}
donde el sombrero $\hat{}$ denota el pullback de los
campos de fondo sobre el volumen de mundo de la brana, como
en~\eqref{pullback}, $\xi$ son coordenadas gen\'ericas del volumen de
mundo, no necesariamente identificadas con las coordenadas del
espacio-tiempo. La acci\'on anteriormente escrita se conoce con el
nombre de acci\'on de \emph{Dirac--Born--Infeld}, y describe la
din\'amica de baja energ\ii a del volumen de mundo de la D$p$-brana
para valores arbitrarios de los campos de fondo en el sector de
$NS-NS$.

Como vimos, el acoplamiento en el sector de $R-R$ est\'a dado de la
forma siguiente~\eqref{mincoup}:
\begin{equation}\label{wv3}
	 \mu_p \int_{\mathcal{M}_{p+1}} \hat{C}_{p+1}\,,
\end{equation}
donde $\mu_p$ es la carga de R-R de la D-brana en las unidades
apropiadas, y donde nuevamente el sombrero denota el pullback de los
campos de fondo $C$ sobre el volumen de mundo. El
t\'ermino~\eqref{wv3} no es el \'unico; las dem\'as contribuciones
pueden otra vez ser obtenidas via T-dualidad. Consideremos ahora una
D1-brana inclinada en el plano $(x^1,x^2)$. El c\'alculo del pull-back
de $C_2$ dar\'a la siguiente expresi\'on para el
acoplamiento~\eqref{wv3}:
\begin{equation}
	\int dx^0 dx^1 \left( (C_2)_{01} +\partial_1 x^2 (C_2)_{02}
        \right)\,,
\end{equation}
que ante una T-dualidad en la direcci\'on $x^2$,
usando~\eqref{Tduality}, deviene en un t\'ermino de la forma:
\begin{equation}
	\int dx^0 dx^1 dx^2 \left( (C_3)_{012} + 2\pi\ls^2 F_{12}
        (C_1)_{0} \right)\,.
\end{equation}
Para branas de mayor dimensi\'on, este procedimiento puede ser
generalizado a potenciales de rango menor. Recordando la forma
invariante que hab\ii amos obtenido para la combinaci\'on de $F$ y $B_2$,
podemos intuir que la forma final del t\'ermino de la de
\emph{Wess--Zumino} ~\cite{Li:1996pq,Douglas:1995bn}:
\begin{equation}\label{WZ}
	S_{\text{WZ}} = \mu_p \int_{\mathcal{M}_{p+1}} \sum_q
        \hat{C}_q \wedge e^{\hat{B}+2\pi\ls^2 F}\,,
\end{equation}
donde la suma selecciona los t\'erminos en la expansi\'on de
$(p+1)$-forma, que aporta con un valor no nulo de su integral sobre la
hoja de mundo.

La acci\'on de volumen de mundo de la D$p$-brana est\'a dada, entonces, por
la suma de Dirac--Born--Infeld ~\eqref{DBI} m\'as la parte de
Wess--Zumino~\eqref{WZ}:
\begin{equation}\label{Dpwv}
\begin{split}
	S_{\text{D}p} = &- \tau_p \int_{\mathcal{M}_{p+1}}
        d^{p+1}\xi\ e^{-\Phi} \sqrt{- \det \left(\hat{G}_{ab} +
          \hat{B}_{ab} + 2\pi\ls^2 F_{ab}\right) }\\ &+ \mu_p
        \int_{\mathcal{M}_{p+1}} \sum_q \hat{C}_q \wedge
        e^{\hat{B}+2\pi\ls^2 F}\,,
\end{split}
\end{equation}
a la que hay que agregarle la parte supersim\'etrica que describe el
acoplamiento de los fermiones, que no discutiremos ac\'a. Por conveniencia, la expresamos en el Einstein frame:
\begin{equation}\label{wvE}
\begin{split}
	S_{\text{D}p}^{\text{(E)}} = &- \tau_p
        \int_{\mathcal{M}_{p+1}} d^{p+1}\xi\ e^{\frac{p-3}{4}\Phi}
        \sqrt{- \det \left[\hat{G}_{ab} +
            e^{-\Phi/2}\left(\hat{B}_{ab} + 2\pi\ls^2
            F_{ab}\right)\right] }\\ &+ \mu_p \int_{\mathcal{M}_{p+1}}
        \sum_q \hat{C}_q \wedge e^{\hat{B}+2\pi\ls^2 F}\,.
\end{split}
\end{equation}

Adem\'as de los t\'erminos aqu{\'\i} descriptos, pueden agregarse
t\'erminos adicionales. Sin embargo, por razones de
simplicidad, no ser\'an considerados aqu{\'\i}. Hasta ahora, todo
lo estudiado ha sido para el caso de una sola D$p$-brana. Cabe
entonces la pregunta de que suceder\ii a si en lugar de una
tuviesemos $N$ D$p$-branas, una encima de la otra. La repuesta a
esta pregunta a\'un no est\'a clara. Los campos del volumen de mundo
$A_a$ y $X^i$ que representan los modos colectivos del movimiento de
las branas se vuelven ahora matrices que toman valores en la
representaci\'on adjunta de U(N). En particular, esto significa que
las coordenadas del espacio transverso deben ser tratadas como
matrices de $N \times N$, y esto trae una serie de consecuencias no
triviales, como por ejemplo, que los pull-backs de los campos de fondo
deben ser realizados usando derivadas covariantes, o que la acci\'on
requiera t\'erminos de la forma $[X^i,X^j]$ y $[A_a,X^i]$. Otro
problema es saber cual es la prescripci\'on precisa que hay que
utilizar para calcular la traza, que es necesaria para obtener
cantidades invariantes de gauge en la acci\'on. Una prescripci\'on
posible es usar la ``traza simetrizada''~\cite{Tseytlin:1997cs}, pero
ha sido mostrado que este procedimiento trae aparejadas algunas
ambig\"uedades.

Sin entrar en detalles de las extensiones no abelianas
de~\eqref{Dpwv}, hay una observaci\'on importante que puede
realizarse, que se deduce sin ambig\"uedades del uso de la traza
simetrizada.  Desarrollando cada extensi\'on no abeliana
de~\eqref{Dpwv} en el background de espacio plano hasta el segundo
orden en los campos de gauge, se ve que el orden dominante reproduce
la acci\'on bos\'onica en $(p+1)$ dimensiones de una teor\ii a de
Super Yang-Mills (SYM)
\begin{equation}
	S_{\text{SYM}} = -\frac{1}{\gym} \int d^{p+1}\xi\ \tr \left[
          \frac{1}{2} F_{ab}F_{ab} + D_a \Phi^i D_a \Phi^i +
          \frac{1}{2}[\Phi^i,\Phi^j]^2 \right]\,.
\end{equation}

Se puede mostrar que los valores para la carga y la tension de la
brana resultan:

\begin{equation}\label{taup}
	\tau_p = \frac{\sqrt{\pi}(2\pi\ls)^{3-p}}{\kappa} =
        \frac{1}{(2\pi)^p \gs \ls^{p+1}}\,.
\end{equation}

\begin{equation}\label{mup}
	\mu_p = \tau_p\,.
\end{equation}
Esto completa la determinaci\'on de la acci\'on de volumen de mundo,
ya que tenemos expresadas todas las cantidades en t\'erminos de las
cantidades de la teor\ii a de cuerdas $\gs$ y $\ls$.

\section{La geometr{\'\i}a de las D-branas}\label{s:classol}

\begin{figure}
\begin{center}
\includegraphics[scale=.7]{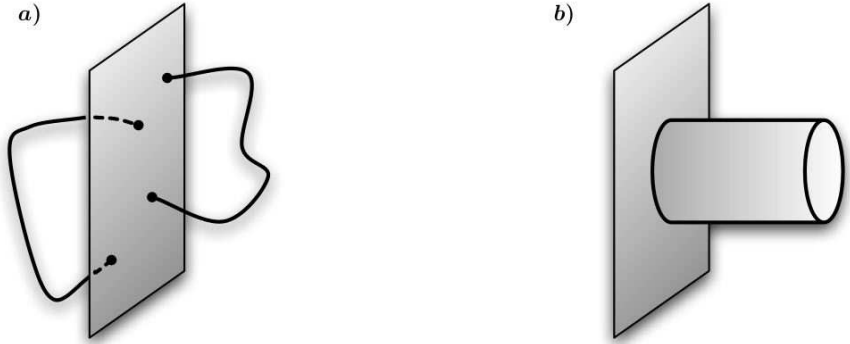}
\caption[Doble interpretaci\'on de las D-branas.]{Doble
  interpretaci\'on de las D-branas. a) Las D-branas son hiperplanos
  donde terminan las cuerdas abiertas; b) Las D-branas son bordes de
  la teor{\'\i}a superconforme de cuerdas cerradas, cargadas bajo
  campos de Ramond-Ramond.}
\label{f:openclosed}
\end{center}
\end{figure}

En la secci\'on anterior vimos que, a nivel perturbativo en la teor\ii
a de cuerdas, aparecen nuevos objetos llamados D$p$-branas que son
hipersuperficies donde las cuerdas pueden tener sus extremos. Sin
embargo, tambi\'en hemos visto que estas D$p$-branas son objetos
fundamentales de la teor\ii a cargados frente a campos potenciales de
$R-R$. La energ\ii a de vac\ii o a un loop de dos branas puede ser
interpretada, tambi\'en, como el intercambio a nivel \'arbol de
cuerdas cerradas entre las branas. Esto es la base de lo que se conoce
como \emph{dualidad cuerda abierta cerrada}, que implica que las
branas tambi\'en pueden ser descriptas como estados de borde de una
teor\ii a conforme de cuerdas cerradas. Este enfoque interpreta las
D-branas como estado de frontera de una teor\ii a conforme, que
resulta ser muy \'util para analizar algunos aspectos de las D-branas.
Una de las consecuencias m\'as importantes de la descripci\'on de
cuerda cerradas es que estas D-branas pueden entenderse como solitones
de las teor\ii as de supergravedad, que son la baja energ\ii a de las
teor{\'\i}as de cuerdas. Uno de las contribuciones m\'as importantes
de~\cite{Polchinski:1995mt} es precisamente la identificaci\'on de
estos solitones con las D$p$-branas encontradas como consecuencia de
la T-dualidad. Esta doble interpretaci\'on, que ser\'a fundamental en
los temas que trataremos en adelante, est\'a graficada en la Figura
~\ref{f:openclosed}.

Buscamos ahora las soluciones de las acciones de SUGRA~\eqref{IIBE} y
\eqref{IIAE}. S\'olo deduciremos las soluciones correspondientes a los
casos llamados {\it extremos} (se denominan extremos porque son
soluciones BPS, es decir que saturan la desigualdad de Bogomonly,
Prasad y Sommerfield) Nuestro punto de partida ser\'a entonces la
acci\'on consistentemente truncada de una de las SUGRA tipo II en el
Einstein frame, donde s\'olo dejamos un campo formas que llamaremos
$F_{p+2}$ (que corresponder\'a a la derivada exterior de uno de los
campos potenciales proveniente del sector $NS-NS$ o del sector $R-R$),
la m\'etrica y el dilat\'on:
\begin{equation}\label{II}
	S_{\text{II}} = \frac{1}{2\kappa^2} \Bigg\{ \int d^{10}x
        \sqrt{-\det G}\ R - \frac{1}{2} \int \Big[ d\Phi \wedge \hd{}
          d\Phi + e^{-a\Phi} F_{p+2} \wedge \hd{}F_{p+2} \Big]
        \Bigg\}\,,
\end{equation}
donde $a$ es una constante apropiada que se elige de forma de ajustar
el rango de la forma as{\'\i} como el sector. Dado que estamos buscando
soluciones extendidas en $p$ direcciones, que llamaremos
$x^0,\ldots,x^p$, resulta conveniente dividir las coordenadas en dos
grupos y asumir algunas condiciones generales que nos
permitir\'an fabricar un ans\"atz para la soluci\'on, que simplifique las
ecuaciones:
\begin{itemize}
\item $x^\alpha$, $\alpha=0,\ldots,p$, son las coordenadas
  longitudinales al volumen de mundo de la brana. Asumiendo que la
  brana no posee ning\'un punto preferencial, resulta natural requerir
  que la soluci\'on tenga invarianza Poincar\'e en estas $p+1$
  direcciones;
\item $x^i$, $i=p+1,\ldots,9$, son las coordenadas del espacio
  transverso al volumen de mundo. Dado que en el espacio transverso la
  brana es visualizada como un punto, parece natural requerir sobre
  estas coordenadas invarianza rotacional $SO(9-p)$ en las
  $(9-p)$-direcciones espaciales.
\end{itemize}
Un ans\"atz compatible con estos requerimientos puede ser escrito de la forma:
\begin{equation}\label{braneans}
	ds^2 = e^{2A(r)}\ \eta_{\alpha\beta}dx^\alpha dx^\beta +
        e^{2B(r)}\ \delta_{ij}dx^i dx^j\,,\qquad \Phi = \Phi(r)\,,
\end{equation}
donde $r=(x^ix^i)^{1/2}$ es la coordenada radial en el espacio
transverso. Para la $(p+2)$-forma hemos optado por un ans\"atz
el\'ectrico, produciendo un acoplamiento como~\eqref{mincoup}, o
magn\'etico para su dual Hodge $\hd{}F_{p+2} = (\hd{}F)_{8-p}$. Eso se
corresponde con buscar soluciones de $p$-brana el\'ectrica o para una
$(6-p)$-brana magn\'etica, que es el objeto dual magn\'etico de la
primera.  La forma del ans\"atz el\'ectrico es:
\begin{equation}
	C_{p+1} = (e^{C(r)}-1)\ dx^0 \wedge \dotsm \wedge dx^p\,.
\end{equation}

Sustituimos luego el ans\"atz en las ecuaciones de movimiento.


La soluci\'on de las ecuaciones de movimiento, cuando $F_{p+2}$ es
interpretado como un campo de fuerza $R-R$, lleva a la siguiente
configuraci\'on de campos para la supergravedad de tipo II:
\begin{equation}\label{DpsolE}
\begin{aligned}
	ds^2 &= H_p(r)^{-\frac{7-p}{8}}\ \eta_{\alpha\beta} dx^\alpha
        dx^\beta + H_p(r)^{\frac{p+1}{8}}\ \delta_{ij} dx^i
        dx^j\,,\\ e^{\Phi} &= H_p(r)^{\frac{3-p}{4}}\,,\\ C_{p+1} &=
        (H_p(r)^{-1} - 1)\ dx^0 \wedge \dotsm \wedge dx^p\,,
\end{aligned}
\end{equation}
donde  $H_p(r)$ es una funci\'on arm\'onica en el espacio transverso dada por:
\begin{equation}\label{H}
	H_p(r) = 1 + \frac{Q_p}{r^{7-p}}\,,\qquad Q_p =
        \frac{2\kappa^2\tau_p}{(7-p)\Omega_{8-p}}\,,
\end{equation}
donde $\Omega_{q} = (2\pi)^{(q+1)/2} / \Gamma ((q+1)/2)$ es el volumen
de la $q$-esfera unidad y las expresiones anteriores son v\'alidas
siempre que $p < 7$.  Notar que ``factorizando la constante de
Newton'' en $Q_p$ uno puede identificar r\'apidamente la tensi\'on de
la $p$-brana, que es igual a $\tau_p$, es decir proporcional a
$\gs^{-1}$. Este resultado concuerda perfectamente con el $\tau_p$
coeficiente de la acci\'on de \emph{Dirac--Born--Infeld}.  

Finalmente, la carga de $R-R$ de la brana puede ser obtenida calculando:
\begin{equation}
	\mathcal{Q}_p = \frac{1}{\sqrt{2}\kappa} \int_{S^{8-p}} \hd{}
        dC_{p+1} = \sqrt{2}\kappa\tau_p\,.
\end{equation}

El hecho de que la carga sea igual a la tensi\' on (en la unidades
apropiadas) es una manifestaci\'on de la propiedad BPS de la
soluci\'on de D$p$-brana obtenida.  Se ve tambi\'en que
$\mathcal{Q}_p\mathcal{Q}_{6-p} = 2\pi$, da la relaci\'on correcta de
cuantizaci\'on de Dirac para objetos electromagn\'eticos duales como una
D$p$ y una D$(6-p)$-branas.

Notar que la soluci\'on de la m\'etrica~\eqref{DpsolE} puede ser
escrita de forma simple en el string frame como:
\begin{equation}\label{Dpsolst}
	ds^2 = H_p(r)^{-1/2}\ \eta_{\alpha\beta} dx^\alpha dx^\beta +
        H_p(r)^{1/2}\ \delta_{ij} dx^i dx^j\,.
\end{equation}

Algunas observaciones finales de estas soluciones de D-branas. En
primer lugar, las soluciones reci\'en obtenidas~\eqref{DpsolE} pueden
ser generalizadas al caso de $N$ $p$-branas BPS coincidentes,
simplemente cambiando $Q_p\to NQ_p$ en~\eqref{H}. En segundo lugar, es
importante notar que todas estas soluciones (para $p\neq3$) presentan
un horizonte en $r=0$, que es efectivamente un punto singular de
\'area nula. En cambio, en el caso de $N$ D3-branas, se puede ver que
la potencia inversa $r^{-4}$ que aparece en~\eqref{H} lleva a una
cancelaci\'on entre la divergencia de la m\'etrica y el tama\~no cero
del horizonte, dejando un horizonte finito en $r_{H} = \ls (4\pi\gs
N)^{1/4}$.

 Como comentario final, hacemos notar que la
 expresi\'on~\eqref{DpsolE} no es la adecuada para el caso de las
 D3-branas, ya que no se ha tenido en cuenta la autodualidad
 $\tilde{F}_5$.  Debemos, entonces, reescribir la soluci\' on en la
 forma siguiente:
\begin{equation}\label{D3b}
\begin{aligned}
	ds^2 &= H_3(r)^{-1/2}\ \eta_{\alpha\beta} dx^\alpha dx^\beta +
        H_3(r)^{1/2}\ \delta_{ij} dx^i dx^j\,,\\ e^{\Phi} &=
        1\,,\\ F_5 &= d H_3(r)^{-1} \wedge dx^0 \wedge \dotsm \wedge
        dx^3 + \hd{} ( d H_3(r)^{-1} \wedge dx^0 \wedge \dotsm \wedge
        dx^3 )\,.
\end{aligned}
\end{equation}


%% file: general_ads_cft.tex

\chapter{Ads/CFT la conjetura de Maldacena y su Extensi\'on}\label{c: ads_cft}
\section{La correspondencia}
\label{correspondence}

\begin{figure}[htb]
\begin{center}
\epsfxsize=3.5in\leavevmode\epsfbox{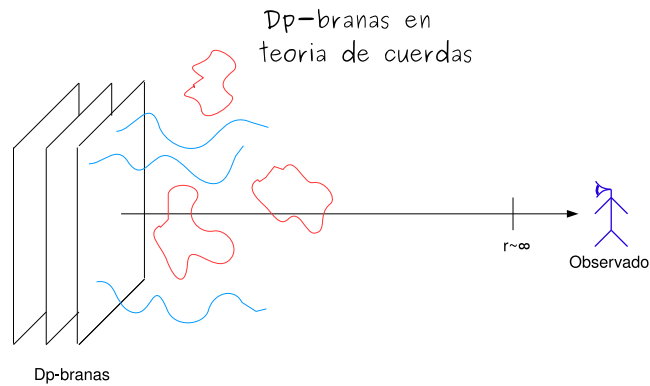}
\end{center}
\caption{$D_p$-branas en teor{\'\i}a de cuerdas.}
\label{branita1}
\end{figure}

En esta secci\'on, repasaremos el argumento que relaciona teor\ii as
tipo IIB compactificadas en $AdS_5\times S^5$ con teor\ii as de
super-Yang-Mills ${\cal N} =4 $. Esta relaci\'on es la base de la
conjetura de Maldacena \cite{Maldacena:1997re} respecto a la dualidad
AdS/CFT.  Comencemos con teor\ii a de cuerdas IIB en un fondo
10-dimensional plano (Minkowski). Consideremos $N$ D3 branas paralelas
puestas juntas. Las D3 branas se extienden a lo largo de un plano
$(3+1)$-dimensional en el espacio-tiempo $(9+1)$-dimensional. Las
excitaciones de teor\ii a de cuerdas sobre este fondo
est\'an dadas tanto por cuerdas abiertas como por cuerdas cerradas (Figura \ref{branita1}).
Las excitaciones de cuerdas cerradas son excitaciones del espacio vac{\'\i}o,
mientras que las excitaciones de cuerdas abiertas (que terminan en las
D-branas) corresponden a excitaciones de las D-branas.

Si analizamos el sistema a bajas energ\ii as, energ\ii as menores que
la escala $1/l_s$, s\'olo los estados no masivos de cuerdas pueden ser
excitados. Podemos entonces escribir una acci\'on efectiva para los
estados sin masa. Los modos de cuerdas cerradas sin masa dan lugar a
un supermultiplete de gravedad en 10 dimensiones, cuya descripci\'on
efectiva de baja energ\ii a est\'a dada por la acci\'on de
supergravedad tipo IIB. Como vimos en el cap\ii tulo anterior, los
modos de cuerdas abiertas sin masa dan lugar a un supermultiplete
vectorial ${\cal N}=4$ en 3+1 dimensiones, cuya descripci\'on efectiva
de baja energ\ii a est\'a dada por el lagrangiano de una teor\ii a
${\cal N}=4$ U(N) Super Yang-Mills (SYM)
\cite{Witten:1996im,Polchinski:1998}.
 
En consecuencia, la acci\'on efectiva total de los modos no masivos
tiene la forma
\begin{equation}
S=S_{bulk}+S_{brana}+S_{int},\label{lowenergy}
\end{equation}
donde $S_{bulk}$ es la acci\'on de supergravedad en 10 dimensiones
m\'as correcciones de orden superior. $S_{brana}$ est\'a definida
sobre el volumen de mundo $p+1$-dimensional, y contiene el lagrangiano
de ${\cal N}=4$ U(N) SYM, m\'as correcciones de orden
superior. Finalmente, $S_{int}$ describe la interacci\'on entre los
modos de la brana y los modos del bulk. Podemos desarrollar la
acci\'on de bulk como la parte cuadr\'atica libre, que describe la
propagaci\'on de los modos libres no masivos (inclu{\'\i}do el
gravit\'on), m\'as interacciones que ser\'an proporcionales a una
potencia positiva de la ra\ii z cuadrada de la constante de Newton.
Esquem\'aticamente, tenemos
\begin{equation}
  S_{buk}\sim \frac{1}{2 \kappa^2} \int \sqrt{g}R \sim \int (\partial h)^2+\;\; \kappa (\partial h)^2 h + \cdot\label{expans}
\end{equation}
donde hemos escrito la m\'etrica como $g = \eta + \kappa h $. Aqu{\ii}
se ha indicado s\'olo la dependencia expl{\'\i}cita en el gravit\'on, pero
los otros t\'erminos del lagrangiano que tienen dependencia en otros
campos pueden ser desarrollados de forma similar. De la
misma manera, la acci\'on de interacci\'on se desarrolla en potencias
de $\kappa$.  El l\'{\i}mite de baja energ\ii a que queremos es $E <<
1/l_s$ y \'este se puede realizar a energ\ii a fija si $\alpha'
\rightarrow 0$, manteniendo todos los par\'ametros dimensionales
fijos. Como consecuencia, dado que $\kappa \sim g_s \alpha'^2$ a $g_s$
constante, se da un desacople de las interacciones y de las
correcciones en derivadas superiores  tanto en la acci\'on bulk como
en la acci\'on de brana.

\begin{figure}[htb]
\begin{center}
\epsfxsize=3.5in\leavevmode\epsfbox{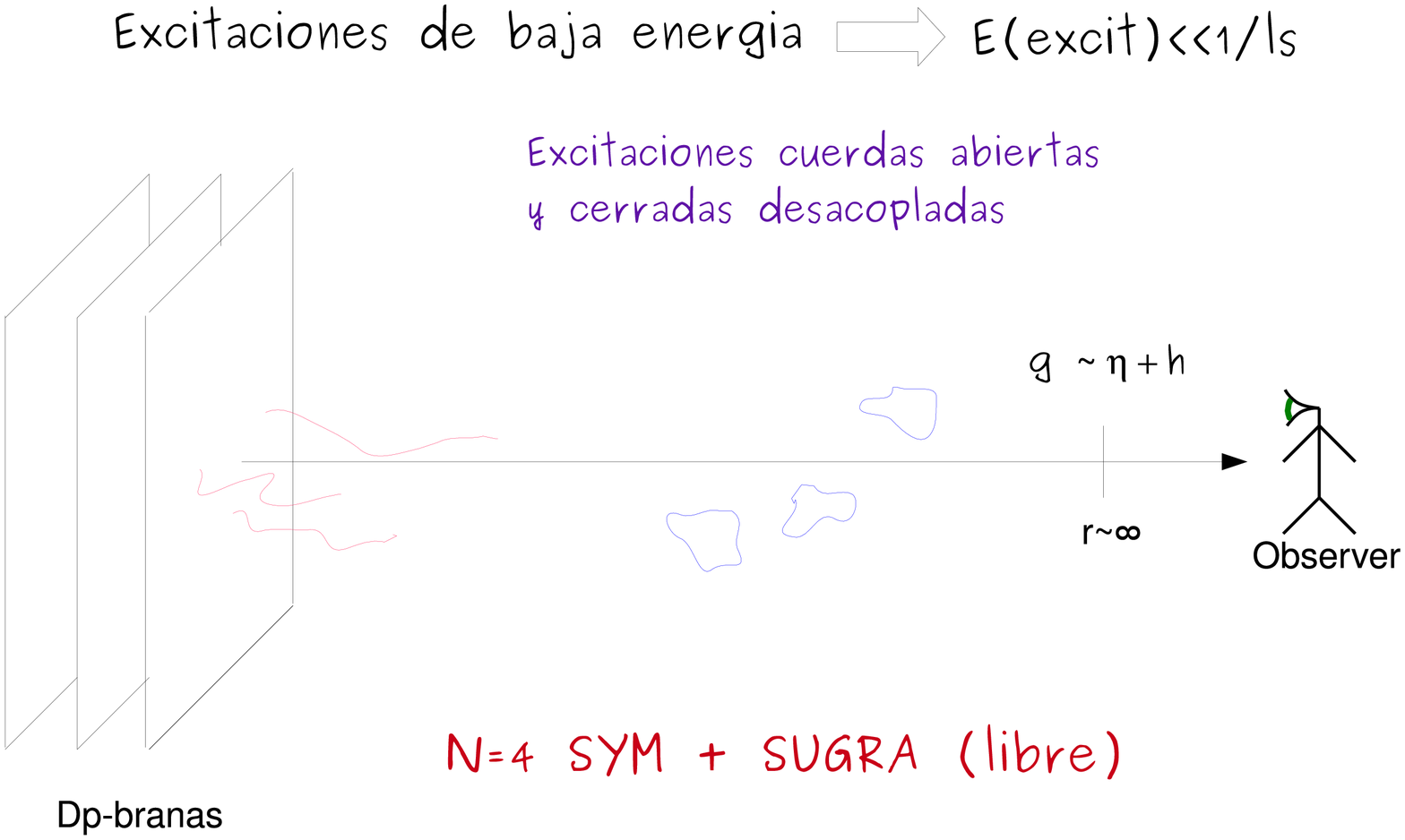}
\end{center}
\caption{Desacople de los modos de gravedad libre y SYM.}
\label{branita2}
\end{figure}

En consecuencia, en este l\ii mite nos quedan dos sistemas
desacoplados; por un lado, una teor\ii a pura ${\cal N}=4$ U(N) SYM en
3+1 dimensiones, y por el otro, gravedad libre sobre un fondo plano (figura \ref{branita2}).

\vspace{1.2cm}

\pagebreak

\begin{figure}[htb]
\begin{center}
\epsfxsize=3.5in\leavevmode\epsfbox{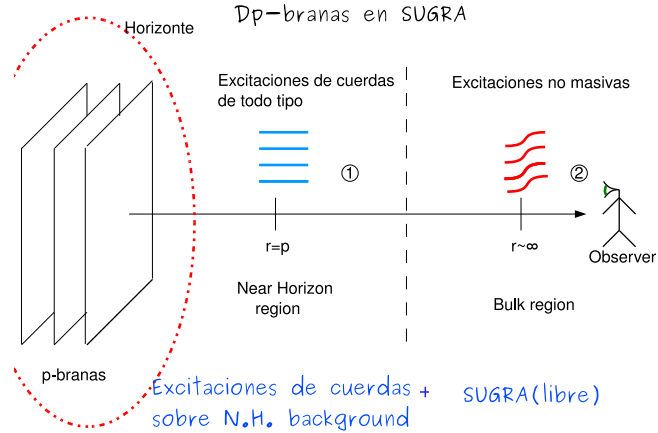}
\end{center}
\caption{Desacople de la regi\'on cercana al horizonte (1) de la regi\'on
  asint\'otica (2).}
\label{branita3}
\end{figure}

Consideremos ahora el mismo sistema pero desde otro punto de vista.
Las D-branas son objetos masivos y cargados que son fuentes de varios
campos de supergravedad. La soluci\'on de D3 brana (\ref{D3b}) en
supergravedad, tiene la forma

\bea  ds^2 &=& f^{-1/2} ( -dt^2 + dx_1^2 + dx_2^2 +
    dx_3^2 ) + f^{1/2} (dr^2 + r^2 d\Omega_5^2 )~, \\
F_5 &=& (1 + * ) dt dx_1 dx_2 dx_3 df^{-1} ~,\\
  f &=& 1 + {\frac{ R^4}{ r^4}}\\
R^4 &\equiv& 4 \pi g_s \alpha'^2 N ~.\label{dthree}\eea Dado que
$g_{tt}$ no es constante, la energ\ii a $E_p$ de un objeto medida por
un observador a una distancia $r_p$ y la energ\ii a $E$
medida por un observador en infinito est\'an relacionadas por el
factor de corrimiento al rojo ({\it redshift}) \be \label{redshift} E =
  f^{-1/4} E_p ~.\ee Esto significa que el mismo objeto acercado m\'as
y m\'as a $r =0$ aparecer\'a como teniendo menor y menor energ\ii a
para un observador que lo mira desde infinito.  Ahora tomamos el l\ii
mite de baja energ\ii a en el fondo descripto por la ecuaci\'on
(\ref{dthree}). Existen aqu{\ii} dos tipos de excitaciones de baja energ\ii a
(desde el punto de vista de un observador en infinito).  Por un lado
tendremos part\ii culas no masivas propag\'andose en la regi\'on del
bulk con longitudes de onda muy grandes, y por otro tendremos
cualquier tipo de excitaci\'on (no necesariamente no masiva) en la
regi\'on cercana a $r=0$. En el l\ii mite de baja energ\ii a, estos
dos tipos de excitaciones se desacoplan unos de otros. Las part\ii
culas no masivas en el bulk se desacoplan de las excitaciones de todo
tipo en la regi\'on cercana al horizonte ({\it the near horizon
  region}, ~ cerca de $r=0$) ya que la secci\'on eficaz de absorci\'on de baja
energ\ii a va como $ \sigma \sim \omega^3 R^8 $
\cite{Klebanov:1997kc,Gubser:1997yh}, donde $\omega $ es la energ\ii
a.  Se puede pensar que en este l\ii mite las longitudes
de onda de las part\ii culas son mucho m\'as grandes que el tama\~no
gravitatorio t\ii pico de una brana (que es del orden de $R$).
Similarmente, a las excitaciones que viven muy cerca de $r = 0$ les
resulta muy dif{\'\i}cil superar el potencial gravitatorio y escapar hacia
la regi\'on asint\'otica. En conclusi\'on, la baja energ\ii a consiste
en dos partes desacopladas, una es la supergravedad bulk libre y la
otra es la regi\'on {\it near horizon} (figura \ref{branita3}).  En esta regi\'on, $r \ll R $,
podemos aproximar $f \sim R^4/r^4$, y la geometr{\'\i}a resulta
\be \label{nearhor} ds^2 = { \frac{r^2}{ R^2} } ( -dt^2 + dx_1^2 + dx_2^2 +
  dx_3^2 ) + R^2 {\frac{ dr^2}{r^2} } + R^2 d\Omega_5^2, \ee que es la
geometr{\'\i}a de $AdS_5 \times S^5$.

\vspace{1.2cm}

\begin{figure}[!]
\begin{center}
\epsfxsize=3.5in\leavevmode\epsfbox{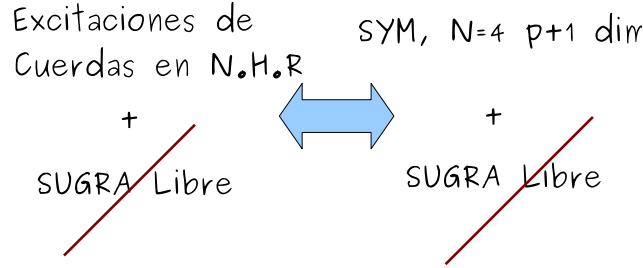}
\end{center}
\caption{Dos descripciones equivalentes del mismo sistema.}
\label{branita4}
\end{figure}

Vemos que tanto desde el punto de vista de la teor\ii a de campos de
cuerdas abiertas viviendo en la brana, como desde el punto de vista de la
descripci\'on de supergravedad, tenemos dos teor\ii as desacopladas en
el l\ii mite de baja energ\ii a. En ambos casos, una de las teor\ii as
desacoplada es supergravedad sobre un fondo plano. As{\ii} que
resulta natural identificar entre s{\'\i} los otros dos sistemas que
aparecen en ambas descripciones. De esta manera, se conjetura que {\it
  ${\cal N} =4 $ $U(N)$ super-Yang-Mills en $3+1$ dimensiones es dual
  a teor\ii a de supercuerdas tipo IIB sobre el fondo $AdS_5\times
  S^5$} \cite{Maldacena:1997re} (figura \ref{branita4}).

\pagebreak
Seamos un poco m\'as precisos respecto del l\ii mite de horizonte
cercano y de c\'omo tomarlo. Tomemos el l\ii mite de baja energ\ii a
$\alpha' \to 0$.  Queremos mantener fija la energ\ii a de los objetos
en la garganta (regi\'on horizonte cercano) en unidades de cuerdas, de
manera de poder considerar estados de excitaciones arbitrarias
all{\ii}. Esto implica que $\sqrt{\alpha'} E_p \sim {\rm fijo}$.
Para $\alpha'$ peque\~no, (\ref{redshift}) se reduce a $E \sim E_p
r/\sqrt{\alpha'} $. Ya que nosotros queremos mantener fija la energ\ii
a medida desde infinito, que es la forma en la que se mide la energ\ii
a en teor\ii as de campos, necesitamos tomar $r \to 0$ manteniendo $r /
\alpha' $ fijo. Es entonces conveniente definir la nueva variable $U
\equiv r / \alpha'$, de tal forma que la m\'etrica resulta
\be \label{metricu} ds^2 = \alpha' \left[ \frac{U^2} {\sqrt{4 \pi g_s
      N} } ( - dt^2 + dx_1^2 + dx_2^2 + dx_3^2 ) + \sqrt{4 \pi g_s N}
  { \frac{d U^2} {U^2}} + \sqrt{4 \pi g_s N} d \Omega_5^2 \right].  \ee

Esto tambi\'en puede ser visto de la siguiente manera. Supongamos que
apartamos alguna D3 brana del origen y la ponemos en $\vec r$.
Esto corresponde a dar un valor de expectaci\'on de vac\ii o a uno de
los escalares de la teor\ii a de Yang-Mills. Ahora tomamos el l\ii
mite de baja energ\ii a $\alpha' \to 0$ manteniendo la masa del
``bos\'on $W$'' fija. Esta masa, que es la masa de la cuerda extendida
entre dos branas, una puesta en $\vec r =0$ y la otra en $\vec r$, es
proporcional a $ U = r/\alpha'$, as{\ii} que \'esta es la cantidad que
debe permanecer fija en el l\ii mite de desacople.
\\~\\

Hasta aqu{\ii} la conjetura es puramente una construcci\'on l\'ogica,
sin embargo, algunas verificaciones son inmediatas. El espacio anti-de-Sitter
tiene un gran grupo de isometr\ii as; el $SO(4,2)$ para el caso que
nos interesa. \'Este es el mismo grupo que el grupo conforme en $3+1$
dimensiones.  Luego, el hecho de que la teor\ii a a bajas energ\ii as sobre
la brana sea conforme se refleja en que la geometr\ii a de la regi\'on
cercana al horizonte sea $AdS$. Tambi\'en tenemos algunas supersimetr\ii
as. El n\'umero de supersimetr\ii as es el doble de los que tiene la
soluci\'on total (\ref{dthree}) \cite{Gibbons:1993sv}.  Este aumento
al doble de las supersimetr\ii as se ve en el lado de la teor\ii a de
campos como consecuencia de la invarianza superconforme, ya que el
\'algebra superconforme tiene el doble de fermiones que el
correspondiente super\'algebra de Poincar\`e.  Por otro lado, tambi\'en
tenemos simetr\ii a $SO(6)$ por la $S^5$. \'Esta est\'a identificada con el
grupo $SU(4)_R$ de la simetr\ii a R de la teor\ii a de campos. En
efecto, todo el subgrupo es el mismo para la teor\ii a de campos
$\cn=4$ y la geometr\ii a $AdS_5 \times S^5 $. De esta manera, ambos
lados de la conjetura tienen las mismas simetr{\'\i}as espacio-temporales.

Estudiemos ahora la validez de nuestras aproximaciones. El estudio de
los diagramas de loops en la teor\ii a de campos muestra que el
an\'alisis perturbativo de la teor\ii a Yang-Mills es v\'alido cuando
\be \label{pert} g^2_{YM} N \sim g_s N \sim \frac{ R^4} {l_s^4 } \ll 1. \ee 
N\'otese que no s\'olo es necesario que $g_{YM}^2$ sea peque\~no
sino que tambi\'en debe serlo $g_{YM}^2 N$. Por otro lado, la
descripci\'on cl\'asica de gravedad resulta confiable cuando el radio
de curvatura $R$ de $AdS$ y de $S^5$ se vuelve grande comparado con la
longitud de la cuerda \be \label{gravity} \frac{ R^4} { l_s^4 } \sim
g_s N \sim g^2_{YM} N \gg 1. \ee Vemos que el r\'egimen de
gravedad (\ref{gravity})\ y la teor\ii a de campos
perturbativa (\ref{pert})\ son totalmente incompatibles. Es por
esta raz\'on que esta correspondencia es llamada ``dualidad''. Se
conjetura que las dos teor\ii as son exactamente las mismas, pero
cuando una est\'a acoplada d\'ebilmente, la otra lo est\'a fuertemente y
viceversa. Obviamente, esto hace que la correspondencia sea dif{\'\i}cil de
probar pero simult\'aneamente, la vuelve muy \'util.  N\'otese que en
(\ref{pert}) (\ref{gravity})\ hemos supesto impl{\'\i}citamente que $g_s <1$. Si $g_s >
1$ podemos hacer una transformaci\'on de dualidad $SL(2,\IZ)$ y
obtener condiciones similares a (\ref{pert})( \ref{gravity})\ pero con $g_s \to
1/g_s$. Es as{\ii} que no podemos ir al r\'egimen de gravedad
(\ref{gravity})\ tomando $N$ peque\~no ($N=1,2,..$) y $g_s$ muy grande, ya
que en este caso las D-string se vuelven muy livianas y hacen que la
aproximaci\'on de supergravedad no sea v\'alida. Otra forma de ver
esto es notando que el radio de curvatura en unidades de Planck es
$R^4/l_p^4 \sim N$.  As{\ii} que es siempre necesario, pero no
suficiente, tomar $N$ grande para tener una descripci\'on de
supergravedad acoplada d\'ebilmente.

Existen varias formas de establecer la conjetura dependiendo de la
regi\'on en la que situemos los par\'ametros $g_s$ y $N$. Sin embargo,
en lo que sigue, trabajaremos con su forma m\'as fuerte, la cual establece
que ambas teor{\'\i}as son exactamente las mismas para todos los valores de
$g_s$ y $N$.  En esta conjetura, el espacio-tiempo s\'olo necesita ser
asint\'oticamente $AdS_5\times S^5$ a medida que nos acercamos al
borde. En el interior, podemos tener todo tipo de procesos; gravitones,
estados de cuerda fundamental muy excitados, D-branas, agujeros
negros, etc. A\'un la topolog\ii a en el interior del espacio-tiempo
puede cambiar. La teor\ii a de Yang-Mills, se supone, es una suma
efectiva sobre todos los espacio-tiempos asint\'oticamente $AdS_5\times
S^5$.

\section{Mapeo entre campos y operadores}
Como acabamos de ver, la correspondencia AdS/CFT establece que la
f\ii sica descripta por una teor\ii a de gauge ${\cal N}=4$ $SYM$ es la
misma que la f\ii sica descripta por teor\ii as de cuerdas sobre un fondo
$AdS$. Sin embargo, no hemos dado a\'un un mapeo preciso entre estas
dos teor\ii as.

Consideremos el siguiente ejemplo para guiarnos en el camino hacia un
mapeo preciso entre estados de la teor\ii a de cuerdas y operadores en
la teor\ii a de gauge conforme dual. Imaginemos que en la teor\ii a
${\cal N} =4 $ super-Yang-Mills tomamos una deformaci\'on marginal por
un operador ${\cal O}$ cambiando el valor de la constante de
acoplamiento. Dado que los acoplamientos de ambas teor\ii as est\'an
relacionados, un cambio en la constante de acoplamiento de la teor\ii
a de gauge implica un cambio en la constante de acoplamiento de la
teor\ii a de cuerdas, es decir, un cambio en el valor de expectaci\'on del
dilat\'on. Ahora bien, el valor de expectaci\'on del dilat\'on est\'a
dado por la condici\'on de contorno que \'este satisface en
infinito. As{\ii} que un cambio en la constante de acoplamiento de la
teor\ii a de gauge nos lleva directamente a un cambio en la condici\'on
de contorno del dilat\'on.  M\'as precisamente, supongamos que
agregamos a la teor\ii a de gauge el siguiente t\'ermino: $\int d^4 x
\phi_0(\vec x) {\cal O}(\vec x)$. De acuerdo a la discusi\'on
previa, es natural suponer que esto cambiar\'a la condici\'on de contorno
del dilat\'on en el borde de $AdS$ (en las coordenadas en las que el borde de
$AdS$ est\'a en $z=0$),  $ \phi(\vec x, z )|_{z = 0} =
\phi_0(\vec x)$.  Es decir, como se sugiere en \cite{Gubser:1998bc,Witten:1998qj}, resulta natural proponer que
\be \label{genera} \langle e^{\int d^4 x \phi_0(\vec x) {\cal O}(\vec x) }
  \rangle_{CFT} = {\cal Z}_{string} \Bigg[ \phi(\vec x, z)\Big|_{z = 0
    } = \phi_0(\vec x) \Bigg], \ee donde el lado izquierdo es la
funci\'on generatriz de las funciones de correlaci\'on de la
teor\ii a de campos. Es decir, de ese lado $\phi_0$ es una funci\'on
arbitraria y podemos calcular las funciones de correlaci\'on de ${\cal
  O}$ tomando las derivadas funcionales necesarias respecto de
$\phi_0$ y poniendo luego $\phi_0 =0$. El lado derecho de la igualdad
es la funci\'on de partici\'on completa de teor\ii as de cuerdas con
la particularidad de que el campo $\phi$ tiene como condici\'on de
contorno el valor $\phi_0$ en el borde de $AdS$. 

La f\'ormula (\ref{genera})\ es v\'alida en general para cualquier campo
$\phi$.  Por lo tanto, cada campo que se propaga en $AdS$ est\'a en
correspondencia uno a uno con un operador de la teor\ii a de campos.

A\'un m\'as, podemos establecer una relaci\'on precisa entre la masa
del campo escalar libre $\phi$ y la dimensi\'on $\Delta$ del operador
de la teor\ii a conforme. No es dif{\'\i}cil ver que si $\Delta$ es la
dimensi\'on de ${\cal O}$, \'esta se relaciona con la masa $m$ de
$\phi$ como \be\label{dimenmass} \Delta = \frac{d} {2} + \sqrt{
  \frac{d^2} {4} + R^2 m^2}.\ee

\begin{figure}[htb]
\begin{center}
\epsfxsize=3.5in\leavevmode\epsfbox{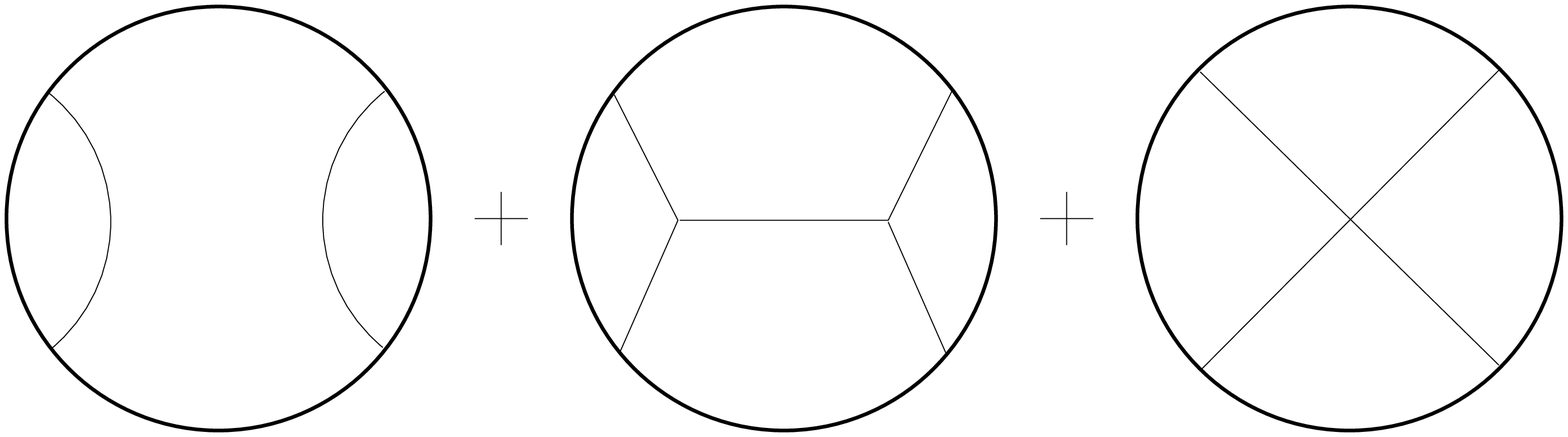}
\end{center}
\caption[Funciones de correlaci\'on y diagramas de Feynman de
  supergravedad.]{ Las funciones de correlaci\'on pueden ser calculadas (en el
  l\ii mite $g_s N$ grande) en t\'erminos de diagramas de Feynman de
  supergravedad. Aqu{\ii} vemos la contribuci\'on principal proveniente
  del diagrama desconectado m\'as los conexos que involucran campos
  de supergravedad en el interior del espacio $AdS$.  Al nivel \'arbol,
  \'estos son los \'unicos diagramas que contribuyen a la funci\'on de
  4 puntos \cite{Aharony:1999ti}. }
\label{dia}
\end{figure}

Como ya dijimos, las funciones de correlaci\'on en la teor\ii a de
gauge pueden calcularse a partir de (\ref{genera})\ derivando respecto
de $\phi_0$.  Cada derivada corresponde a una inserci\'on de un operador
${\cal O}$, que ``env\ii a'' una part\ii cula $\phi$ (una cuerda
cerrada) hacia el interior del espacio.  De esta manera, los
diagramas de Feynman de la teor\ii a de campos pueden ser computados
(a acoplamiento fuerte) calculando diagramas en el interior del
espacio $AdS$.  En el l\ii mite en que supergravedad es aplicable, los
\'unicos diagramas que contribuyen son los que aparecen a nivel \'arbol \cite{Aharony:1999ti}(figura~\ref{dia}).

En el l\ii mite que nos interesa trabajar, podemos aproximar
la funci\'on de partici\'on de teor\ii a de cuerdas por
$e^{-I_{SUGRA}}$, donde $I_{SUGRA}$ es la acci\'on de supergravedad
evaluada en $AdS_5 \times S^5$ (o como ya veremos, evaluada sobre
alg\'un fondo de gravedad conveniente).  Esta aproximaci\'on
corresponde a ignorar todas las correcciones de cuerdas en $\alpha'$
as{\ii} como tambi\'en todas las correcciones en loops que son
controladas por el acoplamiento gravitatorio $\kappa \sim g_{st}
\alpha'^2$.  Del lado de la teor\ii a de gauge, esta aproximaci\'on
corresponde a tomar los l\ii mites $N$ y $g_{YM}^2 N$ grandes, y
entonces la relaci\'on entre las funciones generatrices resulta
\be \label{StringGauge} e^{-I_{SUGRA}} \simeq Z_{\rm string} = Z_{\rm
  gauge} = e^{-W} \ ,\ee donde $W$ es la funci\'on generatriz de
funciones de Green conexas en la teor\ii a de gauge.

El procedimiento consiste ahora en resolver las ecuaciones de movimiento en el
interior del espacio $AdS$, sujetas a condiciones de contorno de
Dirichlet en el borde del espacio, y evaluar la acci\'on en la
soluci\'on.  En este sentido, la aproximaci\'on antes mencionada puede
pensarse como un {\it saddle point} en la funci\'on  de partici\'on de cuerdas.
De esta manera, podemos escribir

\def\extremum{\mathop{\rm extremum}} \be \label{WvsS} W_{\rm
  gauge}[\phi_0] = -\log \left\langle e^{\int d^4 x \, \phi_0(x) {\cal
    O}(x)} \right\rangle_{CFT} \simeq
\extremum_{\phi\big|_{z=\epsilon} = \phi_0} I_{SUGRA}[\phi] \ . \ee Es
decir, la funcional generatriz de funciones de Green conexas en la
teor\ii a de gauge, en el l\ii mite $N, \,g_{YM}^2 N$ grandes, resulta
ser la acci\'on de supergravedad.

Como el lado derecho de la ecuaci\'on es la acci\'on cl\'asica
extremizada, la serie de potencias est\'a representada gr\'aficamente
por diagramas de Feynman a nivel \'arbol para los campos de
supergravedad.

\section{QCD}

Dado el valor y la potencia de c\'alculo de la conjetura, resulta
inevitable tratar de hacer contacto con QCD, sin embargo hasta
aqu{\ii} la conjetura relaciona teor\ii as de cuerdas sobre $AdS$ con
teor\ii as de campos supersim\'etricas y conformes.  Un mecanismo que
permite romper supersimetr\ii a es el propuesto por Witten en
\cite{Witten:1998zw}.

La idea consiste en empezar con una teor\ii a de gauge maximalmente
supersim\'etrica $(p+1)$-dimensional en el volumen de mundo de $N$ $Dp$
branas. Luego, la teor\ii a se compactifica en un c\ii rculo de radio
$R_0$ y se impone sobre \'este condiciones antiperi\'odicas para los
fermiones. En consecuencia, \'estos adquieren masa del orden $m_f \sim
1/R_0$.  Los escalares adquieren masa a un loop y luego, a energ\ii as
menores que $1/R_0$, ambos se desacoplan del sistema. De esta manera,
a grandes distancias comparadas con el radio de $R_0$, la teor\ii a
efectiva deber\ii a representar QCD pura en $p$
dimensiones. Usualmente, la direcci\'on escogida para tal fin es la
direcci\'on temporal, lo que lleva a una teor\ii a de gauge a
temperatura finita $T$, donde el radio de compactificaci\'on es
proporcional a $1/T$.  De esta manera, el l{\'\i}mite de alta temperatura
de una teor\ii a de gauge supersim\'etrica $(p+1)$-dimensional resulta
descripto por una teor\ii a de gauge no supersim\'etrica en $p$
dimensiones. Del lado de gravedad, la descripci\'on corresponde a tomar
teor\ii as de cuerdas sobre un fondo tipo agujero negro de
Schwarzschild en $AdS$, o m\'as generalmente sobre una variedad
simplemente conexa que permita definir una estructura de espin \'unica
tal que s\'olo pueda contribuir al ensamble t\'ermico est\'andar
$Tr\,e^{-\beta H}$ y no a $Tr\,(-1)^F e^{-\beta H}$
\cite{Witten:1998zw}.

En este nuevo contexto, resulta de particular inter\'es el estudio dual
gravitatorio de operadores locales invariantes de gauge, como los lazos
de Wilson, ya que los polos de las funciones de correlaci\'on de estos
operadores deber\ii an darnos el espectro de glueballs de QCD. Para
esto, es fundamental determinar si la descripci\'on dual corresponde a
la fase confinante de la teor\ii a de gauge.

\section{Lazos de Wilson}

En 3 dimensiones el potencial de Coulomb ya es confinante. Es un
confinamiento logar\ii tmico $V(r) \sim ln(r)$, sin embargo las
simulaciones provistas por {\it Lattice QCD} indican que el
confinamiento para QCD$_3$ a grandes distancias es lineal $V(r) \sim \sigma r$.

\begin{figure}[htb]
\begin{center}
\epsfxsize=1.5in\leavevmode\epsfbox{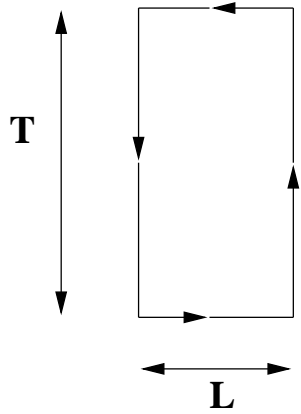}
\end{center}
\caption[Confinamiento lineal para un par quark-antiquark.]{El
  confinamiento lineal $V(L) \sim \sigma L$ para un par
  quark-antiquark puede ser deducido del comportamiento tipo \'area
  del lazo de Wilson $\langle W(C) \rangle \sim \exp(-\sigma TL)$.  }
\label{arealaw}
\end{figure}

Para ver el confinamiento en la descripci\'on dual, resulta \'util
estudiar lazos de Wilson espaciales. En una teor\ii a confinante, el
valor de expectaci\'on de vac\ii o de un operador de lazo de Wilson muestra
comportamiento de \'area \cite{Wilson:1974co}
\be
\langle W(C) \rangle \simeq \exp(-\sigma A(C)) \ ,
\label{area}
\ee donde $A(C)$ es el \'area encerrada por el lazo $C$. La constante
$\sigma$ es llamada, en este contexto, la tensi\'on de la cuerda.  El
comportamiento de \'area (\ref{area}) es equivalente a un potencial de
interacci\'on quark-antiquark que confina linealmente, $V(L) \sim
\sigma L$.  Esto puede ser visto considerando un lazo rectangular $C$
con lados de longitud $T$ y $L$ en espacio eucl{\'\i}deo, como se muestra
en la Figura (\ref{arealaw}).  En el caso en que $T$ sea la direcci\'on
temporal y en el l\ii mite $T \rightarrow \infty$, resulta \be \langle
W(C) \rangle \sim \exp(-TV(L)) \sim \exp(-\sigma A(C)) \ .  \ee

Si pensamos que el tubo de flujo que mantiene unido al par de quarks
se comporta como una cuerda, la descripci\'on dual aparece casi de
manera obvia. Una cuerda con sus extremos pegados a los quarks que
viven en la teor\ii a del borde, no est\'a necesariamente confinada a
vivir sobre \'este. M\'as bien, la cuerda naturalmente se meter\'a
hacia el interior del espacio y $\langle W(C) \rangle$ corresponder\'a a la
suma sobre todas las configuraciones de cuerdas que tengan por
contorno el lazo de Wilson $C$.

Es decir, la prescripci\'on est\'a dada por
 \be \langle W(C) \rangle = \int \exp(-\mu(D)) \ ,
\label{sumarea}
\ee
donde  $\mu(D)$ es el \'area regularizada del volumen de mundo de la cuerda 
$D$ restringida en el borde a $C$. 

En la aproximaci\'on natural, supergravedad, (\ref{sumarea})
se reduce a  
\be
\langle W(C) \rangle = \exp(-\mu_{\textrm{m\ii n}}(D)) \ ,
\label{minarea}
\ee donde el volumen de mundo $\mu_{\textrm{m\ii n}}(D)$ ser\'a aquel que minimice el
\'area con la condici\'on de contorno dictada por el lazo de Wilson del
borde, calculado sobre el fondo de supergravedad correspondiente.

\section{Espectro de Glueballs}

\begin{figure}[htb]
\begin{center}
\epsfxsize=3.5in\leavevmode\epsfbox{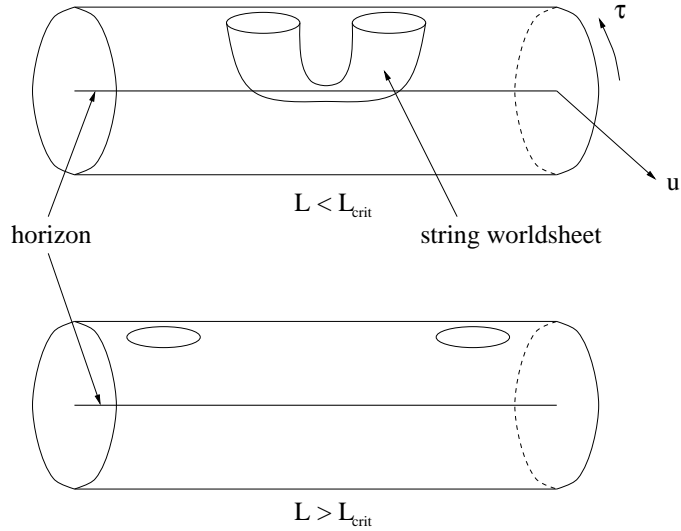}
\end{center}
\caption[$L > L_{\rm crit}$, no existe la hoja de mundo que
  conecta ambos lazos de Wilson.]{Para $L > L_{\rm crit}$, no existe
  la hoja de mundo que conecta ambos lazos. La distancia $L_{\rm
    crit}$ queda determinada por el tama\~no del
  lazo.} \label{disconected}
\end{figure}

La correlaci\'on entre lazos de Wilson en QCD es mediada por el
intercambio de glueballs entre ellos. En la descripci\'on dual, esta
correlaci\'on es descripta a trav\'es de una cuerda que une ambos lazos de
Wilson. Sin embargo, a medida que estos lazos se separan, la hoja
de mundo de la cuerda que los une se vuelve cada vez m\'as
delgada. Para un dado valor de separaci\'on $L$, ($L\,>\,L_{crit}$, que
depende del tama\~no de los lazos) la hoja de mundo se vuelve inestable,
siendo mas favorable energ\'eticamente dos hojas de mundo, cada una con
su respectivo lazo de Wilson (Figura \ref{disconected}). Esto parece ir en
contradicci\'on con la descripci\'on del lado de teor\ii a de gauge, en
la que a grandes distancias los lazos de Wilson son mediados por
glueballs.

\begin{figure}[htb]
\begin{center}
\epsfxsize=3.5in\leavevmode\epsfbox{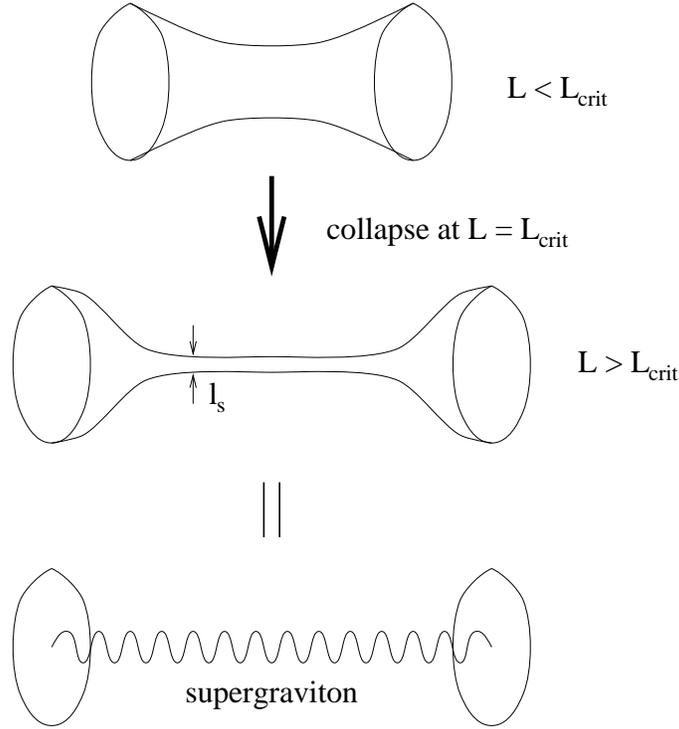}
\end{center}
\caption[$L > L_{\rm crit}$, correlaci\'on soportada por intercambio
  de gravitones.]{Para $L > L_{\rm crit}$, la correlaci\'on entre los
  lazos de Wilson est\'a soportada por el intercambio de
  gravitones.} \label{exchange}
\end{figure}

 Este fen\'omeno se explica en \cite{Gross:1998gk} de la siguiente
 manera. Cuando la distancia entre ambos lazos $C_1$ y $C_2$ es menor
 que la distancia cr\ii tica ($L< L_{\rm crit}$), la contribuci\'on
 principal a la parte conexa de la funci\'on de correlaci\'on $\langle
 W(C_1) W(C_2) \rangle$ proviene de la cuerda cl\'asica que conecta
 $C_1$ con $C_2$. A $L = L_{\rm crit}$, la hoja de mundo de esta
 cuerda se vuelve inestable y comienza a colapsar.  Antes que la
 superficie se desconecte, la aproximaci\'on de supergravedad deja de
 ser v\'alida, cuando el radio de la hoja de mundo se vuelve del orden
 de la longitud caracter\ii stica $l_s$. Luego de eso, las fluctuaciones
 cu\'anticas comienzan soportar la superficie y las dos hojas de mundo
 contin\'uan conectadas por un tubo delgado del orden de la longitud
 caracter\ii stica de la cuerda $l_s$. Para $L$ grande, el tubo delgado
 puede ser representado por el intercambio de gravitones entre las dos
 hojas de mundo (Figura \ref{exchange}). Por lo tanto, la correlaci\'on
 entre los lazos de Wilson no desaparece completamente, sino que es
 mediada por el intercambio de gravitones. Esto indica que los
 supergravitones que se propagan sobre el fondo de agujero negro
 ($p+1$)-dimensional deben ser identificados con los glueballs de
 QCD$_p$.
\\~\\ 

Siendo un poco m\'as precisos, podemos conjeturar que los polos
de las funciones de correlaci\'on entre lazos de Wilson, que
corresponden a las masas de los glueballs, deber\'an corresponder a las
masas de los supergravitones que se propaguen sobre el fondo de
gravedad. Esto \'ultimo es consecuencia de que del lado de gravedad
s\'olo contribuyen diagramas a nivel \'arbol, por lo que la masa
cl\'asica y la masa f\ii sica (polo del propagador del gravit\'on) es la
misma. 

Para ver de manera m\'as expl\ii cita la relaci\'on entre las fluctuaciones
de los campos de supergravedad y el espectro de glueballs, desarrollemos
un ejemplo particular de QCD$_4$, siguiendo \cite{Brower:2000rp}. Una
forma de obtener una teor\ii a de gauge que pueda representar QCD$_4$
es a trav\'es de la reducci\'on dimensional de la teor\ii a de gauge
5-dimensional, que aparece como l\ii mite de baja energ\ii a de la
din\'amica de D4-branas en teor\ii a de cuerdas tipo IIA en 10
dimensiones.

Dado que podemos pensar en los campos de supergravedad como las
constantes de acoplamiento de los operadores de la teor\ii a de gauge,
sus n\'umeros cu\'anticos pueden ser asignados requiriendo invarianza
ante conjugaci\'on de carga y paridad de toda la acci\'on, (es decir,
de todos los t\'eminos con campos de supergravedad acoplados a
operadores compuestos de la teor\ii a de gauge).  Para estudiar de que
manera los campos de supergravedad se acoplan a los campos de la
teor\ii a de gauge que vive sobre las D4-branas, volvamos a la acci\'on
de Born-Infeld m\'as el t\'ermino de Wess-Zumino para este caso,
\be
S=\int d^5x\,
\det[G_{\mu\nu}+e^{-\phi/2}(B_{\mu\nu}+F_{\mu\nu})]+\int d^4x (C_1
F\wedge F+ C_3\wedge F+ C_5)\; ,
\ee
donde hemos considerado la acci\'on de una sola D4-brana, $\mu, \nu =
1,2, 3, 4,\tau$. La asignamci\'on de los n\'umeros cu\'anticos en el caso
de N branas coincidentes ser\'a el mismo.  En la teor\ii a de gauge
5-dimensional, las coordenadas del volumen de mundo son $x_1, x_2,
x_3,x_4,\tau$ donde $\tau$ est\'a compactificada en una $S^1$. Tomemos
$x_4$ como la coordenada temporal eucl\ii dea.  Luego de la reducci\'on dimensional, los campos pueden ser caracterizados por su representaci\'on bajo el peque\~no grupo de $SO(3)$ correspondiente a las rotaciones en las coordenadas espaciales $x_i $, $i=1,2,3$, en la teor\ii a 4-D.

Definamos para los campos de la teor\ii a de gauge 5-D la operaci\'on
de paridad de la siguiente manera:
\bea
P&:& A_i(x_i,x_4,\tau)\rightarrow  -A_i(-x_i,x_4,\tau),\cr
P&:& A_4(x_i,x_4,\tau)\rightarrow \> \>\>\>A_4(-x_i,x_4,\tau),\cr
P&:& A_{\tau}(x_i,x_4,\tau)  \rightarrow \>\>\>\> A_{\tau}(-x_i,x_4,\tau) \; ,
\eea
para $x_i\rightarrow -x_i$, $x_4\rightarrow x_4$, y $\tau\rightarrow
\tau$. Luego de la compactificaci\'on
${\bf R^4 \times S^1}$, podemos definir otra operaci\'on de paridad,
invirtiendo la coordenada $\tau$ en la ${\bf S^1}$. Entonces definimos
otra transformaci\'on de paridad (no relacionada con la anterior). Esta
transformaci\'on de paridad en $\tau-$, $P_{\tau}:$ $\tau \rightarrow
-\tau$, act\'ua de la siguiente manera:
\bea
P_\tau&:& A_i(x_i,x_4,\tau)\rightarrow \> \>\>\>A_i(x_i,x_4,-\tau),\cr
P_\tau&:& A_4(x_i,x_4,\tau)\rightarrow \> \>\>\>A_4(x_i,x_4,-\tau),\cr
P_\tau&:& A_{\tau}(x_i,x_4,\tau) \rightarrow -A_{\tau}(x_i,x_4,-\tau) \; .
\eea
\\
~\\
La conjugaci\'on de carga para un campo no abeliano se define como 
\be
C:{\scriptstyle \frac{1}{2}} T_a A^a_\mu(x) \rightarrow - {\scriptstyle \frac{1}{2}} T^*_a A^a_\mu(x)
\ee
donde $T^a$ es el generador del grupo.  En t\'erminos de los campos
matriciales: ($A \equiv {\scriptstyle \frac{1}{2}} T_a A^a$), $ C: A_\mu(x) \rightarrow -
A^T_\mu(x).$ Esto lleva a una sutileza. Por ejemplo, consideremos esta
transformaci\'on actuando sobre un invariante trilineal de la teor\ii a,
\be C: Tr[ F_{\mu_1 \nu_1} F_{\mu_2 \nu_2} F_{\mu_3 \nu_3}]
\rightarrow - Tr[ F_{\mu_3 \nu_3} F_{\mu_2 \nu_2} F_{\mu_1 \nu_1} ] \; .
\ee
Notar que el \'orden de los campos ha cambiado. Luego, productos
sim\'etricos, $d^{abc} F^a_1 F^b_2 F^c_3$, tendr\'an $C = -1$ y
productos antisim\'etricos, $f^{abc} F^a_1 F^b_2 F^c_3$, tendr\'an $C
= +1$. Es posible mostrar que s\'olo aparecen trazas sim\'etricas,
denotadas como $Sym~\tr[F_{\mu\nu} \cdots]$. De esta manera, los
polinomios pares tendr\'an $C = +1$ y los polinomios impares $C = -1$.
\\~\\

Desarrollando la acci\'on de Born-Infeld, es posible obtener los
operadores a los que las pertubaciones de los campos de fondo se
acoplan y a partir de \'estos, determinar los n\'umeros cu\'anticos $J^{PC}
(P_\tau)$ de los estados de glueballs representados por \'estas:

Naturalmente, el acoplamiento de la m\'etrica resulta,
$G_{\mu\nu}T^{\mu\nu}\sim G_{\mu\nu}\tr(F_{\mu\lambda}F^\lambda_\nu )
    + \cdots  \; ,$
de donde se obtiene
\be
G_{ij}~\rightarrow ~2^{++}\;\; (P_\tau = +) ,\quad\quad
G_{i\tau}~\rightarrow ~1^{-+}\;\; (P_\tau = +) \quad\quad
G_{\tau\tau}\rightarrow ~0^{++}\;\; (P_\tau = +) \; . \ee
Luego de la compactificaci\'on de supergravedad 11-D, el nuevo campo
$G_{\mu ,11}$ resulta una 1-forma $C_{\mu}$ de Ramond-Ramond, que se
acopla como $\sim\epsilon^{\mu\nu\lambda\kappa\eta} C_\mu\;
Sym~\tr[F_{\nu\lambda}F_{\kappa\eta} W]\;,$ donde $W$ es alguna
potencia par del campo $F$.  En consecuencia, el acoplamiento $
\epsilon^{ijk}C_{i}\;Sym~\tr[F_{\tau j}F_{k4}W]+\cdots$ lleva a \be
C_i ~\rightarrow ~1^{++}\;\; (P_\tau = -) \;.\ee Similarmente, $
\epsilon^{ijk}C_{\tau }\tr(F_{ij}F_{4k}W)+\cdots$ da  \be C_\tau
~\rightarrow ~0^{-+}\;\; (P_\tau = +)\; , \ee y  $G_{11, 11}$ lleva a 
un dilat\'on $\phi$ con constante de acoplamiento $\phi \tr{F^2} \; ,$ \be \phi
~\rightarrow ~0^{++}\;\; (P_\tau = +) .  \ee

\noindent Un an\'alisis similar puede realizarse para los otros campos.

%% file: sol_nocritica.tex
\chapter{Soluciones No Cr\ii ticas}\label{nos_sol}
\section{Introducci\'on}
Las teor\ii as de cuerdas no cr\ii ticas $D$ dimensionales est\'an
formuladas en dimensiones $D \neq 26$ en el caso de cuerdas
bos\'onicas, o $D \neq 10$ en el caso de supercuerdas. Estas teor\ii
as incluyen entre sus grados de libertad asociados a la hoja de mundo,
un campo escalar con din\'amica propia. Este campo es llamado {\it
  modo de Liouville} y se combina con las $D-1$ coordenadas restantes
\cite{Polyakov:1981rd}, \cite{Polyakov:1981re}.  Estas teor\ii as han
sido ampliamente estudiadas en el pasado, en particular en $D\leq 2$
\cite{Ginsparg:1993is}, ya que el problema de la llamada barrera
``$c=1$'' imped\ii a formularlas en dimensiones mayores
\cite{Seiberg:1990eb}.  Esta situaci\'on cambi\'o dr\'asticamente con la
introducci\'on de la simetr\ii a superconforme $({\cal N}_L,{\cal
  N}_R)=(2,2)$ en la hoja de mundo. Kutasov y Seiberg
\cite{Kutasov:1990ua}, mostraron que es posible formular teor\ii as no
cr\ii ticas Tipo II en espacio-tiempos de dimensiones $d=2n$, con $
n=0,1,\dots 4$, y describir consistentemente soluciones con
supersimetr\ii a espacio-temporal con al menos $2^{n+1}$ supercargas.
Sobre la hoja de mundo, estas teor\ii as presentan, adem\'as del modo de
Liouville $\phi$ antes mencionado, un bos\'on compacto $X$, dando al
espacio de fondo la forma general \be {\cal M}^d\times \Re_\phi\times
S^1 \times M/\Gamma\qquad , \ee donde ${\cal M}^d$ es el
espacio-tiempo Minkowski $d$-dimensional, $M$ es una teor\ii a de
campos $(2,2)$ superconforme arbitraria, y $\Gamma$ es un subgrupo
discreto que act\'ua sobre $S^1 \times M$. La teor\ii a de Liouville
incluye un fondo de dilat\'on lineal que lleva a una singularidad a
acoplamiento fuerte.  Sin embargo, la existencia de un bos\'on compacto
permite resolver esta singularidad reemplazando la parte del fondo
$\Re_\phi\times S^1$ por el supercoset ${\cal N}=2$ de Kazama y
Suzuki: $SL(2,\Re)_k/U(1)$.  Este espacio tiene la forma geom\'etrica
de un cigarro, con una escala natural dada por $\sqrt{k\,\alpha'}$
\cite{Witten:1991yr}, \cite{Mandal:1991tz}. Esto provee un regulador
geom\'etrico para la singularidad a acoplamiento fuerte, mientras que
coincide con la soluci\'on de dilat\'on lineal en la regi\'on de
acoplamiento d\'ebil.  Se sabe que esta soluci\'on es exacta a todo
orden en teor\ii a Tipo II, a menos de un corrimiento trivial
$k\rightarrow k-2$ \cite{Bars:1992sr}.  Cuando $d=8$, se recupera
supercuerdas cr\ii ticas sobre Minkowski 10-dimensional.

En vista de esto, en este cap\ii tulo, estudiaremos soluciones de
fondo de teor\ii as no cr\ii ticas Tipo II, que incluyan un sector no
trivial que pueda ser parametrizado por una coordenada
radial y una coordenada compacta $S^1$. La coordenada radial
podr\'a ser interpretada como una escala de energ\ii a \cite{Polyakov:1998ju}, \cite{Maldacena:1997re},
\cite{Itzhaki:1998dd}.  


 Presentaremos aqu{\ii}, la soluci\'on correspondiente a la
 cuerda fundamental doblemente localizada (una de las pocas conocidas)
 en el vac\ii o de Minkowski veces el cigarro, soluciones de vac\ii o y
 soluciones cargadas que representan $p$-branas que llenan todo el
 espacio Minkowski no transverso. Todas las soluciones presentadas
 aqu{\ii} se hallan publicadas en \cite{Lugo:2005yf} y
 \cite{Lugo:2006vz}.

El problema de encontrar soluciones a la acci\'on de baja energ\ii a
de las cuerdas no cr\ii ticas ya ha sido abordado en otros trabajos; las
referencias m\'as notables corresponden a \cite{Kuperstein:2004yk}
(ver tambi\'en \cite{Klebanov:2004ya}, \cite{Alishahiha:2004yv},
\cite{Bigazzi:2006ix}).  En nuestro caso, en lugar de resolver un
conjunto de ecuaciones BPS, hemos preferido encarar el problema
resolviendo directamente el conjunto completo de ecuaciones acopladas
de segundo orden derivadas de dicha acci\'on de baja energ\ii a.

Un punto que es importante remarcar es el siguiente. Muchas de las
soluciones encontradas en el caso no cr\ii tico resultan confiables
s\'olo en una (gran) regi\'on del espacio, como pasa en el caso
\cite{Lugo:2005yf} y como pasa para algunas de las soluciones
presentadas aqu{\ii}, de la misma manera que sucede para el caso de
cuerdas cr\ii ticas con las soluciones de $p$-branas. Pero en otros
casos, como sucede con las soluciones tipo $AdS$, la curvatura es del
orden de la unidad, y por tanto no queda claro si estas soluciones
no recibir\'an correcciones importantes de los ordenes superiores de la
acci\'on efectiva. A pesar de esto adoptaremos la postura seguida en
\cite{Polyakov:1998ju}, \cite{Klebanov:2004ya},
\cite{Kuperstein:2004yk}, \cite{Kuperstein:2004yf},
\cite{Casero:2005se}, que asume que para el caso de este tipo de
fondos, la estructura conforme de los mismos no ser\'a modificada por
las contribuciones de orden superior en la curvatura, a menos de
corrimientos en los par\'ametros caracter\ii sticos.

Este cap\ii tulo est\'a organizado como sigue.  En la secci\'on $2$,
presentamos la acci\'on efectiva para cuerdas no cr\ii ticas. En la
secci\'on $3$, presentamos la soluci\'on de cuerda fundamental en el
vac\ii o Minkowski $\times$ el cigarro. Vemos aqu{\ii} que esta
soluci\'on resulta doblemente localizada en el origen del espacio
Minkowski y en el tip del cigarro. En la secci\'on $4$, presentamos,
adem\'as, el ans\"atz general de $p$-branas que llenan todo el
espacio-tiempo. Reducimos, en este caso, el sistema completo de
ecuaciones acopladas de segundo orden a un par de ecuaciones acopladas
m\'as un v\ii nculo, ``condici\'on de energ\ii a cero''; ecuaciones
(\ref{system}) y (\ref{constraint}). En la secci\'on $5$, obtenemos
todas las posibles soluciones de vac\ii o para esta teor\ii a de baja
energ\ii a. \'Estas consisten en la soluci\'on dada por el
espacio-tiempo de Minkowski veces el dilat\'on lineal veces una $S^1$,
y una familia de soluciones en tres par\'ametros, asint\'otica al
vac\ii o antes mencionado. Estas \'ultimas resultan singulares en
general, a excepci\'on de una sub-familia que incluye la bien conocida
soluci\'on Minkowski $\times$ el cigarro.  En la secci\'on $6$,
resolvemos completamente el problema de encontrar soluciones cargadas
NSNS que llenen todo Minkowski.  Por otro lado, adem\'as de la
soluci\'on bien conocida $AdS_{1,2}\times S^1$ (representada por la
teor\ii a exacta $SL(2,\Re)\times U(1)$), encontramos varias familias
de soluciones que tienen a \'esta por l\ii mite
asint\'otico. Notablemente, encontramos adem\'as una soluci\'on
interpretable como la cuerda fundamental no cr\ii tica embebida en el
vac\ii o de dilat\'on lineal, y una familia 2-param\'etrica de
soluciones regulares. M\'as a\'un, recobramos la soluci\'on
previamente hallada en la seccion $3$ de la cuerda fundamental en
el vac\ii o del cigarro y tres familias de soluciones singulares y
oscilantes de dudosa interpretaci\'on f\ii sica.  En la secci\'on $7$,
consideramos el sistema de ecuaciones en el caso de soluciones
cargadas de RR. En este caso, el problema no ha podido ser resuelto
completamente debido a que hemos introducido un anz\"atz particular,
que nos lleva a una familia de soluciones 2-param\'etricas
asint\'oticas al espacio tiempo T-dual a $AdS_{1,p+2}$. \'Estas
resultan singulares en el IR, excepto para dos familias de soluciones
que son regulares.  Via T-dualidad, \'estas pueden ser mapeadas en
familias de espacios de Einstein conformes, con dilat\'on constante,
que incluyen al agujero negro $AdS$ encontrado en
\cite{Kuperstein:2004yk}. Una generalizaci\'on de estas \'ultimas
constituir\'a el tema central del cap\ii tulo que sigue.  Por
\'ultimo, en la secci\'on $8$, se incluyen algunas conclusiones de
este cap\ii tulo.


\section{La acci\'on de baja energ\ii a para las teor\ii as no cr\ii ticas}
Como ya mencion\'aramos, es posible apartarnos de la dimensi\'on cr\ii
tica de manera consistente si admitimos la existencia de campos de
fondo no triviales. En esta secci\'on, repasaremos la formulaci\'on de
teor\ii as de cuerdas sobre fondos curvos y en presencia de campos de
fondo. Por brevedad en la exposici\'on, s\'olo trataremos el caso bos\'onico.
\\~\\
Comencemos recordando la acci\'on de Polyakov para la parte bos\'onica
de la cuerda en D dimensiones planas

\be
S_{\rm p}=-\frac{1}{4\pi \alpha'}\int_{\cal M} d^2 \sigma \sqrt{-h(\sigma)}h^{ab}\eta_{\mu \nu}\partial_a X^{\mu}\partial_b X^{\nu},
\ee
donde $\eta_{\mu \nu}$ es la m\'etrica plana con signatura Minkowski
$(-,+,...,+)$, y $h^{ab}$ es la m\'etrica en la hoja de mundo con
signatura $(-,+)$. De relatividad general, sabemos que si pretendemos
reformular la teor\ii a sobre un espacio curvo debemos reemplazar la
m\'etrica plana $\eta_{\mu \nu}$ por una m\'etrica general $G_{\mu
  \nu}(X)$ que describa tal espacio curvo. Esto lleva naturalmente a
la siguiente acci\'on para la cuerda sobre dicho espacio

\be S_{\sigma}=-\frac{1}{4\pi \alpha'}\int_{\cal M} d^2 \sigma
\sqrt{-h(\sigma)}h^{ab}G_{\mu \nu}(X)\partial_a X^{\mu}\partial_b
X^{\nu}.
\label{f_curvo}
\ee Esta teor\ii a se conoce como modelo {\it sigma no lineal}. Sin
embargo, debemos recordar que la teor\ii a de cuerdas incluye en su
espectro una part\ii cula que puede ser identificada con el
gravit\'on, por lo que cabe preguntarse si es consistente esta
manera de incluir un fondo curvo (estado coherente de
gravitones). Para ver que esta forma es consistente, consideremos el
caso en que la m\'etrica de nuestro espacio curvo puede describirse
como la suma de una m\'etrica plana m\'as una peque\~na perturbaci\'on

\be
G_{\mu \nu}(X)=\eta_{\mu \nu} + {\ell}_{\mu \nu}(X).
\ee

Luego, la expresi\'on (\ref{f_curvo}) resulta

\be S_{\sigma}=-\frac{1}{4\pi \alpha'}\int_{\cal M} d^2 \sigma
\sqrt{-h(\sigma)}h^{ab}(\eta_{\mu \nu}+\ell_{\mu \nu}(X))\partial_a X^{\mu}\partial_b
X^{\nu}.
\ee

Esta expresi\'on es la que entra en la exponencial de la integral de
camino, cuando estudiamos la teor\ii a cu\'antica \bea
\exp(-S_{\sigma})&=&\exp[-\frac{1}{4\pi \alpha'}\int_{\cal M} d^2
  \sigma \sqrt{-h(\sigma)}h^{ab}(\eta_{\mu \nu}+\ell_{\mu
    \nu})\partial_a X^{\mu}\partial_b X^{\nu}]\cr &=&\exp(-S_{\rm
  p})\exp[-\frac{1}{4\pi \alpha'}\int_{\cal M} d^2 \sigma
  \sqrt{-h(\sigma)}h^{ab}\ell_{\mu \nu}\partial_a X^{\mu}\partial_b
  X^{\nu}]\cr &\approx &\exp(-S_{\rm p})[1-\frac{1}{4\pi
    \alpha'}\int_{\cal M} d^2 \sigma \sqrt{-h(\sigma)}h^{ab}\ell_{\mu
    \nu}\partial_a X^{\mu}\partial_b
  X^{\nu}], \label{graviton_vertex}\eea donde al final hemos usado el
desarrollo de la segunda exponencial a primer orden.

De la \'ultima igualdad en (\ref{graviton_vertex}), identificando $
\ell(X)_{\mu \nu}=-4 \pi g_s e^{i k\cdot X}\zeta_{\mu \nu}$, vemos que
una peque\~na perturbaci\'on sobre el espacio plano puede ser
interpretada como la inserci\'on del operador de v\'ertice del
gravit\'on \cite{Polchinski:1998}. Es decir, una peque\~na deformaci\'on del
espacio plano corresponde a la existencia de un estado de
gravit\'on. Luego, reemplazar la m\'etrica plana por $G_{\mu \nu}$
corresponde a exponenciar el operador de v\'ertice del gravit\'on; es
decir, corresponde a un fondo coherente de gravitones.  Naturalmente
podemos incluir de manera similar los otros dos campos que encontramos
cuando estudiamos el sector no masivo NSNS de cuerdas cerradas en la
secci\'on \ref{cuarda_cauntica}, el tensor antisim\'etrico de Kalb-Ramond $B_{\mu
  \nu}$y el dilat\'on $\phi$ \cite{Johnson:2000ch},

\be S_{\sigma}=-\frac{1}{4\pi \alpha'}\int_{\cal M} d^2 \sigma
\sqrt{-h(\sigma)}[(h^{ab}G_{\mu \nu}(X)+i \epsilon^{ab}B_{\mu \nu}(X))\partial_a X^{\mu}\partial_b
X^{\nu} +\alpha' R \phi(X)].
\ee

Claramente, la exigencia de la invarianza de Weyl de la teor\ii a a
nivel cu\'antico se traduce ahora en nuevas funciones $\beta$ (una
por cada campo) que deben anularse. Es decir, dado que la
invarianza de Weyl requiere que la traza del tensor energ\ii a momento
sea nula,

\begin{equation}
T^a_{\phantom{a}a}=-{1\over2\alpha^\prime}\beta_{\mu\nu}^G
h^{ab}\partial_aX^\mu\partial_bX^\nu-{i\over2\alpha^\prime}
\beta_{\mu\nu}^B
\epsilon^{ab}\partial_aX^\mu\partial_bX^\nu-{1\over2}\beta^\Phi R\ ,
\end{equation}

es necesario  que $\beta_{\mu\nu}^G,\beta_{\mu\nu}^B,\beta^\Phi$ sean nulas.

El c\'alculo a segundo orden en las perturbaciones de los campos
(primer orden en $\alpha'$) muestra que \cite{cfmp} \cite{Polchinski:1998}:

\begin{eqnarray}
\beta_{\mu\nu}^G
&=&\alpha^\prime\left(R_{\mu\nu}+2\nabla_\mu\nabla_\nu\Phi
-{1\over4}H_{\mu\kappa\sigma}H_\nu^{\phantom{\nu}\kappa\sigma}\right)
+O(\alpha^{\prime2}),\nonumber\\ 
\beta_{\mu\nu}^B
&=&\alpha^\prime\left(-{1\over2}\nabla^\kappa
H_{\kappa\mu\nu}+\nabla^\kappa \Phi H_{\kappa\mu\nu}
\right)+O(\alpha^{\prime2}),\nonumber\\
\beta^\Phi
&=&\alpha^\prime\left({D-26\over6\alpha^\prime}
-{1\over2}\nabla^2\Phi+\nabla_\kappa\Phi\nabla^\kappa \Phi 
-{1\over24}H_{\kappa\mu\nu}H^{\kappa\mu\nu}
\right)+O(\alpha^{\prime2})\ ,
\label{betas_campos}
\end{eqnarray}
donde $H_{\mu\nu\kappa}\equiv\partial_{\mu}
B_{\nu\kappa}+\partial_{\nu} B_{\kappa\mu}+\partial_{\kappa}
B_{\mu\nu}$.

Es interesante notar que, dado que diferentes \'ordenes en $\alpha'$
corresponden a diferentes niveles de energ\ii a, a mayor orden en
$\alpha'$, mayor nivel de energ\ii a. Luego, si bien s\'olo podemos
conocer orden por orden las ecuaciones para las {\it betas}
(\ref{betas_campos}), anular estas funciones al orden $\alpha'$ tiene
que representar la f\ii sica de esta teor\ii a a bajas energ\ii as. Por
otro lado, es notable que anular las funciones {\it beta} lleve a
ecuaciones para los campos de fondo que est\'an escritas en forma
covariante respecto del espacio-tiempo. De esta manera, contamos con un
conjunto de ecuaciones que deben satisfacer los campos de fondo y que
representan la f\ii sica de baja energ\ii a de la teor\ii a de cuerdas
(sector no masivo).
\\~\\ Es posible mostrar que estas
ecuaciones pueden ser obtenidas a partir de la siguiente acci\'on para
los campos de fondo

 \begin{eqnarray}
{\rm S}&=&{1\over2\kappa^2_0}\int
d^DX(-G)^{1/2}e^{-2\Phi}\left[R+4\nabla_\mu
\Phi\nabla^\mu\Phi-{1\over12}H_{\mu\nu\lambda}
H^{\mu\nu\lambda}\right.\nonumber\\
&&\hskip5cm
\left.-{2(D-26)\over3\alpha^\prime}+O(\alpha^\prime)\right]\ ,
\label{ac_ no_critica}
\end{eqnarray}
en el caso de la cuerda bos\'onica. Para la supercuerda, la parte
bos\'onica de la acci\'on de baja energ\ii a es la misma, donde la
constante cosmol\'ogica $2(D-26)\over3\alpha^\prime$, debe ser
reemplazada por $(D-10)\over\alpha^\prime$. De aqu{\ii} en m\'as, nos
referiremos a esta acci\'on como la acci\'on de baja energ\ii a para
la cuerda no cr\ii tica en el {\it string frame}.

Un an\'alisis r\'apido de las ecuaciones (\ref{betas_campos}) muestra
que \'estas son compatibles con un dilat\'on y campo antisim\'etrico
constante y una m\'etrica plana, si la dimensi\'on del espacio tiempo
es 26 (caso bos\'onico). Sin embargo, si nos apartamos de esta
dimensi\'on cr\ii tica, las ecuaciones s\'olo pueden ser resueltas por
configuraciones de campos no triviales. En las secciones que siguen,
nos enfrentaremos al problema de hallar dichas configuraciones.

\section{La cuerda Fundamental no cr\ii tica}\label{fund_no_critica}

Como vimos, la f\ii sica de baja energ\ii a para la cuerda no cr\ii
tica puede ser obtenida de la acci\'on (\ref{ac_ no_critica}), que
corresponde a considerar los campos que representan los modos sin masa
de la cuerda movi\'endose en una variedad $D$-dimensional $M_D$; ellos
son la m\'etrica $G_{mn}$, el dilat\'on $\Phi$, y el tensor
antisim\'etrico de Kalb-Ramond $B_{mn}$. La parte bos\'onica de la
acci\'on efectiva de baja energ\ii a para la supercuerda resulta
exactamente la misma que presentamos en la secci\'on anterior
\cite{Klebanov:2004ya},
\begin{eqnarray}
S[G,B,\Phi]&=& \int \epsilon_G\; e^{-2\Phi}\left(R[G]+4\left(D\Phi\right)^2\right.
+\left.\Lambda^2-{\frac{1}{2}}\; H^2 \right)\label{acciongra}
\end{eqnarray}
donde $H=dB$ es la 3-forma intensidad de campo, $H^2 \equiv
\frac{1}{2}\, H_{mn}\,H^{mn}$, y $\epsilon_G=d^D x\,\sqrt{-\det G}\, $
es la forma de volumen del espacio-tiempo $D$-dimensional.  No
consideraremos por ahora taquiones o campos de Ramond-Ramond en
(\ref{acciongra}), porque no juegan ning\'un rol en el problema que
queremos estudiar.  A menos que se diga lo contrario, trabajaremos en
esta secci\'on en el {\it string frame}.  La constante cosmol\'ogica
resulta $\Lambda^2 = \frac{2 (26-D)}{3\alpha'}$ en el caso bos\'onico,
y $\Lambda^2 = \frac{10- D}{\alpha'}=\frac{4}{r_0{}^2}\,$ en el caso
de supercuerdas.

Consideremos entonces una cuerda fundamental embebida en el
espacio-tiempo de fondo de la siguiente manera ${\bar X ^m (\sigma)}$
, donde $\sigma\equiv (\sigma^0 ,\sigma^1 )$ son las coordenadas de la
hoja de mundo $\Sigma$.  La cuerda fundamental se acopla con la carga
$Q$ al campo $B$ de acuerdo con el t\'ermino de fuente \be
S_{F_1}[\bar X; B] = Q\;\int_{\Sigma}\; B|_{pull-back} =
\int_{M_D}B\wedge *J\;\;\;\;, \ee donde la 2-forma de corriente  est\'a definida por 
\footnote{
La delta covariante se define como 
\be
\delta^D_G(x-\bar{X})\equiv{1\over\sqrt{-\det G}}\;\delta^D(x-\bar{X})
\;\;\;\;\;,\;\;\;\;\; \int_{M_D}\;\epsilon_G\; \delta^D_G(x-\bar{X}) = 1\;\;\;,
\ee
y los \ii ndices son bajados y subidos con la m\'etrica $G$.}
\bea
J &=& \frac{1}{2}\, J_{mn}(x;\bar X)\, dx^m\wedge dx^n\cr
J^{mn}(x;\bar{X}) &\equiv&  Q\;\int_{\Sigma}d\bar{X}^m\wedge d\bar{X}^n\;
\delta^D_G (x-\bar{X}(\sigma))\;\;\;\;,\label{current}
\eea
De la acci\'on

\be
S = S[G,B,\Phi] + S_{F_1}[\bar X; B]\;\;\;\;,
\ee
las ecuaciones de movimiento resultan,
\begin{eqnarray}
R_{mn}&=&{1\over2}\, H_{mp}\,H_n{}^p - 2D_m D_n\Phi\cr
\Lambda^2&=&e^{2\Phi}\; D^2(e^{-2\Phi})- H^2\cr
d\left(e^{-2\Phi}*H\right)&=& -*J\label{eqnmot}
\end{eqnarray}

En ausencia de fuentes, estas ecuaciones de movimiento admiten como
soluci\'on el producto directo del espacio-tiempo de Minkowski
$d$-dimensional y el cigarro, es decir, el espacio \bea G_0
&=& \eta_{1,d-1} + \tilde g\cr \Phi_0(\tilde\rho) &=& \tilde\Phi_0 +
\tilde\Phi(\tilde\rho)\cr H_0 &=&
0\;\;\;\;\;\;\;\;\;\;\;\qquad,\label{snvac} \eea donde $\,(\tilde g
,\tilde\Phi)$ es la soluci\'on del cigarro ($\tilde\rho\in\Re^+\,,\,
\tilde\theta\in S^1$), \bea \tilde g &=& r_0{}^2 \; \left(
d^2\tilde\rho + \tanh ^2 \tilde\rho\; d^2\tilde\theta \right)\cr
e^{-\tilde\Phi(\tilde\rho)} &=& \cosh\tilde\rho\qquad.\label{sncigar}
\eea $\tilde\Phi_0$ es el valor del dilat\'on en el tip del cigarro
$\tilde\rho =0$, y $r_0=\sqrt{k\,\alpha'}$ es la escala de curvatura,
$R[\tilde g]= \frac{4}{r_0{}^2}\, (\cosh\tilde\rho)^{-2}$.  La
constante de acoplamiento efectiva de la cuerda, $g_s \equiv
e^{\Phi_0(\tilde\rho)}$, est\'a acotada en $\tilde\rho =0$ por
$e^{\tilde\Phi_0}$, que es un par\'ametro libre de la teor\ii a.

Notemos aqu{\ii} que la soluci\'on del cigarro es conforme al espacio
plano (con coordenadas cartesianas $\vec z$), \be \delta \equiv d^2
z^1 + d^2 z^2 = e^{-2\,\tilde\Phi(\tilde\rho) }\; \tilde g = d^2\rho +
\rho^2\; d^2\theta \;\;\;\;,\;\;\;\; \left\{\begin{array}{rcl} \rho
&=& r_0\,\sinh\tilde\rho\cr\theta &=& \tilde\theta
\end{array}\right.\label{flat2dmetric}
\ee

\subsection{El ans\"atz y la ecuaci\'on de movimiento}

Consideremos el fondo (\ref{snvac})-(\ref{sncigar}) como vac\ii o, y
consideremos sobre \'el un objeto unidimensional cargado frente al
campo $B$ a lo largo de las coordenadas $(x^0,x^1)\,$ en ${\cal
  M}^d$. Supongamos que este objeto est\'a localizado en el origen
$r\equiv|\vec y|=0$ del espacio transverso $\Re^{d-2}$ (de coordenads
cartesianas $\vec y$), y en el tip del cigarro $\tilde\rho = 0$.

Un ans\"atz natural para la soluci\'on del problema que
intentamos estudiar, es uno que respete: Poincar\`e en la hoja de mundo,
ya que en principio sobre \'esta no hay ninguna inhomogeneidad que la
rompa; $SO(d-2)$ en el espacio tranverso $\Re^{d-2}$, ya que en este
espacio estar\ii a representada por un punto; y por el mismo motivo
$SO(2)$ sobre el cigarro.  En estas condiciones, el ans\"atz puede
escribirse como \bea G &=& A^2(r;\tilde\rho)\; \eta_{1,1} +
B^2(r;\tilde\rho)\; (dr^2+ r^2\; d^2\Omega_{d-3} ) +
C^2(r;\tilde\rho)\; \tilde g \cr \Phi &=& \Phi(r; \tilde\rho)\cr B&=&
\left( E(r;\tilde\rho) - 1\right)\, dx^0\wedge dx^1
\;\;\longrightarrow\;\; H = dx^0\wedge dx^1 \wedge
dE\qquad.\label{ansatz_0} \eea Usando este ans\"atz, se puede verificar que
la corriente (\ref{current}) resulta \be J= -Q\,\delta^{d-2}_\delta
(\vec y)\;\delta^2_{\delta}(\vec z)\; \frac{
  e^{-2\tilde\Phi(\tilde\rho)}\,A^2}{B^{d-2}\,C^2}\;dx^0 \wedge
dx^1\qquad . \ee Luego, podemos reescribir las ecuaciones
(\ref{eqnmot}) en t\'erminos de las funciones desconocidas del
ans\"atz $(A, B, C, \Phi, E)$.  En este punto, en lugar de escribir
las ecuaciones generales en este ans\"atz, haremos primero las
siguientes suposiciones, \be B=C=1 \qquad,\qquad e^{2\Phi}=e^{2\Phi_0}\,
A^2\qquad,\qquad \epsilon\, E = A^2 - 1\qquad,\label{assump} \ee donde
$\epsilon^2 = 1$.  Resulta entonces que nos hemos quedado con una
\'unica funci\'on desconocida $A$, y un signo.  Definiendo ahora $U
\equiv A^{-2}$, usando la ecuaciones de vac\ii o\footnote{
  Denotamos con ``$\tilde{~}$", las variables, derivadas, campos, en
  el cigarro.  } \bea \tilde{R}_{\tilde a\tilde b}
&=&-2\tilde{D}_{\tilde a}\tilde{D}_{\tilde b}\tilde{\Phi}\cr
\Lambda^2&=& e^{2\tilde\Phi}\; \tilde D^2(e^{-2\tilde\Phi})\;\;\;\;,
\eea y las relaciones \bea
*d(e^{-2\Phi}*H) &=&-A^4\, dx^0\wedge dx^1 \,
D^m\left({e^{-2\Phi}\over A^4}D_m(E)\right)\cr e^{2\Phi} D^m\left(
e^{-2\Phi}\,D_m (U)\right) &=& e^{2\tilde\Phi} D^m_{(0)}\left(
e^{-2\tilde\Phi}\,{D_{(0)}}_m(U)\right)\qquad ,
\label{relations}
\eea es posible mostrar que las ecuaciones (\ref{eqnmot}) se reducen a
la ecuaci\'on diferencial \be -
e^{2\tilde{\Phi}}D^m_{(0)}(e^{-2\tilde{\Phi}}D_{(0)m}(U)) = \epsilon\,
Q\,e^{2\tilde\Phi_0}\, \delta^{d-2}_\delta(\vec y)\;
\delta^2_\delta(\vec z).\label{eqnU} \ee En las ecuaciones
(\ref{relations}) y (\ref{eqnU}), el sufijo $``{(0)}"$ denota
derivada covariante con respecto a la m\'etrica de vac\ii o
(\ref{snvac}); expl{\'\i}citamente, \be l.h.s.\,\textrm{(\ref{eqnU})} = -{1\over
  r^{d-3}}(r^{d-3}U^{'})^{'}- e^{2\tilde{\Phi}} \tilde{D}_{\tilde
  c}\left( e^{-2\tilde{\Phi}}\tilde{D}_{\tilde c}(U)\right) \equiv
\left( \Delta + \tilde{L}_0 \right)(U)\qquad ,  \ee donde es posible
reconocer en el primer t\'ermino el laplaciano en espacio plano
$(d-2)$-dimensional. En el segundo t\'ermino hemos introducido el
operador \be \tilde{L}_0 \equiv -e^{2\tilde{\Phi}}\;\tilde{D}_{\tilde
  c}\;e^{-2\tilde{\Phi}}\; \tilde{D}_{\tilde c}=
-\frac{1}{r_0{}^2}\;\left( \partial^2_{\tilde\rho} +
2\,\coth(2\tilde\rho)\;\partial_{\tilde\rho} +
\coth^2\tilde\rho\;\partial^2_{\tilde\theta} \right)\qquad .\label{L0}
\ee Resumiendo, la funci\'on $U$ est\'a determinada por la ecuaci\'on
diferencial \be \left( \Delta + \tilde{L}_0 \right)(U) =
\epsilon\,Q\,e^{2\tilde\Phi_0}\, \delta^{d-2}_\delta(\vec y)\;
\delta^2_\delta(\vec z),\label{eqnUbis} \ee donde las $\delta$'s que
aparecen del lado derecho de las ecuaciones (\ref{eqnU})
(\ref{eqnUbis}) son respecto a la m\'etrica plana.  

Dado que la ecuaci\'on que intentamos resolver es una ecuaci\'on
diferencial inhomog\'enea en dos variables, repasemos para este caso la soluci\'on
formal de este tipo de ecuaciones en t\'erminos de funciones de Green.

\subsection{La funci\'on de Green}

Sea un espacio m\'etrico $(M,g)$ producto directo de otros dos,
$(M_1,g_1)$ (con coordenadas gen\'ericas $x$) y $(M_2,g_2)$ (con
coordenadas $y$).  Es decir $M =M_1 \times M_2$, con m\'etrica $g=g_1
+ g_2$.  Sean $\hat{A}_1 , \hat{A}_2$ dos operadores lineales sobre el
espacio de funciones en $M_1$ y $M_2$, respectivamente. Consideremos la
siguiente ecuaci\'on diferencial \be \hat{A}\,G(x,y;x_0 ,y_0) \equiv
\left( \hat{A}_1+\hat{A}_2\right)G(x,y;x_0 ,y_0)
=\delta_{g_1}(x-x_0)\;\delta_{g_2}(y-y_0).\label{ecfgreen} \ee La
funci\'on $G(x,y;x_0 ,y_0)$, soluci\'on de esta ecuaci\'on, es la
funci\'on de Green del operador $\hat{A}$ con respecto a la m\'etrica $g$.

Asumiendo que existen conjuntos de auto-funciones $\{ u_m\}$ y $\{ v_i\}$,
con autovalores $\{ \lambda_m\}$ y $\{ \mu_i\}$
\footnote{ Por simplicidad, asumiremos que no hay  modos
  cero presentes. En principio, para la existencia de (\ref{fgreen}), es
  necesario que $\hat A$ no tenga modos cero.} , \bea \hat{A}_1\,
u_m(x)&=&\lambda_m\; u_m (x)\cr \hat{A}_2\, v_i(y)&=& \mu_i\;
v_i(y)\;\;\;\;, \eea donde los \ii ndices $m$  e $i$  corren sobre
conjuntos discretos, continuos, o de ambos tipos.  Si los conjuntos de
auto-funciones son ortonormales y completos, \bea
\int_{M_1}\;\epsilon_{g_1}\; u^*_m\; u_n &=&
\delta_{mn}\;\;\;\;,\;\;\;\; \delta_{g_1}(x - x_0)=\sum_m\;
u^*_m(x_0)\; u_m(x)\cr \int_{M_2}\;\epsilon_{g_2}\; v^*_i\; v_j &=&
\delta_{ij}\;\;\;\;\;\;,\;\;\;\; \delta_{g_2}(y - y_0)=\sum_i\;
v^*_i(y_0)\; v_i(y), \eea entonces las funciones de Green para los
operadores $\hat{A}_1\, ,\hat{A}_2\,$ pueden ser escritas como \bea
G^{(1)}(x,x_0)&\equiv& \sum_m\; {u^*_m(x_0)\; u_m(x)\over
  \lambda_m}\;\;\;\;,\;\;\;\; \hat{A}_1\, G^{(1)}(x,x_0)=
\delta_{g_1}(x- x_0)\cr G^{(2)}(y,y_0)&\equiv& \sum_i\; {v^*_i(y_0)\;
  v_i(y)\over \mu_i}\;\;\;\;\;\;\;\;,\;\;\;\; \hat{A}_2\,
G^{(2)}(y,y_0)= \delta_{g_2}(y- y_0). \eea Definamos ahora \bea
G(x,y;x_0,y_0)&\equiv&\sum_{m,i}{u^*_m(x_0)\; v^*_i(y_0)\; u_m(x)\;
  v_i(y)\over \lambda_m+\mu_i}.\label{fgreen} \eea Es directo mostrar
que \be (\hat{A}_1+\hat{A}_2)\, G(x,y; x_0,y_0)=\delta_{g_1}(x -
x_0)\; \delta_{g_2}(y-y_0). \ee Es decir, (\ref{fgreen}) resulta, de acuerdo a (\ref{ecfgreen}), la funci\'on de Green para el operador $\hat A\equiv
\hat{A}_1+\hat{A}_2$.

Vayamos en particular al caso que nos interesa. Primero identificamos
$\hat{A}_1\equiv \Delta$, el laplaciano sobre el espacio plano
$\Re^{d-2}$. Un conjunto completo de auto-funciones ortonormales con
respecto a la m\'etrica plana es 
\be u_{\vec p }(\vec y) = {e^{i\vec p \cdot \vec y }\over
  (2\pi)^{d-2\over 2}}\qquad,\qquad \int_{\Re^{d-2}}\, d^{d-2} \vec
y\;u_{{\vec p}'}^*(\vec y)\; u_{\vec p}(\vec y) = \delta^{d-2}({\vec
  p}' - \vec p)\label{u} \ee con autovalores $\lambda_{\vec p}={\vec
  p}^2\;,\;\vec p\in\Re^{d-2}$.

Por otro lado, identifiquemos $\hat{A}_2\equiv \tilde L_0$, con
$\tilde L_0$ definido como en (\ref{L0}); se puede mostrar que un
conjunto completo de auto-funciones de este operador corresponde a
\footnote{ Estas son esencialmente funciones de Jacobi $\;{\cal
    P}^{-\frac{1}{2} +i\, s}_{-\frac{m}{2}\,\frac{m}{2}}\;$, ver
  \cite{Dijkgraaf:1991ba} y las referencias que all{\ii} se citan.  } \bea v_{s m}
(\vec z) &=& a_{sm}\; \left(\frac{\rho}{r_0}\right)^{|m|}\;
F\left(\frac{|m| + 1}{2} + i\,s, \frac{|m| + 1}{2} -i\,s; |m| +1;
-\left(\frac{\rho}{r_0}\right)^2\right)\;
\frac{e^{i\,m\,\theta}}{\sqrt{2\,\pi}}\cr
a_{sm} &=& \left(\frac{4}{\pi\,r_0{}^2}\;
\rho_m(s)\right)^\frac{1}{2}\; \frac{i^{-|m|}}{|m|!}\;
\frac{\Gamma(\frac{|m| + 1}{2} + i\,s)}{\Gamma(\frac{-|m| + 1}{2} +
  i\,s)}\cr &=&
\left(\frac{4}{r_0{}^2}\;s\;Re\tanh\left(\pi\,(s+i\,\frac{m}{2})\right)\right)^\frac{1}{2}\;
\frac{i^{-|m|}}{|m|!}\; \frac{\Gamma( \frac{|m| + 1}{2} + i\,s)}{
  \Gamma(\frac{-|m| + 1}{2} + i\,s)} \label{vcompleto}\eea donde $F(a,b;c;z)$ es la
funci\'on hipergeom\'etrica, y $\rho_m(s)$ es la medida de Plancherel.
\'Estas son ortonormales con respecto a la m\'etrica plana $\;\delta =
d\vec z\cdot d\vec z = d^2\rho + \rho^2\,d^2\theta\,$, introducida en
(\ref{flat2dmetric}), \be \int_{\Re^2}\; d^2 \vec z\; v_{s' m'}^*
(\vec z)\; v_{s m} (\vec z) = \delta(s' -s)\;\delta_{m' m}\;\;\;\;,
\ee y los autovalores de $\tilde L_0$ est\'an dados por \be \mu_{sm} =
\frac{1}{r_0{}^2}\,\left( 4\, s^2 + m^2 +1 \right) \qquad,\qquad
s\in\Re^+\;\;,\;\; m\in{\bf z} \ee A partir de este resultado,
podemos construir la funci\'on de Green (\ref{fgreen}).

\subsection{La soluci\'on}

Volvamos entonces a la ecuaci\'on (\ref{eqnUbis}). A partir del
resultado de la secci\'on anterior, vemos que la funci\'on de Green
$U(r,\rho)$ contru{\'\i}da como en (\ref{fgreen}) a partir de (\ref{u}) y
de (\ref{vcompleto}) (en este caso, independiente de $\theta$) admite,
para $d>2$, la siguiente representaci\'on integral \bea U(r,\rho)
&=&1 + \epsilon\,Q\, e^{2\tilde\Phi_0}\, a_d\, \int_0^\infty d\alpha
\; \alpha^{-\frac{d-2}{2}} \;e^{-\alpha - {r^2\over 4r_0^2\alpha}}\;
I(\alpha ; \rho)\cr I(\alpha;\rho)&=& \int_0^\infty ds \;\rho_0(s)\;
e^{-4\alpha s^2} F\left(\frac{1}{2} + is,
\frac{1}{2}-is;1;-{\rho^2\over r_0^2}\right)\label{U}\cr a_d{}^{-1}&
=& 2^{d-3}\pi^{d+2\over 2} r_0{}^{d-2}\qquad .  \eea Por lo tanto,
hemos hallado la configuraci\'on unidimensional de un objeto cargado
ante $B$, embebido en el vac\ii o (\ref{snvac}). Este objeto est\'a
localizado en el origen de coordenadas del espacio transverso
$\Re^{d-2}$, y en el tip del cigarro. Naturalmente, debemos identificar
esta soluci\'on con la cuerda fundamental no cr\ii tica. La soluci\'on
est\'a dada entonces por \bea G &=& U^{-1}\; \eta_{1,1} + dr^2+ r^2\;
d^2\Omega_{d-3} + \tilde g \cr e^{2\Phi}&=&e^{2\tilde{\Phi}}\,
U^{-1}\cr H &=& dx^0\wedge dx^1 \wedge dU^{-1}, \label{funda}\eea donde la funci\'on $U(r,\rho)$ est\'a dada en
(\ref{U}).

En el caso $d=2$, que corresponde a la cuerda que llena todo el
espacio Minkowski, una forma expl\ii cita para $U(\tilde\rho)$ puede
obtenerse, \be U(\tilde\rho)=1 + \beta\; \ln \left(1+{1 \over
  \sinh^2\tilde{\rho}} \right)
\label{Ud2}
\ee donde $\beta= \frac{|Q|\, e^{2\tilde\phi_0}}{4\pi}$ es una
constante num\'erica.

\section{El Tratamiento general}\label{trat_general}

\subsection{La acci\'on no-cr\ii tica y el ans\"atz de branas} 

Dado que hemos sido capaces de encontrar la soluci\'on de cuerda
fundamental no cr\ii tica embebida en el vac\ii o Minkowski $\times$
el cigarro (\ref{funda}), en esta secci\'on nos proponemos realizar un
estudio sistem\'atico de la acci\'on de baja energ\ii a (\ref{ac_
  no_critica}), pero ahora permitiendo incluir tambi\'en campos de R-R,
para poder explorar la posibilidad de soluciones de branas. Escribamos
 nuevamente la parte bos\'onica de la acci\'on de baja
energ\ii a de las teor\ii as de (super)cuerdas no cr\ii ticas en $D$
dimensiones en el string frame, en presencia de campos de formas
generales $A_{q+1}$ y en la base de vielbeins $\omega^m$:
\begin{eqnarray}
S[\Psi]&=&\frac{1}{2 \kappa_D{}^2}\;\int \epsilon_G
\;e^{-2\Phi}\left( R[G]+4\left(D\Phi\right)^2+\Lambda^2-{1\over
  2}\sum_q e^{\left({b_q}+2\right)\Phi}\left(F_{q+2}\right)^2
\right)  \label{sugraction}
\end{eqnarray}
donde $\Psi$ representa el conjunto de campos $\{G_{mn},\Phi,
A_{q+1}\}$, $F_{q+2} = dA_{q+1}$ es el tensor intensidad de campo
asociado al campo de forma $A_{q+1}$, y $\epsilon_G=
\omega^0\wedge\dots\wedge\omega^{D-1}=d^D x\,\sqrt{-\det G}\, $ es el
elemento de volumen invariante. La constante $b_q$ es igual a $-2$
para campos de formas NSNS e igual a $0$ para campos de RR,
$\Lambda^2$ es la constante cosmol\'ogica, que suponemos positiva y que
identificamos en teor\ii a de cuerdas con
$\frac{2\,(26-D)}{3\,\alpha'}$ y en el caso de supercuerdas con
($\frac{10-D}{\alpha'}$), y $\;\kappa_D{}^2$ es la constante de Newton
$D$-dimensional. Consideraremos, adem\'as, un t\'ermino fuente de la
forma
\be 
S^{fuente}[\Psi]= \sum_q\; \mu_q\, \int A_{q+1}\wedge*
J_{q+1}\label{source} 
\ee
donde $\mu_q$ es la carga ante el campo $A_{q+1}$.  Las ecuaciones
de movimiento resultan ser:
\begin{eqnarray}
R_{mn}&=&-2\,D_mD_n\Phi + T^{A}_{mn}\cr
\Lambda^2 &=& e^{2\Phi}D^2(e^{-2\Phi})+
\sum_q {D\,(2+b_q)-2\,b_q-4(q+2)\over 8}\;e^{(2+b_q)\Phi}\; (F_{q+2})^2\cr
d\left(e^{b_q\,\Phi}*F_{q+2}\right)&=& (-)^q \; Q_q\; *J_{q+1}\;\;\;\;,\;\;\;\;
Q_q\equiv 2\,\kappa_D{}^2\, \mu_q \label{ecformal}
\end{eqnarray}
donde el tensor de energ\ii a-momento, el tensor intensidad de campo y
sus contracciones antes se\~naladas son
\footnote{ En teor\ii a de cuerdas, $\kappa_D{}^2 \sim
  \,\alpha'^{\frac{D-2}{2}}$; la carga de $N_q$ $q$-branas es $|\mu_q|
  \sim N_q\,\alpha'^{-\frac{q+1}{2}}$, y $|Q_q|\sim N_q\,
  \alpha'^{\frac{D-q-3}{2}}$.  La normalizaci\'on no ha sido fijada,
  ya que no es trivial en el contexto de teor\ii as de cuerdas no
  cr\ii ticas.}:

\bea 
T^{A}_{mn}&\equiv&\sum_q {1\over2}\, e^{(2+b_q )\Phi}\left(
\left(F_{q+2}\right)^2_{mn} -\frac{2+b_q}{4}\; G_{mn}
\;\left(F_{q+2}\right)^2\right)\cr (F_{q+2})^2_{mn} &\equiv&
\frac{1}{(q+1)!}\, G^{m_1 n_1}\dots G^{m_{q+1}n_{q+1}}\;
     {F_{q+2}}_{mm_1\dots m_{q+1}}{F_{q+2}}_{nn_1\dots n_{q+1}}\cr
     (F_{q+2})^2&\equiv& \frac{1}{(q+2)!}G^{m_1 n_1}\dots
     G^{m_{q+2}n_{q+2}}\; {F_{q+2}}_{m_1\dots
       m_{q+2}}{F_{q+2}}_{n_1\dots n_{q+2}}.\cr
& &\label{emt}
\eea

Consideremos el siguiente ans\"atz para los campos:
\bea G &=& -A^2\;
dx^0{}^2+\tilde{A}^2\;d{\vec x}{}^2 + C^2\;d\rho^2 +\tilde C^2\;
dz^2\cr A_{p+1} &=& dx^0\wedge\dots\wedge dx^p\; E(\rho)\cr \Phi &=&
\Phi(\rho),
\label{fieldansatz}
\eea donde hemos supuesto que hay un s\'olo tipo de carga presente,
que corresponde a la forma $A_{p+1}$ (s\'olo consideramos $q=p$). Este
ans\"atz corresponde presumiblemente a una $p$-brana negra (Black
$p$-brane) extendida a lo largo de las direcciones $( x^0,\vec x)$,
localizada a lo largo de la coordenada radial $\rho$ en el espacio
transverso 2-dimensional eucl{\'\i}deo con coordenadas ($\rho,
z$). Este ans\"atz tiene simetr\ii a $SO(2)$ y la variable $z$ puede
estar compactificada con un radio de compactificaci\'on $R$, $z \sim z
+ 2\,\pi\, R $, que es usualmente determinado por la carga.  En el
ap\'endice \ref{appendI} se dan las f\'ormulas m\'as relevantes del
c\'alculo, en relaci\'on con este ans\"atz.

\subsection{Las ecuaciones de movimiento y la soluci\'on general}

En el caso de $p$-branas que llenan todo el espacio de Minkowski, el
espacio transverso 2-dimensional admite invarianza $SO(2)$, por lo que
puede asumirse, para las funciones del ans\"atz, una dependencia
s\'olo en la ``coordenada radial'' $\rho$ del espacio transverso de
fondo deformado (dilat\'on lineal, cigarro, etc; ver secci\'on \ref{no_carg}).
Notar que, por lo tanto, $C(\rho)$ resultar\'a definido a menos de
reparametrizaciones en la coordenada $\rho$ (en efecto, \'esta
transforma como una 1-forma, mientras que el resto de las funciones de
la m\'etrica lo hacen como escalares). Resulta natural entonces pasar
a una variable que se comporte como escalar ante estos difeomorfismos.
Esto se logra introduciendo la coordenada $x$, definida de la
siguiente manera
\footnote{En lo que sigue $x$ ser\'a considerada no negativa, a menos
  que se indique lo contrario.}:

\bea x &=& \int^{\rho}\; \frac{d\rho}{H(\rho)}
\;\;\;\;,\;\;\;\;\partial_x=H(\rho)\;\partial_{\rho}\cr H&\equiv&
\frac{F_1\; e^{-2\Phi} }{C^2} = A\;\tilde A^p\; \tilde
C\;e^{-2\Phi}\;C^{-1}
\label{xdef}
\eea Notar que $H$ es un vector y $F_1$ es otra $1$-forma, por lo tanto
$H\,C$ es un escalar.  Luego
 \be 
C(\rho)\; d\rho = H\;C\;dx
\ee 
resulta una relaci\'on \'util para escribir la m\'etrica. En t\'erminos
de esta coordenada y denotando con prima las derivadas respecto de
esta variable, las ecuaciones de movimiento (\ref{1})-(\ref{6})
resultan: 
\bea 
(\ln A)''&=& \frac{2-b_p}{8}\;{e^{(2+b_p)\Phi}\over
  (A{\tilde A}^p)^2}\; E'^2\label{A}\\ (\ln \tilde A)''&=&
\frac{2-b_p}{8}\;{e^{(2+b_p)\Phi}\over (A{\tilde A}^p)^2}\; E'^2
\label{tildeA}\\
(\ln \tilde C)''&=& -\frac{2+b_p}{8}\;{e^{(2+b_p)\Phi}\over (A{\tilde A}^p)^2}\; E'^2
\label{tildeC}\\
(\ln e^{-2\Phi})''&=& \Lambda^2\; (H\;C)^2 - \frac{2-b_p}{8}\,(p+1)\;{e^{(2+b_p)\Phi}\over (
A{\tilde A}^p)^2}\; E'^2\label{Phi}\\
\left(\frac{e^{(2+b_p)\Phi}}{(A{\tilde A}^p)^2}\; E'\right)'&=&
(-)^{p+1} \,Q_p\;\frac{(H\;C)^2}{A{\tilde A}^p}\; e^{2\Phi}\;\delta^2_{G^\perp}
\label{E}\\
-(\ln H\,C)'' + (\ln H\,C)'^2 &=& (\ln A)'^2 + p\, (\ln \tilde A)'^2 + (\ln \tilde C)'^2 -
\frac{2-b_p}{8}\;{e^{(2+b_p)\Phi}\over (A{\tilde A}^p)^2}\; E'^2\label{C}\cr
& &
\label{equations}
\eea

Hemos escrito, al final, la ecuaci\'on para la funci\'on $C$, que
tomaremos como v\ii nculo proveniente de la necesidad de fijar el
gauge cuando introducimos la variable $x$.

Para comenzar, resolveremos estas ecuaciones en la regi\'on exterior a
las fuentes. La ecuaci\'on (\ref{E}) en este caso tiene como
soluci\'on a 
 \be E'= q\; \left(A\;\tilde
A^p\right)^2\, e^{-(2+b_p) \Phi} \ee donde $q$ est\'a relacionada con
$Q_p$ por la expresi\'on (obtenida por integraci\'on de
(\ref{ecformal})): \be q = \frac{Q_p}{V_z}\qquad,\qquad V_z
=\int\,dz.\label{qQ} \ee La soluci\'on de (\ref{A}) es \be
A=e^{\alpha\,x}\; \tilde A, \ee mientras que la soluci\'on de
(\ref{tildeC}) resulta \be \tilde C = e^{\gamma\,x}\;\tilde
A^{-\frac{2+b_p}{2-b_p}}, \ee donde $\alpha$ y $\gamma$ son constantes
arbitrarias.  Nos quedan entonces dos ecuaciones, (\ref{tildeA}) y
(\ref{Phi}), m\'as un v\ii nculo (\ref{C}).  Para resolver,
introduciremos las siguientes funciones
\footnote{
Hemos descartado las constantes que pueden ser absorbidas via redefiniciones o re-escaleos de las coordenadas $(x^\mu , z)$.
}
\bea
f_1(x)&\equiv& (H C)^2 = \left(A\;\tilde
A^p\; \tilde C\;e^{-2\Phi}\right)^2 = \left(
e^{(\alpha+\gamma)\,x}\; \tilde
A^{p+1-\frac{2+b_p}{2-b_p}}\;e^{-2\Phi}\right)^2\cr
V_z{}^2\; f_2(x)&\equiv&\left({H\,C \over \tilde C}\right)^2\;e^{(2-b_p)\Phi}
= \left(A\;\tilde A^p\right)^2\;e^{-(2+b_p)\Phi}=
\left(e^{\alpha\,x}\;\tilde A^{p+1}\right)^2\;e^{-(2+b_p)\Phi}\cr .
& &\label{f12def}
\eea
En t\'erminos de estas funciones, la soluci\'on para los campos est\'a dada por
\bea G &=&
\tilde A^2\;\left(-e^{2\alpha\,x}\; dx^0{}^2 + d\vec x ^2\right) + f_1(x)\;dx^2 + e^{2\gamma\,x}\;\tilde
A^{-2\frac{2+b_p}{2-b_p}}\; dz^2\cr
e^{2\sigma\Phi} &=&
e^{2\left(\frac{2+b_p}{2-b_p}\alpha+(p+1)\gamma\right)\,x}\;
\frac{ (V_z{}^2\;f_2)^{p+1 -\frac{2+b_p}{2-b_p}} }{f_1{}^{p+1}}\cr
F_{p+2}&=& V_z\; Q_p\;f_2(x)\; dx\wedge dx^0\wedge\dots\wedge dx^p,\label{gralsolution}
\eea
donde
\bea \tilde A^{2\sigma} &=&
V_z{}^4\;e^{
(-(2-b_p)\alpha+(2+b_p)\gamma)\,x}\;\frac{f_2{}^2}{f_1{}^{\frac{2+b_p}{2}}}\cr
\sigma
&\equiv&\frac{1}{2}\left((2-b_p)(p+1)+\frac{(2+b_p)^2}{2-b_p}\right).
\eea
No es dif\ii cil ver que las ecuaciones (\ref{tildeA})
y (\ref{Phi}) pueden ser reescritas en t\'erminos de $f_1, f_2$ en la
 forma siguiente
\bea
(\ln f_1)''&=&2\,\Lambda^2\; f_1 -{2 +b_p\over
4}\, Q_p{}^2\; f_2\cr (\ln f_2)''&=& \frac{2+b_p}{2}\,\Lambda^2\; f_1 +
\frac{(2-b_p)^2}{16}\,(p+1)\,Q_p^2\;f_2, \label{system}
\eea
mientras que el v\ii nculo (\ref{C}) resulta
\bea
-\frac{1}{2}\,(\ln
f_1)''+\frac{1}{4}\,(\ln f_1)^{'\:2} &=& \left((\ln\tilde A)' +
\alpha\right)^2 +p\,(\ln\tilde A)'{}^2
+\left(-\frac{2+b_p}{2-b_p}\,(\ln\tilde A)' + \gamma\right)^2\cr
&-& \frac{2-b_p}{8}\, Q_p{}^2\; f_2\cr
2\,\sigma\,(\ln\tilde A)' &=& -
(2-b_p)\,\alpha + (2+b_p)\,\gamma + \left(
\ln\frac{f_2{}^2}{f_1{}^{\frac{2+b_p}{2}}}\right)'\qquad.\label{constraint}
\eea

En conclusi\'on, la soluci\'on general (\ref{gralsolution}) para
$p$-branas que llenan todo el espacio-tiempo en teor\ii as no cr\ii
ticas ( space-time filling, non critical $p$-branes), est\'a determinada
por el sistema (\ref{system}) y el v\ii nculo (\ref{constraint}).

En lo que sigue, haremos una b\'usqueda sistem\'atica de las posibles
soluciones, empezando por el caso m\'as simple.

\section{Soluciones no cargadas $Q_p=0$; el vac{\'\i}o}\label{no_carg}

Las soluciones no cargadas corresponden a las soluciones de vac\ii o
de las teor\ii as no cr\ii ticas.
En este caso, el sistema (\ref{system}) se reduce a
\bea
(\ln f_1)''&=&2\,\Lambda^2\; f_1 \cr
(\ln f_2)''&=& \frac{2+b_p}{2}\,\Lambda^2\; f_1
\eea
Reemplazando la primer soluci\'on en la segunda,
 se obtiene $f_2$ en t\'erminos de $f_1$
\be
f_2(x) = e^{\epsilon_0 +\epsilon\, x}\; f_1(x)^{\frac{2+b_p}{4}}
\ee
donde $\epsilon_0, \epsilon$ son constantes arbitrarias.
Para resolver $f_1$, reescribimos la primer ecuaci\'on en la forma siguiente
\be
(\ln f_1)''=2\,\Lambda^2\; f_1\;\;\longleftrightarrow\;\;
{1\over f_1}{d\over d x}\left({1\over f_1}{d\over d x}f_1(x)\right)=2 \,\Lambda ^2
\;\;\longleftrightarrow\;\; \frac{d^2 f_1}{dy^2}=2 \,\Lambda ^2\label{uncharf1eq}
\ee
donde hemos introducido formalmente la primitiva $y(x)$ 
(relevante a menos de una constante)
\be
y(x)=\int^x \;dx\; f_1(x)\;\;\longleftrightarrow\;\; f_1(x) = y'(x)
\ee
Integrando trivialmente en (\ref{uncharf1eq}) obtenemos
\be
f_1(x)= y'(x) = \Lambda^{\:2}\; y(x)^{\:2}+ 2\, \beta \; y(x)+\gamma\label{f1yprime}
\ee
donde $\beta, \gamma$ son constantes de integraci\'on arbitrarias, y las 
soluciones generales son
\be
x - x_0= \int^{y(x)}{dy\over \Lambda^{\:2}\; y^{\:2}+ 2 \,\beta\; y+\gamma}
= \left\{\begin{array}{lcl}
{1\over 2\,\sqrt{-\Delta}}\,\ln\left|{\Lambda^2 y+\beta-\sqrt{-\Delta}\over
\Lambda^2 y+\beta+\sqrt{-\Delta}}\right| \;\;\; &,&\;\;\;\Delta < 0\cr
-{1\over \Lambda^2 y+ \beta}\;\;\; &,&\;\;\;\Delta=0\cr
{1\over \sqrt{\Delta}}\, \arctan{\Lambda ^2 y+ \beta \over \sqrt{\Delta}},
\;\;\; &,&\;\;\;\Delta> 0
\end{array}\right.
\ee
donde $\Delta=\Lambda^2\,\gamma - \beta^2$.
Vemos entonces que hay tres ramas posibles para las soluciones, dependiendo del signo de $\Delta$.
Los correspondientes valores de $y(x)$ son:
\be
\Lambda^2\,y(x) +\beta =
\left\{\begin{array}{lcl}
\begin{array}{lcl}
-\sqrt{-\Delta}\:\coth\left(\sqrt{-\Delta}\,(x-x_0)\right)
\;\;&,&\;\; |\Lambda^2\,y(x) +\beta|>\sqrt{-\Delta}\cr
-\sqrt{-\Delta}\:\tanh\left(\sqrt{-\Delta}\,(x-x_0)\right)
\;\;&,&\;\; |\Lambda^2\,y(x) +\beta|<\sqrt{-\Delta}\cr
\end{array}&,&\Delta < 0\cr
-\frac{1}{x-x_0}&,&\Delta=0\cr
\sqrt{\Delta}\, \tan\left(\sqrt{\Delta}\,(x-x_0)\right)&,&\Delta> 0
\end{array}\right.
\ee
que llevan, de acuerdo con (\ref{f1yprime}), a
\be
f_1(x) =
\left\{\begin{array}{lcl}
\begin{array}{lcl}
-{\Delta \over \Lambda^2} \:{1\over \sinh^2\left(\sqrt{-\Delta}\,(x-x_0)\right)}
\;\;&,&\;\; |\Lambda^2\,y(x) +\beta|>\sqrt{-\Delta}\cr
+{\Delta \over \Lambda^2} \:{1\over \cosh^2\left(\sqrt{-\Delta}\,(x-x_0)\right)}
\;\;&,&\;\; |\Lambda^2\,y(x) +\beta|<\sqrt{-\Delta}\cr
\end{array}\;\;&,&\;\;\Delta < 0\cr
\frac{1}{\Lambda^2\,(x-x_0)^2}\;\;&,&\;\;\Delta=0\cr
{\Delta\over \Lambda^2}\:{1\over \cos^2\left(\sqrt{\Delta}\,(x-x_0)\right)}
\;\;&,&\;\;\Delta> 0\qquad .
\end{array}\right.\label{f1sn}
\ee 
Vemos que $f_1$ depende de dos par\'ametros libres, $x_0$ y $\Delta$, mientras que la soluci\'on total (\ref{gralsolution})
depende adem\'as en forma no trivial (ver m\'as adelante) de $\alpha,
\epsilon, \gamma$.  
Pasemos a analizar cada una de estas ramas por separado.

\subsection{Soluciones con $\Delta=0$; El dilat\'on lineal}

En este caso, la ecuaci\'on del v\ii nculo (\ref{constraint}) impone
la condici\'on $\alpha=\epsilon=\gamma=0$, y de (\ref{f1sn}) tenemos
\bea f_1&=&{1\over \Lambda^2\, (x-x_0)^2}\cr f_2&=&
e^{\epsilon_0}\;\left({1\over \Lambda^2\,
  (x-x_0)^2}\right)^\frac{2+b_p}{4}. 
\eea 

Introduciendo la variable $Y
= -\Lambda^{-1}\,\ln(\Lambda\,|x-x_0|)$, y luego de reescaleos y
redefiniciones triviales, obtenemos
\bea G&=& \eta_{1,p}+d Y^2+
r'_0{}^2\; d\theta^2\cr \Phi&=&\Phi_0 -\frac{1}{r_0}\,Y\qquad,\qquad
r_0 \equiv \frac{2}{\Lambda}. 
\eea 
Esta soluci\'on es el producto directo del espacio-tiempo de Minkowski
($p+1$)-dimensional y la soluci\'on de dilat\'on lineal veces una $S^1$
(asumiendo $\theta\sim\theta+2\,\pi$) de radio arbitrario $r'_0$. El
dilat\'on lineal es una soluci\'on ampliamente conocida y discutida en
la literatura (ver \cite{Myers:1987fv}, \cite{polcho1}). Se conoce
adem\'as que una rotaci\'on de Wick sobre esta soluci\'on lleva a una
nueva soluci\'on de tipo cosmol\'ogica \cite{Antoniadis:1990uu}
\footnote{ Sobre modelos cosmol\'ogicos de teor\ii as de cuerdas, ver
  \cite{Tseytlin:1991xk}, \cite{Bergshoeff:2005bt}.}.

\subsection{Soluciones con $\Delta<0$; El cigarro y m\'as}

De la definici\'on (\ref{f12def}), $f_1= (H \,C)^2 >0$, y por lo tanto
la soluci\'on f\ii sicamente relevante en (\ref{f1sn}) es la primera,
luego \bea f_1 (x) &=& {-\Delta \over \Lambda^2} \:{1\over
  \sinh^2\left(\sqrt{-\Delta}\,(x-x_0)\right)}\cr f_2(x)&=&
e^{\epsilon_0 +\epsilon\, x}\; \left({-\Delta \over \Lambda^2}
\:{1\over \sinh^2\left(\sqrt{-\Delta}\,(x-x_0)\right)}
\right)^{\frac{2+b_p}{4}} \eea despu\'es de redefiniciones varias, la
soluci\'on general que se obtiene es:
\bea G &=& -e^{-2a\,x}\; dx^0{}^2 +e^{-2e\,x}\;d\vec x{}^2
+\frac{1}{\Lambda^2}\;\frac{dx^2}{\sinh^2 x}+e^{-2g\,x}\;d z^2\cr
e^{2\Phi}&=& 2\, e^{2\,\Phi_0}\; e^{-\varphi\, x}\; |\sinh
x|\qquad,\qquad \varphi\equiv a + p\,e + g \eea mientras que la
ecuaci\'on de v\ii nculo (\ref{constraint}) resulta \be a{}^2 +
p\,e{}^2 + g{}^2 = 1.\label{consq0} \ee Podemos poner la soluci\'on en
una forma m\'as familiar introduciendo las variables $\rho$ y
$\theta$ \be e^{-x} \equiv \tanh\rho\qquad,\qquad \theta
\equiv\frac{\Lambda}{2}\, z, \ee en t\'erminos de las cuales, y despu\'es
de re-escalear las coordenadas $x^\mu$, la soluci\'on se
escribe como \bea G&=& -(\tanh \rho)^{2a}\;dx^0{}^2+(\tanh
\rho)^{2e}\;d\vec x{}^2 + {4\over \Lambda^2}\;\left(d\rho^2+(\tanh
\rho)^{2g}\;d\theta^2\right)\cr e^{2 \Phi}&=&e^{2
  \Phi_0}\:\:{(\sinh\rho)^{\varphi-1}\over
  (\cosh\rho)^{\varphi+1}}\qquad,\qquad \varphi\equiv a+p\,e+g\cr
1&=&a^2 + p\,e^2 + g^2.\label{vacuadelta<0} \eea

La soluci\'on queda determinada, adem\'as de por la constante
$\Phi_0$, por tres par\'ametros $a , e, g,$ que deber\'an satisfacer la
ecuaci\'on del v\ii nculo en (\ref{vacuadelta<0}). Toda esta familia
de soluciones va asint\'oticamente, para $\rho$ grande, a la
soluci\'on de dilat\'on lineal de la secci\'on anterior.  De esta
familia, hay algunas soluciones particulares que son bien conocidas.
\bigskip

\noindent\underline{$a =e =0\;,g =1\:\:\:$}
\bigskip

En este caso, la soluci\'on de vac\ii o que se obtiene corresponde al
producto directo del espacio-tiempo de Minkowski ($p+1$)-dimensional
con el cigarro con escala $r_0 =\frac{2}{\Lambda}$ \bea
G&=&\eta_{1,p}+r_0^2\;\left( d^2\rho+ \tanh^2\rho\;d\theta^2\right)\cr
e^{2\Phi}&=&e^{2 \Phi_0}\:\:{1\over \cosh^2\rho} \eea Aqu{\ii}, se
impone periodicidad, $\theta\sim \theta + 2\,\pi$, para evitar la
singularidad c\'onica en el origen $\rho=0$.
\bigskip

\noindent\underline{$a =e =0\;,g =-1\:\:\:$}
\bigskip

Obtenemos, en este caso, la soluci\'on de vac\ii o que corresponde al
producto directo del espacio-tiempo ($p+1$)-dimensional de Minkowski
veces la trompeta

\bea G
&=&\eta_{1,p}+r_0^2\;\left(d^2\rho+\coth^2\rho\;d\theta^2\right)\cr
e^{2\Phi}&=&e^{2\Phi_0}\:\:{1\over\sinh^2\rho}.
 \eea 

Esta soluci\'on es singular en el origen $\rho=0$. Sin embargo, es bien
sabido que el cigarro y la trompeta son soluciones T-duales una de la
otra. Ambas soluciones corresponden a soluciones exactas de teor\ii as CFT
en la hoja de mundo, conocidas como modelos de
Wess-Zumino-Witten-Novikov (WZWN) {\it gaugeados} $SL(2,\Re)/U(1)$, con
gauge vectorial y axial, respectivamente \cite{Witten:1991yr},\cite{Dijkgraaf:1991ba}.

Es interesante notar que todas las soluciones presentan un dilat\'on no
trivial. Sin embargo, para aquellas en las que $\varphi\geq 1$, el
acoplamiento de la cuerda $\,g_s\equiv e^\Phi\,$ est\'a acotado en todos
lados, lo que asegura que el set-up perturbativo de la teor\ii a de
cuerdas est\'a bajo control.  Adem\'as de estas soluciones, hay una
familia 2-param\'etrica de soluciones, aquellas que saturan la
desigualdad $\varphi=1 $, con caracter\ii sticas especiales. En primer
lugar, \'estas presentan el {\it mismo} dilat\'on que la soluci\'on del
cigarro,
y lo que es m\'as, del escalar de Ricci de estas soluciones, 
\be R =
\frac{\Lambda^2}{4}\; \frac{4\,\sinh^2\rho -(\varphi -1)^2 +1 -
  a^2-p\,e^2-g^2}{\sinh^2\rho\,\cosh^2\rho} =
\frac{1}{r_0{}^2}\;\frac{4\,\sinh^2\rho -(\varphi -1)^2
}{\sinh^2\rho\,\cosh^2\rho} \ee 
donde, en el \'ultimo paso, hemos usado el v\ii nculo
(\ref{vacuadelta<0}). Se ve que estas soluciones son las \'unicas
regulares, su escalar de curvatura es el mismo que el del cigarro.
En contraste con la soluci\'on del cigarro, las otras soluciones
presentan un factor de wrapping que diverge en $\rho=0$.  
Para la familia 2-param\'etrica $\varphi = 1$, el tensor de Ricci resulta
\bea
R_{00} = -\frac{2\,a}{r_0{}^2}\; \frac{1}{\cosh^2\rho}\qquad&,&\qquad
R_{IJ} = \frac{2\,e}{r_0{}^2}\; \frac{1}{\cosh^2\rho}\;\delta_{IJ}\cr
R_{p+1p+1} = \frac{2}{r_0{}^2}\; \frac{1}{\cosh^2\rho}\qquad&,&\qquad
R_{p+2p+2} = \frac{2\,g}{r_0{}^2}\; \frac{1}{\cosh^2\rho}. \eea

Ser\ii a interesante poder determinar si alguna de estas soluciones
(\ref{vacuadelta<0}) posee una descripci\'on como CFT exacta, como
ocurre con la trompeta y el cigarro.  Cabe mencionar que en la
referencia \cite{Alvarez:2000it} el espacio de Einstein
correspondiente a la soluci\'on $a=e=g=\frac{1}{\sqrt{p+2}}$ ($\varphi
=\sqrt{p+2}$) ha sido considerado, mientras que la sub-familia de vac\ii os
con $a = e$ ha aparecido recientemente \cite{Kuperstein:2004yk}.

\subsection{Soluciones con $\Delta>0$}

La familia de soluciones est\'a dada por
\bea
G &=& -e^{-2\,a\,x}\; dx^0{}^2 +e^{-2\,e\,x}\;d\vec x{}^2
+\frac{1}{\Lambda^2}\;\frac{dx^2}{\cos^2 x}+e^{-2\,g\,x}\;dz^2\cr
e^{2\Phi}&=&2\, e^{2\,\Phi_0}\;e^{ - \varphi\, x}\;|\cos x|\qquad,\qquad \varphi \equiv a + p\,e + g
\eea
donde el v\ii nculo (\ref{consq0}) es ahora reemplazado por
\be
a{}^2 + p\,e{}^2 + g{}^2 = -1.
\ee
Conclu{\'\i}mos que, en este caso, no existe soluci\'on para la rama $\Delta>0$.
\bigskip

Con esto, hemos terminado el an\'alisis de todas las posibles soluciones no cargadas. Pasemos entonces al caso de las soluciones cargadas.

\section{Soluciones cargadas NSNS, $Q_p\neq 0\;,\;b_p =-2$}

Desde el punto de vista de teor\ii a de cuerdas, estas soluciones son
s\'olo relevantes para el caso particular de $p=1$, con la
identificaci\'on del campo de gauge $A_2$ con la 2-forma de gauge de
Kalb-Ramond usual, $B$, bajo la cual se carga la cuerda fundamental.
Por completitud, dejamos libre $p$, que ser\'a fijado luego para los
casos f\ii sicamente relevantes.

En este caso, las ecuaciones (\ref{system}) para las $f_i$'s se desacoplan,
\bea
(\ln f_1)''&=&2\,\Lambda^2\; f_1 \cr
(\ln f_2)''&=& (p+1)\,Q_p{}^2\;f_2.
\eea
Las soluciones para estas ecuaciones ya fueron encontradas en la
secci\'on anterior. De (\ref{f1sn}) y de la positividad de las
funciones $f_i$
\be
f_1(x) =
\left\{\begin{array}{lcl}
{-\Delta_1 \over \Lambda^2} \:{1\over \sinh^2\left(\sqrt{-\Delta_1}\,(x-x_1)\right)}
\;\;&,&\;\;\Delta_1<0\cr
\frac{1}{\Lambda^2\,(x-x_1)^2}\;\;&,&\;\;\Delta_1=0\cr
{\Delta_1\over \Lambda^2}\:{1\over \cos^2\left(\sqrt{\Delta_1}\,(x-x_1)\right)}
\;\;&,&\;\;\Delta_1> 0
\end{array}\right.
\ee
\be
f_2(x) = \left\{\begin{array}{lcl}
{-2\,\Delta_2 \over (p+1)\,Q_p{}^2} \:{1\over \sinh^2\left(\sqrt{-\Delta_2}\,(x-x_2)\right)}
\;\;&,&\;\;\Delta_2<0\cr
\frac{2}{(p+1)\,Q_p{}^2}\,\frac{1}{(x-x_2)^2}\;\;&,&\;\;\Delta_2=0\cr
{2\,\Delta_2\over (p+1)\,Q_p{}^2}\:{1\over \cos^2\left(\sqrt{\Delta_2}\,(x-x_2)\right)}
\;\;&,&\;\;\Delta_2> 0
\end{array}\right.
\ee donde $\Delta_i , x_i,$ son constantes arbitrarias.  Los campos
son expresados como en (\ref{gralsolution}):
\bea G
&=&e^{-\frac{2\alpha}{p+1}\,x}\; f_2{}^\frac{1}{p+1}\;
\left(-e^{2\alpha\,x}\;dx^0{}^2 +\;d\vec x{}^2\right)
+f_1(x)\;dx^2+e^{2\gamma\,x}\;dz^2\cr
e^{4\Phi}&=&V_z{}^2\,e^{2\gamma\,x}\;\frac{f_2}{f_1}\cr F_{p+2} &=&
Q_p\, f_2(x)\; dx\wedge dx^0\wedge\dots\wedge dx^p 
\eea
 y, entonces,
quedan determinados por las elecciones posibles de $f_1$ y $f_2$, y la
imposici\'on del v\ii nculo (\ref{constraint}), que queda expresado de esta manera
\be
\frac{\Delta_2}{p+1} - \Delta_1 = \frac{p}{p+1}\;\alpha{}^2 +
\gamma{}^2. 
\ee 
Las soluciones dependen de los par\'ametros $\alpha, \gamma,
\Delta_i , x_i$.  

Estas soluciones se detallan a continuaci\'on \footnote{Alertamos al
  lector que las expresiones de los campos se obtuvieron luego de
  redefiniciones, etc, como en la secci\'on 3, y por lo tanto, la
  coordenada $x$ no est\'a definida en (\ref{xdef}).}.
\bigskip

\begin{enumerate}
\item \underline{$\Delta_1 = \Delta_2 = 0$}
\bea
G &=& |x-x_0|^{-\frac{2}{p+1}}\;\eta_{1,p}
+\frac{1}{\Lambda^2}\;\frac{dx^2}{x^2}+ R_1{}^2\;d\theta^2\cr
e^{2\,\Phi}&=&e^{2\,\Phi_0}\; \left|1-\frac{x_0}{x}\right|^{-1}\cr
F_{p+2} &=& s(q)\;\sqrt{\frac{2}{p+1}}\;\frac{1}{(x-x_0)^2}\;
dx\wedge dx^0\wedge\dots\wedge dx^p
\eea
donde $s(q) \equiv sign (q)$, y $\theta\sim\theta +2\,\pi$, y donde el
radio de la $S^1$ es fijado a
\be
R_1 = \sqrt{\frac{p+1}{2}}\;\frac{|Q_p|\,e^{2\Phi_0}}{2\,\pi\,\Lambda}\qquad.\label{rs1}
\ee

Dependiendo del valor de  $x_0$, tenemos 3 posibilidades.

\bigskip

Si $x_0 = 0$, con $x= (r_0\,u)^{-p-1}$, la soluci\'on resulta
\bea
G &=& r_0{}^2\; \left( \frac{d u^2}{u^2} + u^2\;\eta_{1,p}\right) +R_1{}^2\; d\theta^2
\qquad,\qquad r_0\equiv\frac{p+1}{\Lambda}\cr
\Phi&=& \Phi_0\cr
F_{p+2} &=& -s(q)\sqrt{2\,(p+1)}\;r_0{}^{p+1}\; u^p\;du\wedge dx^0\wedge\dots\wedge dx^p.
\label{adss1}
\eea

\bigskip

Se trata de un espacio $AdS_{1,p+1}|_{r_0}\times S^1|_{R_1}$, con
dilat\'on constante.  Este fondo (para el caso de $p=1$) es bien
conocido; es una soluci\'on exacta (super) CFT definida por el modelo
de WZWN $Sl(2,\Re)_{-k}\times U(1)|_{R_1}$ con nivel $(k) k-2 =
\frac{4}{\alpha'\,\Lambda^2}$ \cite{Maldacena:2000hw}.

Si $x_0<0$, despues de re-escalear $x^\mu, x\rightarrow |x_0|^\frac{1}{p+1}\,
x^\mu\, ,|x_0|\,x$, e introducir la variable radial $r$
\be
\frac{1}{1 + x} = 1- \left(\frac{r_h}{r}\right)^{p+1} = f(r) ,\qquad,\qquad r_h <r<\infty
\ee

\bigskip

la soluci\'on toma la forma,

\bea
G &=& f(r)^{\frac{2}{p+1}}\;\eta_{1,p} + r_0{}^2\; \frac{dr^2}{r^2\,f(r)^2}
+ R_1{}^2\; d\theta^2\cr
e^{2\Phi}&=& e^{2\Phi_0}\;\left(\frac{r_h}{r}\right)^{p+1}\cr
F_{p+2} &=& s(q)\;\sqrt{2\,(p+1)}\;\frac{r_h{}^{p+1}}{r^{p+2}}\;
\;dr\wedge dx^0\wedge\dots\wedge dx^p. \label{f1ld}
\eea Se ve que esta soluci\'on interpola entre el vac{\'\i}o de dilat\'on
lineal de la secci\'on 4.2 (para $r$ grande, $r\gg r_h$) y el
espacio $AdS_{1,p+1}|_{r_0}\times S^1|_{R_1}$ de (\ref{adss1}) (para
$r\rightarrow r_h{}^+$). Nos vemos en este punto tentados de
identificar esta soluci\'on (para $p$=1) con la cuerda fundamental
($F1$) sobre un vac\ii o de dilat\'on lineal, el espacio $AdS$ ser\ii a el
near horizon limit que borra la regi\'on del vac\ii o, como pasa en
teor\ii as de cuerdas cr\ii ticas con las soluciones de branas usuales
\footnote{Este {\it near horizon limit} puede hacerse formalmente
  introduciendo la variable $u$, \be (r_0\, u)^{p+1} =
  \left(\frac{r}{r_h}\right)^{p+1} - 1 \ee y tomando el l\ii mite de
  baja energ\ii a $r_0\rightarrow 0$, a $u$ fijo.  Una relaci\'on
  similar es considerada en la referencia \cite{Alvarez:2001ta}.}.  La curvatura escalar se muestra en la Figura~\ref{alpha}.

\begin{figure}[!ht]
\centering
\includegraphics[scale=0.5,angle=-90]{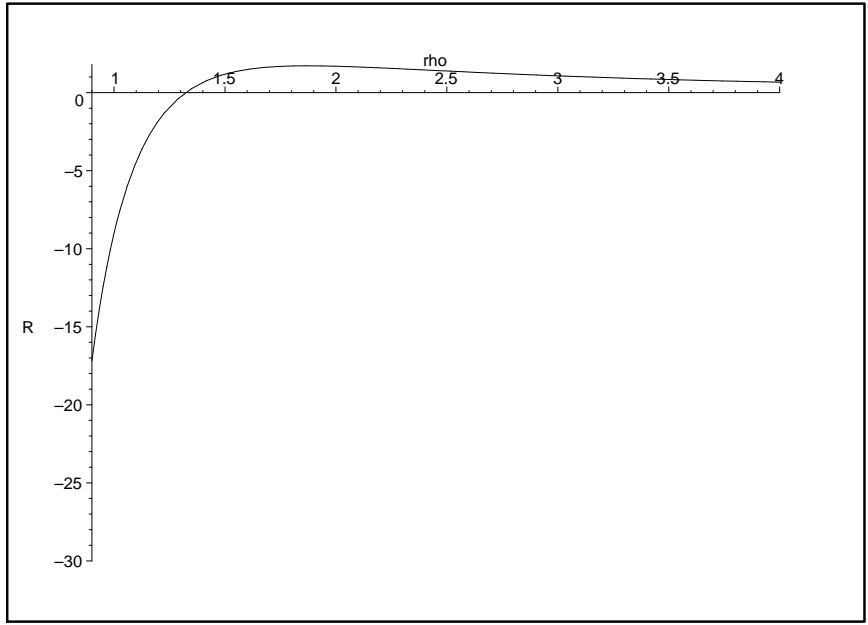}
\caption[Curvatura escalar correspondiente a (\ref{f1ld}).]{La curva muestra $\alpha'\,R$ como funci\'on de $\rho=r/r_h$,
donde $R$ es la curvatura escalar correspondiente a (\ref{f1ld}) y $1< \rho < \infty$.}
\label{alpha}
\end{figure}

\bigskip

Finalmente, si $x_0>0$, obtenemos
\bea
G &=& r_0{}^2\;\left(u^2\;\eta_{1,p} +\left(1 + \left(r_0\,u\right)^{p+1}\right)^{-2}\;
\frac{d u^2}{u^2}\right) + R_1{}^2\;d\theta^2 \cr
e^{2\Phi}&=& e^{2\Phi_0}\;\left(1 + \left(r_0\,u\right)^{p+1}\right)\cr
F_{p+2} &=& -s(q)\;\sqrt{2\,(p+1)}\;r_0{}^{p+1}\;\;u^p\;du\wedge dx^0\wedge\dots\wedge dx^p,
\label{5.10}
\eea
que va a  un espacio  $AdS_{1,p+1}|_{r_0}\times S^1|_{R_1}$ para $r_0\,u\rightarrow 0$,
pero que es singular en la regi\'on $u$ grande, $r_0\,u\rightarrow \infty$.
La curvatura escalar se muestra en la Figura~\ref{R510}.

\bigskip

\begin{figure}[!ht]
\centering
\includegraphics[scale=0.5,angle=-90]{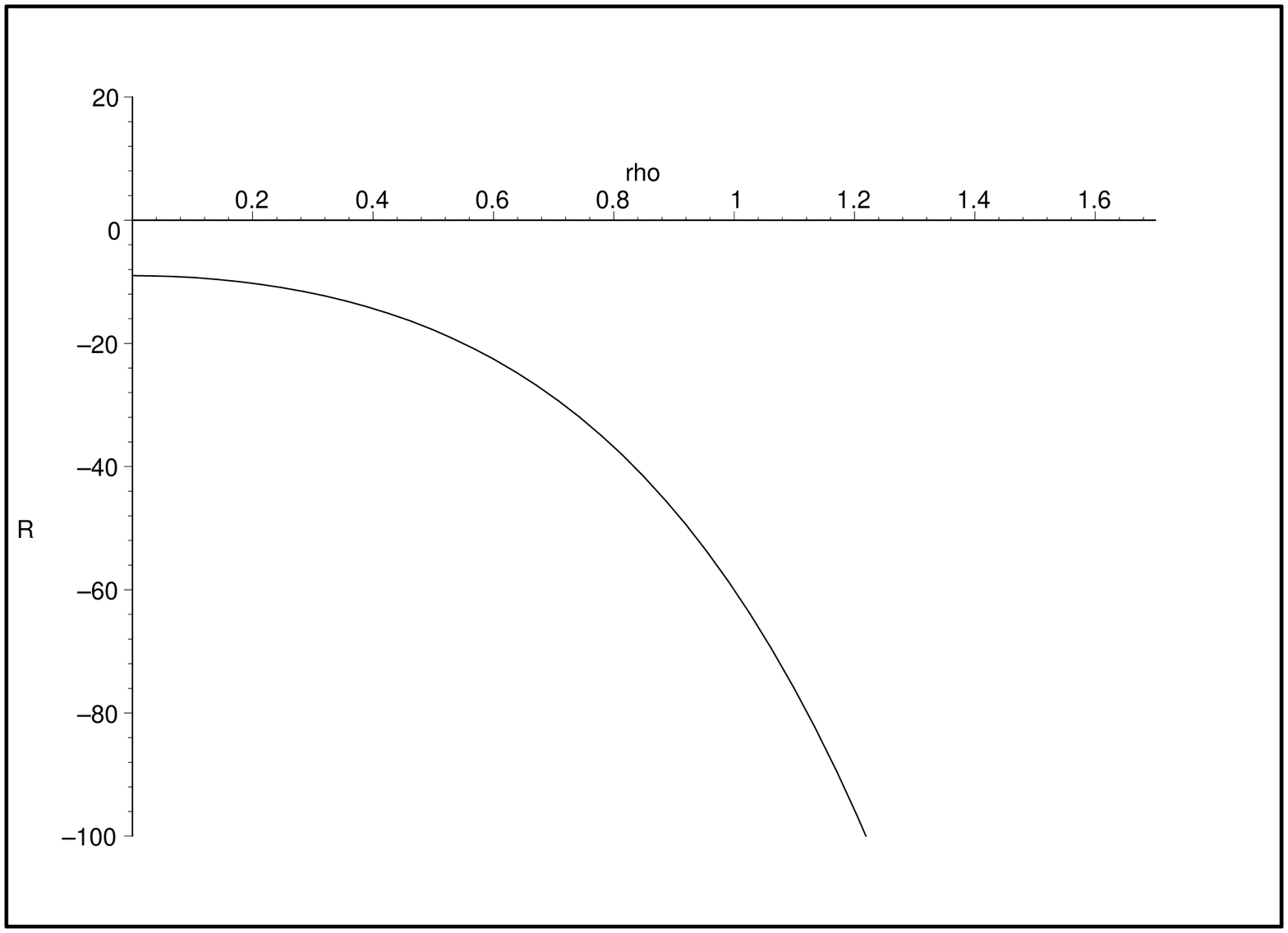}
\caption[Curvatura escalar correspondiente a (\ref{5.10}).]{La curva muestra $\alpha'\,R$ como funci\'on de $\rho=r_0\,u$,
donde $R$ es el escalar de curvatura correspondiente a (\ref{5.10}) y $0< \rho < \infty$.}
\label{R510}
\end{figure}

\item \underline{$\Delta_1 <0\;,\; \Delta_2 = 0$} \bea G &=&
  \frac{e^{-\frac{2\,a}{p+1}\,x}}{|x-x_0|^\frac{2}{p+1}}\;
  \left(-e^{2\,a\, x}\, dx^0{}^2 + d{\bf x}^2\right) +
  \frac{1}{\Lambda^2}\, \frac{dx^2}{\sinh^2x} + R_1{}^2\,e^{2\,g\,
    x}\,d\theta^2\cr e^{2\,\Phi}&=& e^{2\,\Phi_0}\, e^{g\, x}\;
  \left|\frac{\sinh x}{x-x_0}\right|\cr F_{p+2} &=&
  s(q)\;\sqrt{\frac{2}{p+1}}\; \frac{1}{(x-x_0)^2}\; dx\wedge
  dx^0\wedge\dots\wedge dx^p\cr 1&=&\frac{p}{p+1}\,a^2 + g^2 \eea
  donde $R_1$ es el dado en (\ref{rs1}).  Se trata de una familia
  3-param\'etrica de soluciones, con 3 ramas diferentes, dependiendo
  del signo de $x_0$.  Nos concentramos en la soluci\'on con $a=0,
  g=-1$, que tiene un dilat\'on acotado para $x$ grande.  Hacemos un
  cambio de variables conveniente $e^{-x} = \tanh\rho$.

Si $x_0=0$, tenemos
\bea
G&=&|\ln\tanh\rho|^{-\frac{2}{p+1}}\;\eta_{1,p} +
\frac{4}{\Lambda^2}\;d\rho^2 + R_1{}^2\;\tanh^2\rho\;d\theta^2\cr
e^{-2\Phi}&=& 2\;e^{-2\Phi_0}\; \cosh^2\rho\;|\ln\tanh\rho|\cr
F_{p+2} &=&
-s(q)\;\sqrt{\frac{2}{p+1}}\;\frac{1}{\sinh\rho\;\cosh\rho\;(\ln\tanh\rho)^2}
\;d\rho\wedge dx^0\wedge\dots\wedge dx^p\cr \label{5.12}
& & \eea

La soluci\'on resulta, para $\rho$ grande, asint\'otica a
$AdS_{1,p+1}|_\frac{p+1}{\Lambda}\times S^1{}|_{R_1}$ soluci\'on de
(\ref{adss1}), pero es singular en $\rho=0$, donde el volumen de mundo
de la brana se reduce a cero, y el espacio transverso $\Re^2$ presenta
una singularidad c\'onica en el origen. \'Esta puede salvarse imponiendo
la relaci\'on $R_1=\frac{2}{\Lambda}$ (y entonces,
$\,|Q_p|\,e^{2\Phi_0}\sim 1$), en cuyo caso el espacio transverso es
justo el cigarro. El escalar de curvatura se muestra en la figura~\ref{R512}.

\begin{figure}[!ht]
\centering
\includegraphics[scale=0.5,angle=-90]{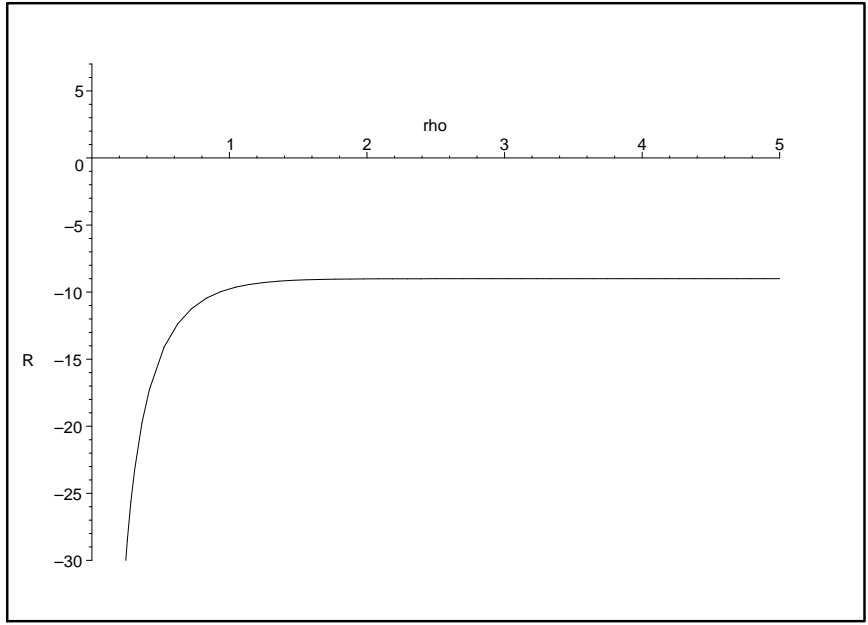}
\caption[Curvatura escalar correspondiente a (\ref{5.12}).]{La curva
  muestra $\alpha'\,R$ como funci\'on de $\rho$, donde $R$ es el
  escalar de curvatura correspondiente a (\ref{5.12}),
  $r_0=2/\Lambda$, y $0< \rho < \infty$.} \label{R512}
\end{figure}

Si $x_0\neq 0$, luego de reescalear $x^\mu\rightarrow
|x_0|^\frac{1}{p+1}\,x^\mu$, y de redefinir 
$e^{2\Phi_0}\rightarrow 2\,|x_0|\,e^{2\Phi_0}$, la soluci\'on queda,
\bea
G &=&|U(\rho)|^{-\frac{2}{p+1}}\; \;\eta_{1,p} +
\frac{4}{\Lambda^2}\, \left(d\rho^2 +\tanh^2\rho\;d\theta^2\right)\cr
e^{-2\Phi}&=& e^{-2\Phi_0}\,\cosh^2\rho\; |U(\rho)|\cr F_{p+2}
&=& s(q\,x_0)\;\sqrt{\frac{2}{p+1}}\; d U(\rho)^{-1}\wedge
dx^0\wedge\dots\wedge dx^p, \label{eq<=0}
\eea
donde
\footnote{ Esta elecci\'on de $|x_0|$ lleva al cigarro como espacio
  transverso.  } \be U(\rho) = 1 +
\frac{1}{x_0}\,\ln\tanh\rho\qquad,\qquad |x_0|\equiv
\sqrt{\frac{2}{p+1}}\;\frac{2\pi}{|Q_p|}\,e^{-2\Phi_0}<0.\label{f1cigar}
\ee Si $x_0<0$, identificamos la soluci\'on (para $p=1$) con una
cuerda fundamental no cr{\'\i}tica sobre el vac\ii o del cigarro.  Por
otro lado, las soluciones con $x_0 >0$ presentan singularidad en
$\rho=\rho_0,\; \tanh\rho_0 = e^{-\frac{2\pi}{|Q_p|e^{2\Phi_0}}}$.  La
solucion para $x_0<0$ es la que encontramos en la secci\'on
\ref{fund_no_critica} para el caso $d=2$. La curvatura
escalar de estas soluciones se muestra en la Figura~\ref{Rmen0}.

\begin{figure}[!ht]
\centering
\includegraphics[scale=0.5,angle=-90]{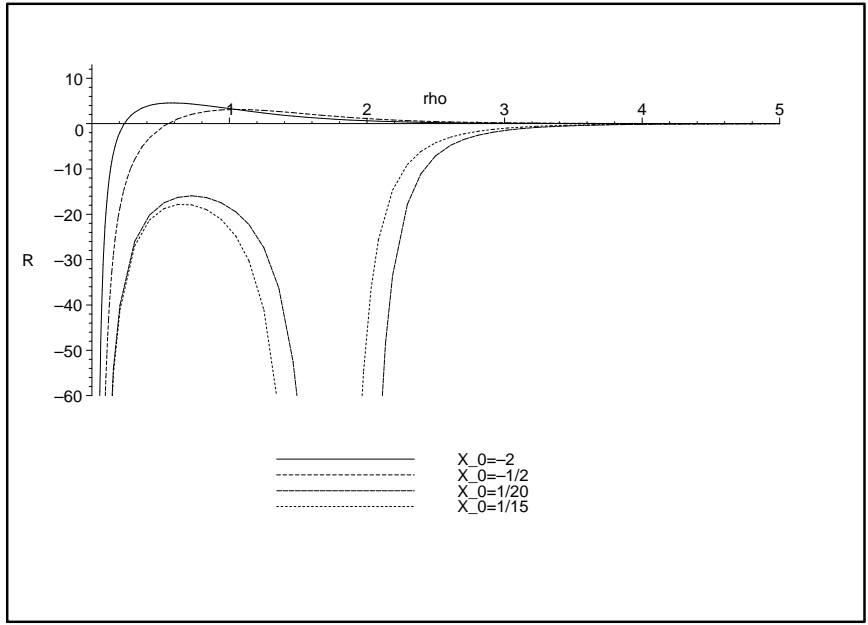}
\caption[Curvatura escalar correspondiente a (\ref{eq<=0}).]{Las
  curvas muestran $\alpha'\,R$ como funci\'on de $\rho$ para
  diferentes valores de $x_0$, donde $R$ es el escalar de curvatura
  correspondiente a (\ref{eq<=0}).}
\label{Rmen0}
\end{figure}

\item \underline{$\Delta_1 <0\;,\; \Delta_2 < 0$}
\bea G &=& \left(\frac{e^{-a\,x}}{|\sinh\left(k\,(x-x_0)\right)|}\right)^{\frac{2}{p+1}}
\left(-e^{2\, a\, x}dx^0{}^2 +d\vec x{}^2\right)
+ \frac{1}{\Lambda^2}\, \frac{dx^2}{\sinh^2x}+R_k{}^2\,e^{2\,g\,x}\;d\theta^2\cr
e^{2\Phi}&=&e^{2\Phi_0}\,\frac{e^{g\,x}\; |\sinh x|}{|\sinh (k\,(x-x_0))|}\cr
F_{p+2}&=&s(q)\;\sqrt{\frac{2}{p+1}}\;\frac{k}{\sinh^2
(k\,(x-x_0))}\; dx\wedge dx^0\wedge\dots\wedge dx^p\cr
1&=&\frac{p}{p+1}\;a^2 +g^2 + \frac{k^2}{p+1}\qquad,\qquad0<k \leq \sqrt{p+1}\label{nsns3}
\eea
La familia de soluciones depende adem\'as de $\Phi_0$ y $x_0$,
de los par\'ametros $a, g, k$, que deben satisfacer el v\ii nculo de (\ref{nsns3}),
mientras que el radio de la $S^1$ est\'a dado por
\be
R_k=\sqrt{\frac{p+1}{2}}\;\frac{|Q_p|\,e^{2\Phi_0}}{2\,\pi\,k\,\Lambda}\qquad.
\label{Rk}
\ee Las soluciones con $g +1-k\leq 0\;$ son especiales en el sentido
que, como en la secci\'on 3.2, tienen el string coupling acotado en
todos lados. M\'as a\'un, su curvatura escalar toma la forma\bea
\frac{1}{\Lambda^{2}}\,R &=& 1 - \frac{5}{2}\;\left(\frac{k\,\sinh
  x}{\sinh(k(x-x_0))}\right)^2 - \left( g\,\sinh x +\cosh x -
k\,\frac{\sinh x}{\tanh(k(x-x_0))} \right)^2\cr & &\label{eq<<0} \eea
de donde se sigue que, si la condici\'on $\,k-1 = g \geq 0\;$ se
cumple, no s\'olo el string coupling est\'a acotado en todos lados, sino
que adem\'as la curvatura es no singular en $x=\infty$. Nos
concentraremos en estas sub-familias.

Si $x_0 = 0$, toda la familia resulta asint\'otica a
$AdS_{1,p+1}|_\frac{p+1}{\Lambda}\times S^1|_{R_k}$ cuando
$x\rightarrow 0$.  En nuestro contexto de cuerdas, esto significa que
la sub-familia de 2 par\'ametros con $g=k-1\geq 0$ y $k =
\frac{2}{3}\,(1+\sqrt{1-\frac{3}{4}a^2})$ determinados por el
v\ii nculo, presenta curvatura regular en todos lados y es
asint\'otica a $AdS_{1,2}|_\frac{2}{\Lambda}\times S^1|_{R_k}$.

Si $x_0>0$, las soluciones desarrollan una singularidad en $x=x_0$ y
no son extensibles a $x=0$.  Por otro lado, si $x_0<0$, las soluciones
resultan asint\'oticas al dilat\'on lineal, cuando $x\rightarrow 0$.  La
curvatura escalar de varios de los casos considerados aqu{\'\i} se muestra
en la Figura~\ref{R1}.

\begin{figure}[ht!]
\centering\includegraphics[scale=0.5,angle=-90]{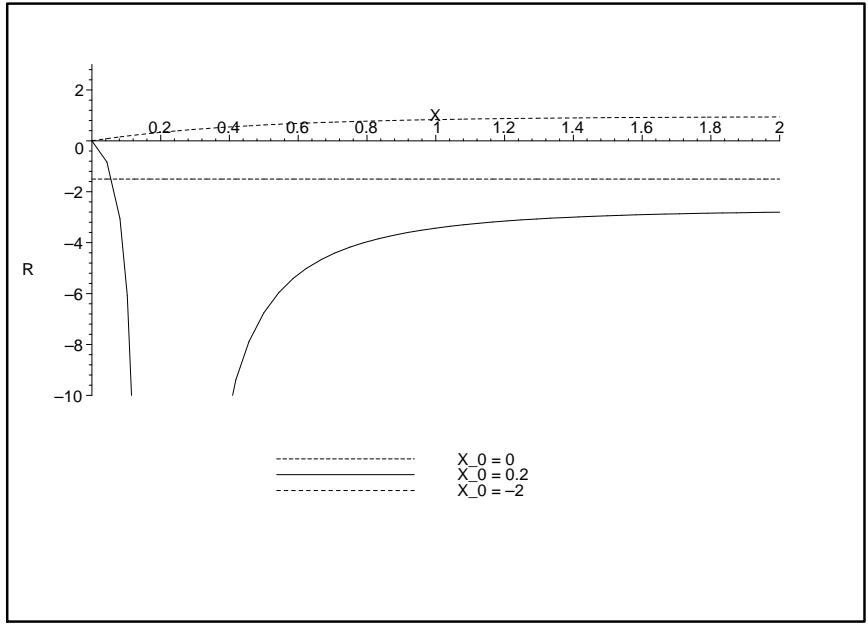}
\caption[Curvatura escalar correspondiente a (\ref{eq<<0}).]{La curva
  muestra $\alpha'\,R$ como funci\'on de $x$ para diferentes valores
  de $x_0$, y para $g=0\,, a=k=1$; $R$ es el escalar de curvatura
  (\ref{eq<<0}).}\label{R1}
\end{figure}

\bigskip

\item \underline{$\Delta_1 = 0\;,\; \Delta_2 > 0$}.
\bea G &=& \left(\frac{e^{-a\, x}}{|\cos(x-x_0)|}\right)^{\frac{2}{p+1}}\;
\left(-e^{2\,a\, x}\,dx^0{}^2 + d{\vec x}^2 \right)
+\frac{1}{\Lambda^2}\;\frac{dx^2}{x^2}+ R_1{}^2\; e^{2g\,x}\,d\theta^2\cr
e^{2\Phi}&=& e^{2\Phi_0}\;\frac{e^{g\,x}\,|x|}{|\cos(x-x_0)|}\cr
F_{p+2} &=&s(q)\;\sqrt{\frac{2}{p+1}}\;\frac{1}{\cos^2(x-x_0)}\;
dx\wedge dx^0\wedge\dots\wedge dx^p\cr
\frac{1}{p+1}&=&\frac{p}{p+1}\,a^2 + g^2\label{eq=>0}
\eea
donde $R_1$ est\'a dado por (\ref{Rk}).
Todos los miembros de la familia tienen el string coupling  acotado y son peri\'odicamente singulares.
No queda claro si \'estas tienen alg\'un sentido f\ii sico.
La curvatura escalar es mostrada en la Figura~\ref{R2} para valores particulares de $a, g$.

\begin{figure}[!ht]
\centering\includegraphics[scale=0.5,angle=-90]{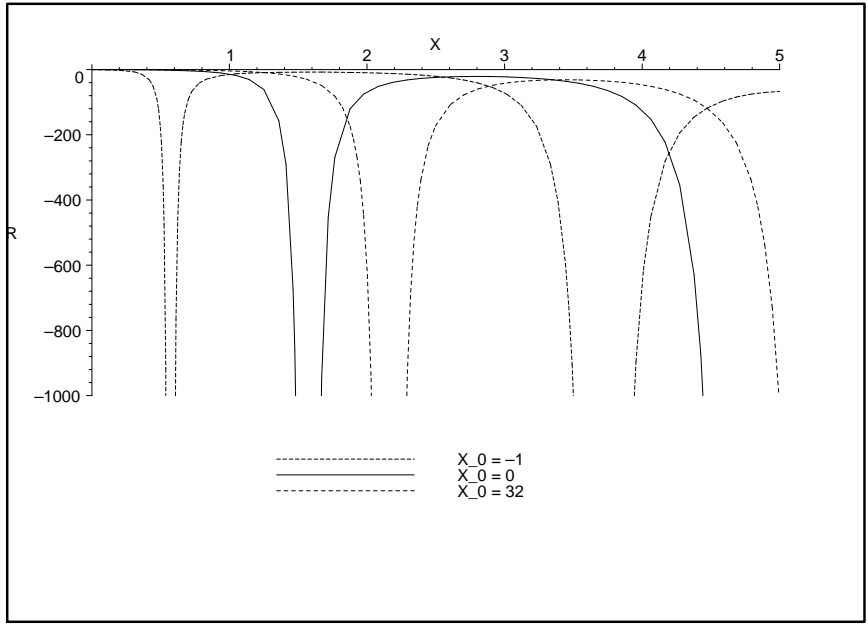}
\caption[Curvatura escalar correspondiente a (\ref{eq=>0}) para $a=1$
  y $g=0$.]{Las curvas muestran $\alpha'\,R$ como funci\'on de $x$
  para diferentes valores de $x_0$, donde $R$ es la curvatura escalar
  correspondiente a (\ref{eq=>0}) para $a=1$ y $g=0$.} \label{R2}
\end{figure}

\item \underline{$\Delta_1 < 0\;,\; \Delta_2 > 0$}.
\bea G &=&\left(\frac{e^{-a\, x}}{|\cos\left(k\,(x-x_0)\right)|}\right)^{\frac{2}{p+1}}
\left(-e^{2\,a\,x} dx^0{}^2 + d\vec x{}^2\right)
+\frac{1}{\Lambda^2}\,\frac{dx^2}{\sinh^2x}+R_k{}^2 e^{2g\,x} d\theta^2\cr
e^{2\Phi}&=&e^{2\Phi_0}\;\frac{e^{g\,x}\;|\sinh x|}{|\cos\left(k\,(x-x_0)\right)|}\cr
F_{p+2} &=&s(q)\;\sqrt{\frac{2}{p+1}}\;\frac{k}{\cos^2\left(k\,(x-x_0)\right)}\;
dx\wedge dx^0\wedge\dots\wedge dx^p\cr
\frac{k^2}{p+1} &=& \frac{p}{p+1}\;a^2 +g^2 -1\qquad,\qquad k>0 \label{eq<>0}
\eea
Como para la soluci\'on (\ref{eq=>0}), \'estas dif{\'\i}cilmente tengan sentido f{\'\i}sico.
El escalar de curvatura se muestra en la Figura~\ref{R3}.

\begin{figure}[!ht]
\centering
\includegraphics[scale=0.5,angle=-90]{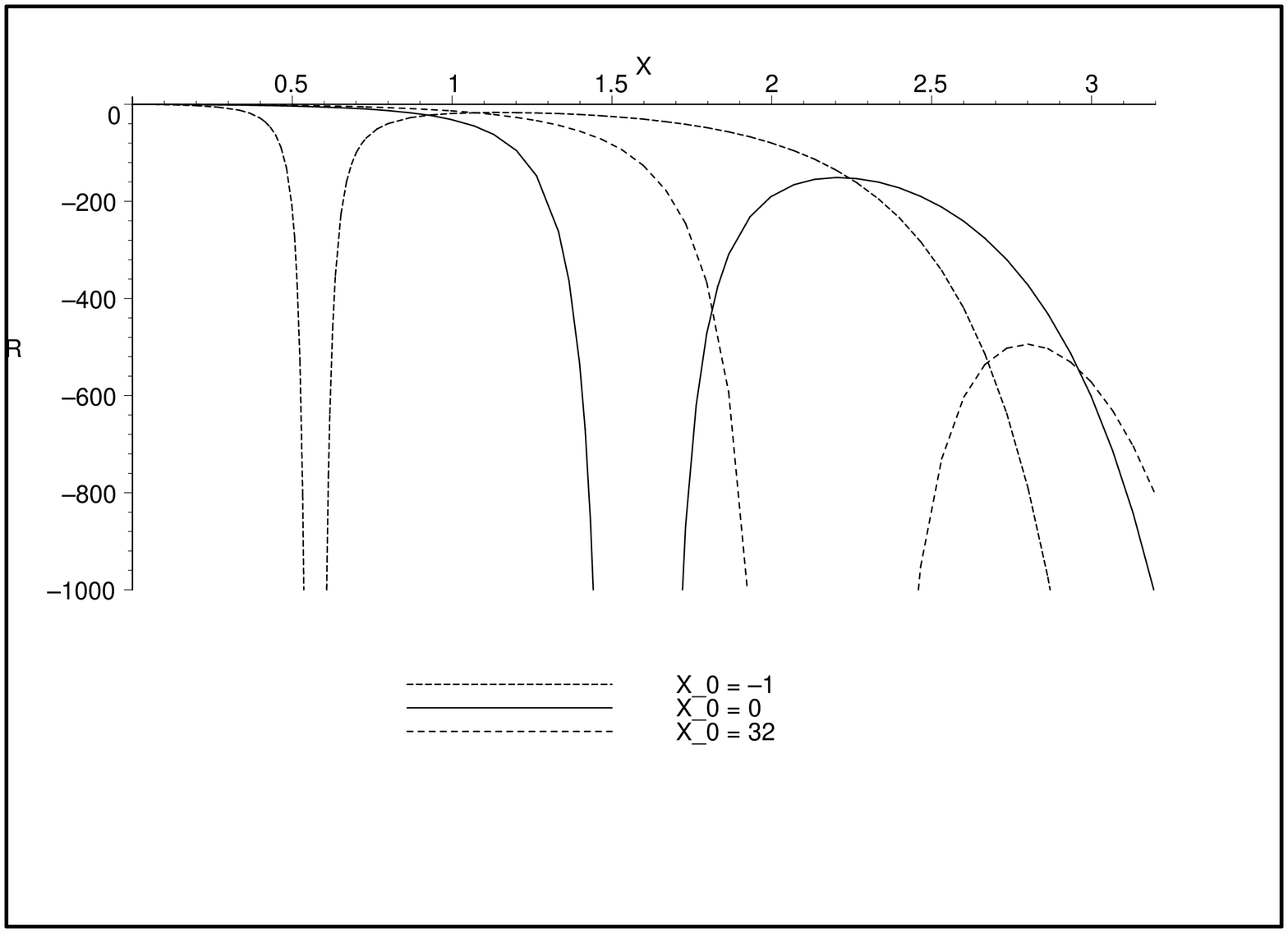}
\caption[Curvatura escalar correspondiente a (\ref{eq<>0}).]{Las
  curvas muestran $\alpha'\,R$ como funci\'on de $x$ para diferentes
  valores de $x_0$ y $k=a=g=1$, donde $R$ es el escalar de curvatura
  correspondiente a (\ref{eq<>0}).}\label{R3}
\end{figure}

\item \underline{$\Delta_1 > 0\;,\; \Delta_2 > 0$}.  \bea G
  &=&\left(\frac{e^{-\,a\,x}}{|\cos\left(k\,(x-x_0)\right)|}\right)^{\frac{2}{p+1}}
  \left(-e^{2\,a\,x}\,dx^0{}^2 +d\vec x{}^2\right)
  +\frac{1}{\Lambda^2}\,\frac{dx^2}{\cos^2x}+R_k{}^2\;e^{2g\,x}\,d\theta^2\cr
  e^{2\Phi}&=&e^{2\Phi_0}\;\frac{e^{g\,x}\;|\cos
    x|}{|\cos\left(k\,(x-x_0)\right)|}\cr F_{p+2}
  &=&s(q)\;\sqrt{\frac{2}{p+1}}\;\frac{k}{\cos^2\left(k\,(x-x_0)\right)}\;
  dx\wedge dx^0\wedge\dots\wedge dx^p\cr \frac{k^2}{p+1} &=&
  \frac{p}{p+1}\;a^2 + g^2 +1\qquad,\qquad k\geq \sqrt{p+1}. \eea El
  comentario hecho para las familias (\ref{eq=>0}) y (\ref{eq<>0})
  tambi\'en se aplica aqu\ii. El comportamiento de la curvatura
  escalar de \'estas es similar a las anteriores y por eso no las
  mostramos.
\end{enumerate}

\section{Soluciones cargadas RR, $b_p = 0$}

En lugar de trabajar con $b_p \neq -2$ arbitrario, nos focalizaremos
en el caso $b_p=0$, que corresponde al caso en que el dilat\'on se
desacopla en el string frame del campo de R-R, $A_{p+1}$.\\ 
De (\ref{gralsolution}) tenemos
\bea G&=& \tilde
A^2\;\left(-e^{2\alpha\,x}\; dx^0{}^2 + {d\vec x}^2\right) +
f_1(x)\;dx^2 + e^{2\gamma\,x}\;\tilde A^{-2}\; dz^2\cr e^{2(p+2)\Phi}
&=&V_z{}^{2p}\; e^{2\left(\alpha+(p+1)\gamma\right)\,x}\;\frac{f_2{}^p
}{f_1{}^{p+1}}\cr F_{p+2} &=& V_z\, Q_p \,f_2(x)\;dx\wedge
dx^0\wedge\dots\wedge dx^p \eea donde \be \tilde A^{2(p+2)} =
V_z{}^4\, e^{2\,(-\alpha+\gamma)\,x}\;\frac{f_2{}^2}{f_1} 
\ee 
El v\ii nculo (\ref{constraint}) deviene
\bea -\frac{1}{2}\,(\ln
f_1)''+\frac{1}{4}\,(\ln f_1)^{'\:2} &=& \left((\ln\tilde A)' +
\alpha\right)^2 +p\,(\ln\tilde A)'{}^2 +\left((\ln\tilde A)'
-\gamma\right)^2 -\frac{Q_p{}^2}{4}\; f_2\cr 2\,(p+2)\,(\ln\tilde A)'
&=& - 2\,\alpha + 2\,\gamma + \left(
\ln\frac{f_2{}^2}{f_1}\right)'\label{consrr} 
\eea 

En contraste con los casos tratados en las secciones anteriores $3$ y
$4$, no hemos sido capaces de resolver en forma completamente general
el conjunto de ecuaciones (\ref{system}). S\'olo hemos logrado
resolverlas a trav\'es del siguiente ans\"atz, 
\be f_1(x) =
f(x)\quad,\qquad f_2(x) = \frac{4}{p+3}\;\frac{\Lambda^2}{Q_p{}^2}\;
f(x).\label{ansatzII} 
\ee 
En este caso, las ecuaciones (\ref{system}) se reducen a
\be 
(\ln f)''=2\;\frac{p+2}{p+3}\,\Lambda^2\;f.
\ee 

Las posibles soluciones para valores no negativos de $f$ 
fueron analizadas en las secciones anteriores (ecuaciones
(\ref{f1sn})). 
Tenemos nuevamente 3 casos, sin embargo, el v\ii nculo (\ref{consrr}) resulta
\be
 -\Delta =\alpha{}^2 +\gamma{}^2
+\frac{2}{p+1}\,\alpha\,\gamma, \label{consrr1} 
\ee 
 que deja afuera autom\'aticamente la opci\'on $\Delta>0$.

\subsection{Soluciones con $\Delta=0$}

Las funciones $f_1 , f_2$ en (\ref{ansatzII}) son \be
f_1(x)=\frac{p+3}{p+2}\,\frac{1}{\Lambda^2}\,
\frac{1}{(x-x_0)^2}\qquad,\qquad f_2(x) =
\frac{4}{p+2}\,\frac{1}{Q_p{}^2}\, \frac{1}{(x-x_0)^2} 
\ee 
y, teniendo en cuenta (\ref{consrr1}), obtenemos la siguiente
soluci\'on
\bea G &=& l_0{}^2\;\left(u^2\;\eta_{1,p} +\frac{d
  u^2}{u^2}\right) + \frac{R_0{}^2}{(l_0\,u)^2}\;d\theta^2 \cr
e^{-2\Phi}&=&e^{-2\Phi_0}\;(l_0\,u)^{2}\cr F_{p+2} &=& s(q)\;
2\sqrt{p+2}\;e^{-\Phi_0}\; l_0{}^{p+2}\; u^{p+1}\;du\wedge
dx^0\wedge\dots\wedge dx^p\label{delta0} 
\eea 
donde la escala $l_0$ y el radio $R_0$ de la $S^1$ son
\be 
l_0 =\sqrt{(p+2)(p+3)}\,\Lambda^{-1}\qquad,\qquad R_0 = \frac{|Q_p|\,
  e^{\Phi_0}}{4\,\pi\,\sqrt{p+2}}\;l_0.\label{l0R0} 
\ee

\'Este es un espacio $AdS_{1,p+1}$ con escala $l_0$, veces una $S^1$ con
un radio que depende de la coordenada $u$. Esto fuerza a un dilat\'on que
corre con dicha variable y que, por lo tanto, lleva a una soluci\'on no
conforme. Para $u$ grande, $u\gg l_0^{-1}$, la $S^1$ toma tama\~no
nulo, y la soluci\'on se reduce a $AdS_{1,p+1}|_{l_0}$; en cambio para
$u\ll l_0{}^{-1}$ es el volumen de mundo de la $Dp$ el que va a cero,
dejando como espacio transverso $AdS_2$, con la misma escala $l_0$.
El tensor de Ricci resulta
\be
R_{\mu\nu} = -\frac{p}{l_0{}^2}\,\eta_{\mu\nu}\qquad,\qquad
R_{p+1,p+1}= -\frac{p+2}{l_0{}^2}\,\qquad,\qquad
R_{p+2,p+2}=\frac{p}{l_0{}^2} 
\ee 
 de donde la curvatura escalar resulta constante
\be
 R = -(p^2+p+2)\;\frac{1}{l_0{}^2}. 
\ee 

 Si bien, como se ve, la soluci\'on es regular en todos lados, el
 dilat\'on  diverge cuando $u\rightarrow 0^+$.  Es
 interesante notar que esta soluci\'on corresponde a realizar una
 T-dualidad (en la direcci\'on $\theta$) de la soluci\'on
 $AdS_{1,p+2}$ con dilat\'on constante (ver al final de esta
 secci\'on). \'Esta ha sido obtenida recientemente en la referencia
 \cite{Kuperstein:2004yk} como el near horizon limit de una
 soluci\'on BPS, que resulta asint\'otica al linear dilaton background.

\subsection{Soluciones con $\Delta<0$}

\begin{figure}[!ht]
\centering
\includegraphics[scale=1]{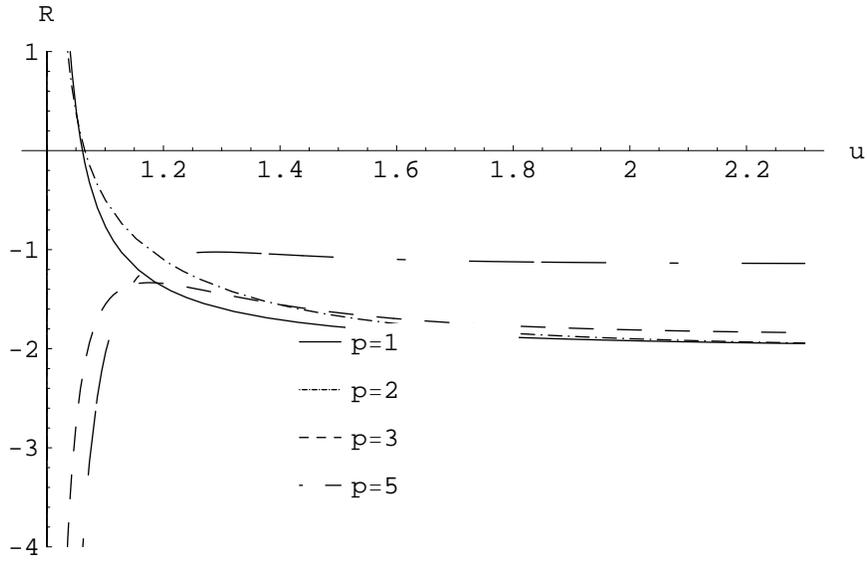}
\caption[Curvatura escalar de (\ref{adsbhsngral}) para $a=1/5$ y
  $e=-3/10$.]{Las curvas muestran $\alpha'\,R$ como funci\'on de
  $\tilde{u}={u \over u_0}$ para diferentes valores de $p$, donde $R$
  es el escalar de curvatura de (\ref{adsbhsngral}) para $a=1/5$ y
  $e=-3/10$.}
\label{R4}
\end{figure}

\begin{figure}[!ht]
\centering
\includegraphics[scale=1]{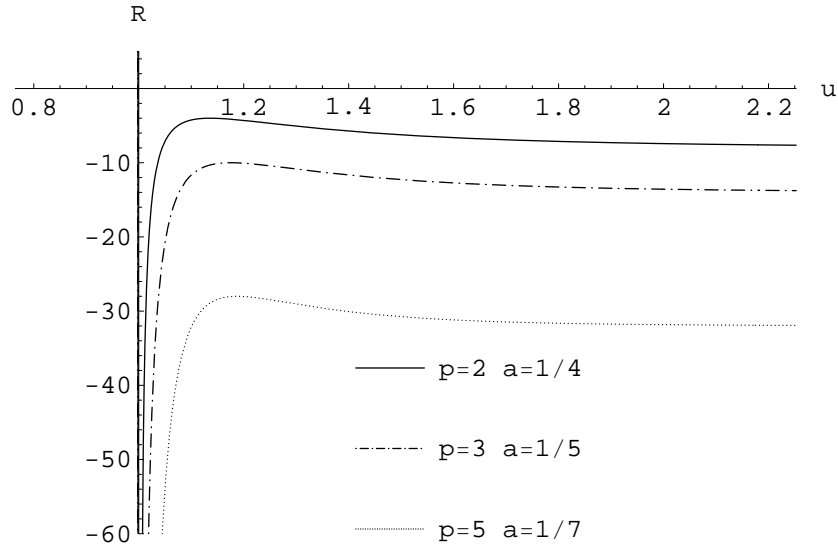}
\caption[Curvatura escalar de (\ref{adsbhsngral}) para distintos
  valores de $p$.]{Las curvas muestran $\alpha'\,R$ para
  (\ref{adsbhsngral}) como funci\'on de $\tilde{u}={u \over u_0}$ para
  valores diferentes de $p$.  Los valores de $a$ y $e$ son tales que
  las soluciones satisfacen el criterio de Maldacena-Nu\~nez y tienen
  acotado el string coupling.}
\label{R5}
\end{figure}

De (\ref{ansatzII}), (\ref{f1sn}), tenemos en este caso,
\bea
f_1(x)&=&\frac{-\Delta}{\Lambda^2}\; \frac{p+3}{p+2}\;\frac{1}{\sinh^2(\sqrt{-\Delta}\,(x-x_0)}
\cr
f_2(x)&=&\frac{-\Delta}{Q_p{}^2}\;\frac{4}{p+2}\;
\frac{1}{\sinh^2(\sqrt{-\Delta}\,(x-x_0))}
\eea
Re-escaleando, la familia de soluciones puede ser reescrita como
\bea
G &=& |\sinh x|^{-\frac{2}{p+2}}\;\left(-e^{2a\,x}\; d^2 x^0 + e^{2e\,x}\; d\vec x{}^2\right)
+ \frac{l_0{}^2}{(p+2)^2}\;\frac{d^2 x}{\sinh^2 x}\cr
&+& R_0{}^2\; e^{2(a+p\,e)\,x}\;|2\,\sinh x|^\frac{2}{p+2}\; d\theta^2\cr
e^{2\Phi} &=& e^{2\Phi_0}\;e^{2(a+p\,e)\,x}\;|2\,\sinh x|^\frac{2}{p+2}\cr
F_{p+2} &=& s(q)\;\frac{2^\frac{p+1}{p+2}}{\sqrt{p+2}}\;e^{-\Phi_0}\;\frac{1}{\sinh^2 x}
\;dx\wedge dx^0\wedge\dots\wedge dx^p\cr
\frac{p+1}{p+2}&=&2\,a^2 + p\,(p+1)\,e^2 + 2\,p\,a\,e
\label{deltaless0}
\eea
donde $R_0$ y $l_0,$ est\'an dados en (\ref{l0R0}).
Ante el cambio de variables,
\be
e^{-2x} = 1 - \left(\frac{u_0}{u}\right)^{p+2}\equiv f(u)
\ee
con $u_0$ una constante, que pone las soluciones, en la forma siguiente
\bea
G &=& l_0{}^2\left(-\frac{u^2}{f(u)^{a-\frac{1}{p+2}}}\;dx^0{}^2 +
\frac{u^2}{f(u)^{e-\frac{1}{p+2}}}\;d\vec x{}^2+ \frac{1}{f(u)}\;\frac{d u^2}{u^2}\right)\cr
&+& R_0{}^2\;\frac{ f(u)^{-a-p\,e-\frac{1}{p+2}}}{(l_0\,u)^2}\;d\theta^2\cr
e^{-2\Phi} &=& e^{-2\Phi_0}\; f(u)^{a+p\,e+\frac{1}{p+2}}\; (l_0\,u)^2\cr
F_{p+2} &=& s(q)\; 2\,\sqrt{p+2}\; e^{-\Phi_0}\;l_0{}^{p+2}\; u^{p+1}\;
du\wedge dx^0\wedge\dots\wedge dx^p\cr
\frac{p+1}{p+2}&=&2\,a^2 + p\,(p+1)\,e^2 + 2\,p\,a\,e\label{adsbhsngral}
\eea
con $l_0$ y $R_0$ como en (\ref{l0R0}).  El l\ii mite $u_0 =0$
corresponde a la soluci\'on $\Delta=0$ (\ref{delta0}), que resulta ser
el l\ii mite asint\'otico UV $\,u\gg u_0\,$, de toda la familia. Por
otro lado, para $\,u\rightarrow u_0{}^+\,$, el comportamiento es
fuertemente dependiente de los exponentes $a$ y $ e$.  La curvatura
escalar, para algunos valores particulares de los par\'ametros, se
muestra en las Figuras~\ref{R4} y \ref{R5}.

Entre los miembros de la familia, aquellos con $a + p\,e
+\frac{1}{p+2}\leq 0$ tienen acotado el string coupling en todos lados.
Y, como pasara con la familia de soluciones de la secci\'on 3.2,
cuando la desigualdad se satura, \'estos se vuelven tambi\'en regulares.  En efecto, 
el escalar de curvatura de (\ref{adsbhsngral}) resulta
\bea l_0{}^2\,R &=&
-\left(1+(p+2)\,(a+pe)\right)^2\;\frac{1}{f(u)} - (p-1)\,(p+2) + 2 -
2\,(p+2)^2\;(a+pe)^2\cr &-& \left(1-(p+2)\,(a+pe)\right)^2\;f(u) 
\eea
mostrando expl\ii citamente que la curvatura es finita en  $u\rightarrow u_0$
si y s\'olo si la condici\'on $a + p\,e +\frac{1}{p+2} = 0$ se satisface.  En ese caso, la siguiente soluci\'on aparece
\bigskip

\noindent\underline{Soluci\'on 1}. $a =-\frac{p+1}{p+2}\;\;,\;\;e =\frac{1}{p+2}$

\bea G &=& l_0{}^2\; \left( -u^2\;f(u)\;dx^0{}^2+ u^2\;d^2 \vec x +
\frac{1}{f(u)}\; \frac{d u^2}{u^2}\right)+
\frac{R_0{}^2}{(l_0\,u)^2}\;d\theta^2\cr e^{-2\Phi} &=&
e^{-2\Phi_0}\;(l_0\,u)^2\cr F_{p+2} &=& s(q)\; 2\,\sqrt{p+2}\;
e^{-\Phi_0}\;l_0{}^{p+2}\; u^{p+1}\; du\wedge dx^0\wedge\dots\wedge
dx^p\label{adsbhsn1} 
\eea 
\'Esta es una soluci\'on tipo $AdS$ black
hole, con dilat\'on que corre.  A\'un m\'as, no es dif\ii cil ver, 
siguiendo la receta usual que, en el l\ii mite $u\rightarrow u_0$, el
espacio eucl{\'\i}deo obtenido por rotaci\'on de Wick $\tau\equiv i\,x^0$ es
regular si imponemos la condici\'on de periodicidad siguiente 
\be 
\tau \sim \tau +\beta\qquad,\qquad\beta\equiv \frac{4\,\pi}{p+2}\,\frac{1}{u_0} .
\ee

Este es un hecho usual en la clase de soluciones que se asocian via la
conjetura con teor\ii as de campos a temperatura finita $\beta^{-1}$
\cite{Witten:1998zw}.

\bigskip

\noindent\underline{Soluci\'on 2}.
$a =\frac{p^2+2p-1}{(p+1)\,(p+2)}\;\;,\;\;e = -\frac{p+3}{(p+1)\,(p+2)}$
\bea
G &=& l_0{}^2\; \left( -\frac{u^2}{f(u)^{\frac{p-1}{p+1}}}\;dx^0{}^2+ u^2\;
f(u)^{\frac{2}{p+1}}\;d^2 \vec x+ \frac{1}{f(u)}\;\frac{d u^2}{u^2}\right)
+\frac{R_0{}^2}{(l_0\,u)^2}\;d\theta^2\cr
e^{-2\Phi} &=& e^{-2\Phi_0}\;(l_0\,u)^2\cr
F_{p+2} &=& s(q)\; 2\,\sqrt{p+2}\; e^{-\Phi_0}\;l_0{}^{p+2}\; u^{p+1}\;
du\wedge dx^0\wedge\dots\wedge dx^p\label{adsbhsn2}
\eea

\begin{figure}[!ht]
\centering
\includegraphics[scale=0.5,angle=-90]{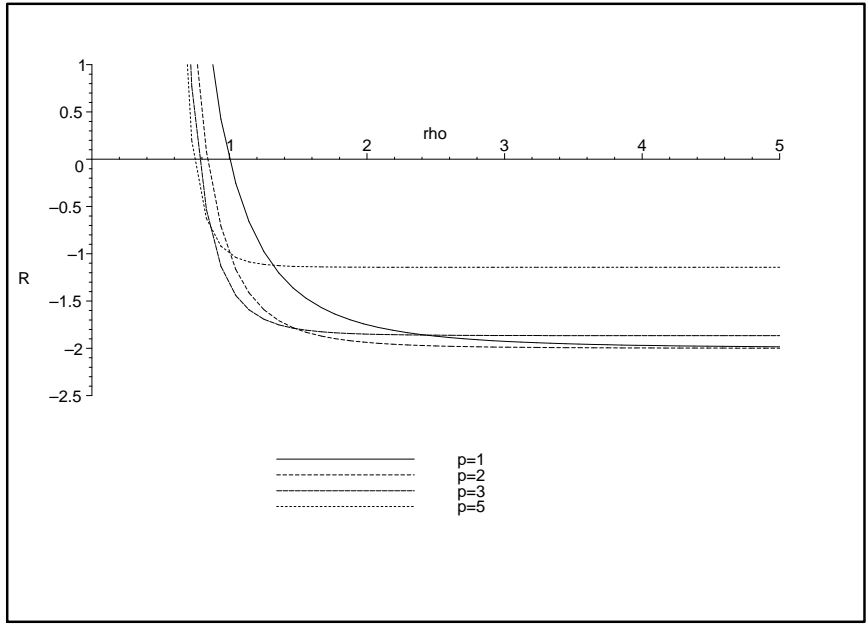}
\caption[Curvatura escalar de (\ref{adsbhsn1}) y (\ref{adsbhsn2}),
  para distintos valores de $p$.]{Las curvas muestran $\alpha'\,R$
  como funci\'on de $\tilde{u}={u \over u_0}$ para valores diferentes
  de $p$, donde $R$ es la curvatura escalar de (\ref{adsbhsn1}) y
  (\ref{adsbhsn2}).}\label{R6}
\end{figure}

El caso de la $D1$-brana se incluye dentro de la Soluci\'on $1$ via
una doble rotaci\'on de Wick, $x^0\rightarrow
-i\,x^1\;,\;x^1\rightarrow i\,x^0$.  Notar que ambas soluciones son soportadas por el mismo dilat\'on 
y el mismo campo de gauge de $RR$, y poseen el mismo escalar de curvatura, que se muestra en la Figura~\ref{R6}.


\subsection{$T$-dualidad}

Aplicando las reglas para transformaciones de $T$-dualidad (ver por
ejemplo \cite{Johnson:2000ch} y sus referencias) a lo largo de la
coordenada $\theta$, podemos generar de (\ref{adsbhsngral}) una nueva
familia de soluciones.  El hecho de haber asumido la proporcionalidad
entre las funciones $f_1$ y $f_2$ (ecuaciones (\ref{ansatzII})), lleva a
la relaci\'on $e^{2\Phi}\propto G_{\theta\theta}$ que impone, al
T-dualizar las soluciones, que el dilat\'on deba ser constante. 
Tenemos
\bea G &=&
l_0{}^2\left(-\frac{u^2}{f(u)^{a-\frac{1}{p+2}}}\;dx^0{}^2 +
\frac{u^2}{f(u)^{e-\frac{1}{p+2}}}\;d\vec x{}^2+
\frac{1}{f(u)}\;\frac{d u^2}{u^2}\right.\cr
&+&\left.\left(\frac{\alpha'}{R_0}\right)^2\;
f(u)^{a+p\,e+\frac{1}{p+2}}\; u^2\;d\theta^2\right)\cr e^{2\Phi} &=&
\frac{16\,\pi^2\,\alpha'\,\Lambda^2}{(p+3)\,Q_p{}^2}\cr F_{p+3} &=&
s(q)\; 2\,\sqrt{p+2}\; e^{-\Phi_0}\;\sqrt{\alpha'}\;l_0{}^{p+2}\;
u^{p+1}\; du\wedge dx^0\wedge\dots\wedge dx^p\wedge d\theta\cr
\frac{p+1}{p+2}&=&2\,a^2 + p\,(p+1)\,e^2 +
2\,p\,a\,e\label{T-adsbhsngral} \eea 

Se puede mostrar que \'esta es una familia de soluciones de espacios de
Einstein con tensor de Ricci 
\be R_{mn} =
-\frac{p+2}{l_0{}^2}\;G_{mn}. 
\ee 
Todos los miembros de la familia son
asint\'oticos para $u$ grande, $u\gg u_0$, a un espacio $AdS_{1,p+2}$
(despu\'es de identificar $x^{p+1}\equiv \frac{\alpha'}{R_0}\,\theta$).
En particular, la soluci\'on T-dual a (\ref{adsbhsn1}) resulta un
$AdS_{1,p+2}$ Schwarzchild black hole, recientemente descubierto en
\cite{Kuperstein:2004yk}, y usado en  \cite{Kuperstein:2004yf},
\cite{Casero:2005se}, como modelo de teor\ii a de YM 
\footnote{Hemos expl{\'\i}citamente verificado que la familia (\ref{T-adsbhsngral})  es soluci\'on de la ecuaci\'on de movimiento (\ref{ecformal}).
En particular, la carga de la  $D(p+1)$ brana (dimensional) debe ser identificada con la carga de la  $Dp$ brana original de la siguiente manera,
\be
Q_{p+1}\equiv \frac{Q_p}{2\,\pi\,\sqrt{\alpha'}}.
\ee
M\'as a\'un, la arbitrariedad, a trav\'es de  $\Phi_0$, en el par\'ametro $\frac{R_0}{l_0}$ dada en (\ref{l0R0})
se traduce aqu{\'\i} como una arbitrariedad en el radio de compactificaci\'on $x_{p+1}$.}
.
\bigskip

\section{Conclusiones}

En este cap\ii tulo, hemos repasado la formulaci\'on de la teor\ii a de
cuerdas en presencia de campos de fondo que lleva, en la aproximaci\'on
de baja energ\ii a, a la que hemos llamado la ``acci\'on efectiva de
cuerdas no cr\ii ticas''. El resto del cap\ii tulo est\'a dedicado a la
presentaci\'on de todas las soluciones que hemos hallado a dicha
acci\'on de baja energ\ii a de teor\ii as de cuerdas no cr\'iticas
Tipo II.\\~\\ En primer lugar, hemos presentado la soluci\'on
doblemente localizada, en el origen del espacio Minkowski y en el tip
del cigarro, que interpretamos como la cuerda fundamental embebida en
dicho vac\ii o. El hecho de que esta soluci\'on aparezca como
doblemente localizada le da un valor extra a la soluci\'on, ya que no
se conocen muchas soluciones con estas caracter\ii sticas.\\~\\ Hemos
agotado completamente el problema de encontrar soluciones de vac\ii
o. Entre las soluciones encontradas, adem\'as de Minkowski
($p+1$)-dimensional veces el dilat\'on lineal veces $S^1$, encontramos
tambi\'en una familia 3-param\'etrica (\ref{vacuadelta<0}), que
incluye entre sus miembros a la conocida soluci\'on Minkowski ($p+1$)
veces el cigarro, y su T-dual Minkowski ($p+1$) veces la trompeta. La
constante de acoplamiento de la cuerda se mantiene acotada para gran
parte del espacio de par\'ametros, encontr\'andose adem\'as una
subfamilia 2-param\'etrica en la que tambi\'en la curvatura escalar se
mantiene acotada.  \\~\\ Tambi\'en presentamos todas las soluciones
posibles al problema de encontrar soluciones de fondo cargadas NSNS
que llenan todo Minkowski, que para $p=1$ representan (el {\it near
  horizon} de) las soluciones de la cuerda fundamental.  Entre la gran
cantidad de soluciones encontradas, se destacan la soluci\'on de
cuerda fundamental embebida en el vac\ii o del cigarro previamente
obtenida, una nueva soluci\'on (\ref{f1ld}) interpretable como la F1
embebida en vac\ii o de dilat\'on lineal, y una familia
2-param\'etrica de soluciones regulares asint\'oticas a $AdS_{1,2}\times
S^1$.  \\~\\ Por \'ultimo, hemos abordado el problema de hallar
soluciones cargadas RR, \'este es el \'unico caso en que no hemos
podido resolver completamente el problema de encontrar soluciones de
la acci\'on de baja energ\ii a no cr\ii tica. Sin embargo, encontramos
una gran cantidad de soluciones anal\ii ticas que satisfacen un v\ii
nculo particular. Entre \'estas se incluyen soluciones de curvatura
constante asint\'oticas en el UV a $AdS_{1,p+1}$, que en el l\ii mite
IR se reducen a $AdS_2$. \'Estas son T-duales a la soluci\'on
$AdS_{1,p+2}$ encontrada por primera vez en \cite{Kuperstein:2004yk}
como el l\ii mite de horizonte cercano (near horizon) de una
soluci\'on num\'erica asint\'otica al vac\ii o de dilat\'on lineal.
M\'as a\'un, encontramos una familia de soluciones 3-param\'etricas
(\ref{adsbhsngral}) que corresponden, presumiblemente, al l\ii mite de
horizonte cercano de una familia de $Dp$-branas negras embebidas en el
vac\ii o de dilat\'on lineal.  Estas \'ultimas no obedecer\ii an
nuestro ans\"atz particular (\ref{ansatz_0}), en alguna regi\'on
particular del espacio-tiempo. Por otro lado, nosotros conjeturamos
que deben existir soluciones de $Dp$- branas embebidas en el vac\ii o
del cigarro que son soluciones de (\ref{system}), pero que no obedecen
nuestro ans\"atz en ninguna regi\'on y que corresponden a aquellas
soluciones constru\ii das como estados de frontera en las referencias
\cite{Ashok:2005py}, \cite{Fotopoulos:2005cn}, \cite{Murthy:2006xt}.

%% file: sym.tex
\chapter{Modelos No Cr\ii ticos de teor\ii as de Yang-Mills ($p+1$)-dimensionales}
\label{sym}
\section{ Introducci\'on }


 En este cap\ii tulo, nos proponemos utilizar algunas de las soluciones
encontradas en el cap\ii tulo anterior, como fondos sobre los cuales
construir teor\ii as duales cercanas a QCD, en el sentido planteado por
Witten en \cite{Witten:1998zw} y que hemos explicado en el cap\ii tulo
\ref{c: ads_cft}.

Estudiaremos con particular inter\'es modelos de teor\ii as de gauge en 3
y 4 dimensiones. Luego de analizar bajo que circunstancias nuestros
fondos de gravedad representan teor\ii as de gauge duales en su fase
confinante, dedicaremos nuestro esfuerzo al c\'alculo del espectro de
glueballs correspondiente a cada una de estas teor\ii as.

En particular, estudiaremos los glueballs asociados con la
perturbaci\'on del campo del dilat\'on, con la perturbaci\'on del
gravit\'on y con la perturbaci\'on de la 1-forma de RR. Hemos
preferido, en lugar del c\'omputo usual de perturbaciones, desarrollar
el c\'alculo num\'erico de las ecuaciones acopladas de segundo orden,
en un marco invariante de gauge. El hecho de que la familia de
soluciones de fondos no cr\ii ticos utilizada est\'e caracterizada por
un conjunto de par\'ametros libres, hace que la teor\ii a dual cuente
al mismo tiempo con dichos par\'ametros tambi\'en como libres. De esta
manera, disponemos de una libertad en la teor\ii a dual que puede
utilizarse para obtener un mejor ajuste con la teor\ii a esperada.
Durante el desarrollo de este cap\ii tulo pondremos \'enfasis en
estudiar la dependencia de nuestros resultados con los par\'ametros
libres de nuestra teor\ii a.

El presente cap\ii tulo se organiza de la siguiente manera:

En la secci\'on \ref{fs} presentamos la familia de soluciones no cr\ii
ticas que analizaremos, y discutiremos las caracter\ii sticas m\'as
sobresalientes de \'estas. En la secci\'on \ref{wlc} describiremos
brevemente el c\'alculo de Lazos de Wilson, que ya fuera presentado en
\ref{c: ads_cft}, y discutiremos las condiciones de confinamiento
sobre nuestros fondos. La secci\'on \ref{setup} est\'a dedicada a
presentar el formalismo de perturbaciones invariantes de gauge. En la
secci\'on \ref{holomodels} discutimos los modelos duales que ser\'an
considerados.  En \ref{3dgs} presentamos los resultados obtenidos en
c\'alculo del espectro de glueballs para la teor\ii a de gauge en 3
dimensiones en el contexto de teor\ii as no cr\ii ticas. En la
secci\'on \ref{4dgs} analizamos de manera similar el modelo para
teor\ii as de gauge en 4 dimensiones. Finalmente, en la secci\'on
\ref{discusion} realizamos un an\'alisis de los espectros obtenidos y
realizamos una comparaci\'on con los valores predichos por {\it
  Lattice QCD}. A\'un cuando los c\'aculos de {\it Laticce} y de
supergravedad son v\'alidos en reg\ii menes opuestos (acoplamiento d\'ebil,
acoplamiento fuerte, respectivamente) veremos que el acuerdo entre
ambos conjuntos de datos es bueno.


\section{La familia de soluciones}\label{fs}

Consideremos una generalizaci\'on de la familia de soluciones a
dilat\'on constante (\ref{T-adsbhsngral}), encontrada en el cap\ii tulo
anterior realizando una transformacion de T-dualidad sobre (\ref{adsbhsngral}).




\bea 
l_0{}^{-2}\,G
&=&\sum_{a=0}^{D-2}\;\,u^2\,f(u)^{a_a}\;dx^a{}^2 +
\frac{du^2}{u^2\,f(u)} \qquad;\qquad f(u)\equiv 1 - \left(
\frac{u_0}{u}\right)^{D-1}\label{f}\cr e^\Phi &=& \frac{2}{\sqrt{D}}\,
\frac{\Lambda}{|Q_{D-2}|}\cr
F_D&=&(-)^D\,Q_{D-2}\;\epsilon_G\;\;\Longleftrightarrow\;\; *F_D=
(-)^{D-1}\;Q_{D-2}\label{Ssn} 
\eea 
donde $l_0 = \sqrt{D\,(D-1)}\,\Lambda^{-1}\,$, $u_0$ es una escala
arbitraria, y donde los exponentes deben satisfacer las siguientes
relaciones
\be
\sum_{a=0}^{D-2}a_a =1\qquad,\qquad \sum_{a=0}^{D-2}a_a{}^2 = 1.\label{constraints}
\ee
Una derivaci\'on de estas relaciones se muestra en el ap\'endice \ref{AA}.


En este cap\ii tulo, nos enfocaremos en esta familia de
soluciones con dilat\'on constante. Un an\'alisis de las soluciones
con dilat\'on no trivial es presentada en el Ap\'endice \ref{AB}.

Las soluciones que estudiaremos pueden ser interpretadas como el {\it
  near horizon limit} de una D-$(D-2)$ black-brane extendida a lo
largo de las coordenadas $x$.  Estas soluciones resultan ser espacios
de Einstein con tensor de Ricci, \be R_{AB} =
-\frac{D-1}{l_0{}^2}\;G_{AB}\qquad,\qquad R = -\Lambda^2. \ee
Mas a\'un, todas ellas son tales que en el l\ii mite $u$ grande, ($u\gg
u_0$) resultan asint\'oticas a espacios $AdS_{1,D-1}$.  Sin embargo, de
todas estas soluciones s\'olo una es estrictamente regular a\'un en
la regi\'on IR, $u\rightarrow u_0{}^+$. \'Esta corresponde a tomar uno
de los exponentes $a_a$ igual a uno, y los otros igual a cero.  La
forma expl{\'\i}cita que toma esta soluci\'on de fondo es
\be
l_0{}^{-2}\,G = u^2\,\left(\eta_{1,D-3} + \,f(u)\; d\tau^2\right) +
\frac{du^2}{u^2\,f(u)}.\label{ks} \ee

A partir del c\'alculo del siguiente invariante de cuarto orden
\bea & &\left(\frac{l_0}{D-1}\right)^4\,
\Re^{ABCD}\,\Re_{ABCD} = \frac{1}{8}\,\left( 4\,(1-s_3) - 1+s_4
\right)\,f(u)^{-2}\cr &+& \frac{1}{2}\left( -\frac{2D}{D-1}\,(1-s_3) +
1-s_4 \right)\,f(u)^{-1}\cr &+&\frac{9}{(D-1)^2} +
\frac{1}{2}\left(1-\frac{12}{D-1}+ \frac{6}{D-1}(1-s_3)
-\frac{3}{2}(1-s_4)\right) + o(f(u)) \eea 
donde $s_n\equiv \sum_a a_a{}^n$,
resulta claro que la \'unica forma de cancelar los t\'erminos
 peligrosos es imponer
 $s_3 =s_4=1$. Esta restricci\'on, sumada a las restricciones 
provenientes de los v\ii nculos $s_1=s_2=1$, deja como
\'unica posibilidad la soluci\'on (\ref{ks}). Esta soluci\'on
corresponde al agujero negro de Schwarzchild  $AdS_{1,D-1}$ hallado
recientemente en \cite{Kuperstein:2004yk}, y usado en
\cite{Kuperstein:2004yf} como modelo, cuando $D=6$, de una teor\ii a
de Y-M cuadrimensional. Cuando la singularidad en el IR es
de tipo c\'onica, se aplica sobre las
soluciones una condici\'on de periodicidad que permita evitarla,
\be \tau \sim \tau +
\beta\qquad,\qquad\beta\equiv
\frac{4\,\pi}{D-1}\,\frac{1}{u_0}.\label{period} \ee

Las soluciones sobre las que se imponen estas condiciones de
periodicidad est\'an usualmente asociadas con teor\ii as de campos a
temperatura finita $\beta^{-1}$ \cite{Witten:1998zw}.

El comportamiento de la curvatura del resto de las soluciones en el IR
hace que \'estas no est\'en completamente bajo control. Esto no es inusual,
ni exclusivo de las teor\ii as no cr\ii ticas, ya que lo mismo sucede
para las soluciones de $D_p$-branas en teor\ii as de cuerdas cr\ii
ticas cuando $p\neq 3$. Esto es consecuencia, en ambos casos, de que
la acci\'on de baja energ\ii a que hemos utilizado para obtener tales
soluciones proviene de anular las funciones $\beta$ de la teor\ii a al
orden m\'as bajo de su desarrollo perturbativo. Es decir, hemos supuesto,
entre otras cosas, que la curvatura se mantiene finita y que es
posible despreciar los \'ordenes superiores de la misma. Argumentos de
supersimetr\ii a permiten asegurar, en el caso de teor\ii as cr\ii
ticas, que tales soluciones existen y que dichas singularidades son
s\'olo manifestaciones de baja energ\ii a. En el caso de las teor\ii as
no cr\ii ticas, es ampliamente conjeturado que sucede lo
mismo. Llamativamente, como veremos m\'as adelante,
las teor\ii as duales que nos interesa estudiar en este cap\ii tulo
parecen no tomar nota de la existencia de dichas singularidades.

Por otro lado, la constante de acoplamiento de cuerdas de cada miembro
de la familia con dilat\'on constante resulta \be g_s = e^\Phi =
\sqrt{\frac{10}{D}-1}\;\frac{4\,\pi}{|Q_p|} \sim
\frac{1}{N},\label{dilaton} \ee donde $N$ es el n\'umero de
D$(D-2)$-branas. Puesto que $N$ es un n\'umero arbitrario, la
constante de acoplamiento de cuerdas puede hacerse tan peque\~na como
se quiera, y por lo tanto, la teor\ii a perturbativa ser\' a v\'alida
para cualquier miembro de la familia, con s\'olo tomar $N$ suficientemente
grande.

A\'un m\'as, siguiendo \cite{Kuperstein:2004yk}, observamos que el
c\'alculo del lado de gravedad del n\'umero de grados de libertad
(``entrop{\'\i}a'') en la regi\'on UV lleva, \be S_{gravity} \sim
\frac{N^2}{\delta^{D-2}}, \ee
donde $\delta$ es un regulador IR. \'Este es exactamente el resultado
esperado para teor\ii as de gauge $U(N)$ ($D-1$)-dimensionales con
regulador UV $\delta^{-1}$ \cite{Susskind:1998dq}.
Esto nos anima a estudiar  algunos aspectos de las posibles
teor\ii as duales a nuestras soluciones. Consideraremos de esta manera
el caso general con exponentes arbitrarios sujetos a las relaciones
(\ref{constraints}). Esto nos permitir\'a encontrar una familia de
teor\ii as duales cuyos par\'ametros libres puedan ajustarse a discreci\'on.
 El objetivo principal de este cap\ii tulo ser\'a estudiar la
dependencia del espectro de esta teor\ii a con los exponentes
mencionados.

\bigskip

\section{Lazos de Wilson y confinamiento}
\label{wlc}
Se conoce desde hace ya bastante tiempo (\cite{Maldacena:1998im},
\cite{Rey:1998ik}) que es posible hacer una descripci\'on en el
contexto de teor\ii a de cuerdas de los lazos de Wilson (Wilson loops)
como cuerdas cuyos extremos unen un par de puntos del borde de un
espacio tipo agujero negro AdS. Estos puntos representan, en la teor\ii
a de gauge que vive en dicho borde, un par quark-antiquark no
din\'amico.  Nos interesa, entonces, estudiar cuerdas que vivan sobre
nuestros fondos y que tengan sus extremos en $u=u_\infty
(\rightarrow\infty), X^1=\pm L/2$, donde $X^1$ denota una de las $p$
direcciones espaciales.  En \cite{Kinar:1998vq}, se deduce la energ\ii
a cl\'asica de un lazo de Wilson asociado con una m\'etrica de fondo
de la forma \be G = G_{00}\; dx^0{}^2+ \sum_{i,j=1}^p
G_{ii}\,\delta_{ij}\;dx^i\,dx^j + C(u)^2\;du^2 + G^\perp
\ee con una dependencia radial gen\'erica, y donde $G^\perp$ es
ortogonal a las direcciones ($x^0, x^i, u$).  A continuaci\'on,
repasaremos la parte sustancial del c\'alculo.

\begin{figure}[!ht]
\centering
\includegraphics[scale=0.95,angle=0]{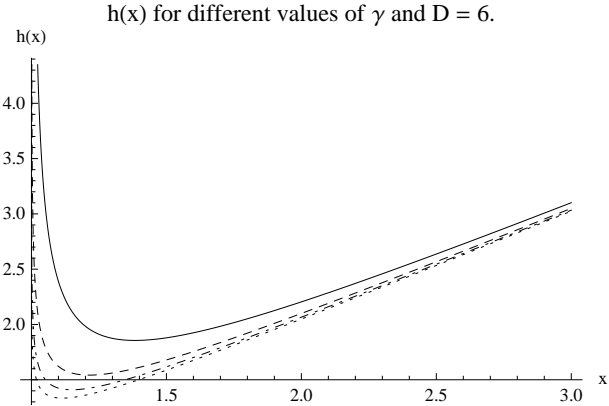}
\caption[Gr\'afico de $h(x)$, para $\gamma$ entre $1/8$ y $1/2$.]{El gr\'afico muestra $h(x)$ como funci\'on de $x$, donde $\gamma$ toma valores entre $1/8$ y $1/2$.
La l{\'\i}nea punteada corresponde a $\gamma=1/8$, mientras que la l{\'\i}nea s\'olida corresponde a $\gamma=1/2$.}
\label{h}
\end{figure}

Sea $(\sigma^\alpha)= (t,\sigma)$ las coordenadas que parametrizan la
hoja de mundo.  En el gauge est\'atico, $X^0(t,\sigma)= t \in \Re\,,\;
X^1(t ,\sigma)= \sigma \in [-\frac{L}{2},\frac{L}{2}]\,$. Consideremos
la configuraci\'on est\'atica de una cuerda definida por $u(t,\sigma)=
u(\sigma) = u(-\sigma) \in[u_0, u_\infty]$, y el resto de las
coordenadas fijas.  El langragiano de Nambu-Goto viene dado por \be
L[X] = T_s\;\int_{-\frac{L}{2}}^{\frac{L}{2}}\,d\sigma\;\sqrt{-\det
  h_{\alpha\beta}(\sigma) } =
T_s\;\int_{-\frac{L}{2}}^{\frac{L}{2}}\,d\sigma\; \sqrt{ F(u)^2 +
  G(u)^2\;u'(\sigma)^2},\label{ng} \ee donde
$h_{\alpha\beta}(\sigma)\equiv G_{MN}(X)\,\partial_\alpha
X^M(\sigma)\partial_\beta X^N(\sigma)\;$ es la m\'etrica inducida, y
las funciones $F$ y $G$ est\'an definidas por \be F(u)\equiv
|G_{00}\,G_{ii}|^\frac{1}{2} \qquad,\qquad G(u)\equiv
|G_{00}\,C|^\frac{1}{2}. \label{FG} \ee La configuraci\'on que
minimiza la acci\'on satisface \be \frac{F(u)\partial_u F(u) +
  G(u)\partial_u G(u)\,u' (\sigma)} {\sqrt{ F(u)^2 +
    G(u)^2\;u'(\sigma)^2}} = \left(\frac{G(u)^2\,u' (\sigma)}{\sqrt{
    F(u)^2 + G(u)^2\; u'(\sigma)^2}}\right)'.\label{eqngeo} \ee De las
ecuaciones anteriores (\ref{ng}) y (\ref{eqngeo}), resulta que la
separaci\'on entre los quarks y la energ\ii a, viene dada por \bea L
&=& 2\,\int_{u_m}^{u_\infty}\;du\;
\frac{G(u)}{F(u)}\;\left(\frac{F(u)^2}{F(u_m)^2}
-1\right)^{-\frac{1}{2}}\cr E&=& T_s\,\left( F(u_m)\, L +
2\,\int_{u_m}^{u_\infty}\;du\; G(u)\;\sqrt{1-\frac{F(u_m)^2}{F(u)^2}}
\right)\label{LE} \eea donde $u_m=u(\sigma_m)\geq u_0$, para
$\sigma_m$ (admitiendo que es \'unica) tal que $u'(\sigma_m) = 0$ y
$F(u)\geq F(u_m)$. En nuestro caso, $\sigma_m=0$ y $u_m=u(0)$.  Sin
embargo, hay que ser cuidadosos, ya que el valor resultante para la
energ\ii a es divergente cuando $u_\infty\rightarrow\infty$. Es
aceptado que esta divergencia se debe a las contribuciones de las
autoenerg\ii as (masas) de los quarks \cite{Maldacena:1998im}. Cada
una de \'estas puede ser representada, en este contexto, por una larga
cuerda con un extremo en el borde $u=u_\infty$ en $x^1=\pm\frac{L}{2}$
que se extiende a lo largo de la direcci\'on $u$ hasta el horizonte
\be m_q =T_s\,\int_{u_0}^{u_\infty}\;du\; G(u). \ee Sustrayendo estas
masas, es posible obtener una energ\ii a de ligadura finita \bea
V(L)&\equiv& E -2\,m_q = T_s\,\left( F(u_m)\, L - 2\, K(L)\right)\cr
K(L)&=& \int_{u_m}^{u_\infty}\;du\; G(u)\;\left( 1-\sqrt{
  1-\frac{F(u_m)^2}{F(u)^2}}\right) + \int_{u_0}^{u_m}\;du\;
G(u).\label{qbarqpot} \eea Del an\'alisis de esta expresi\'on
\cite{Sonnenschein:1999if}, vemos que una condici\'on suficiente para
un comportamiento tipo \'area es que $F(u)$ tenga un m\ii nimo en
$u=u_{min}\geq u_0\,$ y $F(u_{min})>0$, o que $G(u)$ diverja en
alg\'un punto $u_{div}\geq u_0$ y $F(u_{div})>0$
\footnote{En \cite{Sonnenschein:1999if} se enuncia el teorema
  completo, que incluye otras hip\'otesis acerca del comportamiento de
  $F$ y $G$ (que son satisfechas por nuestras familias de
  soluciones). En (\ref{LE}), la separaci\'on $L$ debe ser considerada
  como fija, y $u_m$ pensada como una funci\'on de \'esta. M\'as
  a\'un, $u_{min}$ no depende de $u_m$, pero $u_m\geq u_{min}$ debe
  cumplirse para que exista la configuraci\'on. No es dif{\'\i}cil
  mostrar, para (\ref{FG}), que tomar $u_m \rightarrow u_{min}{}^+$ es
  equivalente a tomar el l{\'\i}mite $L\rightarrow\infty$, el cual es
  la condici\'on que lleva a la definici\'on de la tensi\'on de la
  cuerda $\sigma_s$ en (\ref{stension}).
}.  En este caso, el potencial de quark-antiquark
de la teor\ii a de gauge dual resulta lineal en la distancia de
separaci\'on $L$; se sigue de (\ref{qbarqpot}) que, para $L$ grande,
\be \lim_{L\rightarrow\infty} \frac{V(L)}{L} = \sigma_s \qquad,\qquad
\sigma_s= T_s\,F(u_{min}) \;\;\;\textrm{or}\;\;\;
T_s\,F(u_{div}),\label{stension} \ee donde $\sigma_s$ es la tensi\'on
(YM) de la cuerda.
En particular, sobre nuestra familia de soluciones, no es dif\ii cil ver que
$F(u)= (l_0\,u_0)^2\,h(x)|_{x=\frac{u^2}{u_0{}^2}}$ presenta un m\ii
nimo en $u = u_{min}$ si s\'olo si
\be
\gamma\equiv -\frac{1}{2}\,(a+\tilde a)>0 \Longrightarrow u_{min}= u_0\,
\left(1 + (1 +\frac{D-1}{2})\,\gamma\right)^\frac{1}{D-1}.
\label{p+1confcond}
\ee
La funci\'on $h(x)= x\,\left(1-x^{-\frac{D-1}{2}}\right)^{-\gamma}$ se
muestra en la Figura (\ref{h}) para diferentes valores de $\gamma$
y $D$.
\bigskip

Si restringimos el espacio de par\'ametros a la regi\'on $\gamma>0$,
entonces la teor\ii a de gauge dual deber\'a ser (cl\'asicamente)
confinante, con la tensi\'on de la cuerda dada por
\be
\sigma_s \equiv\tilde\sigma_s\;u_0{}^2\qquad,\qquad
\tilde\sigma_s= \left(\frac{D-1}{2}\right)^\frac{D+1}{D-1}\,\frac{D}{\pi\,(10-D)}\,
\frac{\left(\gamma +\frac{2}{D-1}\right)^{\gamma +\frac{2}{D-1}}}{\gamma^\gamma}.
\ee
\bigskip



Sin embargo, la condici\'on (\ref{p+1confcond}) corresponde a
considerar confinamiento en una teor\ii a ($p+1$)-dimensional a
temperatura finita.  Dado que estamos interesados en una teor\ii a
$p$-dimensional a temperatura cero, debemos considerar ambas
direcciones de entre las $p$. Reemplazando entonces $a$ por $\tilde a$
en la expresi\'on (\ref{FG}), la condici\'on de confinamiento
(\ref{p+1confcond}) se reduce a: \be \gamma\equiv -\tilde a >0
\Longrightarrow u_{min}= u_0\, \left(1 + \frac{D-1}{2}\,|\tilde
a|\right)^\frac{1}{D-1}.
\label{pconfcond}
\ee
\\
En el caso en que $\tilde a =0$, tenemos que $u_{div}= u_0$ y
tambi\'en da confinante, con una tensi\'on para la cuerda dada por \be
\sigma_s = T_s\, F(u_{div})= T_s\,l_0{}^2\,u_0{}^2 =
\frac{D\,(D-1)}{2\,\pi\,(10-D)}\, u_0{}^2, \ee \'este es el caso
analizado por K-S.  
\\
Finalmente, para $\tilde a>0$ no hay
confinamiento.

Es importante notar que, a diferencia de lo que pasa en modelos
provenientes de supercuerdas cr\ii ticas, la tensi\'on de la cuerda
est\'a dada por la escala $u_0{}^2$ a menos de una constante
num\'erica. M\'as a\'un, en teor\ii as no cr\ii ticas $u_0$ siempre
resulta fijado a la escala de la teor\ii a \be u_0\sim
\Lambda_{QCD}.\label{uoscale} \ee
\bigskip

\section{El marco perturbativo invariante de gauge}
\label{setup}

Para poder determinar el espectro de las masas de los glueballs  de
acuerdo a la correspondencia gauge/gravedad, debemos resolver las
ecuaciones de movimientos linearizadas de supergravedad sobre la
soluci\'on de fondo (\ref{Ssn}) \cite{Witten:1998zw},
\cite{Gross:1998gk}.  Los campos presentes en la acci\'on efectiva de
baja energ\ii a son la m\'etrica $G$, el tensor antisim\'etrico de
Kalb-Ramond, el dilat\'on $\Phi$ y las ($q+1$)-formas de RR
$A_{q+1}$, donde el valor de $q$ depende de la teor\ii a.

Analizaremos las perturbaciones consistentes que llevan a las
ecuaciones que ser\'an consideradas en la pr\'ima secci\'on. Como se
menciona en \cite{Kuperstein:2004yf}, en el {\it string frame} las
ecuaciones resultan muy complejas debido a que el gravit\'on aparece
acoplado al resto de las perturbaciones. Sin embargo, 
esto no ocurre en el {\it Einstein frame} para la familia (\ref{Ssn}),
donde en la mayor{\'\i}a de los casos estas fluctuaciones se
desacoplan. S\'olo los modos escalares de la perturbaci\'on de la
m\'etrica contin\'uan acoplados al sistema. Pasaremos entonces a este
frame para poder realizar los c\'alculos.  La relaci\'on para la
m\'etrica en ambos frames es la siguiente $G|_\textrm{s-f} =
e^{\frac{4}{D-2}\,\Phi}G|_\textrm{e-f}$.

La acci\'on de baja energ\ii a para las teor\ii as no cr\ii ticas en
el {\it Einstein frame} resulta
\footnote{ Usaremos la siguiente notaci\'on compacta
\be (\Omega\cdot
  \Lambda)_{A_1\dots A_p ;B_1\dots B_p} \equiv \frac{1}{q!}\; G^{C_1
    D_1}\dots G^{C_q D_q}\, \Omega_{C_1 \dots C_q}{}_{A_1\dots
    A_p}\,\Lambda_{D_1 \dots D_q B_1\dots B_p} \ee donde $\,\Omega$
  y $\Lambda$ son $(p+q)-$formas arbitrarias.}
\begin{eqnarray}
S[G,\Phi, A_{q+1}]&=& \frac{1}{2\,\kappa_D{}^2}\int\epsilon_G\left(
R[G]- \frac{4}{D-2}\left(D(\Phi)\right)^2
+\Lambda^2\,e^{\frac{4}{D-2}\,\Phi} -\frac{1}{2}\sum_q\,
e^{2\,\alpha_q\,\Phi}\left(F_{q+2}\right)^2\right)\cr \alpha_q&=&
\frac{D-2\,q-4}{D-2} - 0 \,(1) \qquad,\qquad \rm{RR\; (NSNS)\;
  formas}\label{Esugraction}
\end{eqnarray}
donde $\, F_{q+2} = dA_{q+1}$ es la intensidad del campo de gauge
correspondiente a la forma $A_{q+1}$, y los posibles valores de $q$
dependen de la teor\ii a. El elemento de volumen es
$\,\epsilon_G=\omega^0\wedge\dots\wedge\omega^{D-1}=d^D x\,E\, $, donde
$\{\omega^A\}$ es el vielbein.

Las ecuaciones de movimiento que se deducen de  (\ref{Esugraction}) son
\begin{eqnarray}
R_{AB}&=&\frac{4}{D-2}\,D_A\Phi\, D_B\Phi -\frac{\Lambda^2}{D-2}\,e^{\frac{4}{D-2}\Phi}\;G_{AB}\cr
&+&\frac{1}{2}\;\sum_q\,e^{2\,\alpha_q\,\Phi}\;\left(  (F_{q+2})^2_{A;B}
-\frac{q+1}{D-2}\;(F_{q+2})^2\;G_{AB}\right)\cr
0&=&D^2(\Phi) +\frac{\Lambda^2}{2}\,e^{\frac{4}{D-2}\Phi}
-\frac{D-2}{8}\,\sum_q\,\alpha_q\,e^{2\,\alpha_q\,\Phi}\;\left(F_{q+2}\right)^2\cr
d\left(e^{2\,\alpha_q\,\Phi}*F_{q+2}\right)&=& (-)^q \; Q_q\;
*J_{q+1}\qquad,\qquad Q_q\equiv 2\,\kappa_D{}^2\,\mu_q\label{efeq}
\end{eqnarray}
donde hemos introducido la corriente $J_{q+1}$ de $q$-branas fuente
con tensi\'on $\mu_q$.

Las perturbaciones en torno a la soluci\'on cl\'asica ($G, \Phi,A_q$)
las escribiremos como \be \textrm{metrica}\rightarrow G_{AB} + h_{AB}
\qquad,\qquad \textrm{dilaton}\rightarrow\Phi +\xi\qquad,\qquad
\textrm{q-forma}\rightarrow A_{q} +a_{q}. \ee Al orden lineal en las
perturbaciones $h,
\xi, a_q$ las ecuaciones resultantes que provienen de (\ref{efeq}) son,
\noindent\underline{ecuaciones-h}
\bea
0&=&A_{AB}(h)-\left(\frac{2\,\Lambda^2}{D-2}\, e^{\frac{4}{D-2}\Phi}+\sum_q\,\frac{q+1}{D-2}\,e^{2\,\alpha_q\,\Phi}
\left(F_{q+2}\right)^2\right) h_{AB}\cr
&+&\sum_q\,e^{2\,\alpha_q\,\Phi}\left(\left(-(F_{q+2})^2{}_{CA;DB} +
\frac{q+1}{D-2}\,G_{AB}\,(F_{q+2})^2{}_{C;D}\,\right) h^{CD}\right.\cr
&+& \left.(F_{q+2}\cdot f_{q+2})_{A;B} + (F_{q+2}\cdot f_{q+2})_{B;A} -
2\,\frac{q+1}{D-2}\,G_{AB}\, F_{q+2}\cdot f_{q+2}\right)\cr
&+&\left(\sum_q\,2\,\alpha_q\,e^{2\,\alpha_q\,\Phi}\left( \left(F_{q+2}\right)^2_{A;B} -\frac{q+1}{D-2}\,\left(
F_{q+2}\right)^2\,G_{AB}\right) -\frac{8\,\Lambda^2}{(D-2)^2}\,e^{\frac{4}{D-2}\Phi}\,G_{AB}\right)\xi\cr
&+&\frac{8}{D-2}\,\left(D_A\Phi\, D_B\xi +D_B\Phi\, D_A\xi\right),\label{ep1}
\eea
donde $A_{AB}(h)$ se deriva en el Ap\'endice \ref{AA}.

\noindent\underline{ecuaci\'on-$\xi$}

\bea
0&=&D^2(\xi) + \left(\frac{2\Lambda^2}{D-2}\,e^{\frac{4}{D-2}\Phi} -
\frac{D-2}{4}\,\sum_q\,\alpha_q{}^2\,e^{2\,\alpha_q\Phi}\,\left(F_{q+2}\right)^2\right)\,\xi\cr
&+&\left(-D_A D_B(\Phi) + \frac{D-2}{16}\,\sum_q\,\alpha_q\,e^{\alpha_q\Phi}\,(F_{q+2})^2{}_{A;B}\right)
h^{AB}\cr
&-&D^C(\Phi)\,\left( D^D h_{CD} -\frac{1}{2}D_C h^D_D\right)
-\frac{D-2}{4}\,\sum_q\,\alpha_q\,e^{2\,\alpha_q\Phi}\,F_{q+2}\cdot f_{q+2}\label{ep2}
\eea

\noindent\underline{ecuaci\'on-$a_{q+1}$}

\bea
0&=& -e^{-2\,\alpha_q\Phi}\,D^B\left(e^{2\,\alpha_q\Phi}\,(f_{q+2})_{A_1\dots A_{q+1}B}\right)
-2\,\alpha_q\,e^{-2\,\alpha_q\Phi}\,D^B\left(e^{2\,\alpha_q\Phi}\,(F_{q+2})_{A_1\dots A_{q+1}B}\;\xi\right)\cr
&+&e^{-\alpha_q\Phi}\,D^B\left(e^{\alpha_q\Phi}\,(F_{q+2})_{A_1\dots A_{q+1}}{}^C\, h_{BC}\right)
-\frac{1}{2}\,(F_{q+2})_{A_1\dots A_{q+1}}{}^B \, D_B h^C_C\cr
&+& \left( (F_{q+2})_{A_1A_2\dots A_{q}}{}^{BC}D_C h_{BA_{q+1}}+ \dots -
(F_{q+2})_{A_{q+1}A_2\dots A_{q}}{}^{BC}D_C h_{BA_1}\right)\cr
& &\label{ep3}
\eea

Como se observa de las ecuaciones anteriores, la diagonalizaci\'on del
sistema para fondos gen\'ericos no es posible. Sin embargo, para las
soluciones de fondo (\ref{Ssn}) (y en el {\it Einstein frame}), \'esta
resulta m\'as sencilla.

El sistema de ecuaciones (\ref{efeq}) resulta invariante ante
reparametrizaciones, en particular, ante reparametrizaciones
infinitesimales $x^M\rightarrow x^M - \epsilon^M(x) + o(\epsilon^2)$,
transformando los campos como tensores.  Por otro lado, dado que las
ecuaciones anteriores pueden interpretarse como un sistema de
ecuaciones para los campos $(h , \xi , a_{q+1})$ en la soluci\'on de
fondo $(G,\Phi, A_{q+1})$, la invarianza ante difeomorfismos se
traduce en invarianza ante \be h \rightarrow h +
L_\epsilon(G)\qquad,\qquad \xi \rightarrow \xi +
L_\epsilon(\Phi)\qquad,\qquad a_{q}\rightarrow a_{q} + L_\epsilon(A_q),
\ee
donde $L_\epsilon(...)$ denota derivada de Lie con respecto al campo
vectorial $\epsilon = \epsilon^M(x)\partial_M = \epsilon^A(x)e_A.$ M\'as
expl{\'\i}citamente, no es dif{\'\i}cil mostrar que (\ref{ep1})-(\ref{ep3})
resultan invariantes ante la transformaci\'on de los
campos \footnote{Esta es una generalizaci\'on del tratamiento
  perturbativo usual alrededor de un fondo plano en el contexto de
  Relatividad General, ver por ejemplo \cite{Carroll:1997ar}, cap{\'\i}tulo
  6.}, \bea ^\epsilon h_{AB} &=& h_{AB} + D_A\epsilon_B +
D_B\epsilon_A\cr ^\epsilon\xi &=& \xi + \epsilon^A D_A\Phi\cr
^\epsilon a_q{}_{A_1\dots A_q}&=& a_q{}_{A_1\dots A_q}+
\epsilon^B\,D_B A_q{}_{A_1\dots A_q} +
D_{A_1}\epsilon^B\,A_q{}_{BA_2\dots A_q} + \dots
+D_{A_q}\epsilon^B\,A_q{}_{A_1\dots A_{q-1}B}.\cr & &\label{gaugetrans}
\eea Como el sistema es lineal, se sigue que
$(L_\epsilon(G),L_\epsilon(\Phi),L_\epsilon(A_q))$ es soluci\'on para
cualquier $\epsilon$, (la soluci\'on trivial puro gauge). A
diferencia de las teor\ii as de gauge ordinarias, donde las
transformaciones no son lineales, en este caso deber\ii a ser posible
definir cantidades invariantes de gauge y expresar las ecuaciones para
las perturbaciones (\ref{ep1}) en t\'erminos de \'estos de manera
expl{\'\i}citamente invariante. Es importante remarcar aqu{\ii} que desde el
trabajo pionero de J.M. Bardeen \cite{Bardeen:1980kt}, la teor\ii a de
perturbaciones invariantes de gauge se ha desarrollado en las \'ultimas
d\'ecadas esencialmente en contextos cosmol\'ogicos
\footnote{Para una extensi\'on a teor\ii a de perturbaciones de
  segundo orden, ver \cite{Nakamura:2004rm}.}. La implementaci\'on de
perturbaciones invariantes de gauge en el contexto de Gauge/Gravedad
es un aporte novedoso de \cite{adrian}. Para la familia que nos
interesa en este trabajo, podemos hacer esto como sigue.  Primero,
introducimos la fluctuaci\'on $\chi$ de acuerdo con \be f_D \equiv
\chi\, F_D\qquad,\qquad ^\epsilon\chi=\chi + D_A\epsilon^A, \ee donde
la transformaci\'on de gauge de $\chi$ se sigue de (\ref{gaugetrans}).
Luego, observemos que \be I_\xi\equiv\xi\qquad,\qquad I_\chi\equiv
\chi - \frac{1}{2}\,h^A_A \ee son ambas invariantes de gauge, junto
con los campos de ($q+1$)-formas $a_{q+1}\,,\,q\neq D-2$.  En
t\'erminos de \'estas, las ecuaciones para las fluctuaciones
(\ref{ep1})-(\ref{ep3}) son escritas de forma manifiestamente
invariante \bea 0&=&A_{AB}(h) -
\frac{2\Lambda^2}{D}\,e^{\frac{4}{D-2}\Phi}\,h_{AB} - \frac{8\Lambda^2
  e^{\frac{4}{D-2}\Phi}}{D(D-2)}\,\,G_{AB}\,\left(\frac{2D}{D-2}\,I_\xi-I_\chi\right)\cr
0&=&D^2(I_\xi) +
\frac{D+2}{D-2}\,\Lambda^2\,e^{\frac{4}{D-2}\Phi}\,I_\xi
-\Lambda^2\,e^{\frac{4}{D-2}\Phi}\, I_\chi\cr 0&=& D^B\left(
e^{2\,\alpha_q\,\Phi}\,(f_{q+2})_{A_1\dots
  A_{q+1}B}\right)\qquad,\qquad q\neq D-2\cr 0 &=&
D_A\left(\frac{2D}{D-2}\,I_\xi-I_\chi\right)\label{ep4} \eea De la
\'ultima ecuaci\'on resulta, \be I_\chi = \frac{2D}{D-2}\,I_\xi. \ee
De esta manera, nos queda un sistema parcialmente desacoplado \bea
0&=&A_{AB}(h) - \frac{2\Lambda^2}{D}\,e^{\frac{4}{D-2}\Phi}\,h_{AB}\cr
0&=&D^2(I_\xi) -
\frac{\Lambda^2}{D-2}\,e^{\frac{4}{D-2}\Phi}\,I_\xi\cr 0&=&
D^B\left(e^{2\,\alpha_q\,\Phi}\,(f_{q+2})_{A_1\dots
  A_{q+1}B}\right)\qquad,\qquad q\neq D-2.\label{ep5} \eea Todo el
espectro perturbativo proviene de estas ecuaciones. Vale la pena
se\~nalar que la ecuaci\'on para la m\'etrica es invariante de gauge,
como puede ser verificado usando la propiedad general \be
A_{AB}(^\epsilon h)-A_{AB}(h) = -2\,\left( D_C R_{AB}\,\epsilon^C +
R_{AC}\,D_B\epsilon^C+R_{BC}\,D_A\epsilon^C \right) \ee y las
ecuaciones de movimiento para el fondo (\ref{efeq}). Luego, en la
subsecci\'on \ref{Smetricfluct}, ser\'an constru{\'\i}dos nuevos
invariantes a partir de las fluctuaciones de la m\'etrica, para el
anz\"atz particular considerado.

\subsection{La ecuaci\'on para las fluctuaciones dilat\'onicas}

Las fluctuaciones dilat\'onicas corresponden a tomar
\footnote{Con la siguiente ecuaci\'on no invariante de gauge $h_{AB}
  =0$, queremos expresar que ponemos a cero toda posible fluctuaci\'on
  de la m\'etrica constru{\'\i}da a partir de $h_{AB}$,  ver subsecci\'on
  \ref{Smetricfluct}.} \be h_{AB} =0\qquad;\qquad f_{q+2}\equiv
da_{q+1}= \left\{
\begin{array}{ll}
0 &,\;\; q\neq D-2\cr
(-)^{p+1}\,\frac{2D}{D-2}\,Q_{D-2}\, e^{\frac{2D}{D-2}\,\Phi}\;\epsilon_G\, I_\xi&,\;\; q= D-2,
\end{array}\right.
\ee donde $I_\xi$ satisface la segunda ecuaci\'on en (\ref{ep5}).
Descomponemos la perturbaci\'on en modos de Fourier,
\be I_\xi(x, u)=\chi(u)\;e^{ip_a\,x^a}.\label{xians} \ee Por simetr\ii
a translacional en las coordenadas $x^a$ de las soluciones de fondo,
los modos no se mezclan.  En la secci\'on \ref{holomodels},
consideraremos algunas de ellas compactificadas.

Reemplazando (\ref{xians}) en (\ref{ep5}), ecuaci\'on $I_\xi$,
tenemos \be \frac{1}{E}\;\partial_u\left( \frac{E}{C^2}\,\partial_u
\chi(u) \right) - \left( \sum_a \frac{p^a\,p_a}{A_a{}^2} + \Lambda^2
\right)\,\chi(u) = 0\label{xieq} \ee donde hemos usado la forma
expl\ii cita de la m\'etrica en el {\it string frame} (\ref{Ssn}).  El
siguiente cambio de variables y la definici\'on, \bea
\left(\frac{u}{u_0}\right)^{D-1} &=& 1 + e^x \equiv g(x)\qquad,\qquad
x\in\Re\cr \chi(u)&\equiv& g(x)^{-\frac{1}{2}}\; H(x)\label{cv-red}
\eea ponen (\ref{xieq}) en la forma t\ii pica de una ecuaci\'on de
Schr\"odinger para $H(x)$, \be 0=-H''(x) + V(x)\,H(x)\label{schro} \ee
con el potencial dado por \be V(x) = \frac{1}{4} -
\frac{1}{4\,g(x)^2} + \sum_a {\hat p}^a {\hat p} _a\;
\frac{e^{(1-a_a)x}}{g(x)^{1-a_a +\frac{2}{D-1}}}+
\frac{D}{D-1}\,(1+e^{-x})^{-1}\label{dilatoneq} \ee donde ${\hat p}_a
\equiv \frac{1}{(D-1)\,u_0}\, p_a$.

\subsection{La ecuaci\'on para las fluctuaciones de la RR 1-forma}

Entre las posibles fluctuaciones de las $(q+1)$-formas, nosotros
consideraremos campos $a_1$ de RR que siempre est\'an presentes en
teor\ii as D-dimensionales tipo IIA NCST\footnote{ En $D=8$ puede
  estar presente tambi\'en $a_3$, y por supuesto, en cualquier
  dimensi\'on el campo de Kalb-Ramond $B_2{}_{AB}$ proveniente del
  sector NS-NS, pero no ser\'an considerados en este trabajo.}.  De
acuerdo a (\ref{ep5}), las fluctuaciones obtenidas prendiendo s\'olo
$a_{q+1}$, para cualquier $q\neq D-2$ es consistente si $a_{q+1}$
satisface las ecuaciones de Maxwell generalizadas en la m\'etrica de
fondo.
En particular, para la forma $a_1$,
\be
D^B\left( e^{2\,\alpha_q\,\Phi}\, D_A(a_1{}_B)\right) - D^B\left(
e^{2\,\alpha_q\,\Phi}\,D_B(a_1{}_A)\right)= 0.\label{1form} \ee
Nuevamente, desarrollando Fourier \be a_A(x, u)= \chi_A(u)\;e^{i p_a\,
  x^a},\label{a1ans} \ee y usando los resultados del Ap\'endice A tenemos
\bea D^B F_{aB}\;e^{-i p_c\, x^c} &=& \sum_c \frac{p^c p_c}{A_c{}^2}\;
P_a{}^b \chi_b + \frac{C\,A_a}{E}\,e_n\left( \frac{E}{C\,A_a}\left(
i\,\frac{p^a}{A_a}\,\chi_n - e_n(\chi_a)
-\sigma_a\,\chi_a\right)\right)\cr D^B F_{nB}\;e^{-i p_c\, x^c} &=&
\sum_b \frac{p^b p_b}{A_b{}^2}\; \chi_n + i\,\sum_b
\frac{p^b}{A_b}\,\left( e_n(\chi_b) +\sigma_b\,\chi_b\right), \eea
donde $\,P_a{}^b\equiv \delta_a{}^b - \left(\sum_c\frac{p^c
  p_c}{A_c{}^2}\right)^{-1} \frac{p_a}{A_a}\frac{p^b}{A_b}$.  La
segunda ecuaci\'on 
es un v\ii nculo que da $\chi_n$ en t\'erminos de las funciones
$\chi_b$. Poniendo \'esta en la primer ecuaci\'on resulta \be
\frac{C\,A_a}{E}\,e_n\left( \frac{E}{C\,A_a}\;
P_a{}^b\left(e_n(\chi_b) +\sigma_b\,\chi_b\right)\right) - \sum_c
\frac{p^c p_c}{A_c{}^2}\; P_a{}^b \chi_b = 0\label{a1eq} \ee Este
sistema acoplado ser\'a reducido en la secci\'on siguiente usando el
ans\"atz usual en cada uno de los modelos considerados.

\subsection{La ecuaci\'on para las fluctuaciones de la m\'etrica}
\label{Smetricfluct}
De (\ref{ep5}),
\be
I_\xi = I_\chi= 0\qquad;\qquad f_{q+2}\equiv da_{q+1}=
\left\{
\begin{array}{ll}
0 &,\;\; q\neq D-2\cr
\frac{(-)^D}{2}\,Q_{D-2}\, e^{\frac{2D}{D-2}\,\Phi}\;\epsilon_G\,h^C_C&,\;\; q= D-2
\end{array}\right.
\ee es consistente si la perturbaci\'on de la m\'etrica satisface
\bea A_{AB}(h)\equiv D_A D_B h^C_C + D^2 h_{AB} - D^C D_A h_{CB} -D^C
D_B h_{AC} = \frac{2\,\Lambda^2}{D}\,
e^{\frac{4}{D-2}\Phi}\,h_{AB}.\label{3dmetric} \eea Para escribirla en
forma manifiestamente invariante de gauge, introducimos los campos
($g, g_a, I_{ab}$) \bea e_n(g)&\equiv& h_{nn} \cr
A_a\,e_n\left(\frac{g_a}{A_a}\right)&\equiv& h_{an} - \frac{1}{2}\,
e_a(g)\cr I_{ab}&\equiv& h_{ab} - e_a(g_b) - e_b(g_a) -
\eta_{ab}\,\sigma_a\, g,\label{ls-bdgduals} \eea que ante una
transformaci\'on de gauge resultan \be \delta_\epsilon g =
2\,\epsilon_n\qquad,\qquad \delta_\epsilon g_a = \epsilon_a \qquad
,\qquad \delta_\epsilon I_{ab}= 0. \ee Las ecuaciones que  siguen de
(\ref{3dmetric}) y (\ref{AAB}) son \bea 0 &=& e^A e_A (I_{ab})+ e_a
e_b(I^c_c) - e^c e_a (I_{bc})-e^c e_b (I_{ac})
+\sigma\,e_n(I_{ab})-\left( \sigma_a-\sigma_b \right)^2\,I_{ab}\cr &+&
\eta_{ab}\,\sigma_a\,e_n(I^c_c)\cr 0 &=& e^c e_n (I_{ac}) - e_a e_n
(I^c_c) + (\sigma_c - \sigma_a)\,\left( e^c(I_{ac}) -
e_a(I^c_c)\right)\cr 0 &=& e_n{}^2 (I^c_c)+2\,\sigma_c\,
e_n(I^c_c),\label{metric1} \eea donde en las \'ultimas dos ecuaciones
se suma sobre el \ii ndice $``c"$.  La dependencia en $g$ y $g_a$ ha
desaparecido dejando todas las ecuaciones expresadas en t\'erminos de
los campos de fluctuaciones invariantes de gauge $I_{ab}$.
Introduciendo los modos de Fourier de la manera
usual, \be I_{ab}(x, u)=\chi_{ab}(u)\;e^{ip_a\,x^a},\label{Iab} \ee las
ecuaciones para los $\chi_{AB}(u)$ que provienen de (\ref{metric1})
resultan \bea 0 &=& e_n{}^2(\chi_{ab})+ \sigma\,e_n(\chi_{ab}) +
\eta_{ab}\,\sigma_a\,e_n(\chi^c_c) -
\frac{p_a\,p_b}{A_a\,A_b}\,\chi^c_c + \left(-
\frac{p^c\,p_c}{A_c{}^2}- \left(\sigma_a-\sigma_b
\right)^2\right)\,\chi_{ab}\cr &+&
\frac{p_a}{A_a}\,\frac{p^c}{A_c}\,\chi_{bc} +
\frac{p_b}{A_b}\,\frac{p^c}{A_c}\,\chi_{ac}\cr 0 &=& \frac{p^c}{A_c}\,
e_n (\chi_{ac}) - \frac{p_a}{A_a}\, e_n (\chi^c_c) + (\sigma_c -
\sigma_a)\,\left( \frac{p^c}{A_c}\, \chi_{ac} -
\frac{p_a}{A_a}\,\chi^c_c\right)\cr 0 &=& \sum_c
\frac{1}{A_c{}^2}\,e_n\left( A_c{}^2\,
e_n(\chi^c_c)\right).\label{chiABeq} \eea


\section{Modelos hologr\'aficos de teor{\'\i}as de Yang-Mills  $d$-dimensionales}
\label{holomodels}

Tomemos de entre las coordenadas $x^a$, $d$ coordenadas $x^\mu$
equivalentes no compactas, con $\mu=0,1,\dots, d-1,\;$ y $D-d-1$
coordenadas compactas no equivalentes $\tau^i,\;$ con $i=1,\dots
D-d-1, \;\;\tau_i\equiv \tau_i +2\,\pi\,R_i$.  Denotaremos con
``$\;\,\tilde{} \,\;$'' las cantidades asociadas con las direcciones
no compactas ($A_\mu = \tilde A\,,\, a_\mu=\tilde a\, ,\,\sigma_\mu =
\tilde \sigma\,,\,$, etc).  La m\'etrica y los v\ii nculos
(\ref{constraints}) resultan \bea l_0{}^{-2}\;G &=& u^2\left(
f(u)^{\tilde a}\;\eta_{\mu\nu}\,dx^\mu\,dx^\nu +
\sum_i\,f(u)^{a_i}\;d\tau^i{}^2\right) + \frac{du^2}{u^2\,f(u)}\cr &
&d\,\tilde a + \sum_i a_i =1\qquad,\qquad d\,\tilde a^2 + \sum_i
a_i{}^2 =1. \eea Notar que $D-d-2$ exponentes contin\'uan siendo libres.

\subsection{Fluctuaciones dilat\'onicas}

De (\ref{dilatoneq}), la ecuaci\'on a resolver es \bea 0&=&-H''(x) +
V(x)\,H(x)\cr V(x) &=& \frac{1}{4} - \frac{1}{4\,g(x)^2}+
\frac{D}{D-1}\,\frac{e^x}{g(x)} - {\hat M}^2\; \frac{e^{(1-\tilde
    a)x}}{g(x)^{1-\tilde a +\frac{2}{D-1}}}+ \sum_i {\hat p} _i{}^2\;
\frac{e^{(1-a_i)x}}{g(x)^{1-a_i +\frac{2}{D-1}}} \cr&
&\label{eqdilaton} \eea donde $M \equiv (D-1)\,u_0\,\hat M$ es la masa
$d$-dimensional.  Los t\'erminos con momentos en las direcciones
compactas son cuantizados en unidades de $R_i{}^{-1}$ y representan
modos de Kaluza-Klein, \'estos se desacoplan para $R_i\rightarrow 0$;

\subsection{Campo de gauge RR: fluctuaciones transversas}

El ans\"atz consistente incluye la condici\'on transversa,
\be
\chi_\mu(u)= \epsilon_\mu(p)\;\chi(u)\qquad;\qquad \epsilon_\mu (p)\,p^\mu = 0.
\label{a1anstrans}
\ee
De (\ref{a1eq}), (\ref{cv-red}),obtenemos
\bea
0&=&-H''(x) + V(x)\,H(x)\cr
V(x) &=& \frac{1}{4\,g(x)^2} \left( \left(\frac{D-3}{D-1}\,e^x -\tilde a\right)^2 + 2\,\left( \frac{D-3}{D-1} +
\tilde a\right)\,e^x\right)- \hat M^2\, \frac{e^{(1-\tilde a)x}}{g(x)^{1-\tilde a +\frac{2}{D-1}}}\cr
&+& \sum_i \hat p_i{}^2\, \frac{e^{(1-a_i)x}}{g(x)^{1-a_i +\frac{2}{D-1}}}.\label{a1transpot2}
\eea

\subsection{Campo de gauge RR: fluctuaciones longitudinales}

El ans\" atz consistente, a $i$ fijo (pero arbitrario $i=1,\dots,D-d-1)$ es
\be
\chi_i(u)= \chi(u)\qquad;\qquad p_i =0.\label{a1anslongtau}
\ee
De (\ref{a1eq}), (\ref{cv-red}), tenemos
\bea
0&=&-H''(x) + V(x)\,H(x)\cr
V(x) &=& \frac{1}{4\,g(x)^2} \left( \left(\frac{D-3}{D-1}\,e^x -a_i\right)^2 + 2\,\left( \frac{D-3}{D-1} +
a_i\right)\,e^x\right)- \hat M^2\, \frac{e^{(1-\tilde a)x}}{g(x)^{1-\tilde a +\frac{2}{D-1}}}\cr
&+& \sum_{j\neq i}\hat p_j{}^2\, \frac{e^{(1-a_j)x}}{g(x)^{1-a_j +\frac{2}{D-1}}}\label{a1longtaupot}
\eea

\subsection{M\'etrica: fluctuaciones   transversas}

\'Estas corresponden a tomar el ans\" atz \be \chi_{\mu\nu}(u)=
\epsilon_{\mu\nu}(p)\;\chi(u)\qquad;\qquad
\epsilon_\rho^\rho=0\;\;,\;\;\epsilon_{\mu\nu}\,p^\nu =
0.\label{hmnanstrans} \ee y poner el resto a cero.  Las ecuaciones
(\ref{chiABeq}) se satisfacen si luego de hacer el cambio
(\ref{cv-red}), $H$ obedece la ecuaci\'on \bea 0&=&-H''(x) +
V(x)\,H(x)\cr V(x) &=& \frac{1}{4} - \frac{1}{4\,g(x)^2} - {\hat
  M}^2\; \frac{e^{(1-\tilde a)x}}{g(x)^{1-\tilde a +\frac{2}{D-1}}}+
\sum_i {\hat p} _i{}^2\; \frac{e^{(1-a_i)x}}{g(x)^{1-a_i
    +\frac{2}{D-1}}}.\label{eqmetrictrans} \eea

\subsection{M\'etrica: fluctuaciones  longitudinales}

Estas perturbaciones son las correspondientes a tomar ans\" atz, a $i$
fijo (arbitrario) \be \chi_{i\mu}(u)=
\epsilon_\mu(p)\;\chi(u)\qquad;\qquad
p_i=0\;\;,\;\;\epsilon_\mu\,p^\mu = 0\label{hmnanslong} \ee y el resto
cero.  \'Este obedece las ecuaciones (\ref{chiABeq}) si luego del cambio
(\ref{cv-red}), H satisface la ecuaci\'on \bea 0&=&-H''(x) +
V(x)\,H(x)\cr V(x) &=& \frac{1}{4} - \frac{1 - (\tilde a -
  a_i)^2}{4\,g(x)^2} - {\hat M}^2\; \frac{e^{(1-\tilde
    a)x}}{g(x)^{1-\tilde a +\frac{2}{D-1}}}+ \sum_{j\neq i} {\hat p}
_j{}^2\; \frac{e^{(1-a_j)x}}{g(x)^{1-a_j +\frac{2}{D-1}}}.
\label{eqmetriclong}
\eea

\subsection{M\'etrica: fluctuaciones escalares}

\'Este es el caso m\'as complicado, debido a que involucra, en
general, un sistema de ecuaciones acoplado.  El sistema puede ser
reducido a una ecuaci\'on del tipo (\ref{dilatoneq}) para alg\'un
potencial particular, como sucede en los casos de las perturbaciones
analizadas anteriormente, s\'olo en el caso analizado por
\cite{Constable:1999gb}. Por lo tanto, lo mejor ser\'a hacer aqu{\'\i}
un an\'alisis detallado para estas fluctuaciones.


Consideremos el siguiente ans\"atz \be \chi_{\mu\nu}(u) =
a(u)\,\eta_{\mu\nu} + b(u)\,p_\mu\,p_\nu \qquad,\qquad \chi_{ij}(u) =
a_i(u)\,\delta_{ij} \qquad,\qquad\chi_{\mu i}(u) = 0. \ee Notemos que
este ans\"atz depende de $D-d+1$ funciones invariantes $(a, a_i, b)$.
A diferencia de \cite{Constable:1999gb}, todas estas funciones son
relevantes por ser invariantes de gauge. Resulta conveniente entonces
introducir las siguientes fluctuaciones invariantes \bea F &=& a -
\tilde\sigma\, \tilde A{}^2\,e_n(b)\cr F_i &=& a_i - \sigma_i\, \tilde
A{}^2\,e_n(b)\cr F_n &=& -e_n\left(\tilde A{}^2\,e_n(b)\right). \eea En
t\'erminos de \'estas, las ecuaciones (\ref{chiABeq}) se escriben como
\bea 0 &=& e_n{}^2(F)+ \sigma\,e_n(F) + \tilde\sigma\,e_n\left(d\,F +
F_\tau - F_n\right) + \frac{M^2}{\tilde A{}^2}\, F -
\frac{2\,\Lambda^2}{D}\, e^{\frac{4}{D-2}\Phi}\, F_n\cr 0 &=&
e_n{}^2(F_i)+ \sigma\,e_n(F_i) + \sigma_i\,e_n\left(d\,F + F_\tau -
F_n\right) + \frac{M^2}{\tilde A{}^2}\, F_i - \frac{2\,\Lambda^2}{D}\,
e^{\frac{4}{D-2}\Phi} F_n\cr 0 &=& e_n{}^2\left(d\,F + F_\tau\right)+
2\,d\,\tilde\sigma\, e_n(F) + 2\, \sum_i\sigma_i\,e_n(F_i) -
\sigma\,e_n(F_n) + \left( \frac{M^2}{\tilde A{}^2}-
\frac{2\,\Lambda^2}{D}\,e^{\frac{4}{D-2}\Phi}\right)\,F_n\cr 0 &=&
e_n\left( (d-1)\,F + F_\tau\right) + \sum_i (\sigma_i -\tilde\sigma)\,
F_i -(\sigma -\tilde\sigma)\,F_n\cr 0 &=& (d-2)\, F + F_\tau +
F_n,\label{Feq1} \eea donde $F_\tau\equiv \sum_i F_i\,$. La \'ultima
ecuaci\'on es claramente un v\ii nculo, que se resuelve trivialmente
para $\,F_n = - (d-2)\, F - F_\tau\,$.  El resto de las ecuaciones
toman la forma \bea 0 &=& e_n{}^2(F)+ \sigma\,e_n(F) +
2\,\tilde\sigma\,e_n\left((d-1)\,F + F_\tau\right) + \frac{M^2}{\tilde
  A{}^2}\,F+\frac{2\,\Lambda^2}{D}\,
e^{\frac{4}{D-2}\Phi}\,\left((d-2)\,F + F_\tau\right)\cr 0 &=&
e_n{}^2(F_i)+ \sigma\,e_n(F_i) + 2\,\sigma_i\,e_n\left((d-1)\,F +
F_\tau\right) + \frac{M^2}{\tilde A{}^2}\, F_i +
\frac{2\,\Lambda^2}{D}\, e^{\frac{4}{D-2}\Phi}\, \left((d-2)\,F +
F_\tau\right)\cr 0 &=& e_n{}^2\left(d\,F + F_\tau\right)+
\sigma\,e_n\left(d\,F + F_\tau\right) + 2\, \sum_i\sigma_i\,e_n(F_i
-F)\cr &+&\left( -\frac{M^2}{\tilde A{}^2} + \frac{2\,\Lambda^2}{D}\,
e^{\frac{4}{D-2}\Phi}\right)\,\left((d-2)\,F + F_\tau\right)\cr 0 &=&
e_n\left( (d-1)\,F + F_\tau\right) + (d-2)\,(\sigma-\tilde\sigma)\,F +
\sum_i (\sigma_i +\sigma - 2\,\tilde\sigma)\, F_i.\label{Feq2} \eea En
este punto, 3 hechos son notorios:
\begin{itemize}
\item Hay $D-d$ inc\'ognitas $(F , F_i)$ y $D-d+2$ ecuaciones
  diferenciales; obviamente esto est\'a directamente relacionado con el
  fijado de la invarianza de gauge;
\item Los t\'erminos $e_n\left( (d-1)\,F + F_\tau\right)$ en las dos
  primeras ecuaciones en (\ref{Feq1}) pueden ser eliminados usando la
  \'ultima;
\item La tercer ecuaci\'on puede ser transformada en una de primer orden
  usando las dos primeras ecuaciones.
\end{itemize}

Teniendo en cuenta todo esto, el sistema (\ref{Feq1}) puede ser dividido
en dos conjuntos, un sistema de ecuaciones de segundo orden con $D-d$
ecuaciones y $D-d$ inc\'ognitas \bea 0 &=& e_n{}^2(F)+ \sigma\,e_n(F) +
\left( \frac{M^2}{\tilde A{}^2}+ 2\, (d-2)\,\left(
\frac{\Lambda^2}{D}\, e^{\frac{4}{D-2}\Phi} - \tilde\sigma\,(\sigma
-\tilde\sigma)\right)\right)\,F\cr &+& 2\,\sum_i
\left(\frac{\Lambda^2}{D}\,
e^{\frac{4}{D-2}\Phi}-\tilde\sigma\,\left(\sigma_i +\sigma
-2\,\tilde\sigma\right)\right)\, F_i\cr 0 &=& e_n{}^2(F_i)+
\sigma\,e_n(F_i) + 2\,(d-2)\,\left( \frac{\Lambda^2}{D}\,
e^{\frac{4}{D-2}\Phi} - \sigma_i\,\left(\sigma
-\tilde\sigma\right)\right)\,F\cr &+& \sum_j\left(\frac{M^2}{\tilde
  A{}^2}\,\delta_{ij}+ \frac{2\,\Lambda^2}{D}\, e^{\frac{4}{D-2}\Phi}
- 2\,\sigma_i\,\left(\sigma_j + \sigma
-2\,\tilde\sigma\right)\right)\, F_j,\label{Feqsecond} \eea y dos
ecuaciones de primer orden \bea 0 &=& \sum_i\sigma_i\, e_n\left(F -
F_i\right) + \left( (d-1)\,\frac{M^2}{\tilde A{}^2} +
(d-2)\,\sum_i\sigma_i\,\left(\tilde\sigma -\sigma_i\right)\right)\,
F\cr &+&\sum_i\left( \frac{M^2}{\tilde A{}^2}+ \sum_j
\sigma_j\,\left(\tilde\sigma -\sigma_j\right) +
\sigma\,\left(\tilde\sigma -\sigma_i\right)\right)\, F_i\cr 0 &=&
e_n\left( (d-1)\,F + F_\tau\right) + (d-2)\,(\sigma-\tilde\sigma)\,F +
\sum_i (\sigma_i +\sigma - 2\,\tilde\sigma)\, F_i.\label{Feqfirst} \eea
Nos concentramos en primer lugar en el sistema de ecuaciones de segundo
orden (\ref{Feqsecond}).  Usando (\ref{cv-red}), introducimos la
variable $x$ y los campos $(H , H_i)$ de la siguiente manera \be
F(u)\equiv g(x)^{-\frac{1}{2}}\; H(x)\qquad,\qquad F_i(u)\equiv
g(x)^{-\frac{1}{2}}\; H_i(x).\label{FH} \ee Luego de algunos c\'alculos,
(\ref{Feqsecond}) puede ser reescrito de la siguiente forma
\bea \vec 0 &=&- \vec H''(x) + {\bf V}(x)\; \vec H(x)\cr {\bf V}(x)
&\equiv& v(x)\, {\bf 1} + \left(\begin{array}{cc}m(x)& \vec
  m^{(1)}{}^t(x)\cr\vec m^{(2)}(x)&{\bf
    m}(x)\end{array}\right),\label{Heq} \eea donde $\vec H(x)\equiv
(H(x), H_1 (x),\dots, H_{D-d-1}(x))$.  Los elementos que definen el
potencial matricial ${\bf V}(x)$ est\'an dados por
\bea
v(x) &=& \frac{1}{4} - \frac{1}{4\,g(x)^2} - \hat M{}^2\;
\frac{ e^{(1-\tilde a)x} }{ g(x)^{1-\tilde a +\frac{2}{D-1}} }\cr
m(x) &=& \frac{D-4}{g(x)^2}\,\left( \frac{\tilde a}{2}\,(1-\tilde a) +
\frac{\,(D-3)\,\tilde a - 1}{D-1}\, e^x - \frac{2}{(D-1)^2}\, e^{2\,x}\right)\cr
m_i^{(1)}(x) &=& \frac{1}{g(x)^2}\,\left( \frac{\tilde a}{2}\,(1+a_i-2\,\tilde a) +
\frac{(D-4)\,\tilde a + a_i - 1}{D-1}\, e^x - \frac{2}{(D-1)^2}\, e^{2\,x}\right)\cr
m_i^{(2)}(x) &=& \frac{D-4}{g(x)^2}\,\left( \frac{a_i}{2}\,(1-\tilde a) +
\frac{(D-2)\,a_i -\tilde a -1}{D-1}\, e^x - \frac{2}{(D-1)^2}\, e^{2\,x}\right)\cr
{\bf m}_{ij}(x) &=& \frac{1}{g(x)^2}\,\left( \frac{a_i}{2}\,(1+a_j-2\,\tilde a) +
\frac{(D-2)\, a_i + a_j -2\,\tilde a -1}{D-1}\, e^x - \frac{2}{(D-1)^2}\, e^{2\,x}\right).\cr
& &
\label{element_potential}
\eea Este sistema ser\'a analizado en profundidad cuando calculemos
el espectro respectivo.
\bigskip

\noindent{\bf El caso $d=D-2$}

> Qu\'e ocurre con las ecuaciones lineales (\ref{Feqfirst})?  Para
$d=D-2$ hay s\'olo una direcci\'on compacta $\tau^i\equiv\tau$, y
consecuentemente introducimos $a_i\equiv a_\tau\,$.  Existen, para
este caso, dos soluciones que corresponden a los siguientes valores de
los exponentes $\,(\tilde a = 0,\, a_\tau = 1)$ y $\,(\tilde a =
\frac{2}{D-1} ,\, a_\tau = -\frac{D-3}{D-1})$.  La primera corresponde
justamente al agujero negro de Schwarzchild AdS en $D$ dimensiones; el
per\ii odo de $\tau$ es usualmente fijado como en (\ref{period})
requiriendo la ausencia de la singularidad c\'onica en el plano
$\tau-u$; de esta manera la soluci\'on se vuelve regular. La segunda,
en cambio, no es regular en el IR. M\'as \'aun, esta soluci\'on no es
confinante seg\'un (\ref{pconfcond}) y por lo tanto no ser\'a
considerada.  En este caso, tenemos s\'olo dos inc\'ognitas $F$ y
$F_\tau$, y las ecuaciones (\ref{Feqfirst}) resultan entonces un
sistema de dos ecuaciones de primer orden con dos
inc\'ognitas. Haciendo el cambio (\ref{FH}), (\ref{Feqfirst}) resulta
\bea 0 &=&\left(\begin{array}{c}H(x)\cr H_\tau(x)\end{array}\right)' -
     {\bf U}(x)\; \left(\begin{array}{c}H(x)\cr H_\tau
       (x)\end{array}\right)\cr {\bf U}(x) &\equiv&
     \frac{e^x}{2\,g(x)}\, 1 + \left(\begin{array}{cc}u_{11}(x)&
       u_{12}(x)\cr u_{21}(x)&u_{22}(x)
\end{array}\right)\label{Heqlinear}
\eea donde los elementos que definen ${\bf U}(x)$ est\'an dados por
\bea u_{11} &=& -\frac{(D-3)\,(D-1)}{D-2}\, \hat M^2\,\frac{
  e^{(1-\tilde a)x}}{ g(x)^{1-\tilde a +\frac{2}{D-1}} }
\,\frac{g(x)}{\frac{D-1}{2}\,a_\tau + e^x}
-\frac{D-4}{D-1}\,\frac{\frac{D-1}{2}\,\tilde a + e^x}{g(x)}\cr u_{12}
&=& -\frac{D-1}{D-2}\, \hat M^2\,\frac{ e^{(1-\tilde a)x}}{
  g(x)^{1-\tilde a +\frac{2}{D-1}} }
\,\frac{g(x)}{\frac{D-1}{2}\,a_\tau + e^x} -
\frac{1}{D-1}\,\frac{\left(\frac{D-1}{2}\,\tilde a +
  e^x\right)^2}{g(x)\,\left( \frac{D-1}{2}\,a_\tau + e^x\right)}\cr
u_{21} &=& \frac{(D-3)^2\,(D-1)}{D-2}\, \hat M^2\,\frac{ e^{(1-\tilde
    a)x} } { g(x)^{1-\tilde a
    +\frac{2}{D-1}}}\,\frac{g(x)}{\frac{D-1}{2}\,a_\tau + e^x}
-\frac{D-4}{D-1}\, \frac{\frac{D-1}{2}\,a_\tau + e^x}{g(x)}\cr u_{22}
&=& \frac{(D-3)\,(D-1)}{D-2}\, \hat M^2\,\frac{ e^{(1-\tilde a)x} } {
  g(x)^{1-\tilde a +\frac{2}{D-1}}
}\,\frac{g(x)}{\frac{D-1}{2}\,a_\tau + e^x}\cr &+&
\frac{1}{g(x)}\,\left(
\frac{D-3}{D-1}\,\frac{\left(\frac{D-1}{2}\,\tilde a + e^x\right)^2}
     {\frac{D-1}{2}\,a_\tau + e^x}+\tilde a - \frac{1+a_\tau}{2}
     -\frac{D-2}{D-1}\,e^x\right). \eea 

Se puede probar sin mayor dificultad que (\ref{Heqlinear})
implica el sistema de segundo orden (\ref{Heq}) si y s\'olo si se
cumple la siguiente identidad \be {\bf V}(x) = {\bf U}'(x) + {\bf
  U}(x)^2\label{U-V} .\ee Esta relaci\'on ha sido verificada por
c\'omputo directo.  M\'as a\'un, la equivalencia del conjunto completo
de ecuaciones con el sistema lineal (\ref{Heqlinear}) nos permite
tambi\'en encarar el problema resolviendo directamente una ecuaci\'on de
segundo orden para alguno de los campos, constru{\'\i}da a partir del
sistema lineal de dos ecuaciones con dos inc\'ognitas (\ref{Heqlinear}).
En el caso general $D-d-1>1$, las ecuaciones lineales act\'uan
presumiblemente como v\'inculos, y debiendo resolver (\ref{Heq}) y
verificar a posteriori (\ref{Heqlinear})
\footnote{ Inversamente, si como haremos, suponemos que se cumple
  (\ref{Heq}), entonces si hallamos una matriz ${\bf U}$ que verifique
  (\ref{U-V}), se sigue que \be ({\vec H}'(x) - {\bf U}(x)\, \vec
  H(x))' + {\bf U}(x)\,(\vec H'(x) - {\bf U}(x)\,\vec H(x)) = 0. \ee
  Creemos que con una elecci\'on conveniente de ${\bf U}$ (y tal vez,
  apropiadas condiciones de contorno), $\vec H'(x) = {\bf U}(x)\,\vec
  H(x)$, hecho que ciertamente ocurre en el caso $d=D-2$.  Esto nos
  dejar\ii a con un sistema lineal $(D-d)$-dimensional que
  presumiblemente implicar\ii a las dos ecuaciones (\ref{Heq}), pero
  no hemos verificado esto debido a la complejidad de (\ref{U-V}).}.
Seguiremos con esta estrategia en la secci\'on siguiente, presentando
tambi\'en resultados relacionados con este caso, compatibles con los
hallados en \cite{Constable:1999gb}.


\section{Espectro de Glue-ball de teor{\'\i}as de Yang-Mills $3D$}\label{3dgs}

En esta secci\'on, consideraremos supercuerdas no cr\ii ticas tipo IIA
en dimensi\'on $D=6$.  Las formas de RR presentes son $A_1$ (con $A_3$
como su dual Hodge) y $A_5$, con intensidades de campo $F_2 = dA_1$ y
$F_6 = dA_5$, respectivamente.  Tomaremos adem\'as $d=3$ direcciones
equivalentes, y $D-d-1=2$ direcciones compactas no equivalentes. La
familia considerada es interpretada como soluciones de $D4$-branas
enrolladas en un 2-toro de radios $(R_1, R_2)$. Las ecuaciones de
v{\'\i}nculos (\ref{constraints}) son \be 3\,\tilde a + a_1+a_2
=1\qquad,\qquad 3\,\tilde a^2 + a_1{}^2+a_2{}^2 =1. \ee Como ya
se\~nal\'aramos antes, la idea aqu{\'\i} es estudiar la dependencia
del espectro con los exponentes.  Resulta \'util resolver el v\ii
nculo en la forma siguiente \bea \sqrt{3}\;\tilde
a&=&\frac{\cos\beta^- -
  \cos\beta^+}{1-\cos\beta^+\,\cos\beta^-}\;\sin\beta^-\cr a_1
&=&\frac{\cos\beta^- -
  \cos\beta^+}{1-\cos\beta^+\,\cos\beta^-}\;\cos\beta^-\cr
a_2&=&-\frac{\sin\beta^+\;\sin\beta^-}{1-\cos\beta^+\,\cos\beta^-},
\eea donde el espacio de soluciones es una $S^1$ parameterizada por
$\beta\sim\beta+\pi\,$, y \be \beta^\pm\equiv
\beta\pm\beta_0\qquad,\qquad \tan\beta_0 \equiv
\frac{1}{\sqrt{d}}\;\;,\;\; \beta_0\in \left[0,\frac{\pi}{2}\right].
\ee Para $\beta=\beta_0$ obtenemos la soluci\'on de KS. Es posible
pensar nuestra familia como una deformaci\'on con par\'ametro $\beta^-
= \beta-\beta_0$ de esa soluci\'on.  En ese sentido, resulta natural
imponer la condici\'on de periodicidad (\ref{period}) al menos sobre
una de las direcciones compactas \cite{Kuperstein:2004yk}
\cite{Lugo:2006vz}, pensando en ella como la que permite romper
SUSY. Recordemos que (\ref{period}) aparece para imponer suavidad a la
soluci\'on a $u=u_0$, sin embargo esto no resulta claro en nuestro
caso dado que nuestras soluciones siguen siendo singulares en ese
punto. Nosotros dejaremos libres ambos radios, por ahora.

Antes de presentar los resultados num\'ericos que hemos obtenido
\footnote{Los espectros fueron calculados tanto con el enfoque WKB,
  como num\'ericamente, pero dado el gran acuerdo entre ellos, s\'olo
  mostramos en las tablas los resultados num\'ericos.}, resulta
instructivo analizar c\'omo funciona en nuestro caso el l\ii mite de
desacoplamiento (decoupling limit), siguiendo el an\'alisis usual.
Primero, para desacoplar toda la torre de estados de cuerdas abiertas
debemos imponer el l\ii mite de desacople de baja energ\ii a,
$l_s\equiv\sqrt{\alpha'}\rightarrow 0$.  Esto nos lleva a una
constante de acoplamiento de Newton 6-dimensional $2\,\kappa_6{}^2
\sim l_s{}^4\,g_s{}^2$\footnote{El factor num\'erico en teor{\'\i}as
  cr{\'\i}ticas de tipo II es $(2\,\pi)^7$; a partir de, por ejemplo,
  la amplitud de dispersi\'on de 4 gravitones, deber{\'\i}a ser
  posible fijarlo tambi\'en en teor{\'\i}as no cr{\'\i}ticas, pero
  hasta donde tenemos conocimiento, este c\'alculo (o cualquier otro
  que permita fijar el acoplamiento) no ha sido realizado.
}, este l\ii mite, a $g_s$ fijo, tambi\'en desacopla
las interacciones {\it bulk-open} y {\it bulk-bulk}.  Por otro lado, de
acuerdo con (\ref{dilaton}), el l\ii mite de $N$ grande nos deja con la
teor\ii a de cuerdas cl\'asica.


Recordemos en primer lugar que un vac\ii o no cr\ii tico tipo IIA
(dilat\'on lineal, cigarro, etc.) preserva $2^\frac{D}{2}= 8$ supercargas
\cite{Kutasov:1990ua}.  Una $Dp$-brana BPS en \'el, preserva la mitad
de estas supercargas.  En efecto, se muestra en la referencia
\cite{Ashok:2005py} que el l\ii mite de baja energ\ii a de una
D3-brana sobre el vac\ii o del cigarro es ${\cal N}= 1$ super Yang
Mills en 4 dimensiones, esto es, preserva $4=\frac{1}{2}\,8$
supercargas.  Ahora, nuestra soluci\'on de D4 con dilat\'on
constante puede pensarse como la versi\'on negra de una D4 BPS ($u_0
=0$) en el {\it near horizon l\ii mit}, que es T-dual a una D3
brana BPS viviendo sobre el vac\ii o de dilat\'on lineal, que corresponde
al l\ii mite $u$ grande del vac\ii o del cigarro. De esta manera,
podemos decir que nuestra familia describe en el UV alguna CFT
(desconocida) en 5 dimensiones, que es el completamiento de la
teor\ii a de YM 5-dimensional con constante de acoplamiento a la
escala $\Lambda_s\equiv l_s{}^{-1}$ dada por \be g_{YM_5}{}^2 =
\frac{T_s{}^2}{T_{D4}}\sim l_s\,g_s.\label{cc5d} \ee

El acoplamiento de t'Hooft a dicha escala, y la constante de
acoplamiento efectiva adimensional a la escala $E$ son \be
\lambda_5{}^2 \equiv g_{YM_5}{}^2\, N \sim l_s\qquad;\qquad
\lambda_5^{eff}(E)^2\equiv \lambda_5{}^2\, E \sim
\frac{E}{\Lambda_s},\label{tof5d} \ee donde hemos usado
(\ref{dilaton}).  De (\ref{tof5d}), dos hechos bien conocidos se deducen:
Para $E<<\Lambda_s$, $\,\lambda_5^{eff}<<1$ y la teor\ii a de
perturbativa de YM es v\'alida a bajas energ\ii as. Por otro lado, es
claro que $g_{YM_5}$ no puede ser fijado por $l_s\rightarrow 0$, y por
lo tanto no existe l\ii mite de desacoplamiento; este hecho puede
interpretarse como una manifestaci\'on de la no renormalizabilidad de
las teor\ii as de YM en dimensiones mayores que 4
\cite{Itzhaki:1998dd}. Sin embargo, nosotros estamos interesados en la
teor\ii a tridimensional obtenida por debajo de la escala de
compactificaci\'on $\Lambda_c \equiv (4\pi^2\,R_1\,R_2)^{-\frac{1}{2}}$;
el acoplamiento de t'Hooft a esta escala es \be \lambda_3{}^2 =
\frac{\lambda_5{}^2}{4\pi^2\,R_1\,R_2}\sim
\frac{\Lambda_c{}^2}{\Lambda_s}. \ee Luego, siguiendo el argumento de
Witten \cite{Witten:1998zw}, la compactificaci\'on deber\ii a romper
supersimetr\ii a, dando masa a los fermiones y a los bosones a nivel
\'arbol y a un loop, respectivamente. Agrandando la escala de
compactificaci\'on, estos deber\ii an desacoplarse dejando YM
tridimensional en el l\ii mite \cite{Aharony:1999ti} \be
\lambda_3{}^2\;\xrightarrow[\Lambda_s\rightarrow\infty]{\Lambda_c\rightarrow\infty}\;
\Lambda_{QCD}\sim
u_0\qquad\longleftrightarrow\qquad\frac{\Lambda_c{}^2}{\Lambda_s}\sim
u_0, \ee donde se ha tenido en cuenta (\ref{uoscale}).

Usaremos la notaci\'on usual que asigna a cada tipo de perturbaci\'on
la notaci\'on dual de glueball correspondiente $J^{PC}$.  El spin $J$,
paridad $P$ y conjugaci\'on de carga $C$, son deducidos de los
n\'umeros cu\'anticos de los operadores de la teor\ii a del borde al
cual se acopla la perturbaci\'on considerada. Los detalles se pueden
encontrar en \cite{Csaki:1998qr,Brower:2000rp}.

En la secciones que siguen, calcularemos detalladamente el espectro
correspondiente para cada tipo de perturbaci\'on.
\bigskip

\subsection{Espectro de fluctuaciones dilat\'onicas: $O^{++}$ glue-balls}

En este caso, la ecuaci\'on de tipo Schr\"odinger con energ\ii a cero
a resolver corresponde al potencial 
\be V(x)=\frac{1}{4} - \frac{1}{4\,g(x)^2}
+ \frac{6}{5}\,\frac{e^x}{g(x)}- \hat M^2\, \frac{e^{(1-\tilde
    a)x}}{g(x)^{\frac{7}{5}-\tilde a}} + \sum_{i=1}^2 \hat p_i{}^2\,
\frac{e^{(1- a_i)x}}{g(x)^{\frac{7}{5}-a_i}}.\label{schrodil} \ee

\begin{center}
\begin{minipage}{5cm}
\begin{tabular}{|c|c|c|}
\hline
 $0$ &$\frac{\pi}{6}$ &$\frac{\pi}{12}$  \\
\hline
 7.59 & 7.59 & 4.80   \\
\hline
 10.40 & 10.40 & 7.81 \\
\hline
 13.08& 13.08 &  10.19 \\
\hline
 15.71 & 15.71 & 12.41 \\
\hline
 18.30 & 18.30 & 14.56 \\
\hline
\end{tabular}
\end{minipage}
\end{center}

\vspace{-0.3cm} {\footnotesize La tabla muestra los valores de las
  masas $M_{0^{++}}$, correspondientes a la perturbaci\'on
  dilat\'onica, para valores $\beta=0,\;\;\beta=\frac{\pi}{6}$ y
  $\beta=\frac{\pi}{12}$ con $d=3$.}

\bigskip

\subsection{Espectro de fluctuaciones de  RR $1$-formas.}

Resulta directo verificar que la perturbaci\'on obtenida cuando s\'olo
se enciende $f_{q+2}$ es consistente para cualquier $q\neq D-2$, y
est\'a gobernada por la ecuaci\'on de Maxwell generalizada de
(\ref{ep4}).  En particular, para $D=6$ nos queda $a_1$ del sector de
RR\footnote{Es importante notar que estos c\'alculos est\'an
  realizados en la base local, y no en una base coordenada, por eso
  nuestras componentes tensoriales son diferentes de aquellas en
  \cite{Kuperstein:2004yf} en factores de la m\'etrica.}.

De la secci\'on 5, analizamos 
\begin{itemize}

\item \underline{Polarizaciones longitudinales: $0^{-+}$ glueballs}

Realizando los mismos pasos que en (\ref{cv-red}),
obtenemos la forma de Schr\"odinger (\ref{dilatoneq}), con potencial
\be
V(x)= \frac{1}{4\,g(x)^2} \left( \frac{9}{25}\,e^{2\,x} + \frac{2}{5}\, (3+ 2\,a_i)\,e^x+ a_i{}^2\right)
- \hat M^2\, \frac{e^{(1-\tilde a)x}}{g(x)^{\frac{7}{5}-\tilde a}}
+\sum_{j\neq i}\hat p_j{}^2\, \frac{e^{(1-a_j)x}}{g(x)^{\frac{7}{5}-a_j}}.\label{a1longtaupot3d}
\ee

\begin{center}
\begin{tabular}{c c}

\begin{minipage}{5cm}
{\small
\begin{tabular}{|c|c|c|}
\hline
 $0$ &$\frac{\pi}{6}$ &$\frac{\pi}{12}$\\
\hline
 2.96 & 4.06  & 3.97  \\
\hline
 5.55 & 6.69  & 6.67  \\
\hline
 8.09 & 9.25  & 9.31  \\
\hline
10.61 & 11.78 & 11.94 \\
\hline
13.13 & 14.30 & 14.57 \\
\hline
15.64 & 16.82 & 17.18 \\
\hline
\end{tabular}}
\end{minipage}
&

\begin{minipage}{5cm}
{\small
\begin{tabular}{|c|c|c|}
\hline
 $0$ &$\frac{\pi}{6}$ &$\frac{\pi}{12}$\\
\hline
 4.06  & 2.96 &   3.79\\
\hline
 6.69  & 5.55 &   6.45 \\
\hline
  9.25 & 8.09 &   9.08 \\
\hline
 11.78 & 10.61 & 11.70  \\
\hline
 14.31 & 13.12 & 14.32  \\
\hline
 16.83 & 15.64 &  16.93 \\
\hline
\end{tabular}}
\end{minipage}
\end{tabular}
\end{center}
{\footnotesize En la tabla de la izquierda, mostramos los valores de
  $M_{0^{-+}}$ correspondientes a la perturbaci\'on de la 1-forma
  longitudinal, polarizada a lo largo de la direcci\'on caracterizada
  por $a_1$. El par\'ametro toma los valores siguientes:
  $\beta=0,\;\;\beta=\frac{\pi}{6}$ y $\beta=\frac{\pi}{12}$. En la
  tabla de la derecha, se muestran los valores obtenidos para la
  perturbaci\'on longitudinal polarizada a lo largo de $a_2$. En ambos
  casos $d=3$.}

\item \underline{Polarizaciones transversales: $1^{++}$ glueballs}

El potencial es
\be
V(x)= \frac{1}{4\,g(x)^2} \left( \frac{9}{25}\,e^{2\,x} + \frac{2}{5}\, (3+ 2\,\tilde a)\,e^x+ \tilde a{}^2\right)
- \hat M^2\, \frac{e^{(1-\tilde a)x}}{g(x)^{\frac{7}{5}-\tilde a}}
+\sum_{j}\hat p_j{}^2\, \frac{e^{(1-a_j)x}}{g(x)^{\frac{7}{5}-a_j}}\label{a1transtaupot3d}
\ee


\end{itemize}

\begin{center}
\begin{minipage}{5cm}
\begin{tabular}{|c|c|c|}
\hline
 $0$ &$\frac{\pi}{6}$ &$\frac{\pi}{12}$  \\
\hline
 2.96 & 2.96  & 3.10 \\
\hline
 5.55 &  5.55 &  5.82\\
\hline
 8.09 & 8.09  &  8.47 \\
\hline
 10.61 & 10.61  & 11.10 \\
\hline
 13.12& 13.12  &  13.72\\
\hline
\end{tabular}
\end{minipage}
\end{center}
\vspace{-0.1cm} {\footnotesize La tabla muestra los valores de
  $M_{1^{++}}$, correspondientes a la 1-forma perturbada
  transversalmente, para $\beta=0,\;\;\beta=\frac{\pi}{6}$ y
  $\beta=\frac{\pi}{12}$ con $d=3$.}
\bigskip

\subsection{Espectros de Glueball de perturbaciones en la m\'etrica}

Del ans\"atz para la m\'etrica propuesto en la secci\'on 5, analizamos
en $d=3$, los siguientes casos:

\begin{itemize}

\item \underline{Polarizaciones transversas: $2^{++}$ glueballs}

El potencial
\be
V(x) = \frac{1}{4}- \frac{1}{4\,g(x)^2}- \hat M^2\, \frac{e^{(1-\tilde a)x}}{g(x)^{\frac{7}{5}-\tilde a}}.
\label{metrictranspot}
\ee
Notemos que no hay degeneraci\'on con el  espectro del $0^{++}$ como s{\'\i} sucede en el caso cr\ii tico; este hecho es notado en \cite{Kuperstein:2004yf} y es v\'alido para todas las soluciones de nuestra familia.
\bigskip

\begin{center}
\begin{minipage}{5cm}
\begin{tabular}{|c|c|c|}
\hline
 $0$ &$\frac{\pi}{6}$ &$\frac{\pi}{12}$  \\
\hline
4.61 & 4.61 &4.28 \\
\hline
6.69 & 6.69 & 7.00 \\
\hline
9.25 & 9.25 & 9.66 \\
\hline
11.79 &11.79& 12.31 \\
\hline
14.31 & 14.3 & 14.94\\
\hline
\end{tabular}\\
\end{minipage}
\end{center}
\vspace{-0.3cm} {\footnotesize En la tabla de arriba, mostramos los
  valores de $M_{2^{++}}$, correspondientes a la perturbaci\'on
  transversa de la m\'etrica, con $\beta=0,\;\;\beta=\frac{\pi}{6}$
  y $\beta=\frac{\pi}{12}$.  Todos ellos son calculados en  $d=3$.}
\bigskip

\item \underline{Polarizaciones longitudinales: $1^{-+}$ glueballs}

El potencial es
\be
V(x) = \frac{1}{4}- \frac{1 - (a_i-\tilde a)^2}{4\,g(x)^2}-
\hat M^2\, \frac{e^{(1-\tilde a)x}}{g(x)^{\frac{7}{5}-\tilde a}}
\label{ametriclongpot}
\ee
\bigskip

\begin{center}
\begin{tabular}{c c}

\begin{minipage}{5cm}
{\small
\begin{tabular}{|c|c|c|}
\hline
 $0$ &$\frac{\pi}{6}$ &$\frac{\pi}{12}$  \\
\hline
4.06& 5.00& 5.06 \\
\hline
 6.69 & 7.73 &  7.88 \\
\hline
 9.25 & 10.34  & 10.59  \\
\hline
 11.79 & 12.91 &  13.25 \\
\hline
 14.31& 15.45 & 15.90 \\
\hline
 16.83  & 17.99 & 18.54 \\
\hline
\end{tabular}}
\end{minipage}
&

\begin{minipage}{5cm}
{\small
\begin{tabular}{|c|c|c|}
\hline
 $0$ &$\frac{\pi}{6}$ &$\frac{\pi}{12}$  \\
\hline
 5.00 &4.06 &4.92\\
\hline
7.73 & 6.69&  7.70\\
\hline
10.34 & 9.25&  10.39\\
\hline
12.91 & 11.79 & 13.051 \\
\hline
15.45& 14.31&  15.69\\
\hline
17.99& 16.83 &  18.33 \\
\hline
\end{tabular}}
\end{minipage}
\end{tabular}
\end{center}
\vspace{-0.5cm} {\footnotesize En la tabla de la izquierda, mostramos
  los valores $M_{1^{-+}}$ correspondientes a la polarizaci\'on
  longitudinal de la m\'etrica a lo largo de la direcci\'on
  caracterizada por $a_1$, para valores del par\'ametro
  $\beta=0,\;\;\beta=\frac{\pi}{6}$ y $\beta=\frac{\pi}{12}$. En la
  tabla de la derecha, mostramos los valores de la polarizaci\'on
  caracterizados por $a_2$. En ambos casos, $d=3$.}

\item \underline{Escalares: $0^{++}$ glueballs}

El sistema (\ref{Heq}) es tal que el elemento
(\ref{element_potential}) del potencial se reduce a
\bea
v(x) &=& \frac{1}{4} - \frac{1}{4\,g(x)^2} - \hat M{}^2\;
\frac{ e^{(1-\tilde a)x} }{ g(x)^{1-\tilde a +\frac{2}{5}} }\cr
m(x) &=& \frac{2}{g(x)^2}\,\left( \frac{\tilde a}{2}\,(1-\tilde a) +
\frac{\,3\,\tilde a - 1}{5}\, e^x - \frac{2}{25}\, e^{2\,x}\right)\cr
m_i^{(1)}(x) &=& \frac{1}{g(x)^2}\,\left( \frac{\tilde a}{2}\,(1+a_i-2\,\tilde a) +
\frac{2\,\tilde a + a_i - 1}{5}\, e^x - \frac{2}{25}\, e^{2\,x}\right)\cr
m_i^{(2)}(x) &=& \frac{2}{g(x)^2}\,\left( \frac{a_i}{2}\,(1-\tilde a) +
\frac{4\,a_i -\tilde a -1}{5}\, e^x - \frac{2}{25}\, e^{2\,x}\right)\cr
{\bf m}_{ij}(x) &=& \frac{1}{g(x)^2}\,\left( \frac{a_i}{2}\,(1+a_j-2\,\tilde a) +
\frac{4\, a_i + a_j -2\,\tilde a -1}{5}\, e^x - \frac{2}{25}\, e^{2\,x}\right)\cr.
& &
\eea

\bigskip

\begin{center}
\begin{minipage}{5cm}
\begin{tabular}{|c|c|c|}
\hline
 $0$ &$\frac{\pi}{6}$ &$\frac{\pi}{12}$  \\
\hline
3.97 &3.97 & 3.22 \\
\hline
6.67 &6.67 & 6.36 \\
\hline
9.26 &9.26 & 9.25   \\
\hline
11.81&11.81 &  12.04 \\
\hline
14.45 &14.45  & 14.78\\
\hline
\end{tabular}\\
\end{minipage}
\end{center}
\vspace{-0.5cm} {\footnotesize La tabla muestra los valores de
  $M_{0^{++}}$, correspondiente a la perturbaci\'on escalar de la
  m\'etrica para $\beta=0,\;\;\beta=\frac{\pi}{6}$ y
  $\beta=\frac{\pi}{12}$, con $d=3$.}
\end{itemize}


\section{Espectros de Glueballs de las teor{\'\i}as de Yang-Mills en $4D$}
\label{4dgs}

Aqu{\ii} consideraremos supercuerdas no cr\ii ticas tipo IIA en
$D=8$, en el fondo (\ref{Ssn}) de D6-branas negras.  Tomaremos $d=4$
direcciones equivalentes, y $D-d-1=3$ como compactas y no
equivalentes.  Las ecuaciones de v\ii nculos (\ref{constraints}) son
\be 4\,\tilde a + a_1+a_2 + a_3=1\qquad,\qquad 4\,\tilde a^2 +
a_1{}^2+a_2{}^2 + a_3{}^2=1. \ee El espacio que resuelve estos v\ii
nculos es una esfera 2-dimensional, como resulta claro de la
soluci\'on presentada aqu{\ii} \bea a_1 &=& \frac{1}{7}\, \left(
\sqrt{21}\,x + \sqrt{15}\, z + 1 \right)\cr a_2 &=& \frac{1}{7}\,
\left( - \sqrt{21}\,x + \sqrt{15}\, z + 1 \right)\cr a_3 &=&
\frac{1}{7}\, \left( -2\, \sqrt{\frac{42}{5}}\,y
-\sqrt{\frac{12}{5}}\, z + 1 \right)\cr 2\,\tilde a &=& \frac{1}{7}\,
\left( \sqrt{\frac{42}{5}}\,y - 2\,\sqrt{\frac{12}{5}}\, z + 2 \right)
\eea donde \be x^2 + y^2 + z^2 = 1 \ee y por tanto el espacio de
par\'ametros est\'a naturalmente caracterizado por dos par\'ametros
angulares $\theta$ y $\phi$.\\

Estudiemos ahora el l\ii mite de desacoplamiento de la misma manera que hicimos en el caso de los modelos tridimensionales.
La constante de Newton en 8  dimensiones es $2\,\kappa_8{}^2 \sim
l_s{}^6\,g_s{}^2$. Luego, para desacoplar la torre de estados de
cuerdas abiertas, as{\ii} como los estados de interacci\'on {\it
  bulk-open} y {\it bulk-bulk}, es necesario tomar el l\ii mite de
baja eneg\ii a $\,l_s\rightarrow 0$. Nuevamente el problema se
focaliza en lo que queda en el volumen de mundo de las D6-branas.
Como ya dijeramos, el vac{\'\i}o no cr\ii tico preserva $2^\frac{D}{2}= 16$
supercargas, y una D6-brana BPS embebida en \'el preservar\'a 8 de
\'estas, de manera similar a lo discutido en la secci\'on \ref{4dgs} a
partir del conocimiento de que el l\ii mite de baja energ\ii a de una
D5-brana en el vac\ii o del cigarro corresponde a una ${\cal N}=
(0,1)$ super Yang Mills m{\'\i}nima en 6 dimensiones \cite{Ashok:2005py}.
De aqu{\ii} podemos argumentar que nuestra familia es hologr\'afica en el
UV a una CFT (desconocida) en 7 dimensiones, que corresponde al
completamiento de la teor\ii a de YM cuyo acoplamiento a la escala
$\Lambda_s\equiv l_s{}^{-1}$ es \be g_{YM_7}{}^2 =
\frac{T_s{}^2}{T_{D6}}\sim l_s{}^3\,g_s.\label{cc7d} \ee El
acoplamiento de t'Hooft a esa escala, y el acoplamiento efectivo
adimensional a la escala $E$ son \be \lambda_7{}^2 \equiv
g_{YM_7}{}^2\, N \sim l_s{}^3\qquad;\qquad \lambda_7^{eff}(E)^2\equiv
\lambda_7{}^2\, E^3 \sim
\left(\frac{E}{\Lambda_s}\right)^3,\label{tof7d} \ee de donde se siguen
la validez de la descripci\'on perturbativa en la regi\'on
$E<<\Lambda_s\,$ y la ausencia de un l\ii mite de desacople
\cite{Itzhaki:1998dd}.  El acoplamiento de t'Hooft a la escala de
compactificaci\'on $\Lambda_c \equiv
(8\pi^3\,R_1\,R_2\,R_1)^{-\frac{1}{3}}\,$ de la teor\ii a
cuadridimensional en la que estamos interesados es \be \lambda_4{}^2
= \frac{\lambda_7{}^2}{8\pi^3\,R_1\,R_2\,R_3}\sim
\left(\frac{\Lambda_c}{\Lambda_s}\right)^3. \ee Para hacer contacto con
la teor\ii a de YM pura en cuatro dimensiones tenemos que tomar el
l\ii mite de escaleo \cite{Witten:1998zw} \cite{Gross:1998gk}, \be
\Lambda_c\, e^{-\frac{1}{B\,\lambda_4(\Lambda_c)}}
\;\xrightarrow[\Lambda_s\rightarrow\infty]{\Lambda_c\rightarrow\infty}\;
\Lambda_{QCD}\sim u_0\qquad\longleftrightarrow\qquad
\ln\frac{\Lambda_c}{u_0}\sim
\left(\frac{\Lambda_s}{\Lambda_c}\right)^\frac{3}{2}, \ee donde $B$ es
el coeficiente de la funci\'on $\beta$ a un loop definida por
$\,\beta(\lambda_4)\equiv\mu\partial_\mu\lambda_4(\mu) =
-B\,\lambda_4{}^2 +\dots\,$ ($B=\frac{23}{48\pi^2}$ para YM $SU(N)$
puro).
\bigskip

No consideraremos perturbaciones del tipo  $A_3^+$.
\subsection{Espectro de fluctuaciones dilat\'onicas: $O^{++}$ glueballs}

La ecuaci\'on tipo Schr\"odinger a resolver da el siguiente potencial
\be
V(x)=\frac{1}{4} - \frac{1}{4\,g(x)^2} + \frac{8}{7}\,\frac{e^x}{g(x)}- \hat M^2\, \frac{e^{(1-\tilde a)x}}{g(x)^{\frac{9}{7}-\tilde a}}
+ \sum_{i=1}^3 \hat p_i{}^2\, \frac{e^{(1- a_i)x}}{g(x)^{\frac{9}{7}-a_i}}\label{schrodil4d}
\ee


\begin{center}
\begin{minipage}{5cm}
\begin{tabular}{|c|c|c|}
\hline
 $P1$ &$P2$ &$P3$  \\
\hline
 10.69 &11.15 & 11.43 \\
\hline
13.80  & 14.96 & 15.17  \\
\hline
 16.78 & 18.4 &  18.68 \\
\hline
19.68 &  21.85& 22.08 \\
\hline
22.54 & 25.18 & 25.67\\
\hline
\end{tabular}
\end{minipage}
\end{center}

{\footnotesize La tabla muestra los valores para las masas
  $M_{0^{++}}$, correspondientes a las perturbaciones dilat\'onicas, en
  los puntos del espacio de par\'ametros
  $P1=(\frac{\pi}{6},\;\;\frac{7\pi}{6}),\;\;P2=(\frac{\pi}{2},\;\;\frac{4\pi}{3})$
  y $P3=(\frac{\pi}{3},\;\;\frac{9\pi}{6})$ del caso $d=4$.}

\bigskip

\subsection{Espectros de fluctuaciones de RR $1$-formas}

Es directo verificar que las perturbaciones definidas prendiendo s\'olo
$f_{q+2}$, para cualquier $q\neq p+1$ son consistentes, dando
ecuaciones de Maxwell generalizadas.  En particular, para $D=6$ tenemos
s\'olo $a_1$ del sector de RR\footnote{Trabajamos en base local, no en una base coordenada, y por
  lo tanto, nuestras componentes tensoriales difieren de aquellas en
  \cite{Kuperstein:2004yf} por factores de la m\'etrica.}.

De las secci\'on  5, analizamos
\begin{itemize}

\item \underline{Polarizaciones longitudinales: $0^{-+}$ glueballs}

Llevando a cabo los mismo pasos que en (\ref{cv-red})
obtenemos la forma de ecuaci\'on de Schr\"odinger (\ref{schro}), con potencial
\be
V(x)= \frac{1}{4\,g(x)^2} \left( \frac{25}{49}\,e^{2\,x} + \frac{2}{7}\, (5+ 2\,a_i)\,e^x+ a_i{}^2\right)
- \hat M^2\, \frac{e^{(1-\tilde a)x}}{g(x)^{\frac{9}{7}-\tilde a}}
+\sum_{j\neq i}\hat p_j{}^2\, \frac{e^{(1-a_j)x}}{g(x)^{\frac{9}{7}-a_j}}\label{a1longtaupot4d}
\ee

\begin{center}
\begin{tabular}{c c c}

\begin{minipage}{5cm}
{\small
\begin{tabular}{|c|c|c|}
\hline
 $P1$ &$P2$ &$P3$\\
\hline
4.84& 5.36&4.58\\
\hline
7.68& 8.25 & 7.43 \\
\hline
10.46& 11.06 &  10.20 \\
\hline
13.21& 13.84 & 12.95 \\
\hline
\end{tabular}}
\end{minipage}
&
\begin{minipage}{5cm}
{\small
\begin{tabular}{|c|c|c|}
\hline
 $P1$ &$P2$ &$P3$\\
\hline
4.44&4.925& 5.26 \\
\hline
7.29 & 7.74 &8.12 \\
\hline
10.04 &  10.47 & 10.89 \\
\hline
12.75& 13.19  & 13.62  \\
\hline
\end{tabular}}
\end{minipage}
&

\begin{minipage}{5cm}
{\small
\begin{tabular}{|c|c|c|}
\hline
  $P1$ &$P2$ &$P3$\\
\hline
4.96&4.96&5.24\\
\hline
7.85&7.85&, 8.17\\
\hline
10.66&10.66& 11.00\\
\hline
13.45&13.45&, 13.80\\
\hline
\end{tabular}}
\end{minipage}
\end{tabular}
\end{center}
 {\footnotesize En la tabla de la izquierda, mostramos los valores de
   las masas $M_{0^{-+}}$, correspondientes a la perturbaci\'on
   longitudinal de la 1-forma polarizada a lo largo de la direcci\'on
   caracterizada por $a_1$. Los par\'ametros toman valores llamados
   $P1, \;\; P2$ and $P3$, del espacio param\'etrico asociado con las
   soluciones en 4 dimensiones. En la tabla del centro y de la
   derecha, mostramos estos valores para las polarizaciones
   longitudinales $a_2$ and $a_3$, respectivamente.}

\item \underline{Polarizaciones transversas: $1^{++}$ glueballs}

El potencial es
\be
V(x)= \frac{1}{4\,g(x)^2} \left( \frac{25}{49}\,e^{2\,x} + \frac{2}{7}\, (5+ 2\,\tilde a)\,e^x+ \tilde a{}^2\right)
- \hat M^2\, \frac{e^{(1-\tilde a)x}}{g(x)^{\frac{9}{7}-\tilde a}}
+\sum_{j}\hat p_j{}^2\, \frac{e^{(1-a_j)x}}{g(x)^{\frac{9}{7}-a_j}}\label{a1transtaupot4d}
\ee

\end{itemize}

\begin{center}
\begin{minipage}{5cm}
\begin{tabular}{|c|c|c|}
\hline
 $P1$ &$P2$ &$P3$  \\
\hline
4.42& 4.31& 4.52\\
\hline
 7.31 & 7.14 & 7.44 \\
\hline
 10.10& 9.89 & 10.27\\
\hline
 12.86& 1260 & 13.06 \\
\hline
 15.61& 15.30 &  15.83\\
\hline
\end{tabular}
\end{minipage}
\end{center}
\vspace{-0.3cm} {\footnotesize La tabla muestra los valores de
  $M_{1^{++}}$ de las perturbaciones transversas de la 1-forma en los
  puntos $P1,\;\;P2$ y $P3$ del espacio de par\'ametros
  correspondiente a $d=4$.}
\bigskip

\subsection{Espectros de Glueball de perturbaciones en la m\'etrica}

Del ans\"atz propuesto para la m\'etrica en la secci\'on 5 analizamos,
en $d=4$, los siguientes casos:

\begin{itemize}

\item \underline{Polarizaciones transversas: $2^{++}$ glueballs}

El potencial
\be
V(x) = \frac{1}{4}- \frac{1}{4\,g(x)^2}- \hat M^2\, \frac{e^{(1-\tilde a)x}}{g(x)^{\frac{9}{7}-\tilde a}}.
\label{metrictranspot2}
\ee Notemos que no hay degeneraci\'on con el espectro $0^{++}$, como
s{\'\i} sucede en el caso de cuerdas cr\ii ticas, tal como es notado por
\cite{Kuperstein:2004yf}, hecho que es v\'alido para todas las
soluciones de nuestra familia.
\bigskip

\begin{center}
\begin{minipage}{5cm}
\begin{tabular}{|c|c|c|}
\hline
 $P1$ &$P2$ &$P3$  \\
\hline
 5.52 & 5.40 &5.60\\
\hline
 8.45 & 8.29 &  8.55\\
\hline
 11.27 & 11.07 & 11.40\\
\hline
 14.05 & 13.80 & 14.21\\
\hline
 16.81 & 16.52  &  17.00\\
\hline
\end{tabular}
\end{minipage}
\end{center}
\vspace{-0.5cm} {\footnotesize En esta tabla mostramos los valores de $M_{2^{++}}$, correspondientes a la m\'etrica transversa. Estos valores corresponden a $P1, \;\;P2$ y $P3$, todos ellos calculados en $d=4$.}
\bigskip

\item \underline{Polarizaciones longitudinales: $1^{-+}$ glueballs}

El potencial es
\be
V(x) = \frac{1}{4}- \frac{1 - (a_i-\tilde a)^2}{4\,g(x)^2}-
\hat M^2\, \frac{e^{(1-\tilde a)x}}{g(x)^{\frac{9}{7}-\tilde a}}.
\label{a1longpot}
\ee
\bigskip

\begin{center}
\begin{tabular}{c c c}

\begin{minipage}{5cm}
{\small
\begin{tabular}{|c|c|c|}
\hline
 $P1$ &$P2$ &$P3$\\
\hline
5.97&6.39 &5.79\\
\hline
  8.93& 9.42 & 8.72\\
\hline
 11.76&  12.30& 11.54 \\
\hline
 14.55 & 15.12 & 14.32 \\
\hline
\end{tabular}}
\end{minipage}
&
\begin{minipage}{5cm}
{\small
\begin{tabular}{|c|c|c|}
\hline
 $P1$ &$P2$ &$P3$\\
\hline
5.54,&5.96 & 6.26 \\
\hline
8.41 &8.87 &  9.23\\
\hline
11.19 &11.66 &  12.06\\
\hline
 13.92& 14.41 &  14.82\\
\hline
\end{tabular}}
\end{minipage}
&

\begin{minipage}{5cm}
{\small
\begin{tabular}{|c|c|c|}
\hline
 $P1$ &$P2$ &$P3$\\
\hline
6.14 &6.14 &6.36 \\
\hline
 9.14& 9.14 & 9.41 \\
\hline
12.00 & 12.00& 12.31 \\
\hline
14.83 & 14.83 & 15.15 \\
\hline
\end{tabular}}
\end{minipage}
\end{tabular}
\end{center}
\vspace{-0.3cm} {\footnotesize En las tablas de arriba mostramos los
  valores para $M_{1^{-+}}$, de la perturbaci\'on longitudinal de la  m\'etrica,
a lo largo de 3 direcciones diferentes  asociadas con $a_1, \;\; a_2$
  y $a_3$. Los valores corresponden a puntos del espacio de par\'ametros que hemos denominado $P1,\;\;P2$, y $P3$ con $d=4$.}

\item \underline{Escalares: $0^{++}$ glueballs}

El sistema (\ref{Heq}) es tal que el elemento
(\ref{element_potential}) del potencial se reduce a
\bea
v(x) &=& \frac{1}{4} - \frac{1}{4\,g(x)^2} - \hat M{}^2\;
\frac{ e^{(1-\tilde a)x} }{ g(x)^{1-\tilde a +\frac{2}{7}} }\cr
m(x) &=& \frac{4}{g(x)^2}\,\left( \frac{\tilde a}{2}\,(1-\tilde a) +
\frac{\,5\,\tilde a - 1}{7}\, e^x - \frac{2}{49}\, e^{2\,x}\right)\cr
m_i^{(1)}(x) &=& \frac{1}{g(x)^2}\,\left( \frac{\tilde a}{2}\,(1+a_i-2\,\tilde a) +
\frac{4\,\tilde a + a_i - 1}{7}\, e^x - \frac{2}{49}\, e^{2\,x}\right)\cr
m_i^{(2)}(x) &=& \frac{4}{g(x)^2}\,\left( \frac{a_i}{2}\,(1-\tilde a) +
\frac{6\,a_i -\tilde a -1}{7}\, e^x - \frac{2}{49}\, e^{2\,x}\right)\cr
{\bf m}_{ij}(x) &=& \frac{1}{g(x)^2}\,\left( \frac{a_i}{2}\,(1+a_j-2\,\tilde a) +
\frac{6\, a_i + a_j -2\,\tilde a -1}{7}\, e^x - \frac{2}{49}\, e^{2\,x}\right)\cr
& &
\eea

\begin{center}
\begin{minipage}{5cm}
\begin{tabular}{|c|c|c|}
\hline
 $P1$ &$P2$ &$P3$  \\
\hline
6.44 &6.37 &6.50\\
\hline
9.12 & 9.00 & 9.20 \\
\hline
 11.62& 11.47 &  11.72\\
\hline
 14.06 & 13.87 & 14.18 \\
\hline
\end{tabular}
\end{minipage}
\end{center}
\vspace{-0.3cm} {\footnotesize La tabla muestra los valores de $M_{0^{++}}$,
  correspondientes a las perturbaciones escalares de la  m\'etrica para tres diferentes valores del espacio de par\'ametros  $P1,\;\;P2$, y $P3$ en $d=4$.}
\end{itemize}

\bigskip


\section{Resumen de resultados y discusi\'on}\label{discusion}


En este punto es interesante resaltar algunos aspectos que conciernen
a los resultados obtenidos en el c\'alculo de los espectros de
glueballs. Bien puede pensarse que el enfoque que hemos utilizado para
el c\'omputo de los espectros es bastante indirecto, puesto que no
proviene de una teor\ii a fenomenol\'ogica ni de una teor\ii a
fundamental, s\'olo se motiva en la dualidad conjeturada entre teor\ii
as de gauge y teor\ii as de gravedad. Sin embargo, los resultados
obtenidos muestran una gran concordancia con los valores esperados
para estos espectros. Dichos valores no son predichos por QCD, ya que
se trata de estados inalcanzables perturbativamente, pero aparecen de
manera natural en {\it Lattice QCD} y se espera que realmente existan
en la naturaleza. A\'un cuando estos estados no han sido confirmados
experimentalmente, el actual grado de confianza de la comunidad
cient\ii fica en los c\'alculos de {\it Lattice QCD} hace que los
glueballs, estados singletes de color, sean parte de la zoolog\ii a
actualmente aceptada de QCD. En este apartado nos proponemos comparar
los resultados obtenidos por nosotros, con los c\'alculos de {\it
  Lattice QCD} en el l\ii mite de N grande. Esto \'ultimo es
fundamental ya que nuestro {\it setup} tiene sentido s\'olo en este
l\ii mite de N. La pregunta de si $N=3$ es lo suficientemente grande
como para esperar alg\'un tipo de contacto con la realidad, puede
responderse, al menos fenomenol\'ogicamente, teniendo en cuenta que
los valores predichos por {\it Lattice QCD} para $N=2,3,4,5$
var{\'\i}an muy poco unos de otros (\cite{Teper:1998te},
\cite{Morningstar:1999rf}). Vale decir entonces que es de esperar que
la teor\ii a simplificada $SU(N)$ en el l\ii mite de t'Hooft, sea
capaz de reproducir con sensibilidad la f\ii sica de glueballs.  A
modo de ejemplo, hemos calculado los espectros en dimensi\'on $D=3$ y
$D=4$. Si bien el inter\'es central de estas teor\ii as de gauge es
describir QCD en el mundo real, $D=4$, grande es el inter\'es que
despiertan estas teor\ii as de YM en dimensiones menores, por eso
hemos realizado el c\'alculo en $D=3$. Cabe destacar, sin embargo, que
mucha de la fenomenolog\ii a esperada en ambas dimensiones es la
misma.

\begin{figure}[!ht]
\centering
\includegraphics[scale=0.95,angle=0]{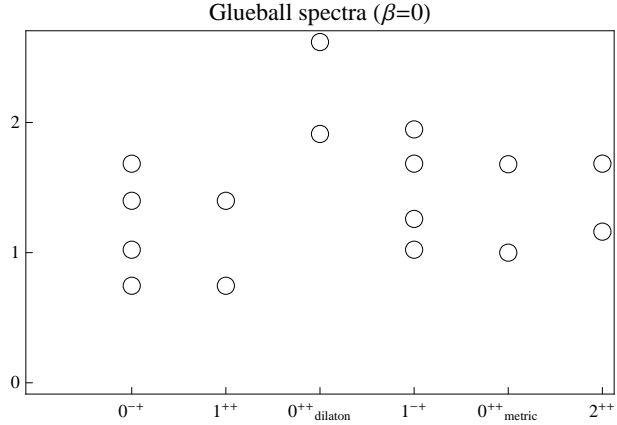}
\caption[Espectro de glueballs, en $d=3$ para $\beta= 0$.]{El
  gr\'afico muestra el espectro de glueballs normalizado por la masa
  m\'as baja del $0^{++}$, en $d=3$ para $\beta= 0$.}\label{3Fig0}
\end{figure}

Pasemos entonces al an\'alisis de los resultados obtenidos.  Es
interesante notar que para $d=3$, en nuestro caso, por cada
perturbaci\'on polarizada longitudinalmente existen dos contribuciones
diferentes, una asociada con la direcci\'on caracterizada por $a_1$ y
la otra con la direcci\'on caracterizada por $a_2$.  Estas
contribuciones no son modos de Kaluza-Klein, sino que son consecuencia
de polarizaciones a lo largo de dos direcciones compactas no
equivalentes. Es por esto que nuestro espectro posee el doble de
estados $0^{-+}$ y $1^{-+}$ que aquel hallado en
\cite{Kuperstein:2004yf}. Estos modos en general est\'an desdoblados y
no contribuyen con una degeneraci\'on superior, excepto para valores
particulares del par\'ametro $\beta$. En general, \'estos est\'an
separados ligeramente en los valores de sus masas, como consecuencia
del v\ii nculo (\ref{constraints}) (Figura (\ref{3Fig0})).

\begin{figure}[!ht]
\centering
\includegraphics[scale=0.95,angle=0]{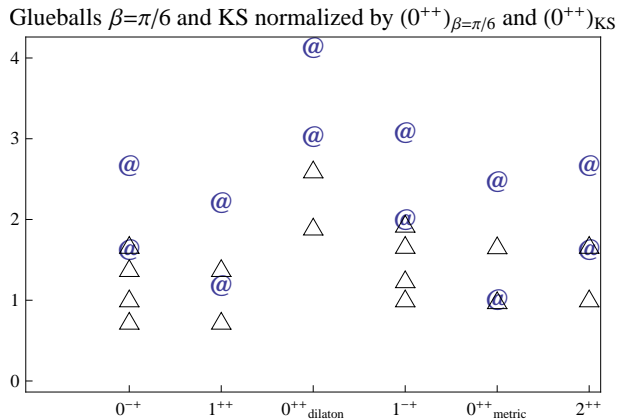} 
\caption[Comparaci\'on entre espectros, normalizados por sus $0^{++}$]{En este
  gr\'afico se comparan los espectros obtenidos por Kuperstein y
  Sonnenschein \cite{Kuperstein:2004yf}(@) y nuestro caso $\beta=
  \frac{\pi}{6}\;$($\bigtriangleup$), cada uno normalizado por su
  correspondiente $0^{++}$ m\'as bajo. En principio, estos dos
  espectros deber\ii an ser iguales, a menos de los desdoblamientos en
  $0^{-+}$ y en $1^{-+}$.}\label{3FigKSP-1-6}
\end{figure}

\begin{figure}[!ht]
\centering
\includegraphics[scale=0.95,angle=0]{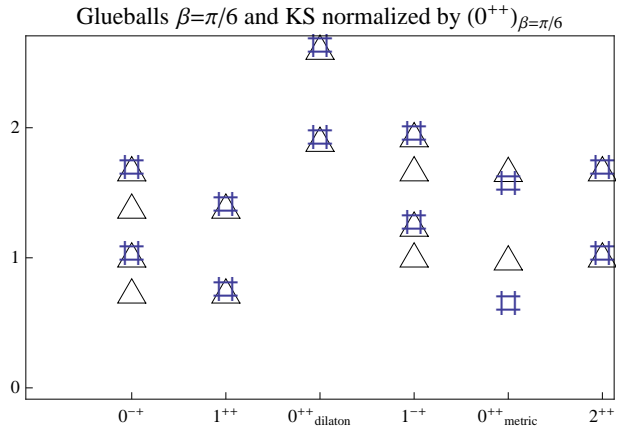} 
\caption[Comparaci\'on entre espectros, normalizados por
  nuestro $0^{++}$.]{En este gr\'afico se comparan los
  espectros obtenidos por Kuperstein y Sonnenschein
  \cite{Kuperstein:2004yf}($\sharp$) y nuestro caso $\beta=
  \frac{\pi}{6}\;$($\bigtriangleup$), ambos normalizados por nuestro
  valor $0^{++}$ m\'as bajo. El acuerdo entre ambos espectros es muy
  bueno, a menos de los desdoblamientos en $0^{-+}$ y en
  $1^{-+}$.}\label{3FigKS-1-6}
\end{figure}

A\'un cuando esperabamos ser capaces de reproducir a la perfecci\'on
el espectro obtenido por \cite{Kuperstein:2004yf} (a menos de los
desdoblamientos), cuando nuestro par\'ametro $\beta$ toma el valor
$\beta=\pi/6$ tal cosa no sucede. Esto se debe a que nuestro espectro
difiere del encontrado en \cite{Kuperstein:2004yf} en los autovalores
m\'as bajos de los estados $0^{++}$ asociados a la perturbaci\'on
m\'etrica. En nuestro caso, la concordancia entre los valores
obtenidos en forma num\'erica y a trav\'es de WKB, es muy buena, no
siendo \'este el caso para \cite{Kuperstein:2004yf}. Es por este
motivo que creemos que el valor correcto para el estado m\'as bajo
del $0^{++}$ es el obtenido por nosotros. A continuaci\'on, se muestra
en el gr\'afico de la Figura \ref{3FigKSP-1-6}, el espectro obtenido
por \cite{Kuperstein:2004yf} y nuestro espectro, ambos normalizados por
el autovalor m\'as bajo del $0^{++}$ respectivo. En el gr\'afico de la
Figura \ref{3FigKS-1-6} los mismos dos espectros, pero ahora ambos
normalizados por el valor m\'as bajo $0^{++}$ obtenido por nosotros. El
acuerdo entre ambos espectros es absoluto.

\begin{figure}[!ht]
\centering
\includegraphics[scale=0.95,angle=0]{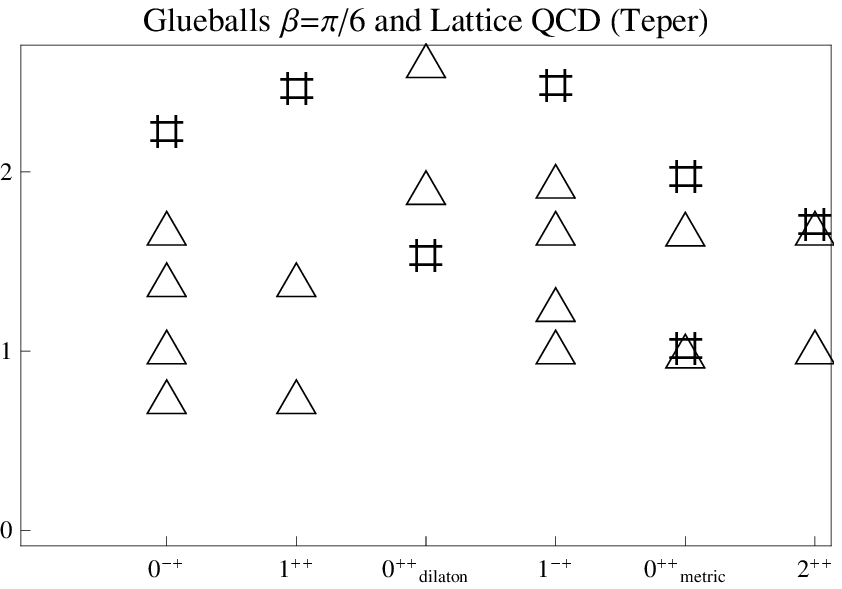}
\caption[Comparaci\'on entre $\beta=\pi/6$ y
  Lattice QCD para $d=3$.]{En este gr\'afico comparamos el espectro
  obtenido por nosotros para $\beta=\pi/6\;$ ($\bigtriangleup$) con el
  espectro obtenido en {\it Lattice QCD} ($\sharp$) para $d=3$ por
  Teper (\cite{Teper:1998te}) .}
\label{3Fig0-1-12}
\end{figure}

\begin{figure}[!ht]
\centering
\includegraphics[scale=0.95,angle=0]{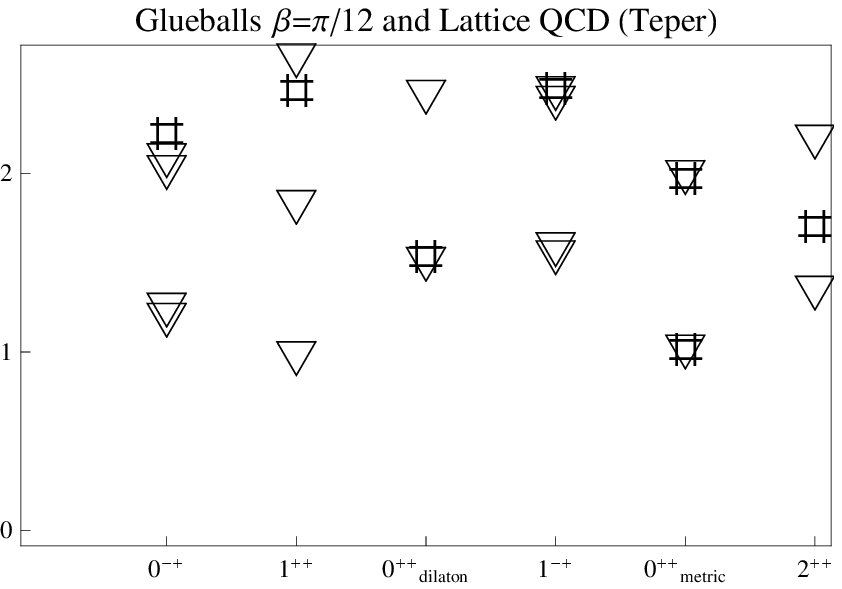}
\caption[Comparaci\'on entre $\beta=\pi/12$ y
  Lattice QCD para $d=3$.]{En este gr\'afico comparamos el espectro
  obtenido por nosotros para $\beta=\pi/12\; (\bigtriangledown)$ con
  el espectro obtenido en {\it Lattice QCD} ($\sharp$) para $d=3$ por
  Teper (\cite{Teper:1998te}).}
\label{3Fig0-1-12-teper}
\end{figure}

Es importante notar aqu{\ii} que la forma cualitativa esperada para el
espectro de glueballs no es exactamente la que observamos para los
valores del par\'ametro $\beta$ mostrados en los gr\'aficos
anteriores. En general es ampliamente asumido, y en algunos casos
verificado en el c\'aculo de {\it Lattice QCD}, que el glueball de
menor masa corresponde al operador $0^{++}$. Esto claramente no es lo
que sucede con nuestro espectro en los gr\'aficos que se muestran en
las Figuras \ref{3Fig0}, \ref{3FigKSP-1-6} y \ref{3FigKS-1-6}.  Sin
embargo, debido a la libertad en la elecci\'on del par\'ametro $\beta$,
es posible ajustar el mismo para obtener un mejor acuerdo con el
espectro esperado.  \\ Si bien durante el transcurso de esta tesis no
hemos realizado una exploraci\'on sistem\'atica del sector confinante del
espacio  parametrizado por $\beta$, se puede obervar, por simple
inspecci\'on, que algunos valores del par\'ametro acuerdan mejor con
los valores obtenido en el c\'alculo de {\it Lattice QCD}
(\cite{Teper:1998te})(Ver Figura \ref{3Fig0-1-12-teper}).

Por \'ultimo, es interesante notar que para el mismo valor de $\beta$
para el cual el espectro obtenido se parece m\'as al esperado, en
cuanto a los cocientes relativos entre las masas de los distintos
glueballs, simult\'aneamente el desdoblamiento de estados debido a las
polarizaciones longitudinales se hace cada vez m\'as peque\~no
(creemos que un valor m\'as preciso de $\beta$ podr{\'\i}a hacer
desaparecer tal desdoblamiento dejando un espectro degenerado). De
esta manera, el espectro resulta cualitativamente y cuantitativamente
m\'as parecido a\'un al predicho por {\it Lattice QCD}.

\begin{figure}[!ht]
\centering
\includegraphics[scale=0.95,angle=0]{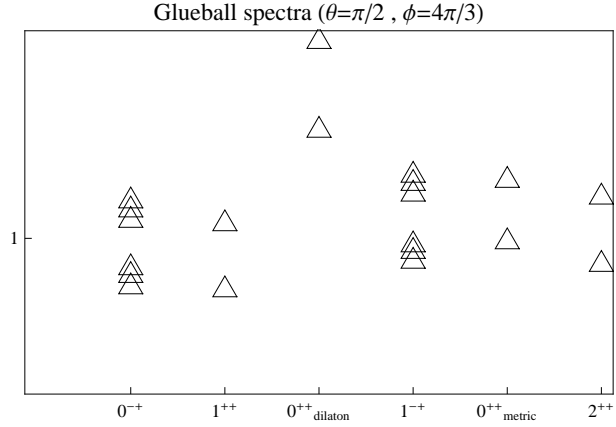}
\caption[Espectro de glueballs para $d=4$, para $\theta=\frac{\pi}{2}$
  y $\beta=\frac{4\pi}{3}$.]{El gr\'afico muestra el espectro de
  glueballs para $d=4$ normalizado por el valor m\'as bajo $0^{++}$
  para el punto del espacio de par\'ametros denominado $P2$, que
  corresponde a $\theta=\frac{\pi}{2}$ y $\beta=\frac{4\pi}{3}$.}
\label{4dFig2}
\end{figure}

\begin{figure}[!ht]
\centering
\includegraphics[scale=0.95,angle=0]{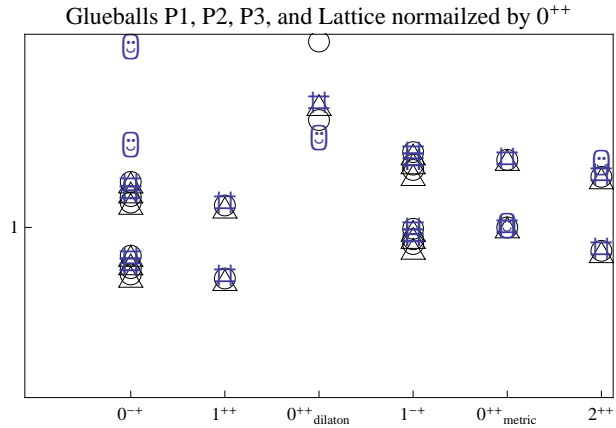}
\caption[Espectro de glueballs para $d=4$ y
  comparaci\'on con Lattice QCD.]{En el gr\'afico se muestra el
  espectro de glueballs para $d=4$, calculado en tres puntos
  diferentes del espacio de par\'ametros P1($\bigcirc$,
  P2($\bigtriangleup$), P3($\sharp$) y se lo compara con el espectro
  obtenido en {\it Lattice QCD} ($\ddot\smile$) por Morningstar and
  Peardon \cite{Morningstar:1999rf}.}
\label{4dFig8}
\end{figure}

En el caso $d=4$ (Figura \ref{4dFig2}), nuestras soluciones poseen
3 direcciones compactas no equivalentes, y por lo tanto, 3
estados con los mismos n\'umeros cu\'anticos pero diferentes masas. Como
en el caso anterior, las polarizaciones longitudinales se ven
desdobladas como consecuencia de la libertad de polarizar a lo largo
de las 3 direcciones no equivalentes.  Nuevamente, la libertad en
la elecci\'on de los par\'ametros (ahora son dos), nos da la
posibilidad de obtener diferentes espectros dependiendo de la
elecci\'on de los mismos. Como ya mencionamos, no ha sido la
intenci\'on de este trabajo realizar una b\'usqueda sistem\'atica de los
par\'ametros que mejor ajustan el espectro deseado. Sin embargo, a
pesar de la dificultad que implica el c\'alculo del espectro
completo y la complejidad de estudiar el espacio de par\'ametros
bidimensional, creemos que es posible hallar un punto en dicho espacio
que d\'e lugar a un mejor ajuste entre los datos de {\it Lattice QCD} y
nuestro espectro (ver Figura (\ref{4dFig8})).

Por otro lado, es muy importante notar que todas nuestras soluciones
llevan a un problema y a un espectro resultante bien definidos, a\'un
cuando resultan singulares en el IR. Es como si alg\'un mecanismo, que
act\'ua de manera natural, impidiera que la f\ii sica del borde est\'e
relacionada con excitaciones en el IR profundo. Ciertamente esto es lo
que sucede cuando estudiamos en la secci\'on \ref{wlc} los lazos de
Wilson, donde la cuerda amarrada a los quarks no din\'amicos s\'olo
pod\ii a penetrar hasta un valor $u=u_c$ con $u_c> u_0$.  M\'as a\'un,
ambas familias (\ref{Ssn}) y (\ref{sns}) llevan exactamente al mismo
espectro, tal como se muestra en (\ref{AB}), hecho que se pod\ii a ya
anticipar por argumentos de T-dualidad. Sin embargo, resulta llamativo
que mientras la aproximaci\'on de cuerdas para toda la familia de
soluciones con dilat\'on constante resulta bajo control en el l\ii
mite de N grande, la familia T-dual a \'esta aparece como restringida a
la regi\'on $a_\theta<0$ debido a que la constante de acoplamiento de
cuerdas efectiva $g_s\equiv e^\Phi$ explota en la regi\'on infrarroja
$u\rightarrow u_0{}^+$.  Esto es una se\~nal m\'as de la irrelevancia
de la singularidad de la soluci\'on en la f\ii sica del borde. Estos
hechos han sido ya notados en la referencia \cite{Gursoy:2007er} (ver
\cite{Kiritsis:2009hu} para un resumen), este comportamiento de las
singularidades de las soluciones, que impide que la f\ii sica del borde
sea captada por excitaciones en la regi\'on singular, es com\'un en
las singularidades llamadas repulsivas.  \\

Finalmente, creemos que es posible que existan soluciones {\it exactas}
que se aproximen a las nuestras en el UV, pero que sean regulares en
el IR, a\'un al nivel de acci\'on efectiva de baja energ\ii a, es
decir, a trav\'es de una posible dependencia no trivial en
$\theta$. Esto sucede con el fondo de Klebanov-Tseytlin
\cite{Klebanov:2000nc}, el cual presenta una singularidad desnuda que
resulta regularizada por la soluci\'on de Klebanov-Strassler
\cite{Klebanov:2000hb}, pero que modifica su espectro de forma suave.



%% file: conclusiones.tex
\chapter{Conclusiones}

En esta tesis, hemos abordado el problema del estudio de la acci\'on
efectiva de baja energ\ii a de las teor\ii as de cuerdas no cr\ii
ticas y sus soluciones. A diferencia de otros trabajos previos, en los
que las soluciones son encontradas a trav\'es de un conjunto de
ecuaciones BPS derivadas de un superpotencial, hemos preferido encarar
el problema directamente, resolviendo de manera sistem\'atica el
conjunto completo de ecuaciones de segundo orden fuertemente acoplado
correspondiente a las ecuaciones de movimiento. Dada la complejidad de
dichas ecuaciones, este problema no es nada trivial. Sin embargo,
hemos podido obtener soluciones a dichas ecuaciones en diferentes
contextos, llegando en algunos casos a agotar todas las posibles
soluciones. En particular, como ya coment\'aramos en la discusi\'on
del cap\ii tulo \ref{nos_sol}, hemos encontrado todas las posibles
soluciones de vac\ii o, y por supuesto entre ellas hemos recuperado
las previamente conocidas en la literatura. Resulta interesante
remarcar que es posible que algunas de ellas tambi\'en se comporten
como CFT exactas, y s\'olo reciban correcciones menores que no
modifiquen sustancialmente sus caracter\ii sticas \footnote{Esta es
  una interesante hip\'otesis que podr\ii a ser abordada en un futuro
  trabajo} . De manera similar, tambi\'en hemos agotado el problema de
hallar soluciones cargadas frente a campos de NSNS que llenan todo
Minkowski, que en el caso de objetos unidimensionales, representar\ii
an cuerdas fundamentales. Hemos encontrado, adem\'as, la soluci\'on
interpretable como una cuerda fundamental doblemente localizada en el
tip del cigarro y en el origen del espacio plano \footnote{Esta es una
  de las pocas soluciones doblemente localizadas que se conocen} . Por
otro lado, encontramos tambi\'en la que corresponder\ii a a una cuerda
fundamental en el vac\ii o del dilat\'on lineal, entre varias otras
soluciones. Por \'ultimo, tambi\'en en el cap\ii tulo \ref{nos_sol},
encontramos familias de soluciones de fondos cargados RR que pueden
ser interpretados como {\it near horizon} de D$_p$-branas en el vac\ii
o del dilat\'on lineal, y tales que sus T-duales son familias de
dilat\'on constante e incluyen de manera particular la soluci\'on
encontrada previamente por Kuperstein y Sonnenschein en
\cite{Kuperstein:2004yk}.

En el Cap\ii tulo \ref{sym} hemos estudiado, via la dualidad
gauge/gravedad, el posible comportamiento de teor\ii as de gauge
duales a una generalizaci\'on de esta familia de soluciones cargadas
frente a RR con dilat\'on constante. En particular, hemos abordado el
estudio de los lazos de Wilson en estas teor\ii as de gauge a trav\'es
de su prescripci\'on dual y hemos sido capaces de determinar la
condici\'on de confinamiento, en t\'erminos de condiciones sobre el
espacio de par\'ametros que caracterizan nuestra familia. En estas
condiciones hemos emprendido el estudio del espectro de glueballs de
estas teor\ii as de gauge, que como se sabe, son los grados de
libertad dominantes a grandes valores de N (r\'egimen en el cual
nuestro estudio es estrictamente v\'alido). A trav\'es de la
identificaci\'on de los grados de libertad de gravedad con los de
operadores de la teor\ii a de gauge, hemos obtenido tal espectro
resolviendo ecuaciones diferenciales acopladas de segundo orden de
manera num\'erica. Durante el transcurso de estos estudios, tambi\'en
hemos calculado estos espectros en la aproximaci\'on WKB, pero dado que
el acuerdo entre ambos c\'alculos es total, por prolijidad en la
presentaci\'on s\'olo hemos mostrado los resultados num\'ericos. Es
importante destacar que el acuerdo logrado con el c\'alculo de {\it
  Laticce QCD} es muy bueno, teniendo en cuenta que se tratan de dos
enfoques totalmente diferentes, y que los datos obtenidos
pertenecer\ii an a reg\ii menes opuestos de la constante de
acoplamiento de la teor\ii a de gauge bajo estudio. El acuerdo es
mejor en el caso de YM en $D=3$ que en $D=4$. Esto posiblemente se
deba a que no hemos hecho un estudio sistem\'atico de los valores de
los par\'ametros que mejor ajustan si no que, a modo ilustrativo,
hemos buscado, a prueba y error, puntos diferentes del espacio de
par\'ametros ajustando lo mejor posible el espectro deseado. En el
caso de $D=3$, el espacio de par\'ametros es unidimensional por lo que
buscar {\it a mano} un valor que ajuste aceptablemente es m\'as
sencillo que en el caso de $D=4$, en el que el espacio de par\'ametros
es bidimensional.

%% file: appendI.tex



\chapter{Algunas f\'ormulas \'utiles I}\label{appendI}
\cleqn

Aqu{\ii} recolectamos algunas propiedades \'utiles relacionadas al
ans\"atz asumido para los campos en la secci\'on \ref{trat_general}, \bea G &=&
-A^2\; dx^0{}^2+\tilde{A}^2\;d{\vec x}{}^2 + C^2\;d\rho^2 + \tilde
C^2\;dz^2\equiv\eta_{mn}\;\omega^m\;\omega^n\cr A_{p+1} &=&
dx^0\wedge\dots\wedge dx^p\; E(\rho)\cr \Phi &=&
\Phi(\rho)\label{ansatzapen} \eea donde $\vec x$ es un vector $p$
dimensional, y donde hemos introducido los vielbein $\omega^n$ y los
campos vectoriales duales $e_m$, $e_m(\omega^n)=\delta_m^n$, definidos
por \bea
\begin{array}{rclcrcl}
\omega^0&=&A(\rho)\;dx^0
\;\;\;\;&,&\;\;\;\; e_0 &=& A(\rho)^{-1}\;\partial_0\cr
\omega^{I}&=&\tilde A(\rho)\;dx^I
\;\;\;\;&,&\;\;\;\; e_I &=& \tilde A (\rho)^{-1}\;\partial_I\qquad,\qquad I=1,\dots,p\cr
\omega^{p+1}&=&C(\rho)\;d\rho
\;\;\;\;&,&\;\;\;\; e_{p+1}&=& C(\rho)^{-1}\;\partial_{\rho}\cr
\omega^{p+2}&=&\tilde C(\rho)\; dz
\;\;\;\;&,&\;\;\;\; e_{p+2} &=& \tilde C(\rho)^{-1}\;\partial_z
\end{array}
\eea
\section{Conexiones}

Como es usual, las conexiones $\omega^m{}_n$ quedan completamente determinadas por las condiciones de 
\begin{itemize}
\item  Torsi\'on nula: \hspace{2cm}$d\omega^m+{\omega^m}_n\wedge \omega^n=0$
\item Metricidad:\hspace{4cm}$\omega_{mn}\equiv\eta_{mp}\,\omega^p{}_n
  =-\omega_{nm}.$
\end{itemize}
Del c\'alculo directo de los coeficientes ${\;\alpha^m}_{nl}$,
\begin{equation}
d\omega^m \equiv {1\over2}{\alpha^m}_{n l}\;\omega^n \wedge
\omega^l\;\;\;,\;\;\; {\alpha^m}_{nl} = - {\alpha^m}_{ln}
\end{equation}
obtenemos las conexiones en la siguiente forma,
\be
\omega_{mn}\equiv\omega_{m n l}\;\omega^l\qquad,\qquad
\omega_{m n l}={1\over2}(\alpha_{mn l}-\alpha_{l m n}+\alpha_{n l m})
\ee

Aqu{\ii} resumimos todas las conexiones no nulas asociaxdas con la m\'etrica (\ref{ansatzapen}),
\begin{eqnarray}
\omega_{0 p+1}&=&{\partial_{\rho}\ln A\over C}\;\omega_0= e_{p+1}(\ln A)\;\omega_0 \cr
\omega_{I p+1}&=&{\partial_{\rho}\ln \tilde A\over C}\;\omega_I
=e_{p+1}(\ln\tilde A)\;\omega_I\cr
\omega_{p+1,p+2}&=&-{\partial_{\rho}\ln\tilde C\over C}\;\omega_{p+2}=
-e_{p+1}(\ln \tilde C)\;\omega_{p+2}\label{conexiones}
\end{eqnarray}

\section{Derivadas covariantes}

En esta secci\'on, presentamos todas las derivadas covariantes de
inter\'es relacionadas con las conexiones mostradas arriba.  Por
definici\'on,
\begin{center}
$D_m(A_n)\equiv e_m(A_n)-\omega_{s n m}\; A_s$
\end{center}
Ya que todas nuestras dependencias son a trav\'es de funciones
escalares de $\rho$, es suficiente conocer s\'olo algunas de ellas. Sea
$\phi(\rho)$ una funci\'on escalar arbitraria de $\rho$; luego se obtienen las siguientes derivadas no nulas (se asume contracci\'on de {\'\i}ndices),
\begin{eqnarray}
D_{p+1}(\phi) &=& e_{p+1}(\phi) = \frac{1}{C}\,\partial_{\rho}\phi\cr
D_0{}^2(\phi)&=&-\frac{1}{C^2}\;\partial_{\rho}\ln A\;\partial_{\rho}\phi\cr
D_I D_J (\phi)&=&\delta_{IJ}\;e_{p+1}(\ln\tilde A)\;e_{p+1}(\phi)\cr
D_{p+1}{}^2(\phi)&=&\partial_{\rho}\left({\partial_{\rho}\phi\over C^2}\right)
+ \frac{1}{C^2}\;\partial_{\rho}\ln C\;\partial_{\rho}\phi\cr
D_{p+2}{}^2(\phi)&=& \frac{1}{C^2}\;\partial_{\rho}\ln \tilde C \;\partial_{\rho}\phi\cr
D^2(\phi) &=& \frac{1}{F_1}\;\partial_{\rho}\left(\frac{F_1}{C^2}\;\partial_{\rho}\phi\right)
=\partial_{\rho}\left(\frac{\partial_{\rho}\phi}{C^2}\right)+
\frac{1}{C^2}\;\partial_{\rho}\ln F_1\;\partial_{\rho}\phi\cr
e^{b\xi}D(e^{-b\xi}\; \psi\; D(\phi)) &=&\frac{1}{F_1\, e^{-b\xi}}\;\partial_{\rho}\left(\frac{F_1\,e^{-b\xi}}{C^2}\;
\psi\;\partial_{\rho}\phi\right)\cr
& & \label{derivadascov}
\end{eqnarray}
donde $F_1 \equiv A\,\tilde A^p\, C\,\tilde C\;$,
$\;\epsilon_G =  dx^0\wedge dx^1\wedge\dots\wedge dx^p\wedge d\rho\wedge
d\theta \;F_1\,$.

\section{Tensor de curvatura}

Por definici\'on,
\be
\Re_{mn} \equiv d\omega_{mn} + \omega_{mp}\wedge\omega^p{}_n =
\frac{1}{2}\,\Re_{mnpq}\;\omega^p\wedge\omega^q
\ee
Del c\'alculo resulta que,
\bea
\Re_{0I}&=& D_0{}^2(\ln \tilde A)\;\omega_0\wedge\omega_I\cr
\Re_{0p+1}&=&-{1\over{A}}{D_{p+1}}^2(A)\;\omega_0\wedge\omega_{p+1}\cr
\Re_{Ip+1}&=&-{1\over{\tilde{A}}}D_{p+1}{}^2(\tilde{A})\;\omega_I\wedge\omega_{p+1}\cr
\Re_{IJ}&=&- D_{p+1}(\ln\tilde A)^2\;\omega_I\wedge\omega_J\cr
\Re_{0p+2}&=&-\frac{1}{A}D_{p+2}{}^2(A)\;\omega_0\wedge\omega_{p+2}\cr
\Re_{Ip+2}&=&-\frac{1}{\tilde A}D_{p+2}{}^2(\tilde{A})\;\omega_I\wedge\omega_{p+2}\cr
\Re_{p+1p+2}&=&-{1\over{\tilde{C}}}D_{p+1}{}^2(\tilde C)\;\omega_{p+1}\wedge\omega_{p+2}
\label{curvatura}
\eea

\section{Tensor de Ricci y escalar de Ricci}

De (\ref{curvatura}), se obtienen las siguientes componentes del tensor de Ricci $R_{mn} \equiv \Re^p{}_{mpn}$,
\bea
R_{00}&=& D^2(\ln A)\cr
R_{IJ}&=&- D^2(\ln \tilde A)\;\delta_{IJ}\cr
R_{p+1p+1}&=&-D^2(\ln C)-{1\over A}\partial_{\rho}\left({\partial_{\rho}A\over
C^2}\right)-{p\over \tilde A}\partial_{\rho}
\left({\partial_{\rho}\tilde A\over C^2}\right)+
{1\over C}\partial_{\rho}\left({\partial_{\tilde
\rho}C\over C^2}\right)- {1\over \tilde C}\partial_{\rho}
\left({\partial_{\rho}\tilde C\over C^2}\right)\cr
&=&-D^2(\ln C) - \frac{D_{p+1}{}^2(A)}{A} - p\,\frac{D_{p+1}{}^2(\tilde A)}{\tilde A}
+\frac{D_{p+1}{}^2(C)}{C} -\frac{D_{p+1}{}^2(\tilde C)}{\tilde C}\cr
&+&\frac{1}{C^2}\;\partial_{\rho}\ln C\;\partial_{\rho}\ln \frac{F_1}{C^2}\cr
R_{p+2p+2}&=&- D^2(\ln \tilde C)
\label{ricci}
\eea
\bigskip
La curvatura escalar resulta,
\bea
R&=&
-{2\over C}\partial_{\rho}^2\left({1\over C}\right)
-D(\ln A)^2-p\,D(\ln \tilde A)^2-D(\ln C)^2-D(\ln \tilde C)^2\cr
&-&2\,D^2(\ln F_1)+D(\ln F_1)^2
\eea
\label{ricciscalar}

\section{Tensor intesidad de campo}

A partir del ans\"atz (\ref{fieldansatz}) para las ($p+1$)-formas se
obtiene,
\begin{eqnarray}
F_{p+2} &=& dx^0\wedge\dots\wedge dx^p\; \partial_\rho E(\rho)\cr
\left(F_{p+2}\right)^2_{\mu\nu}&=&-{(\partial_{\rho}E)^2\over(A\,\tilde{A}^p\,C)^2}\;
\eta_{\mu \nu}\cr
\left(F_{p+2}\right)^2_{p+1p+1}&=&-
{(\partial_{\rho}E)^2\over(A\,\tilde{A}^p\,C)^2}\;
\end{eqnarray}

La contribuci\'on de gauge al tensor intensidad de campo resulta,

\begin{eqnarray}
T^A_{\mu\nu}&=&-\frac{2-b_p}{8}\;\frac{e^{(2+ b_p)\Phi}}{\left(A\,{\tilde A}^p\,C\right)^2}\;
\; (\partial_{\rho}E)^2\;\eta_{\mu\nu}\cr
T^A_{p+1p+1}&=&-\frac{2-b_p}{8}\;\frac{e^{(2+ b_p)\Phi}}{\left(A\,{\tilde A}^p\,C\right)^2}\;
\; (\partial_{\rho}E)^2\cr
T^A_{p+2p+2}&=&\frac{2+b_p}{8}\;\frac{e^{(2+ b_p)\Phi}}{\left(A\,{\tilde A}^p\,C\right)^2}\;
\; (\partial_{\rho}E)^2\label{fieldstrength}
\end{eqnarray}

\section{Las ecuaciones de movimiento}

Utilizando los resultados anteriores para el c\'alculo del tensor de
Ricci, el tensor intensidad de campo, etc., las ecuaciones de
movimiento (\ref{ecformal}) para el ans\"atz (\ref{fieldansatz})
pueden ser reescritas de la siguiente manera,
\bigskip

\noindent\underline{$A$-equation}
\be
e^{2\Phi}D(e^{-2\Phi}D(\ln A))={2-b_p\over 8}\;{e^{(b_p+2)\Phi}\over (A{\tilde A}^p\,C)^2}\;
(\partial_\rho E)^2\label{1}
\ee
\noindent\underline{$\tilde A$-equation}
\be
e^{2\Phi}D(e^{-2\Phi}D(\ln \tilde A))=
{2-b_p\over 8}\;{e^{(b_p+2)\Phi}\over(A{\tilde A}^p\,C)^2}\;(\partial_\rho E)^2\label{2}
\ee
\noindent\underline{$C$-equation}
\bea
&-&e^{2\Phi}D(e^{-2\Phi}D(\ln C))={D_{p+1}{}^2(A)\over A} + p\,{D_{p+1}{}^2(\tilde A)\over
\tilde A}- {D_{p+1}{}^2(C)\over C} + {D_{p+1}{}^2(\tilde C)\over \tilde C}\cr
&-&2\,D_{p+1}{}^2(\Phi)
-\frac{1}{C^2}\; \partial_\rho\ln C\;\partial_\rho\ln\left({F_1\over C^2}e^{-2\Phi}\right)
- \frac{2-b_p}{8}\;{e^{(b_p+2)\Phi}\over (A{\tilde A}^p\,C)^2}\;(\partial_\rho E)^2\label{3}
\eea
\noindent\underline{$\tilde C$-equation}
\be
-e^{2\Phi}D(e^{-2\Phi}D(\ln \tilde C))={2+b_p\over 8}\;
{e^{(b_p+2)\Phi}\over (A{\tilde A}^p\,C)^2}\;(\partial_\rho E)^2\label{4}
\ee
\noindent\underline{$\Phi$-equation}
\be
e^{2\Phi}D(e^{-2\Phi}D(\ln e^{-2\Phi}))=\Lambda^2 - (p+1)\,{2-b_p\over 8}\;
{e^{(b_p+2)\Phi}\over (A\tilde A^p\,C)^2}\;(\partial_\rho E)^2\label{5}
\ee
\noindent\underline{E-equation} \be
-D\left(\frac{e^{b_p\Phi}}{(A{\tilde A}^p)^2}\;D(E)\right)= (-)^p
\,Q_p\;\frac{\delta^2_{\perp}}{F_1}\label{6} \ee En la \'ultima
ecuaci\'on, la correspondiente a $E$, hemos usado la siguiente
relaci\'on, \be *F_{p+2} = (-)^p\;\frac{\tilde
  C^2}{F_1}\;\partial_{\rho}E\;dz \ee y hemos supuesto una fuente de
forma $\,Q_p\,J_{p+1}$, con \be J_{p+1}^{01\dots p} =
\frac{\delta^2_\perp}{\sqrt{-\det G}}=
\frac{\delta^2_{G^\perp}}{A\,\tilde A^p}\;\;\longrightarrow\;\;
J_{p+1} = -\delta^2_{G^\perp}\;\omega^0\wedge\dots\wedge\omega^p \ee
donde usamos $G^\perp$ para referirnos a la m\'etrica del espacio
donde la $p$-brana plana es localizada.


%% file: appendII.tex
\chapter{F\'ormulas \'utiles II}\label{AA}

En este ap\'endice, resumimos convenciones y f\'ormulas relevantes para
los c\'alculos presentados en el cap\ii tulo \ref{sym} de esta
tesis. A menos que sea especificado expl\ii citamente, trabajaremos en
la base local con \ii ndices $A, B, C,\dots=0,1,\dots,D-1$.

Consideremos una m\'etrica de la forma,
\be
G = \eta_{AB}\;\omega^A\,\omega^B = \eta_{ab}\;\omega^a\,\omega^b + \omega^n{}^2\qquad,\qquad a,b = 0,1,\dots,D-2
\label{metrica}
\ee
donde los vielbein $\omega^A$, los vectores duales $e_A ,\, e_A(\omega^B) = \delta_A^B\,$, y el elemento de volumen son,
\bea
\omega^a &=& A_a(u)\,dx^a\qquad,\qquad e_a= A_a(u)^{-1}\,\partial_a\cr
\omega^n &=& C(u)\,du \qquad\;\;\;,\qquad e_n= C(u)^{-1}\,\partial_u\cr
\epsilon_G&=& \omega^0\wedge\dots\wedge\omega^n = dx^0\wedge\dots\wedge dx^{D-2}\wedge du\, F_1
\qquad,\qquad F_1 = \prod_a A_a\;C
\eea

Reescribiremos nuevamente algunas f\'ormulas del ap\'endice anterior y
otras nuevas en estas coordenadas.  Las conexiones
pseudo-riemmanianas, $\,\omega_{AB}=-\omega_{BA}\,:\, d\omega^A+
\omega^A{}_B\wedge \omega^B=0\;$ son, \bea \omega_{ab} = 0\qquad
&,&\qquad \omega_{an} = \sigma_a\,\omega_a\cr \sigma_a\equiv e_n(\ln
A_a)\qquad &,&\qquad \sigma\equiv \sum_a\,\sigma_a =
e_n\left(\ln\frac{F_1}{C}\right)\label{conn} \eea Las derivadas
covariantes sobre un escalar $\phi(x,u)$ son, \bea D_A(\phi) &=&
e_A(\phi)\qquad,\qquad \forall A\cr D_a D_b(\phi)&=& e_a e_b(\phi) +
\sigma_a\,e_n(\phi)\,\eta_{ab}\cr D_a D_n(\phi)&=& e_a e_n(\phi) -
\sigma_a\,e_a(\phi) \cr D_n D_a(\phi)&=& e_n e_a(\phi) = D_a
D_n(\phi)\cr D_n{}^2 (\phi) &=& e_n{}^2 (\phi)\cr D^2(\phi) &=& e^a
e_a(\phi) + e_n{}^2(\phi) + \sigma\,e_n(\phi) = \sum_a
\frac{\partial^a\partial_a\phi}{A_a{}^2} +
\frac{1}{E}\,\partial_u\left(\frac{E}{C^2}\;\partial_u\phi\right)
\eea
con $\sigma$ definido en (\ref{conn}).

Las derivadas covariantes relevantes de la 1-forma son $A =
A_A(x,u)\,\omega^A$ son, \bea D_b A_a&=& e_b (A_a) +
\sigma_b\,A_n\,\eta_{ab}\cr D_n A_a&=& e_n (A_a)\cr D_a A_n&=& e_a
(A_n) - \sigma_a\,A_a\cr D_n A_n&=& e_n(A_n)\cr D_c D_a A_b&=& e_c e_a
(A_b) + \sigma_c\,e_n(A_b)\,\eta_{ca}-\sigma_c\,\sigma_a\,
A_a\,\eta_{cb} + \sigma_b\,\left(e_c(A_n)\,\eta_{ab} +
e_a(A_n)\,\eta_{cb}\right)\cr D_n D_a A_b&=& e_n e_a (A_b) + e_n(
\sigma_a\,A_n)\,\eta_{ab}\cr D_a D_n A_b&=& e_a e_n (A_b) -
\sigma_a\,e_a(A_b) + \sigma_b\,\left(e_n(A_n) -
\sigma_b\,A_n\right)\,\eta_{ab}\cr D_n D_n A_b &=& e_n{}^2(A_b)\cr D_c
D_a A_n&=& e_c e_a (A_n) - \sigma_a\,e_c(A_a)-\sigma_c\,e_a(A_c) +
\sigma_a\,\left(e_n(A_n) - \sigma_a\,A_n\right)\,\eta_{ac}\cr D_n D_a
A_n&=& e_n e_a (A_n) - e_n( \sigma_a\,A_a)\cr D_a D_n A_n&=& e_a e_n
(A_n) - \sigma_a\,\left(e_a(A_n) + e_n(A_a) - \sigma_a\,A_a\right)\cr
D_n D_n A_n &=& e_n{}^2(A_n) \eea Las derivadas covariantes relevantes
sobre un 2-tensor sim\'etrico $h=h_{AB}(x,u)\,\omega^A \,\omega^B$
son, \bea D_c h_{ab} &=& e_c(h_{ab}) +
\sigma_c\,\left(\eta_{ac}\,h_{bn} + \eta_{bc}\,h_{an} \right)\cr D_c
h_{an} &=& e_c(h_{an}) + \sigma_c\,\left(\eta_{ac}\,h_{nn} -
h_{ac}\right)\cr D_c h_{nn} &=& e_c(h_{nn}) - 2\,\sigma_c\,h_{cn}\cr
D_n h_{AB} &=& e_n(h_{AB})\qquad,\qquad \forall\;\; A,B\cr D_d D_c
h_{ab} &=& e_de_c(h_{ab}) + \sigma_c\,\sigma_d\left( (\eta_{bc}\,
\eta_{ad}+\eta_{ac}\, \eta_{bd})\, h_{nn} -
\eta_{ad}\,h_{bc}-\eta_{bd}\,h_{ac}\right)\cr &+&\sigma_c\,\left(
\eta_{cd}\, e_n(h_{ab}) + \eta_{ac}\, e_d(h_{bn}) + \eta_{bc}\,
e_d(h_{an}\right) +\sigma_d\,\left( \eta_{ad}\, e_c(h_{bn}) +
\eta_{bd}\, e_c(h_{an})\right)\cr D_n D_c h_{ab} &=& e_n e_c(h_{ab}) +
\eta_{ac}\, e_n(\sigma_c\,h_{bn}) + \eta_{bc}\,
e_n(\sigma_c\,h_{an})\cr D_c D_n h_{ab} &=& e_c e_n(h_{ab}) +
\sigma_c\left(\eta_{ac}\, e_n(h_{bn})+\eta_{bc}\,
e_n(h_{an})-e_c(h_{ab})\right) - \sigma_c{}^2\,\left(\eta_{ac}\,
h_{bn} +\eta_{bc}\, h_{an}\right)\cr D_d D_c h_{an} &=& e_d
e_c(h_{an}) -\sigma_c\,\sigma_d\,\left(\eta_{ac}\, h_{dn} +
2\,\eta_{ad}\,h_{cn}\right) + \sigma_c\,\left(\eta_{ac}\, e_d(h_{nn})
- e_d(h_{ac})\right)\cr &+& \sigma_d\,\left(\eta_{ad}\, e_c(h_{nn}) -
e_c(h_{ad}\right) + \sigma_d\,\eta_{cd}\,\left(e_n(h_{an}) -
\sigma_c\,h_{an}\right)\cr D_n D_c h_{an} &=& e_n e_c(h_{an}) +
\eta_{ac}\, e_n(\sigma_a\,h_{nn}) - e_n(\sigma_c\,h_{ac})\cr D_c D_n
h_{an} &=& e_c e_n(h_{an}) + \sigma_c\,\left(\eta_{ac}\, e_n(h_{nn})-
e_n(h_{ac})-e_c(h_{an})\right) - \sigma_c{}^2\,\left(\eta_{ac}\,
h_{nn} - h_{ac}\right)\cr D_d D_c h_{nn} &=& e_d e_c (h_{nn})
-2\,\sigma_c\, \sigma_d\,\left(\eta_{cd}\, h_{nn}- h_{cd}\right) +
\sigma_c\,\eta_{cd}\, e_n(h_{nn})\cr &-&2\,\left(\sigma_c\,
e_d(h_{cn})+ \sigma_d\,e_c(h_{dn})\right)\cr D_n D_c h_{nn} &=& e_n
e_c(h_{nn}) -2\, e_n(\sigma_c\,h_{cn})\cr D_c D_n h_{nn} &=& e_c
e_n(h_{nn}) -\sigma_c\,\left(e_c(h_{nn})+
2\,(e_n(h_{cn})-\sigma_c\,h_{cn})\right)\cr D_n D_n h_{AB} &=&
e_n{}^2(h_{AB})\qquad,\qquad \forall\;\; A,B\label{metricderiv} \eea
El tensor de curvatura $\,\Re_{AB} \equiv d\omega_{AB} +
\omega_{AC}\wedge\omega^C{}_B \,$ es, \bea
\Re_{ab}&=&-\sigma_a\,\sigma_b\;\omega_a\wedge\omega_b\cr
\Re_{an}&=&-\frac{1}{A_a}\, e_n{}^2(A_a)\;\omega_a\wedge\omega_n
\label{curvaturaII}
\eea
y el tensor de Ricci y el escalar de Ricci son,
\bea
R_{ab}&=& -D^2(\ln A_a)\;\eta_{ab}\cr
R_{nn}&=&-\sum_a\,\frac{1}{A_a}\,
e_n{}^2(A_a)= - \left(e_n(\sigma) + \sum_a\sigma_a{}^2\right)\cr
R&=&  - 2\, D^2\left(\ln\frac{F_1}{C}\right) +\sigma^2 - \sum_a\sigma_a{}^2\label{riccis}
\eea

\section{C\'alculo del tensor $ A_{AB}$}\label{AA1}

Cuando se tienen en cuenta las perturbaciones de la m\'etrica, el siguiente 2-tensor aparece naturalmente,
\be
A_{AB}(h) \equiv D_A D_B h^C_C + D^2 h_{AB} -  D^C D_A h_{CB} -D^C D_B h_{AC}.
\ee
Con ayuda de (\ref{metricderiv}) obtenemos las expresiones,
\bea
A_{ab}(h) &=&  e^A e_A (h_{ab})+ e_a e_b(h^C_C)
-e_a e_n (h_{bn})-e_b e_n (h_{an}) -e^c e_a (h_{bc})-e^c e_b (h_{ac})\cr
&+&\sigma\,e_n(h_{ab}) +\left(  e_n(\sigma_a + \sigma_b) + \sigma\,(\sigma_a+\sigma_b) -
(\sigma_a-\sigma_b)^2 \right)\,h_{ab}\cr
&+& (2\,\sigma_a - \sigma_b -\sigma)\,e_a(h_{bn}) +
(2\,\sigma_b - \sigma_a -\sigma)\,e_b(h_{an})\cr
&+& \eta_{ab}\,\left(  -2\,\left( e_n(\sigma_a) + \sigma\,\sigma_a\right)\,h_{nn} +
\sigma_a\,\left( e_n(h^c_c) - e_n(h_{nn}) - 2\,\,e^c(h_{cn}\right) \right)\cr
A_{an}(h) &=& e^c e_c(h_{an}) - e^c e_n (h_{ac}) - e^c e_a (h_{cn}) + e_n e_a (h^c_c)
+2\, (e_n (\sigma_a) + \sigma\,\sigma_a )\, h_{an}\cr
&-& (\sigma - \sigma_a)\, e_a(h_{nn})+ (\sigma_a - \sigma_c)\, e^c(h_{ac}) + \sigma_c\,e_a(h^c_c)\cr
A_{nn}(h) &=& e^c e_c(h_{nn}) - \sigma\, e_n(h_{nn}) -2\, e^c e_n (h_{cn}) - 2\,\sigma_c\, e^c(h_{cn})
+ e_n{}^2 (h^c_c)+ 2\,\sigma_c\, e_n(h^c_c)\cr& &\label{AAB}
\eea
Bajo la transformaci\'on de gauge (\ref{gaugetrans}),
\bea
A_{ab}(\delta_\epsilon h) &=&
\left( e_n(\sigma_a) + \sigma\,\sigma_a + e_n(\sigma_b) + \sigma\,\sigma_b\right)\,
\delta_\epsilon h_{ab} + 2\, e_n\left( e_n(\sigma_a) + \sigma\,\sigma_a\right)\,\epsilon_n\,\eta_{ab}\cr
&+&\left( e_n(\sigma_b) + \sigma\,\sigma_b - e_n(\sigma_a) - \sigma\,\sigma_a \right)\,
\left(e_a(\epsilon_b) - e_b(\epsilon_a)\right)\cr
A_{an}(\delta_\epsilon h) &=&
2\,\left( e_n(\sigma_a) + \sigma\,\sigma_a\right)\,\delta_\epsilon h_{an} + 2
\left( \sum_c\left( e_n(\sigma_c) + \sigma_c{}^2\right) - e_n(\sigma_a) - \sigma\,\sigma_a\right)\,
e_a(\epsilon_n)\cr
A_{nn}(\delta_\epsilon h) &=& 2\, \sum_c\left( e_n(\sigma_c) + \sigma_c{}^2\right)\, \delta_\epsilon h_{nn}
+ 2\, e_n\left(\sum_c\left( e_n(\sigma_c) + \sigma_c{}^2\right)\right)\,\epsilon_n
\eea

\section{Una derivaci\'on corta de las soluciones}\label{AA2}

Esbozaremos aqu{\ii } la familia de soluciones (\ref{Ssn})
consideradas.  Supongamos un ans\"atz para la m\'etrica de la forma
(\ref{metrica}), junto con, \bea \Phi(u) &=& \Phi = \rm{constant}\cr
F_D&=&(-)^D\,Q_{D-2}\;\epsilon_G\;\;\Longleftrightarrow\;\; *F_D=
(-)^{D-1}\;Q_{D-2}\label{ansatz_b} \eea El tensor intensidad de campo en
(\ref{ansatz_b}) lleva a, \bea (F_D)^2{}_{AB} =
(F_D)^2\;\eta_{AB}\qquad,\qquad (F_D)^2
=-\left(\frac{C\;e_n(E)}{F_1}\right)^2 \eea Las ecuaciones de
movimiento en el {\it string frame} resultan (consideraremos D-branas,
$b_{D-2}=0$), \bea R_{AB} &=& \frac{1}{4}\,
e^{2\Phi}\;\left(F_D\right)^2\;\eta_{AB}\cr \Lambda^2 &=& -
\frac{D}{4}\;e^{2\Phi}\; (F_D)^2\cr d\left(*F_D\right)&=& (-)^D \;
Q_{D-2}\; *J_{D-1}\;\;\;\;,\;\;\;\; Q_{D-2}\equiv 2\,\kappa_D{}^2\,
\mu_{D-2} \label{ecformalII} \eea donde $\kappa_D{}^2 =8\pi G_D$ es la
constante de acoplamiento gravitatorio y $\mu_{D-2}$ la tensi\'on de
la D$(D-2)$-brana.  Las \'ultimas dos ecuaciones pueden ser resueltas
por, \be \frac{C\; e_n(E)}{F_1}=-Q_{D-2}\qquad,\qquad e^{2\Phi} =
\frac{4}{D}\;\frac{\Lambda^2}{Q_{D-2}{}^2} \ee mientras que las
ecuaciones para la m\'etrica se reducen a, \bea D^2(\ln A_a) \equiv
e_n(\sigma_a) + \sigma\,\sigma_a &=& \frac{\Lambda^2}{D}\,
\qquad,\qquad \forall a\cr
\sum_a\,\left(e_n(\sigma_a) +
\sigma_a{}^2\right)&=&\frac{\Lambda^2}{D}\label{eqmetrica}
\eea
Siguiendo los pasos del cap\ii tulo \ref{nos_sol}, la soluci\'on general a (\ref{eqmetrica}) puede ser escrita como,
\be
A_a(u) = l_0\,u\,f(u)^\frac{a_a}{2}\qquad;\qquad C(u)=l_0\,u^{-1}\,f(u)^{-\frac{1}{2}}
\ee
con $f(u)$ como en (\ref{f}), y el v\ii nculo (\ref{constraints}) sobre los exponentes.


%% file: appendIII.tex
\chapter{Las fluctuaciones en las soluciones T-duales}\label{AB}

\section{La familia de soluciones}

La familia de soluciones (en el {\it string frame}) obtenida haciendo
una T-dualidad en la coordenada $\,x^{D-2}-$ resulta, \bea
l_0{}^{-2}\,G &=& u^2\,f(u)^{a_\mu}\,\eta_{\mu\nu}\,dx^\mu\,dx^\nu +
\frac{d u^2}{u^2\,f(u)} +
u_1{}^2\,\frac{d\theta^2}{u^2\,f(u)^{a_\theta}}\qquad,\qquad
\Lambda^2\,l_0{}^2 = D\,(D-1)\cr e^{\Phi} &=&
\frac{4\,\pi\,\sqrt{D-1}}{|Q_{D-3}|}\,\frac{u_1}{u}\,f(u)^{-\frac{a_\theta}{2}}\cr
F_{D-1} &=&
\frac{Q_{D-3}\,l_0{}^{D-2}}{2\,\pi\,u_1}\,u^{D-2}\,du\wedge
\,dx^0\wedge\dots\wedge dx^{D-3} \;\;\leftrightarrow\;\; *F_{D-1} =
(-)^{D-1}\,Q_{D-3}\,\frac{d\theta}{2\,\pi}\cr & &\label{sns} \eea
donde $\mu,\nu= 0,1,\dots,D-3\,$, la coordenada $\theta$ es
$2\,\pi$-peri\'odica, y \be f(u) = 1 -
\left(\frac{u_0}{u}\right)^{D-1} \ee Las escalas $u_0$ y $u_1$ son arbitrarias, y los exponentes cumplen, \be \sum_\mu a_\mu+ a_\theta = 1\qquad,\quad \sum_\mu
a_\mu{}^2 + a_\theta{}^2 = 1 \ee Near horizon de  D(D-3)-branas sobre un fondo de dilat\'on lineal

Las siguientes son algunas relaciones \'utiles,
\bea
e_n(\sigma_\mu) + \sigma\,\sigma_\mu&=& \frac{\Lambda^2}{D-2}\,e^{\frac{4}{D-2}\,\Phi}\cr
e_n(\sigma_\theta) + \sigma\,\sigma_\theta &=& -\frac{D-4}{D (D-2)}\,\Lambda^2\,e^{\frac{4}{D-2}\,\Phi}\cr
\sum_a \left( e_n(\sigma_a) + \sigma_a{}^2\right)&=& \frac{\Lambda^2}{D-2}\,e^{\frac{4}{D-2}\,\Phi} -
\frac{4\,(D-2)}{(D-4)^2}\,\sigma_\theta{}^2\cr
\sigma^2 - \sum_a \sigma_a{}^2 &=& \frac{D-2}{D}\,\Lambda^2\,e^{\frac{4}{D-2}\,\Phi}
+ \frac{4\,(D-2)}{(D-4)^2}\,\sigma_\theta{}^2\cr
e_n{}^2(\Phi) + \sigma\,e_n(\Phi) &=& -\frac{\Lambda^2}{D}\,e^{\frac{4}{D-2}\,\Phi}\cr
e_n(\Phi)&=& \frac{D-2}{D-4}\,\sigma_\theta
\label{orig.rel}
\eea

\section{Las ecuaciones para las fluctuaciones en el e-frame}

Dado que el sistema es lineal, resulta que
$(L_\epsilon(G),L_\epsilon(\Phi),L_\epsilon(A_q))$ es soluci\'on para
todo $\epsilon$, la soluci\'on puro gauge.  A diferencia de las
teor\ii as de gravedad ordinarias o de gauge, donde las
transformaciones no son lineales, en nuestro caso es posible definir
expl\ii citamente cantidades invariantes de gauge, y expresar las
ecuaciones para las perturbaciones (\ref{ep1}) en t\'erminos de
\'estas, de manera manifiestamente invariante de gauge.  En efecto,
para las familias que nos interesan podemos hacerlo de la siguiente
manera.  Primero, nos restringiremos a las fluctuaciones tales que,
\be f_{D-1} \equiv \chi\, F_{D-1}\qquad,\qquad ^\epsilon\chi=\chi +
D_{\tilde a}\epsilon^{\tilde a} +
\frac{D}{D-4}\,\sigma_\theta\,\epsilon_n \ee Por consistencia con las
identidades de Bianchi, se debe satisfacer que $\chi$ debe ser
independiente de $\theta$, y por consiguiente lo ser\'an todas las
fluctuaciones; la transformaci\'on de gauge para $\chi$ se sigue de
(\ref{gaugetrans}).  En segundo lugar, siguiendo (\ref{ls-bdgduals})
introducimos los campos $(g, g_a, I_{ab}, I_\xi,I_\chi)$ de la
siguiente manera,
\begin{eqnarray}
h_{nn} &\equiv& 2\,e_n(g) \qquad\qquad\qquad\qquad\qquad\qquad\qquad\qquad,\qquad\delta_\epsilon g = \epsilon_n\cr
h_{an} &\equiv& A_a\,e_n\left(\frac{g_a}{A_a}\right) + e_a(g)\;\;\qquad\qquad\qquad\qquad\qquad,\qquad\delta_\epsilon g_a = \epsilon_a\cr
h_{ab}&\equiv& I_{ab} + e_a(g_b) + e_b(g_a) + 2\,\eta_{ab}\,\sigma_a\, g\qquad\qquad\qquad,\qquad\delta_\epsilon I_{ab} = 0\cr
\xi &\equiv& I_\xi + \frac{D-2}{D-4}\,\sigma_\theta\, g\;\;\qquad\qquad\qquad\qquad\qquad\qquad,\qquad\delta_\epsilon I_\xi = 0\cr
\chi &\equiv& I_\chi + e^\mu(g_\mu) + e_n(g) + \left(\sigma + \frac{4}{D-4}\,\sigma_\theta\right)\,g
\qquad,\qquad\delta_\epsilon I_\chi = 0
\end{eqnarray}
donde las transformaciones de gauge se siguen de (\ref{gaugetrans}).
En t\'erminos de estas variables (\ref{AAB}) se escriben como, \bea
A_{ab}(h) &=& A^{(i)}_{ab}(I) + 2\, \left( e_n(\sigma_b) +
\sigma\,\sigma_b\right)\,e_a(g_b) +2\, \left( e_n(\sigma_a) +
\sigma\,\sigma_a\right)\,e_b(g_a)\cr &+& 2\,\eta_{ab}\,\left( e_n
\left( e_n(\sigma_a) + \sigma\,\sigma_a\right)+ 2\,\sigma_a\,\left(
e_n(\sigma_a) + \sigma\,\sigma_a\right)\right) \,g\cr
A^{(i)}_{ab}(I)&\equiv& e^A e_A (I_{ab})+ e_a e_b(I^c_c)- e_ae^c
(I_{bc})- e_be^c (I_{ac})+\sigma\,e_n(I_{ab})\cr &+&\left(
e_n(\sigma_a + \sigma_b) + \sigma\,(\sigma_a+\sigma_b)
-(\sigma_a-\sigma_b)^2 \right)\,I_{ab} + \eta_{ab}\,\sigma_a\,
e_n\left(I^c_c\right)\cr A_{an}(h) &=& A^{(i)}_{an}(I) + 2\, \left(
e_n(\sigma_a) + \sigma\,\sigma_a\right)\, \left( e_n(g_a) -
\sigma_a\,g_a\right)+ 2\,\sum_c \left( e_n(\sigma_c) +
\sigma_c{}^2\right)\; e_a(g)\cr A^{(i)}_{an}(I)&\equiv&- e^c e_n
(I_{ac}) + e_a e_n (I^c_c) + (\sigma_a - \sigma_c)\, \left(
e^c(I_{ac}) - \,e_a(I^c_c)\right)\cr A_{nn}(h) &=&A^{(i)}_{nn}(I)
+4\,\sum_c \left( e_n(\sigma_c) + \sigma_c{}^2\right)\; e_n(g) + 2\,
e_n\left(\sum_c \left( e_n(\sigma_c) + \sigma_c{}^2\right)\right)\;
g\cr A^{(i)}_{nn}(I)&\equiv& e_n{}^2 (I^c_c)+ 2\,\sigma_c\,
e_n(I^c_c)\label{Ainv} \eea Usando estas expresiones, las ecuaciones
(\ref{ep1}) en nuestros fondos resultan, \bea 0 &=& e^A e_A
(I_{\mu\nu})+ e_\mu e_\nu(I^c_c)- e_\mu e^c (I_{\nu c})- e_\nu e^c
(I_{\mu c})+\sigma\,e_n(I_{\mu\nu})\cr &+&\eta_{\mu\nu}\,\sigma_\mu\,
e_n\left(I^c_c\right)- \left(\sigma_\mu-\sigma_\nu \right)^2
\,I_{\mu\nu} - \frac{8\,\Lambda^2}{(D-2)^2}\, \eta_{\mu\nu}\,
e^{\frac{4}{D-2}\,\Phi} \,I_\xi\cr 0 &=& - e^c e_n (I_{\mu c}) + e_\mu
e_n (I^c_c) + (\sigma_\mu - \sigma_c)\, \left( e^c(I_{\mu c}) -
\,e_\mu (I^c_c)\right)+ \frac{8}{D-4}\, \sigma_\theta\, e_\mu (I_\xi)
\cr 0 &=& e_n{}^2 (I^c_c)+ 2\,\sigma_c\, e_n(I^c_c)+ \frac{16}{D-4}\,
\sigma_\theta\, e_n (I_\xi) - \frac{8\,\Lambda^2}{(D-2)^2}\,
e^{\frac{4}{D-2}\,\Phi} \,I_\xi \cr 0 &=& e^A e_A (I_{\theta\theta})+
\sigma\,e_n(I_{\theta\theta}) + \sigma_\theta\,e_n(I_c^c) +
\frac{8\,\Lambda^2}{D}\, e^{\frac{4}{D-2}\,\Phi}\, \left( \,I_\chi
-\frac{1}{2}\, I_\mu^\mu - \frac{D^2 -3\,D
  +4}{(D-2)^2}\,I_\xi\right)\cr 0 &=& e^A e_A (I_{\mu\theta})+
\sigma\,e_n(I_{\mu\theta}) - e_\mu e^c(I_{c\theta}) + \left(
\frac{2\,\Lambda^2}{D}\, e^{\frac{4}{D-2}\,\Phi} - (\sigma_\mu
-\sigma_\theta)^2\right)\,I_{\mu\theta}\cr 0 &=& e^c\left(
e_n(I_{c\theta }) + (\sigma_c -\sigma_\theta)\,I_{c\theta}\right)\cr 0
&=& e^A e_A(I_\xi) + \sigma\, e_n(I_\xi)+
\frac{D-2}{2(D-4)}\,\sigma_\theta\,e_n(I_c^c)\cr &+&
\frac{D-2}{D}\,\Lambda^2\, e^{\frac{4}{D-2}\,\Phi}\, \left( -I_\chi
+\frac{1}{2}\, I_\mu^\mu + \frac{D^2 -2\,D
  +4}{(D-2)^2}\,I_\xi\right)\cr 0&=&
-e^{-\alpha_q\Phi}\,D^B\left(e^{\alpha_q\Phi}\,(f_{q+2})_{A_1\dots
  A_{q+1}B}\right)\qquad,\qquad q\neq D-3
\label{hxieqn}
\eea

\noindent\underline{ h-equations}

\bea
A_{\mu\nu}(h)\, e^{-\frac{4}{D-2}\,\Phi} &=& \frac{2\,\Lambda{}^2}{D-2}\, h_{\mu\nu} +
\frac{8\,\Lambda^2}{(D-2)^2}\,\eta_{\mu\nu}\,\xi\cr
A_{\mu n}(h)\, e^{-\frac{4}{D-2}\,\Phi} &=& \frac{2\,\Lambda{}^2}{D-2}\, h_{\mu n} -
\frac{8}{D-2}\,e^{-\frac{4}{D-2}\,\Phi}\,e_n(\Phi)\,e_\mu(\xi)\cr
A_{nn}(h)\, e^{-\frac{4}{D-2}\,\Phi} &=& \frac{2\,\Lambda{}^2}{D-2}\, h_{nn} -
\frac{16}{D-2}\,e^{-\frac{4}{D-2}\,\Phi}\,e_n(\Phi)\,e_n(\xi) + \frac{8\,\Lambda^2}{(D-2)^2}\,\xi\cr
A_{\theta\theta}(h)\, e^{-\frac{4}{D-2}\,\Phi} &=& -\frac{2\,(D-4)}{D (D-2)}\, \Lambda^2\,h_{\theta\theta} +
\frac{8 (D^2-3D+4)}{D(D-2)^2}\,\Lambda^2\,\xi +
\frac{4}{D}\,\Lambda^2\,\left( h^{\tilde a}_{\tilde a} - 2\,\chi\right)\cr
A_{\tilde a\theta}(h)\, e^{-\frac{4}{D-2}\,\Phi} &=& -\frac{2\,(D-4)}{D (D-2)}\,\Lambda{}^2\, h_{\tilde a\theta} -
\frac{8}{D-2}\, e^{-\frac{4}{D-2}\,\Phi}\, e_{\tilde a}(\Phi)\,e_\theta(\xi)\qquad,\qquad \tilde a\neq \theta\cr & &
\eea

\noindent\underline{$\xi$-equation}

\bea
e^{-\frac{4}{D-2}\,\Phi}\,D^2(\xi) &=& e^{-\frac{4}{D-2}\,\Phi}\,e_n(\Phi)\,
\left( e^A(h_{An}) - \frac{1}{2}\,e_n(h^A_A)\right)+ e^{-\frac{4}{D-2}\,\Phi}\,D^2(\Phi)\,h_{nn}\cr
&-& \frac{D-2}{2\,D}\,\Lambda^2\,h_{\theta\theta}  +\frac{D^2 -6\,D+4}{D (D-2)}\,\Lambda^2\,\xi
\eea

\noindent\underline{$a_{D-2}\,$-equation}
\bigskip

Obtenemos una ecuaci\'on con $(D-2)$ \ii ndices antisim\'etricos, que
lleva a las siguientes ecuaciones \bea 0 &=&e_\mu (I_{\nu\theta})
-e_\nu (I_{\mu\theta})\qquad,\qquad \forall\, \mu,\nu\cr 0&=&
e_n\left(\frac{A_\mu}{A_\theta}\,I_{\mu\theta}\right)\qquad,\qquad
\forall\, \mu\cr 0&=& e_{\tilde a}\left(2\,I_\xi - I_\chi +
\frac{1}{2}\,(I_\mu^\mu - I_{\theta\theta})\right)\qquad,\qquad
\forall\,\tilde a \eea con soluci\'on, \bea
I_{\mu\theta}&=&0\qquad,\qquad \forall\,\mu\cr I_\chi &=& 2\,I_\xi +
\frac{1}{2}\,(I_\mu^\mu - I_{\theta\theta}). \eea Introduciendo estos
valores en (\ref{hxieqn}) obtenemos el siguiente sistema, \bea 0 &=&
e_n{}^2 (I_{\mu\nu})+ \sigma\,e_n(I_{\mu\nu}) + e^\rho e_\rho
(I_{\mu\nu}) + e_\mu e_\nu(I^c_c)- e_\mu e^\rho (I_{\nu\rho})- e_\nu
e^\rho (I_{\mu\rho})- \left(\sigma_\mu-\sigma_\nu \right)^2
\,I_{\mu\nu}\cr &+&\eta_{\mu\nu}\,\sigma_\mu\, e_n\left(I^c_c\right)-
\frac{8\,\Lambda^2}{(D-2)^2}\, \eta_{\mu\nu}\,
e^{\frac{4}{D-2}\,\Phi}\,I_\xi\cr 0 &=& e_n{}^2(I_{\theta\theta})+
\sigma\,e_n(I_{\theta\theta})+ e^\rho e_\rho (I_{\theta\theta})+
\sigma_\theta\,e_n(I_c^c)\cr &+& \frac{8\,\Lambda^2}{D}\,
e^{\frac{4}{D-2}\,\Phi}\, \left( \frac{(D-1)(D-4)}{(D-2)^2}\, I_\xi -
\frac{1}{2}\,I_{\theta\theta}\right)\cr 0 &=& - e^\rho e_n
(I_{\mu\rho}) + e_\mu e_n (I^c_c) + (\sigma_\mu - \sigma_\rho)\,
e^\rho(I_{\mu\rho}) + (\sigma_c - \sigma_\mu) \,e_\mu (I^c_c)+
\frac{8}{D-4}\, \sigma_\theta\, e_\mu (I_\xi) \cr 0 &=& e_n{}^2
(I^c_c)+ 2\,\sigma_c\, e_n(I^c_c)+ \frac{16}{D-4}\, \sigma_\theta\,
e_n (I_\xi) - \frac{8\,\Lambda^2}{(D-2)^2}\, e^{\frac{4}{D-2}\,\Phi}
\,I_\xi \cr 0 &=& D^2(I_\xi) +
\frac{D-2}{2(D-4)}\,\sigma_\theta\,e_n(I_c^c) -
\frac{(D-2)\,\Lambda^2}{D}\, e^{\frac{4}{D-2}\,\Phi}\, \left(\frac{D^2
  -6D+4}{(D-2)^2}\, I_\xi - \frac{1}{2}\,I_{\theta\theta}\right)\cr&
&\label{eqn1} \eea Ahora introducimos los modos en los momentos, \be
I_{\mu\nu} = \chi_{\mu\nu}(u)\,e^{i p\cdot x}\qquad;\qquad
I_{\theta\theta} = \chi_\theta(u)\,e^{i p\cdot x}\qquad;\qquad I_\xi =
\chi_\xi(u) \,e^{i p\cdot x} \ee donde $p\cdot x \equiv
p_\rho\,x^\rho\,$.  Con la definici\'on $p_\mu \equiv A_\mu\,\tilde
p_\mu\,$, obtenemos el sistema en la forma, \bea 0 &=& e_n{}^2
(\chi_{\mu\nu})+ \sigma\,e_n(\chi_{\mu\nu}) - \left(\tilde p^\rho\,
\tilde p_\rho + \left(\sigma_\mu-\sigma_\nu \right)^2
\right)\,\chi_{\mu\nu} + \tilde p_\mu \,\tilde
p^\rho\,\chi_{\nu\rho}+\tilde p_\nu \,\tilde p^\rho\,\chi_{\mu\rho}\cr
&-&\tilde p_\mu\,\tilde p_\nu\, ( \chi^\rho_\rho + \chi_\theta)+
\eta_{\mu\nu}\,\sigma_\mu\, e_n\left(\chi^\rho_\rho +
\chi_\theta\right)- \frac{8\,\Lambda^2}{(D-2)^2}\, \eta_{\mu\nu}\,
e^{\frac{4}{D-2}\,\Phi}\,\chi_\xi\cr 0 &=& e_n{}^2(\chi_\theta)+
\sigma\,e_n(\chi_\theta)- \left( \tilde p^\rho \tilde
p_\rho+\frac{4\,\Lambda^2}{D}\,
e^{\frac{4}{D-2}\,\Phi}\right)\,\chi_\theta +
\sigma_\theta\,e_n(\chi_\rho^\rho + \chi_\theta)\cr &+&
\frac{8(D-1)(D-4)}{D(D-2)^2}\,\Lambda^2\,e^{\frac{4}{D-2}\,\Phi}\,
\chi_\xi \cr 0 &=& e_n{}^2(\chi_\xi)+ \sigma\,e_n(\chi_\xi) - \left(
\tilde p{}^2+ \frac{D^2 -6D+4}{D(D-2)}\,\Lambda^2\,
e^{\frac{4}{D-2}\,\Phi}\, \chi_\xi\right) +
\frac{D-2}{2(D-4)}\,\sigma_\theta\,e_n(\chi^\rho_\rho +
\chi_\theta)\cr &+& \frac{D-2}{2\,D}\,\Lambda^2
e^{\frac{4}{D-2}\,\Phi} \,\chi_\theta\cr 0 &=& \tilde p^\mu\,\left(
e_n (\chi^\rho_\rho + \chi_\theta) + (\sigma_\rho - \sigma_\mu)
\,\chi^\rho_\rho + (\sigma_\theta - \sigma_\mu) \,\chi_\theta +
\frac{8}{D-4}\, \sigma_\theta\,\chi_\xi\right) \cr &-& \tilde
p^\rho\,\left( e_n (\chi_{\mu\rho}) + (\sigma_\rho -
\sigma_\mu)\,\chi_{\mu\rho} \right)\cr 0 &=& e_n{}^2
(\chi^\rho_\rho+\chi_\theta) + 2\,\sigma_\rho\, e_n(\chi^\rho_\rho)+ +
2\,\sigma_\theta\, e_n(\chi_\theta)+\frac{16}{D-4}\, \sigma_\theta\,
e_n (\chi_\xi) - \frac{8\,\Lambda^2}{(D-2)^2}\,
e^{\frac{4}{D-2}\,\Phi} \,\chi_\xi\cr & &
\label{eqn2}
\eea Notamos que nos hemos quedado con las inc\'ognitas
$(\chi_{\mu\nu}, \chi_{\theta\theta}, \chi_\xi)$, cuyos sistemas de
ecuaciones diferenciales de segundo orden acopladas estan dados por
las primeras tres ecuaciones; las \'ultimas dos, deben actuar como
v{\'\i}nculos.

\section{Modelos hologr\'aficos de teor{\'\i}as de Yang-Mills $d$-dimensionales}

Tomemos, de entre las coordenadas $x^\mu$, $d$ coordenadas no
compactas equivalentes $x^\alpha$, $\alpha=0,1,\dots, d-1$, y $D-d-2$
compactas y no equivalentes $\tau^i, i=1,\dots D-d-2, \tau_i\equiv
\tau_i +2\,\pi\,R_i$.  Llamaremos con $\tilde{} $ a las cantidades
asociadas con las direcciones no compactas ($A_\alpha = \tilde A\,,\,
a_\alpha = \tilde a\, ,\,\sigma_\alpha = \tilde \sigma\,,\,$ etc).  La
m\'etrica y los v{\'\i}nculos (\ref{constraints}) son, \bea
l_0{}^{-2}\;G &=& u^2\left( f(u)^{\tilde
  a}\;\eta_{\mu\nu}\,dx^\mu\,dx^\nu +
\sum_i\,f(u)^{a_i}\;d\tau^i{}^2\right) + \frac{du^2}{u^2\,f(u)}+
u_1{}^2\,\frac{d\theta^2}{u^2\,f(u)^{a_\theta}}\cr & &d\,\tilde a +
\sum_i a_i + a_\theta=1\qquad,\qquad d\,\tilde a^2 + \sum_i
a_i{}^2+a_\theta{}^2 =1 \eea Enfatizamos que $D-d-2$ exponentes quedan
libres.

\section{M\'etrica: fluctuaciones transversales}

Corresponden a tomar el  ans\" atz,
\be
\chi_{\alpha\beta}(u)= \epsilon_{\alpha\beta}(p)\;\chi(u)\qquad;\qquad
\epsilon_\alpha^\alpha=0\;\;,\;\;\epsilon_{\alpha\beta}\,p^\beta = 0\label{hmnanstransorig}
\ee
y el resto de las fluctuaciones como cero. Este  ans\" atz resuelve (\ref{eqn2}) si $\chi$ satisface
\be
e_n{}^2 (\chi)+ \sigma\,e_n(\chi) + \left( \frac{M^2}{\tilde A^2} - \sum_i \tilde p _i{}^2 \right)\,\chi=0
\ee
donde $M^2\equiv -p^\alpha p_\alpha$ es la masa $d$-dimensional. 

\section{M\'etrica: fluctuaciones longitudinales}

Corresponden a tomar el ans\" atz, a $i$ fijo (pero arbitrario),
\be
\chi_{i\alpha}(u)= \epsilon_\alpha(p)\;\chi(u)\qquad;\qquad
p_i=0\;\;,\;\;\epsilon_\alpha\,p^\alpha = 0\label{hmnanslongorig}
\ee
y el resto de las fluctuaciones a cero. 
Es consistente con (\ref{eqn2}) si $\chi(u)$ obedece
\be
e_n{}^2 (\chi)+ \sigma\,e_n(\chi) + \left( \frac{M^2}{\tilde A^2} - \sum_i \tilde p _i{}^2
- (\tilde\sigma - \sigma_i{})^2\right)\,\chi=0
\ee

\section{Fluctuaciones escalares}

Esperamos que vengan del dilat\'on y de la m\'etrica.  Sin embargo, a
diferencia de los casos resueltos en \cite{Constable:1999gb},
\cite{Kuperstein:2004yk}, estos no se desacoplan, en el sentido que
poner a cero las fluctuaciones de la m\'etrica no es consistente.  Por
eso, nosotros comenzamos con este ans\" atz para la m\'etrica, \be
\chi_{\mu\nu}(u) = a_\mu(u)\,\eta_{\mu\nu} +
b_{\mu\nu}(u)\,p_\mu\,p_\nu \ee el cual, a diferencia de los casos
tratados anteriormente, no es transversal, ni de traza nula.  Resulta
conveniente introducir las siguientes variables \bea \chi_\mu &\equiv&
a_\mu - \sigma_\mu\, \frac{\chi_\theta}{\sigma_\theta}\cr
\tilde\chi_{\mu\nu} &\equiv& A_\mu\,A_\nu\, b_{\mu\nu} - \frac{1}{2} (
A_\mu{}^2\, b_{\mu\mu} + A_\nu{}^2\, b_{\nu\nu})\cr \tilde\chi_\mu
&\equiv& A_\mu{}^2\, e_n(b_{\mu\mu}) -
\frac{\chi_\theta}{\sigma_\theta}\cr \tilde \chi_\theta &\equiv&
e_n\left(\frac{\chi_\theta}{\sigma_\theta}\right)\cr \tilde
\chi_\xi&\equiv& \chi_\xi -\frac{D-2}{2(D-4)}\, \chi_\theta \eea
Notamos que $b_{\mu\nu}$ es reemplazado por $\tilde\chi_{\mu\nu}$ y
$\tilde\chi_\mu$ (por construcci\'on, $\tilde\chi_{\mu\mu}\equiv 0$
para cualquier $\mu$).  Con ellos, y combinando ecuaciones en
(\ref{eqn2}), podemos reacomodar el sistema de la siguiente manera,
\bea 0 &=& e_n{}^2 (\chi_\mu)+ \sigma\,e_n(\chi_\mu) - \tilde p^\rho\,
\tilde p_\rho\,\chi_\mu + \frac{2\,\Lambda^2}{D-2}\,
e^{\frac{4}{D-2}\,\Phi}\, \left( 1 +
\frac{D-4}{D}\,\frac{\sigma_\mu}{\sigma_\theta}\right)\,\tilde\chi_\theta\cr
&-&\frac{8\,\Lambda^2}{(D-2)^2}\, e^{\frac{4}{D-2}\,\Phi}\, \left( 1 +
(D-1)\frac{D-4}{D}\,\frac{\sigma_\mu}{\sigma_\theta}\right)\,\tilde\chi_\xi\cr
0 &=& e_n{}^2 (\tilde\chi_{\mu\nu})+ \left(\sigma -2(\sigma_\mu
+\sigma_\nu)\right)\,e_n(\tilde\chi_{\mu\nu}) + \sum_\rho \tilde
p^\rho\, \tilde p_\rho \,\left( \tilde\chi_{\mu\rho} +
\tilde\chi_{\nu\rho} - \tilde\chi_{\mu\nu}\right)\cr &+& \left(
4\,\sigma_\mu \,\sigma_\nu -\frac{2\,\Lambda^2}{D-2}\,
e^{\frac{4}{D-2}\,\Phi}\right)\,\tilde\chi_{\mu\nu} +
\tilde\chi_\theta + \chi_\mu + \chi_\nu-\sum_\rho\chi_\rho\cr &+&
\frac{1}{2}\,\left(e_n(\tilde\chi_\mu +\tilde\chi_\nu)
+\sigma\,(\tilde\chi_\mu
+\tilde\chi_\nu)\right)-\sigma_\mu\,\tilde\chi_\nu
-\sigma_\nu\,\tilde\chi_\mu\cr 0 &=& e_n{}^2(\tilde\chi_\xi)+
\sigma\,e_n(\tilde\chi_\xi) - \left( \tilde p{}^2 + \Lambda^2\,
e^{\frac{4}{D-2}\,\Phi}\right)\,\tilde\chi_\xi\cr 0 &=&
\sigma_\theta\,e_n(\tilde\chi_\theta)+ 2\, D^2(\ln
A_\theta)\,\tilde\chi_\theta + \sigma_\theta\,e_n(\sum_\rho\chi_\rho)
+ \sigma_\theta\, \sum_\rho \tilde p^\rho\,\tilde
p_\rho\,\tilde\chi_\rho\cr &+&
\frac{8(D-1)(D-4)}{D(D-2)^2}\,\Lambda^2\,e^{\frac{4}{D-2}\,\Phi}\,\tilde\chi_\xi
\cr 0 &=& \sum_\rho \left( e_n(\chi_\rho) + (\sigma_\rho -
\sigma_\mu)\, \chi_\rho\right) - e_n(\chi_\mu) + (\sigma -
\sigma_\mu)\, \tilde\chi_\theta +\frac{8}{D-4}\, \sigma_\theta\,
\tilde\chi_\xi\cr &-&\sum_\rho \tilde p^\rho\,\tilde p_\rho\,\left(
\frac{1}{2}(\tilde\chi_\mu -\tilde\chi_\rho) +
e_n(\tilde\chi_{\mu\rho}) -
2\,\sigma_\mu\,\tilde\chi_{\mu\rho}\right)\cr 0 &=& \sum_\rho \tilde
p^\rho\,\tilde p_\rho\, \left( \chi_\rho + (\sigma - \sigma_\rho)\,
\tilde\chi_\rho +\sum_\sigma\tilde p^\sigma\,\tilde
p_\sigma\,\tilde\chi_{\rho\sigma}\right)
+\frac{D-2}{D}\,\Lambda^2\,e^{\frac{4}{D-2}\,\Phi}\,\left(\tilde\chi_\theta
- \frac{8}{(D-2)^2}\,\tilde\chi_\xi\right)\cr & &\label{eqn3} \eea

%% file: 4mesc.tex
\chapter{C\'odigo num\'erico}

Por completitud, se incluye uno de los c\'odigos num\'ericos
utilizados para calcular los espectros de glueballs. El mismo est\'a
programado en Fortran 77. En este caso, se trata del c\'odigo
utilizado para computar el espectro correspondiente a la
perturbaci\'on escalar de la m\'etrica. El nombre del archivo es {\sc
  programa.for}.

%% file: tesis.bbl

\begin{thebibliography}{99}


\bibitem{Maldacena:1997re}
  J.~M.~Maldacena,
  Adv.\ Theor.\ Math.\ Phys.\  {\bf 2}, 231 (1998)
  [Int.\ J.\ Theor.\ Phys.\  {\bf 38}, 1113 (1999)]
  [arXiv:hep-th/9711200].

\bibitem{Witten:1998qj}
  E.~Witten,
  Adv.\ Theor.\ Math.\ Phys.\  {\bf 2}, 253 (1998)
  [arXiv:hep-th/9802150].


\bibitem{Witten:1998zw}
  E.~Witten,
   ``Anti-de Sitter space, thermal phase transition, and confinement in  gauge
  theories,''
  Adv.\ Theor.\ Math.\ Phys.\  {\bf 2}, 505 (1998)
  [arXiv:hep-th/9803131].


\bibitem{Kuperstein:2004yk}
  S.~Kuperstein and J.~Sonnenschein,
  JHEP {\bf 0407}, 049 (2004)
  [arXiv:hep-th/0403254].



\bibitem{Polchinski:1998}
  J.~Polchinski, {\em String theory}.
  \newblock Cambridge, UK: Univ. Pr., 1998.

\bibitem{'tHooft:1973jz}
  G.~'t Hooft,
  Nucl.\ Phys.\  B {\bf 72} (1974) 461.



\bibitem{Lugo:2005yf}
  A.~R.~Lugo and M.~B.~Sturla,
  Phys.\ Lett.\ B {\bf 637}, 338 (2006)
  [arXiv:hep-th/0604202].

\bibitem{Lugo:2006vz}
  A.~R.~Lugo and M.~B.~Sturla,
  Nucl.\ Phys.\  B {\bf 792}, 136 (2008)
  [arXiv:0709.0471 [hep-th]].

\bibitem{adrian} A. R. Lugo and M. B. Sturla,``Gauge invariant
  perturbation theory and non-critical string models of Yang-Mills
  theories''. A ser enviado




\bibitem{Imeroni:2003jk}
  E.~Imeroni,
  arXiv:hep-th/0312070.
and references therein.

\bibitem{Green:1987}
  M.~B. Green, J.~H. Schwarz, and E.~Witten, {\em Superstring theory}.
  \newblock Cambridge, UK: Univ. Pr., 1987.

\bibitem{DiVecchia:1999rh}
  P.~Di~Vecchia and A.~Liccardo, ``D-branes in string theory, I,''

\bibitem{DiVecchia:2003ne}
  P.~Di~Vecchia and A.~Liccardo, ``Gauge theories from D-branes,''

\bibitem{Johnson:2003gi}
  C.~V. Johnson, {\em D-branes}.
  \newblock Cambridge, UK: Univ. Pr., 2002.

\bibitem{Bergshoeff:1995as}
  E.~Bergshoeff, C.~M. Hull, and T.~Ortin, ``Duality in the type II superstring
  effective action,'' {\em Nucl. Phys.} {\bf B451} (1995) 547--578,

\bibitem{Polchinski:1995mt}
  J.~Polchinski, ``Dirichlet-Branes and Ramond-Ramond Charges,'' {\em Phys. Rev.
  Lett.} {\bf 75} (1995) 4724--4727,


\bibitem{Li:1996pq}
  M.~Li, ``Boundary States of D-Branes and DY-Strings,'' {\em Nucl. Phys.} {\bf
  B460} (1996) 351--361,

\bibitem{Douglas:1995bn}
  M.~R. Douglas, ``Branes within branes,''

\bibitem{Tseytlin:1997cs}
   A.~A. Tseytlin, ``On non-abelian generalisation of the Born-Infeld action in
  string theory,'' {\em Nucl. Phys.} {\bf B501} (1997) 41--52,



\bibitem{Witten:1996im}
E.~Witten, ``Bound States of Strings and p-Branes,'' {\em Nucl. Phys.} {\bf
  B460} (1996) 335--350, 

\bibitem{Klebanov:1997kc}
I.~R. Klebanov, ``World volume approach to absorption by nondilatonic branes,''
  {\em Nucl. Phys.} {\bf B496} (1997) 231,

\bibitem{Gubser:1997yh}
S.~S. Gubser, I.~R. Klebanov, and A.~A. Tseytlin, ``String theory and classical
  absorption by three-branes,'' {\em Nucl. Phys.} {\bf B499} (1997) 217,

\bibitem{Gibbons:1993sv}
G.~W. Gibbons and P.~K. Townsend, ``Vacuum interpolation in supergravity via
  super p-branes,'' {\em Phys. Rev. Lett.} {\bf 71} (1993) 3754--3757,

\bibitem{Gubser:1998bc}
  S.~S.~Gubser, I.~R.~Klebanov and A.~M.~Polyakov,
  Phys.\ Lett.\  B {\bf 428}, 105 (1998)
  [arXiv:hep-th/9802109].

\bibitem{Aharony:1999ti}
 O.~Aharony, S.~S.~Gubser, J.~M.~Maldacena, H.~Ooguri and Y.~Oz,
  Phys.\ Rept.\  {\bf 323}, 183 (2000)
  [arXiv:hep-th/9905111], and references therein.


\bibitem{Wilson:1974co}
K.~Wilson, ``Confinement of quarks,'' {\em Phys. Rev.} {\bf D10} (1974) 2445.


\bibitem{Gross:1998gk}
  D.~J.~Gross and H.~Ooguri,
  Phys.\ Rev.\  D {\bf 58}, 106002 (1998)
  [arXiv:hep-th/9805129].

\bibitem{Brower:2000rp}
  R.~C.~Brower, S.~D.~Mathur and C.~I.~Tan,
  Nucl.\ Phys.\  B {\bf 587}, 249 (2000)
  [arXiv:hep-th/0003115].





\bibitem{Polyakov:1981rd}
  A.~M.~Polyakov,
  Phys.\ Lett.\ B {\bf 103}, 207 (1981).

\bibitem{Polyakov:1981re}
  A.~M.~Polyakov,
  Phys.\ Lett.\ B {\bf 103}, 211 (1981).

\bibitem{Ginsparg:1993is}
  P.~H.~Ginsparg and G.~W.~Moore,
  arXiv:hep-th/9304011.

\bibitem{Seiberg:1990eb}
  N.~Seiberg,
  Prog.\ Theor.\ Phys.\ Suppl.\  {\bf 102}, 319 (1990).

\bibitem{Kutasov:1990ua}
  D.~Kutasov and N.~Seiberg,
  Phys.\ Lett.\ B {\bf 251}, 67 (1990).


\bibitem{Witten:1991yr}
 E.~Witten,
  Phys.\ Rev.\ D {\bf 44}, 314 (1991).


\bibitem{Mandal:1991tz}
  G.~Mandal, A.~M.~Sengupta and S.~R.~Wadia,
  Mod.\ Phys.\ Lett.\ A {\bf 6}, 1685 (1991).

\bibitem{Bars:1992sr}
  I.~Bars and K.~Sfetsos,
  Phys.\ Rev.\ D {\bf 46}, 4510 (1992)
  [arXiv:hep-th/9206006].

\bibitem{Polyakov:1998ju}
  A.~M.~Polyakov,
  Int.\ J.\ Mod.\ Phys.\ A {\bf 14}, 645 (1999)
  [arXiv:hep-th/9809057].

\bibitem{Itzhaki:1998dd}
  N.~Itzhaki, J.~M.~Maldacena, J.~Sonnenschein and S.~Yankielowicz,
  Phys.\ Rev.\ D {\bf 58}, 046004 (1998)
  [arXiv:hep-th/9802042].



\bibitem{Klebanov:2004ya}
  I.~R.~Klebanov and J.~M.~Maldacena,
  Int.\ J.\ Mod.\ Phys.\ A {\bf 19}, 5003 (2004)
  [arXiv:hep-th/0409133].

\bibitem{Alishahiha:2004yv}
  M.~Alishahiha, A.~Ghodsi and A.~E.~Mosaffa,
  JHEP {\bf 0501}, 017 (2005)
  [arXiv:hep-th/0411087].

\bibitem{Bigazzi:2006ix}
  F.~Bigazzi, R.~Casero, A.~Paredes and A.~L.~Cotrone,
  Fortsch.\ Phys.\  {\bf 54}, 300 (2006).


\bibitem{Kuperstein:2004yf}
  S.~Kuperstein and J.~Sonnenschein,
  JHEP {\bf 0411}, 026 (2004)
  [arXiv:hep-th/0411009].


\bibitem{Casero:2005se}
  R.~Casero, A.~Paredes and J.~Sonnenschein,
  JHEP {\bf 0601}, 127 (2006)
  [arXiv:hep-th/0510110].


\bibitem{Johnson:2000ch}
  C.~V.~Johnson,
  arXiv:hep-th/0007170.


\bibitem{cfmp}
  C.~G.~.~Callan, E.~J.~Martinec, M.~J.~Perry and D.~Friedan,
  Nucl.\ Phys.\ B {\bf 262}, 593 (1985).

\bibitem{Dijkgraaf:1991ba}
  R.~Dijkgraaf, H.~L.~Verlinde and E.~P.~Verlinde,
  Nucl.\ Phys.\  B {\bf 371}, 269 (1992).



\bibitem{Myers:1987fv}
  R.~C.~Myers,
  Phys.\ Lett.\  B {\bf 199}, 371 (1987).

\bibitem{polcho1} J. Polchinski, ``String theory, vol. 1",
Cambridge University Press, Cambridge (1998).

\bibitem{Antoniadis:1990uu}
  I.~Antoniadis, C.~Bachas, J.~R.~Ellis and D.~V.~Nanopoulos,
  Phys.\ Lett.\  B {\bf 257}, 278 (1991).

\bibitem{Tseytlin:1991xk}
  A.~A.~Tseytlin and C.~Vafa,
  Nucl.\ Phys.\ B {\bf 372}, 443 (1992)
  [arXiv:hep-th/9109048].

\bibitem{Bergshoeff:2005bt}
  E.~A.~Bergshoeff, A.~Collinucci, D.~Roest, J.~G.~Russo and P.~K.~Townsend,
  Class.\ Quant.\ Grav.\  {\bf 22}, 4763 (2005)
  [arXiv:hep-th/0507143].


\bibitem{Alvarez:2000it}
  E.~Alvarez, C.~Gomez and L.~Hernandez,
  Nucl.\ Phys.\  B {\bf 600}, 185 (2001)
  [arXiv:hep-th/0011105].

\bibitem{Maldacena:2000hw}
A lot of work in the literature is devoted to the $SL(2,\Re)$ WZW model, see for example,
  J.~M.~Maldacena and H.~Ooguri,
  J.\ Math.\ Phys.\  {\bf 42}, 2929 (2001)
  [arXiv:hep-th/0001053],
and references therein.


\bibitem{Alvarez:2001ta}
  E.~Alvarez, C.~Gomez, L.~Hernandez and P.~Resco,
  Nucl.\ Phys.\  B {\bf 603}, 286 (2001)
  [arXiv:hep-th/0101181].


\bibitem{Ashok:2005py}
  S.~K.~Ashok, S.~Murthy and J.~Troost,
  Nucl.\ Phys.\ B {\bf 749}, 172 (2006)
  [arXiv:hep-th/0504079].

\bibitem{Fotopoulos:2005cn}
  A.~Fotopoulos, V.~Niarchos and N.~Prezas,
  JHEP {\bf 0510} (2005) 081
  [arXiv:hep-th/0504010].


\bibitem{Murthy:2006xt}
  S.~Murthy and J.~Troost,
  JHEP {\bf 0610} (2006) 019
  [arXiv:hep-th/0606203].

\bibitem{Klebanov:2000nc}
  I.~R.~Klebanov and A.~A.~Tseytlin,
  Nucl.\ Phys.\  B {\bf 578}, 123 (2000)
  [arXiv:hep-th/0002159].

\bibitem{Klebanov:2002gr}
  I.~R.~Klebanov, P.~Ouyang and E.~Witten,
  Phys.\ Rev.\  D {\bf 65} (2002) 105007
  [arXiv:hep-th/0202056].

\bibitem{Klebanov:2000hb}
  I.~R.~Klebanov and M.~J.~Strassler,
  JHEP {\bf 0008}, 052 (2000)
  [arXiv:hep-th/0007191].

\bibitem{Maldacena:2000mw}
  J.~M.~Maldacena and C.~Nunez,
  Int.\ J.\ Mod.\ Phys.\  A {\bf 16}, 822 (2001)
  [arXiv:hep-th/0007018].

\bibitem{Gubser:2000nd}
  S.~S.~Gubser,
  Adv.\ Theor.\ Math.\ Phys.\  {\bf 4}, 679 (2000)
  [arXiv:hep-th/0002160].


%
%
\bibitem{Susskind:1998dq}
  L.~Susskind and E.~Witten,
  arXiv:hep-th/9805114.

\bibitem{Maldacena:1998im}
  J.~M.~Maldacena,
  Phys.\ Rev.\ Lett.\  {\bf 80}, 4859 (1998)
  [arXiv:hep-th/9803002].

\bibitem{Rey:1998ik}
  S.~J.~Rey and J.~T.~Yee,
  Eur.\ Phys.\ J.\  C {\bf 22}, 379 (2001)
  [arXiv:hep-th/9803001].


\bibitem{Kinar:1998vq}
  Y.~Kinar, E.~Schreiber and J.~Sonnenschein,
  Nucl.\ Phys.\  B {\bf 566}, 103 (2000)
  [arXiv:hep-th/9811192].

\bibitem{Sonnenschein:1999if}
  J.~Sonnenschein,
  arXiv:hep-th/0003032.


\bibitem{Carroll:1997ar}
  S.~M.~Carroll,
  ``Lecture notes on General Relativity,''
  arXiv:gr-qc/9712019.

\bibitem{Bardeen:1980kt}
  J.~M.~Bardeen,
  Phys.\ Rev.\  D {\bf 22}, 1882 (1980).

\bibitem{Nakamura:2004rm}
  K.~Nakamura,
  Prog.\ Theor.\ Phys.\  {\bf 117}, 17 (2007)
  [arXiv:gr-qc/0605108].

\bibitem{Constable:1999gb}
  N.~R.~Constable and R.~C.~Myers,
  JHEP {\bf 9910}, 037 (1999)
  [arXiv:hep-th/9908175].

\bibitem{polcho2} J. Polchinski, ``String theory, vol. 1 and 2",
Cambridge University Press, Cambridge (1998).


\bibitem{Csaki:1998qr}
  C.~Csaki, H.~Ooguri, Y.~Oz and J.~Terning,
  JHEP {\bf 9901}, 017 (1999)
  [arXiv:hep-th/9806021].

\bibitem{Teper:1998te}
  M.~J.~Teper,
  Phys.\ Rev.\  D {\bf 59}, 014512 (1999)
  [arXiv:hep-lat/9804008].


\bibitem{Morningstar:1999rf}
  C.~J.~Morningstar and M.~J.~Peardon,
  Phys.\ Rev.\  D {\bf 60} (1999) 034509
  [arXiv:hep-lat/9901004].

\bibitem{Gursoy:2007er}
  U.~Gursoy, E.~Kiritsis and F.~Nitti,
  JHEP {\bf 0802}, 019 (2008)
  [arXiv:0707.1349 [hep-th]].

\bibitem{Kiritsis:2009hu}
  E.~Kiritsis,
  arXiv:0901.1772 [hep-th], and references therein.



%
%



































%

  [arXiv:hep-th/9809106].



























\bibitem{CT97}
I.~Chepelev and A.~A. Tseytlin, ``Long distance interactions of D-brane bound
  states and longitudinal five-brane in M(atrix) theory,'' {\em Phys. Rev.}
  {\bf D56} (1997) 3672--3685,



\bibitem{joebook}
J.~Polchinski, ``String Theory,''. Cambridge University Press (1998).

\bibitem{Horowitz:1991cd}
G.~T. Horowitz and A.~Strominger, ``Black strings and P-branes,'' {\em Nucl.
  Phys.} {\bf B360} (1991) 197--209.







\bibitem{pol} A. M. Polyakov, Phys. Lett. 103 {\bf B} (1981) 207, 211;

\bibitem{ginmoo} P. Ginsparg and G. Moore, ``Lectures on $2D$ gravity and $2D$ string
theory", hep-th/9304011;

\bibitem{sei}
N.~Seiberg,
``Notes On Quantum Liouville Theory And Quantum Gravity'',
Prog.\ Theor.\ Phys.\ Suppl.\  {\bf 102}, 319 (1990).

\bibitem{kutasei} D. Kutasov and N. Seiberg, ``Non critical superstrings", Phys. Lett. B.  251
(1990), 67.

\bibitem{bs}
I.~Bars and K.~Sfetsos,
``Conformally exact metric and dilaton in string theory on curved space-time,''
Phys.\ Rev.\ D {\bf 46}, 4510 (1992) [arXiv:hep-th/9206006].

\bibitem{argu}
See, for example, R.~Argurio, ``Brane physics in M-theory'',
arXiv:hep-th/9807171, and references therein.



\bibitem{ks1}
S.~Kuperstein and J.~Sonnenschein,
``Non-critical supergravity (d > 1) and holography,''
JHEP {\bf 0407}, 049 (2004)
[arXiv:hep-th/0403254].

\bibitem{ks2}
S.~Kuperstein and J.~Sonnenschein,
``Non-critical, near extremal AdS(6) background as a holographic laboratory
of four dimensional YM theory,''
JHEP {\bf 0411}, 026 (2004) [arXiv:hep-th/0411009].

\bibitem{hs}
G.~T.~Horowitz and A.~Strominger,
``Black strings and P-branes,''
Nucl.\ Phys.\ B {\bf 360}, 197 (1991).

\bibitem{imsy}
N.~Itzhaki, J.~M.~Maldacena, J.~Sonnenschein and S.~Yankielowicz,
``Supergravity and the large N limit of theories with sixteen
supercharges,''
Phys.\ Rev.\ D {\bf 58}, 046004 (1998) [arXiv:hep-th/9802042];
O.~Aharony, S.~S.~Gubser, J.~M.~Maldacena, H.~Ooguri and Y.~Oz,
``Large N field theories, string theory and gravity,''
Phys.\ Rept.\  {\bf 323}, 183 (2000) [arXiv:hep-th/9905111].

\bibitem{chs}
C.~G.~.~Callan, J.~A.~Harvey and A.~Strominger,
``Supersymmetric string solitons'', arXiv:hep-th/9112030.

\bibitem{cgo}
C.~M.~Chen, D.~V.~Gal'tsov and N.~Ohta,
``Intersecting non-extreme p-branes and linear dilaton background'',
Phys.\ Rev.\ D {\bf 72}, 044029 (2005) [arXiv:hep-th/0506216].

\bibitem{zgw}
B.~Zwiebach, ``Curvature Squared Terms And String Theories,''
Phys.\ Lett.\ B {\bf 156}, 315 (1985);
D.~J.~Gross and E.~Witten,
``Superstring Modifications Of Einstein's Equations,''
Nucl.\ Phys.\ B {\bf 277}, 1 (1986).

\bibitem{fnp}
  A.~Fotopoulos, V.~Niarchos and N.~Prezas,
  JHEP {\bf 0510}, 081 (2005)
  [arXiv:hep-th/0504010].

\bibitem{amt}
  S.~K.~Ashok, S.~Murthy and J.~Troost,
  arXiv:hep-th/0504079.

\bibitem{lsprep} A. Lugo and M. Sturla, in preparation.











\end{thebibliography}














